\documentstyle[12pt,epsf]{ioplppt}
\input bbnrev.sty 
\eqnobysec
\begin{document}
\submitted{}
\review{Big Bang nucleosynthesis and physics beyond the Standard
         Model}[BBN and physics beyond the SM]
         \footnotetext{\em Dedicated to Dennis Sciama on his 67th birthday}

\author{Subir Sarkar}

\address{Department of Physics, University of Oxford, 1 Keble Road,
 Oxford OX1 3NP, U.K.}

\begin{abstract} 
  The Hubble expansion of galaxies, the $2.73\dK$ blackbody radiation
  background and the cosmic abundances of the light elements argue for
  a hot, dense origin of the universe --- the standard Big Bang
  cosmology --- and enable its evolution to be traced back fairly
  reliably to the nucleosynthesis era when the temperature was of
  $\Or(1)$ MeV corresponding to an expansion age of $\Or(1)$ sec. All
  particles, known and hypothetical, would have been created at higher
  temperatures in the early universe and analyses of their possible
  effects on the abundances of the synthesized elements enable many
  interesting constraints to be obtained on particle properties. These
  arguments have usefully complemented laboratory experiments in
  guiding attempts to extend physics beyond the Standard
  $SU(3)_{\c}{\otimes}SU(2)_{\L}{\otimes}U(1)_{Y}$ Model,
  incorporating ideas such as supersymmetry, compositeness and
  unification. We first present a pedagogical account of relativistic
  cosmology and primordial nucleosynthesis, discussing both
  theoretical and observational aspects, and then proceed to examine
  such constraints in detail, in particular those pertaining to new
  massless particles and massive unstable particles. Finally, in a
  section aimed at particle physicists, we illustrate applications of
  such constraints to models of new physics.
\end{abstract}


\vspace{-16cm}\hfill{\small OUTP-95-16P (rev)}\vspace{16cm}

\maketitle

\contents
\entry{1.}{Introduction}{3} 
\entry{2.}{The standard cosmology}{8} 
\subentry{2.1}{The Friedmann-Lema\^{\ii}tre-Robertson-Walker models}{8}
\subentry{2.2}{Thermal history of the early universe}{14} 
\entry{3}{Primordial nucleosynthesis}{26} 
\subentry{3.1}{The standard BBN model}{27} 
\subentry{3.2}{Primordial elemental abundances}{42} 
\subentry{3.3}{Theory versus observations}{54} 
\entry{4.}{Constraints on new physics}{60} 
\subentry{4.1}{Bounds on relativistic relics}{61} 
\subentry{4.2}{Bounds on non-relativistic relics}{67} 
\entry{5.}{Applications}{82} 
\subentry{5.1}{Neutrinos}{82} 
\subentry{5.2}{Technicolour}{93} 
\subentry{5.3}{Supersymmetry and supergravity}{95} 
\subentry{5.4}{Grand unification and cosmic strings}{112} 
\subentry{5.5}{Miscellaneous models}{113} 
\subentry{5.6}{Implications for the dark matter}{114} 
\entry{6.}{Conclusions}{117} 
\entry{}{Acknowledgements}{118}
\entry{}{References}{118}

\newpage
\section{Introduction\label{intro}} 

There has been interest in problems at the interface of cosmology and
particle physics for over thirty years (see Zel'dovich 1965), but it
is only in the past decade or so that the subject has received serious
attention (see B\"orner 1988, Collins \etal 1989, Kolb and Turner
1990, Ellis 1993). Cosmology, once considered to be outside the
mainstream of physics and chiefly of interest to astronomers and
applied mathematicians, has become a {\em physical} subject, largely
due to the advances which have been made on the observational front
(see Weinberg 1972, Peebles 1993). It has become increasingly clear
that particle physicists can no longer afford to ignore the
cosmological ``laboratory'', which offers a powerful probe of new
physical phenomena far beyond the reach of terrestrial laboratories
(see Steigman 1979, Dolgov and Zel'dovich 1981). Cosmological
phenomena have thus come under detailed scrutiny by particle
physicists, prompting deeper theoretical analyses (see Weinberg 1980,
Wilczek 1991) as well as ambitious observational programmes (see
Sadoulet 1992, Kolb and Peccei 1995).

The increasing interaction between particle physics and cosmology has
largely resulted from the establishment of `standard models' in both
fields which satisfactorily describe all known phenomena but whose
very success, paradoxically, establishes them as intrinsically
incomplete pictures of physical reality. Our reconstruction of the
history of the universe in \fref{cosmohist} emphasizes the
interdependence of these models. The familiar physics of
electromagnetism, weak interactions and nuclear reactions provide a
sound basis for the standard Big Bang cosmology up to the beginning of
the nucleosynthesis era, when the universe was about $10^{-2}\sec$
old. The Standard $SU(3)_{\c}{\otimes}SU(2)_{\L}{\otimes}U(1)_{Y}$
Model (SM) of particle physics (see Cheng and Li 1984, Kane 1987),
brilliantly confirmed by all experiments to date (see Burkhardt and
Steinberger 1991, Bethke and Pilcher 1992), allows us to extrapolate
back further, to $t\sim10^{-12}$ sec. Two phase transitions are
believed to have occurred in this interval, although a detailed
understanding of their dynamics is still lacking (see Kapusta 1988,
Yaffe 1995). The first is associated with the confinement of quarks into
hadrons and chiral symmetry breaking by the strong interactions at
$T_{\c}^{\qh}\sim\Lambda_{\QCD}\approx200\MeV$, and the second with
the spontaneous breaking of the unified electroweak symmetry to
electromagnetism, $SU(2)_{\L}{\otimes}U(1)_{Y}\,\rightarrow\,U(1)_{\rm
em}$ at $T_{\c}^{\EW}\sim250\GeV$, when all known particles received
their masses through the Higgs mechanism. To go beyond this point
requires an extension of the SM; indeed, the very success of the model
demands such new physics.


Similarly, the standard cosmological model of an adiabatically
expanding, homogeneous and isotropic universe requires extreme fine
tuning of the initial conditions of the Big Bang, as emphasized by
Dicke and Peebles (1979). The problem essentially consists of
explaining why the universe is as old ($\geqsim3\times10^{17}\sec$) or
as large ($\geqsim10^{28}\cm$) as it is today, relative to the Planck
time ($5.39\times10^{-44}\sec$) or the Planck length
($1.62\times10^{-33}\cm$), which are the appropriate physical scales
governing gravitational dynamics. Although a resolution of this may
have to await progress in our understanding of quantum gravity (see
Penrose 1979, 1989), there has been enthusiastic response to the
simpler solution proposed by Guth (1981), viz. that there was a period
of non-adiabatic accelerated expansion or `inflation', possibly
associated with a phase transition in the early universe (see Linde
1990). This has the additional advantage that it naturally generates a
nearly scale-invariant `Harrison-Zel'dovich' spectrum of scalar
density fluctuations (see Mukhanov \etal 1992) which can seed the
growth of the observed large-scale structure in the expanding universe
(see Efstathiou 1990, Padmanabhan 1993). Another fundamental problem
of the standard cosmology is that the observed abundance of baryonic
matter is $\sim10^{9}$ times greater than the relic abundance expected
from a state of thermal equilibrium, while no antimatter is observed
(see Steigman 1976), thus requiring a primordial asymmetry between
matter and anti-matter. To generate this dynamically requires new
physics to violate baryon number ($B$) and charge-parity ($CP$) at
high temperatures, in an out-of-equilibrium situation to ensure time
asymmetry (Sakharov 1967, see Cohen \etal 1994, Rubakov and
Shaposhnikov 1996). More recently, it has been recognized that baryons
are probably a minor constituent of the universe, since all observed
structures appear to be dominated by dark matter (see Binney and
Tremaine 1987, Peebles 1993) which is probably non-baryonic. The
growing interest in the early universe stems from the realization that
extensions of physics beyond the SM naturally provide the mechanisms
for processes such as inflation and baryogenesis, as well as new
particle candidates for the dark matter. These exciting developments
have been discussed in a number of schools and conferences (see
e.g. Gibbons \etal 1983, Setti and Van Hove 1984, Kolb \etal 1986b,
Piran and Weinberg 1986, Alvarez \etal 1987, Hinchliffe 1987, De
R\'ujula \etal 1987, Unruh and Semenoff 1988, Yoshimura \etal 1988,
Peacock \etal 1990, Nilsson \etal 1991, Sato and Audoze 1991,
Nanopoulos 1991, Sanchez and Zichichi 1992, Akerlof and Srednicki
1993, Astbury \etal 1995).

Such new physics is in fact necessary to address the theoretical
shortcomings of the Standard Model itself (see Ross 1984, Mohapatra
1992). Its phenomenological success requires that the Higgs boson,
which gives masses to all known particles, cannot itself be much more
massive than its vacuum expectation value (VEV) which sets the
electroweak (`Fermi') scale,
$v\equiv(\sqrt{2}G_{\F})^{-1/2}=246\GeV$. This creates the
`naturalness' or `hierarchy' problem, viz. why is the Higgs mass not
pushed up to the Planck mass
($M_{\Pl}\,\equiv\,G_{\rm N}^{-1/2}=1.221\times10^{19}\GeV$) due to the
{\em quadratically} divergent radiative corrections it receives due to
its couplings to all massive particles?\footnote{By contrast, it is
`natural' for fermions to be light relative to the Planck scale since
letting their masses go to zero reveals a chiral symmetry which tames
the radiative corrections to be only logarithmically divergent; there
is no such symmetry to `protect' the mass of a scalar Higgs boson.}
Supersymmetry (SUSY) addresses this problem by imposing a symmetry
between bosons and fermions which makes such radiative corrections
cancel to zero. This requires all known particles (boson/fermion) to
have supersymmetric (fermion/boson) partners distinguished by a new
quantum number called $R$-parity; if it is conserved, the lightest
supersymmetric particle would be stable. Supersymmetry must be broken
in nature since known particles do not have supersymmetric partners of
the same mass.  However the Higgs mass would still be acceptable if
the scale of SUSY breaking (hence the masses of the supersymmetric
partners) is not much beyond the Fermi scale. When such breaking is
realized {\em locally}, as in gauge theories, a link with general
coordinate transformations, i.e. gravity, emerges; this is
supergravity (SUGRA) (see Van Nieuwenhuizen 1981, Wess and Bagger
1993). Technicolour is an alternative approach in which the offending
{\em elementary} Higgs particle is absent (see Farhi and Susskind
1981, Kaul 1983); electroweak symmetry breaking is now seen as a
dynamic phenomenon (see King 1995, Chivukula \etal 1995), akin to the
breaking of chiral symmetry by the strong interactions. However no
realistic technicolour model has been constructed satisfying all
experimental constraints, in particular the small radiative
corrections to SM parameters measured at \LEP (see Lane 1993).

Another conundrum is that $CP$ is known to be well conserved by the
strong interactions, given the stringent experimental upper limit on
the neutron electric dipole moment, whereas QCD, the successful theory
of this interaction, contains an arbitrary $CP$ violating
parameter. An attractive solution is to replace this parameter by a
field which dynamically relaxes to zero --- the axion (see Kim 1987,
Cheng 1988). This is a pseudo-Goldstone boson generated by the
breaking of a new global $U(1)$ `Peccei-Quinn' symmetry at a scale
$f_{a}$ (see Peccei 1989). This symmetry is also explicitly broken by
QCD instanton effects, hence the axion acquires a small mass
$m_{a}\,\sim\,f_{\pi}^2/f_{a}$ when the temperature drops to
$T\sim\Lambda_{\QCD}$. The mixing with the pion makes the axion
unstable against decay into photons; negative experimental searches
for decaying axions then constrain $f_{a}$ to be beyond the Fermi
scale, implying that axions are light enough to be produced in stellar
interiors. Considerations of stellar cooling through axion emission
imply $f_{a}\geqsim10^{10}$ GeV, which {\em requires} the axion (if it
exists!) to have an interesting cosmological relic density (see
Raffelt 1990, Turner 1990).

Yet another motivation for going beyond the Standard Model is the
unification of forces. Grand unified theories (GUTs) of the strong and
electroweak interactions at high energies also provide a physical need
for inflation in order to dilute the embarrassingly large abundance of
magnetic monopoles expected to be created during the breaking of the
unified symmetry (see Preskill 1984). Unification naturally provides
for baryon and lepton number violation (see Langacker 1981, Costa and
Zwirner 1986) which allows for generation of the cosmological baryon
asymmetry (see Kolb and Turner 1983) as well as masses for neutrinos
(see Mohapatra and Pal 1991). Recent data from \LEP on the evolution
of the gauge interaction couplings with energy indicate that such
unification can occur at $M_{\GUT}\approx2\times10^{16}$ GeV, but
only in a (broken) supersymmetric theory with superparticle masses at
around the Fermi scale (see Dimopoulos 1994, Ellis 1995). Moreover in
such unified models, electroweak symmetry breaking via the Higgs
mechanism is driven quite naturally by supersymmetry breaking (see
Ib\'a\~nez and Ross 1993). A dynamical understanding of how
supersymmetry itself is broken is expected to come from the theory of
superstrings, the most ambitious attempt yet towards a finite quantum
theory of gravity and its unification with all other forces (see Green
\etal 1987). Following the initial euphoria over the discovery of the
anomaly-free heterotic superstring, progress has been difficult due to
the problems of relating low energy physics to the higher dimensional
world which is the natural domain of the string. However explicit
examples of compactified four-dimensional strings have been
constructed which reduce to a supersymmetric version of the Standard
Model at low energies and also contain additional gauge bosons and
gauge singlets which have only gravitational couplings to matter (see
Dine 1988, 1990, Ib\'a\~nez 1994). Furthermore, the recent exciting
discoveries of the special `duality' properties of string theories
(see Giveon \etal 1994, Polchinski 1996) have begun to provide
important insights into the issue of supersymmetry breaking (see
Zwirner 1996).

It is thus a common feature of new physics beyond the Fermi scale to
predict the existence of new particles which are unstable in general
but some of which may be stable by virtue of carrying new conserved
quantum numbers. Moreover their generic weak interactions ensure a
cosmologically significant relic density (see Primack \etal 1988,
Turner 1991). In addition, known particles such as neutrinos, although
strictly massless in the Standard Model, may acquire masses from such
new physics, enabling them also to be candidates for dark
matter. Conventionally, particle physicists look for new physics
either by directly attempting to produce the new particles in high
energy collisions at accelerators or by looking for exotic phenomena
such as nucleon instability or neutrino masses. In this context, the
standard cosmology, in particular Big Bang nucleosynthesis (BBN),
provides an important new testing ground for new physics and, indeed,
in many cases, provides the {\em only} ``experimental'' means by which
the properties of new particles may be restricted (see Sarkar 1985,
1986). Whether or not one finds this satisfactory from a philosophical
point of view, it is essential for this enterprise that we have the
best possible understanding of the cosmological laboratory. This
review presents the current status and is primarily aimed at particle
physicists, although it is hoped that astrophysicists and cosmologists
will also find it useful.

A decade or more ago, it was possible for reviewers (e.g. Steigman
1979, Dolgov and Zel'dovich 1981) to give a comprehensive discussion
of all constraints on fundamental physics from cosmological
considerations and many of the key papers could be found in one
collection (Zee 1982). Subsequently several hundred papers on this
subject have been published. For reasons of space we will restrict
ourselves to a discussion of the constraints which follow from BBN
alone but, for completeness, present all the neccessary cosmological
background --- both theory and observations. Rather than engage in a
detailed critique of every published work, we present a pedagogical
discussion of the basic physics, together with a summary of the key
observational inputs, so that readers can assess the reliability of
these constraints. Raffelt (1990) has presented a model review of this
form which deals with astrophysical methods for constraining novel
particle phenomena. A similar discussion of all types of cosmological
constraints, including those deduced from the observed Hubble
expansion and the $2.73\dK$ blackbody radiation as well as other
radiation backgrounds, will appear in Sarkar (1997).

We begin by outlining in \sref{stdcosm} the basic features of the
standard Big Bang cosmological model and then discuss the
thermodynamics of the early radiation-dominated era. In
\sref{primnucl} we present the essential physics of the BBN era and
then discuss the observational data in some detail, highlighting the
sources of uncertainty. We argue for the consistency of the standard
model and briefly mention possible variations. This sets the stage for
deriving general constraints in \sref{cons} on both relativistic and
non-relativistic hypothetical particles which may be present during
nucleosynthesis. Finally, for the benefit of particle physicists we
illustrate in \sref{appl} how such cosmological arguments have
complemented experimental searches for physics beyond the Standard
Model, particularly in the neutrino sector, and also provided entirely
new probes of such physics, e.g. technicolour and supersymmetry. We
also discuss the implications for the nature of the dark matter.

It appears to be a widely held belief that cosmological data are not
particularly accurate and the associated errors uncertain, so that the
derived constraints cannot compare in reliability with those obtained
in the laboratory. Although not entirely incorrect, this view is being
increasingly challenged by modern observations; for example
measurements of the background radiation temperature and anisotropy,
the cosmic abundance of helium \etc are now routinely quoted to
several significant figures. Correspondingly there has been a growing
appreciation of the systematic effects involved in the analysis of
cosmological observations and careful attempts at their estimation.
More importantly, cosmological data, even if more imprecise than
accelerator data, are often much more {\em sensitive} to novel
particle phenomena; for example, even a crude upper limit on the
present energy density of the universe suffices to bound the masses of
relic neutrinos to a level which improves by several orders of
magnitude over precise laboratory experiments. Nevertheless, one
should be cautious about rejecting an interesting theoretical
possibility on the basis of a restrictive cosmological constraint
(e.g. the bound on the number of neutrino-like particles present
during BBN) without a critical appreciation of the underlying
assumptions. We have tried wherever possible to clarify what these
assumptions are and to refer to expert debate on the issues
involved. (In writing down numerical values where errors are not
quoted, the symbols $\sim$, $\approx$ and $\simeq$ indicate equality
to within a factor of 10, factor of 2 and $10\%$, respectively.)

Due to space limitations, the references are not comprehensive but do
include the seminal papers and recent reviews from which the
intervening literature can be traced. We have used `natural' units
($\hbar=c=k_{\B}=1$) although astronomical units such as year,
megaparsec or Solar mass are given where convenient. (For reference,
$1\,{\GeV}^{-1}=1.973\times10^{-14}\cm=6.582\times10^{-25}\sec$,
$1\,\GeV=1.160\times10^{13}\dK=1.783\times10^{-24}\gm$,
$1\,\Mpc=3.086\times10^{24}\cm$, $1\,\yr=3.156\times10^{7}\sec$,
$1M_{\odot}=1.989\times10^{33}\gm$.) Astronomical quantities are
listed in Allen (1973) and Lang (1992), while clarification of
unfamiliar astrophysical terms may be sought in the excellent
textbooks by Shu (1981), Mihalas and Binney (1981) and Longair (1981).

\section{The standard cosmology\label{stdcosm}} 

The standard Big Bang cosmological model assumes that the universe is
spatially homogeneous and isotropic, an assumption originally
dignified as the `Cosmological Principle' (Milne 1935).  Subsequently
cosmological observations have provided empirical justification for
this assumption as reviewed by Peebles (1980). Astronomical
observations in the last decade have required a reappraisal of this
issue with the discovery of cosmic structures on very large spatial
scales. However careful studies of the clustering of galaxies and of
the small angular fluctuations in the $2.73\dK$ cosmic microwave
background (CMB) have established (see Peebles 1993) that the universe
is indeed homogeneous when averaged on scales exceeding a few hundred
Mpc, out to spatial scales comparable to its present ``size''
\eref{hubrad} of several thousand Mpc.

\subsection{The Friedmann-Lema\^{\ii}tre-Robertson-Walker models\label{frw}} 

Homogeneity and isotropy considerably simplify the mathematical
description of the cosmology since all hypersurfaces with constant
cosmic standard time~\footnote{Spatial coordinates may be defined
 through observables such as the apparent brightness or redshift,
 while time may be defined as a definite (decreasing) function of a
 cosmic scalar field such as the proper energy density $\rho$ or the
 blackbody radiation temperature $T$, which are believed to be
 monotonically decreasing everywhere due to the expansion of the
 universe. Knowledge of the function $t=t(T)$ requires further
 assumptions, for example that the expansion is adiabatic.} are
then maximally symmetric subspaces of the whole of space-time
and all cosmic tensors (such as the metric $g_{\mu\nu}$ or
energy-momentum $T_{\mu\nu}$) are form-invariant with respect to
the isometries of these surfaces (see Weinberg 1972). These symmetries
enable a relatively simple and elegant description of the dynamical
evolution of the universe. Although the mathematical complexities of
general relativity do allow of many exotic possibilities (see Hawking
and Ellis 1973), these appear to be largely irrelevant to the physical
universe, except perhaps at very early epochs. There are many
pedagogical accounts of relativistic cosmology; to keep this review
self-contained we reiterate the relevant points.

For a homogeneous and isotropic evolving space-time, we can choose
comoving spherical coordinates (i.e. constant for an observer
expanding with the universe) in which the proper interval between two
space-time events is given by the Robertson-Walker ({\rm R-W}) metric
\begin{equation} 
 \d s^{2} = g_{\mu \nu} \d x^{\mu} \d x^{\nu} = \d t^{2} - R^{2} (t)
  \left[\frac{\d r^{2}}{1 - k r^{2}} + r^{2} (\d \theta^{2} + \sin^{2}
   \theta \d \phi^{2}) \right] .
\end{equation}
Here $R(t)$ is the cosmic scale-factor which evolves in time
describing the expansion (or contraction) of the universe and $k$ is
the 3-space curvature signature which is conventionally scaled (by
tranforming $r\to\vert{k}\vert^{1/2}r$ and
$R\to\vert{k}\vert^{-1/2}R$) to be $-1$, $0$ or $+1$ corresponding to 
an elliptic, euclidean or hyperbolic space.\footnote{This does not
 however fix the global topology; for example Euclidean space may be
 $\bf{R^3}$ and infinite or have the topology of a 3-torus (${\bf
 T^3}$) and be finite in extent; however the latter possibility has
 recently been severely constrained by the non-observation of the
 expected characteristic pattern of fluctuations in the CMB (Starobinsky 
 1993, Stevens \etal 1993).}

The energy-momentum tensor is then required to be of the `perfect
fluid' form
\begin{equation}
 T_{\mu \nu} = p g_{\mu \nu} + (p + \rho) u_{\mu} u_{\nu}\ ,
\end{equation}
in terms of the pressure $p$, the energy density $\rho$ and the
four-velocity $u_{\mu}\equiv{\d}x_{\mu}/{\d}s$. (Here and below, we
follow the sign conventions of Weinberg (1972).) The Einstein field
equations relate $T_{\mu\nu}$ to the space-time curvature
$R_{\lambda\mu\nu\kappa}$:
\begin{equation}
\label{einstein}
 R_{\mu \nu} - \case{1}{2} g_{\mu \nu} R_{c} = -\frac{8 \pi
  T_{\mu\nu}}{M_{\Pl}^{2}}\ ,
\end{equation}
where $R_{\mu\nu}{\equiv}g^{\lambda\kappa}R_{\lambda\mu\kappa\nu}$ is
the Ricci tensor and $R_{c}{\equiv}g^{\mu\nu}R_{\mu\nu}$ is the
curvature scalar. For the present case these equations simplify to
yield the Friedmann-Lema\^{\ii}tre (F-L) equation for the normalized
expansion rate $H$, called the Hubble parameter,
\begin{equation}
\label{fried1}
 H^{2} \equiv \left({\dot R \over R} \right)^{2} = {8 \pi \rho \over
  3 M_{\Pl}^{2}} - {k \over R^{2}}\ ,
\end{equation}
as well as an equation for the acceleration
\begin{equation} 
\label{fried2}
 \ddot R = - {4 \pi \rho \over 3 M_{\Pl}^{2}} (\rho + 3 p) R\ .
\end{equation}
Further, the conservation of energy-momentum
\begin{equation} 
\label{conserv1}
 T^{\mu \nu}_{;\nu} = 0\ ,
\end{equation}
implies~\footnote{This does not imply conservation of the energy
 of matter since $\rho\,R^{3}$ clearly decreases (for positive p) in
 an expanding universe due to work done against the gravitational
 field. In fact, we cannot in general even define a conserved total
 energy for matter plus the gravitational field unless space-time is
 asymptotically Minkowskian, which it is {\em not} for the R-W metric
 (see Witten 1981a).}
\begin{equation}
\label{conserv2}
 {\d (\rho R^{3}) \over \d R} = - 3 p R^{2} .
\end{equation}
This can also be derived from \eref{fried1} and \eref{fried2} since
\eref{einstein} and \eref{conserv1} are related by the Bianchi
identities:
\begin{equation}
 \left(R^{\mu \nu} - \case{1}{2} g^{\mu \nu} R_{c}\right)_{;\mu} = 0\ .
\end{equation}

In principle we can add a cosmological constant, $\Lambda g_{\mu
\nu}$, to the field equation \eref{einstein}, which would appear as an
additive term $\Lambda/3$ on the RHS of the F-L equations
\eref{fried1} and \eref{fried2}. This is equivalent to the freedom
granted by the conservation equation \eref{conserv1} to scale $T_{\mu
\nu}{\to}T_{\mu\nu}+\Lambda g_{\mu \nu}$, so that $\Lambda$ can be
related to the energy-density of the vacuum (see Weinberg 1989):
\begin{equation}
\label{Lambda}
 \langle 0 \mid T_{\mu \nu} \mid 0 \rangle = - \rho_{\vac} g_{\mu \nu}\ , 
 \qquad \Lambda =  {8 \pi \rho_{\vac} \over M_{\Pl}^{2}}\ .
\end{equation}
Empirically $\Lambda$ is consistent with being zero today; in natural
units $\Lambda<10^{-120}M_{\Pl}^{-2}$ (see Carroll \etal
1992). However the present vacuum is known to violate symmetries of
the underlying gauge field theory, e.g. the
$SU(2)_{\L}{\otimes}U(1)_{Y}$ symmetry of the electroweak interaction
and (very probably) the symmetry unifying the $SU(3)_{\c}$ and
electroweak interactions in a GUT (see Ross 1984). These symmetries
would have been presumably restored at sufficiently high temperatures
in the early universe and a finite value of $\Lambda$ associated with
the symmetric or false vacuum (see Linde 1979). (There are also other
ways, not associated with symmetry breaking, in which the universe may
have been trapped in a false vacuum state.) This possibility is
exploited in the inflationary universe model of Guth (1981) and its
successors (see Linde 1984, 1990, Olive 1990a), wherein the
(approximately constant) vacuum energy drives a huge increase of the
scale-factor during the transition to the true vacuum and is then
converted during `reheating' into interacting particles, thus
accounting for the large entropy content of the universe, which is
otherwise unexplained in the standard cosmology.

Knowing the equation of state, $p=p(\rho)$, $\rho$ can now be
specified as a function of $R$. For non-relativistic particles
(`matter' or `dust') with $p/\rho\,\approx\,T/m\,\ll\,1$,
\begin{equation}
  \rho_{\NR} \propto R^{-3} ,
\end{equation}
reflecting the dilution of density due to the increasing proper
volume. For relativistic particles (`radiation') with $p=\rho/3$, an
additional factor of $R^{-1}$ enters due to the redshifting of the
momentum by the expansion:
\begin{equation}
 \rho_{\R} \propto R^{-4} .
\end{equation}
In the modern context, it is also relevant to consider the
contribution of `vacuum energy' (i.e. a cosmological constant) for
which the equation of state, dictated by Lorentz-invariance of the
energy-momentum tensor, is $p=-\rho$, i.e.
\begin{equation}
\label{vac}
 \rho_{\vac} \propto \con .
\end{equation}
This completes the specification of the ensemble of
Friedmann-Lema\^{\ii}tre-Robertson-Walker ({\rm F-L-R-W}) models. (As
a historical note, Friedmann presented the dynamical equation
\eref{fried1} only for the case of pressureless dust, while
Lema\^{\ii}tre extended it to include the case of radiation and also
wrote down the conservation equation \eref{conserv2}.)

Taking $\Lambda=0$, the curvature term $k/R^2$ in \eref{fried1} is
positive, zero or negative according as $\rho$ is greater than, equal
to or less than the critical density
\begin{equation}
\label{rhocrit}
 \rho_{\c} = {3 H^2 M_{\Pl}^{2} \over 8 \pi} \equiv {\rho \over \Omega}\ ,
\end{equation}
where $\Omega$ is the density parameter. The critical density today
is~\footnote{The subscript $_{0}$ on any quantity denotes its present
 value.}
\begin{equation} 
 \rho_{\c_{0}} = (2.999 \times 10^{-12} h^{1/2}\ {\GeV})^4
               = 1.054 \times 10^{-5} h^{2} \GeV {\cm}^{-3}\ ,
\end{equation}
where $h$, the Hubble constant, is defined in terms of the present
expansion rate,
\begin{equation}
  h \equiv {H_{0} \over 100 \km \sec^{-1} \Mpc^{-1}}\ , \qquad 
   H_{0} \equiv \frac{\dot R_{0}}{R_{0}}\ .
\end{equation}
The extragalactic distance scale is set by $H_{0}$ since a measured
redshift
\begin{equation}
 z \equiv \frac{\lambda (t_{0}) - \lambda (t)}{\lambda (t)} = 
   \frac{R(t_{0})}{R(t)} - 1
\end{equation}
is assumed to correspond to the distance $d\simeq\,z/H_{0}$. (This is
an approximate relationship, since it is the recession velocity, not
the redshift, which is truly proportional to distance for the R-W
metric (see Harrison 1993), hence corrections are neccessary (see
Weinberg 1972) for cosmologically large distances.) The major
observational problem in obtaining $H_{0}$ is the uncertainty in
determining cosmological distances (see Rowan-Robinson 1985, 1988,
Jacoby \etal 1992, Huchra 1992, Van den Bergh 1992, 1994, Fukugita
\etal 1993). Different estimates, while often inconsistent within the
stated errors, generally fall in the range
$40-100\km{\sec}^{-1}{\Mpc}^{-1}$, i.e.
\begin{equation}
\label{h}
  0.4 \leqsim h \leqsim 1\ .
\end{equation}
The {\it Hubble Space Telescope} has recently provided the means to
directly calibrate various techniques through observations of Cepheid
variables in distant galaxies (see Kennicutt \etal 1995). According to
the review by Hogan (1996), the central values obtained by the most
reliably calibrated methods lie in the range $0.65-0.85$, although
there is, as yet, no consensus among observers. This issue may soon be
resolved by new techniques such as measurements of time delays between
variations in multiple images of gravitationally lensed quasars
(Kundic \etal 1995, see Blandford and Narayan 1992) or of the
`Sunyaev-Zel'dovich' effect on the CMB by the X-ray emitting plasma in
clusters of galaxies (Birkinshaw and Hughes 1994, see Rephaeli 1990),
which bypass the traditional error-prone construction of the
`cosmological distance ladder'.

Since $(\rho + 3 p)$ is positive for both matter and radiation, $\ddot
R$ is always negative (see \eref{fried2}), hence the present age is
bounded by the Hubble time
\begin{equation}
 t_{0} < H_{0}^{-1} = 9.778 \times 10^9\ h^{-1} \yr\ ,
\end{equation}
corresponding to a present Hubble radius of 
\begin{equation}
\label{hubrad}
 R_{\H} (t_{0}) = H^{-1} (t_{0}) \simeq 3000\ h^{-1} \Mpc\ ,
\end{equation}
which sets the {\em local} spatial scale for the universe. Another
scale, which depends on the past evolutionary history, is set by the
finite propagation velocity of light signals. Consider a ray emitted
at time $t$ which has just reached us at time $t_{0}$:
\begin{eqnarray}
 \int_{0}^{r} \frac{\d r'}{\sqrt{1 - k r'^{2}}} 
 &= \int_{t}^{t_{0}} \frac{\d t'}{R (t')} \\ \nonumber
 &= \int_{R (t)}^{R (t_{0})} \frac{\d R}{R} 
  \left(\frac{8 \pi \rho R^{2}}{3 M_{\Pl}^{2}} - k\right)^{1/2} .
\end{eqnarray}
Since $\rho R^{2}\to\infty$ as $R\to0$, for both non-relativistic and
relativistic particles, the above integral converges as $t\to0$.  This
indicates (see Rindler 1977) that there are sources from which light
has not yet reached us, which are said to lie beyond our particle
horizon, at proper distance
\begin{equation}
\label{parthor}
 d_{\H} (t_{0}) = R (t_{0}) \int_{0}^{t_{0}} \frac{\d t'}{R (t')}
  = \kappa t_{0}\ ,
\end{equation} 
where $\kappa=2,3$ for $\rho=\rho_{\R},\rho_{\NR}$ (taking $k=0$).
This creates a problem for the standard cosmology because looking back
to earlier times we observe regions which were outside each other's
(shrinking) horizons, but which nevertheless appear to be
well-correlated. Consider the photons of the $2.73\dK$ microwave
background radiation which have been propagating freely since
$z\approx1000$; the particle horizon at that epoch subtends only
$\approx1^0$ on the sky, yet we observe the radiation arriving from
all directions to have the same temperature to within 1 part in about
$10^{5}$. This problem too is solved in the inflationary universe
(Guth 1981) where the energy density becomes dominated by a positive
cosmological constant \eref{vac} at early times. The {\em accelerated}
growth of $R(t)$ ($\ddot{R}>0$ for $p=-\rho$) then rapidly blows up a
region small enough to be causally connected at that time into the
very large universe we see today.

Returning to the standard cosmology, the future evolution is
determined by the sign of $k$, or equivalently, the value of $\Omega$
(assuming $\Lambda=0$). For $k=-1$, $\dot{R}^2$ is always positive and
$R{\to}t$ as $t\to\infty$. For $k=0$, $\dot{R}^2$ goes to zero as
$R\to\infty$. For $k=+1$, $\dot{R}^2$ drops to zero at $R_{\max}=(3
M_{\Pl}^{2}/8\pi\rho)^{1/2}$ after which $R$ begins decreasing. Thus
$\Omega<1$ corresponds to an open universe which will expand forever,
$\Omega=1$ is the critical or flat universe which will asymptotically
expand to infinity while $\Omega>1$ corresponds to a closed universe
which will eventually recollapse.

Dynamical measurements of the present energy density in {\em all}
gravitating matter require (see Peebles 1993, Dekel 1994)
\begin{equation}
\label{omegaobs}
 \Omega_{0} \approx 0.1-1\ ,
\end{equation}
although such techniques are insensitive to matter which is not
clustered on the largest scales probed (for example relativistic
particles). The present energy density of visible radiation alone is
better known, since it is dominated by that of the blackbody CMB with
present temperature (Mather \etal 1994)
\begin{equation}
\label{Tcmb}
 T_{0} = 2.726 \pm 0.01 \dK ,
\end{equation}
hence, defining $\Theta\,\equiv\,T_{0}/2.73\dK$,
\begin{equation}
 \rho_{\gamma_{0}} = {\pi^2 T_{0}^4 \over 15} = 2.02 \times 10^{-51} 
  \Theta^4\ {\GeV}^{4} ,
\end{equation} 
and
\begin{equation} 
 \Omega_{\gamma_{0}} = {\rho_{\gamma_{0}} \over \rho_{c_{0}}} =
  2.49 \times 10^{-5} \Theta^4 h^{-2} .
\end{equation} 
A primordial background of (three) massless neutrinos is also believed
to be present (see \eref{gsgrhonow}); this raises the total energy
density in relativistic particles to
\begin{equation}
 \Omega_{\R_{0}} =
  \Omega_{\gamma_{0}} + \Omega_{\nu_{0}} = 1.68\,
  \Omega_{\gamma_{0}} = 4.18 \times 10^{-5} \Theta^4 h^{-2} .
\end{equation}
Since this is a negligible fraction of the total energy density
$\Omega_{0}$,\footnote{There can be a much higher energy density in
 massless particles such as neutrinos or hypothetical Goldstone
 bosons (see Kolb 1980) if these have been created relatively recently 
 rather than being relics of the early universe.} the universe is assumed 
to be matter dominated (MD) today by non-relativistic particles, i.e.
\begin{equation}
 \Omega_{0} \equiv \Omega_{\R_{0}} + \Omega_{\NR_{0}} \simeq \Omega_{\NR_{0}} .
\end{equation}
In F-L-R-W models this has actually been true for most of the age of
the universe, thus a lower bound to the age of the universe implies an
upper bound on its matter content (see Weinberg 1972). Conservatively
taking $t_0>10^{10}\yr$ and $h>0.4$ requires (see Kolb and Turner
1990)
\begin{equation}
\label{omegah2}
 \Omega_{\NR_{0}} h^2 \leqsim 1 \quad \Rightarrow \quad 
 \rho_{\NR_{0}} \leqsim 1.05 \times 10^{-5} \GeV {\cm}^{-3}\ .
\end{equation}
However as $R$ decreases, $\rho_{\R}$ rises faster than $\rho_{\NR}$
so that the universe would have been radiation dominated (RD) by
relativistic particles for
\begin{equation}
\label{matdom}
 {R \over R_{0}} < {R_{\equ} \over R_{0}} = 4.18 \times 10^{-5}
  \Theta^4\ (\Omega_{0} h^2)^{-1} .
\end{equation}
Assuming that the expansion is adiabatic, the scale-factor is related
to the blackbody photon temperature $T\,(\equiv\,T_\gamma)$ as
$RT=\con$ (see \eref{adiabat}). Hence `radiation' overwhelmed `matter'
for
\begin{equation}
\label{radera}
 T > T_{\equ} = 5.63 \times 10^{-9} \GeV\ (\Omega_{0} h^2 \Theta^{-3})\ .
\end{equation}
Cosmological processes of interest to particle physics therefore
occured during the RD era, with which we will be mainly concerned in
subsequent sections.

\subsection{Thermal history of the early universe\label{thermhist}} 

As the temperature rises, all particles are expected to ultimately
achieve thermodynamic equilibrium through rapid interactions,
facilitated by the increasing density. The interaction rate $\Gamma$
typically rises much faster with temperature than the expansion rate
$H$, hence the epoch at which $\Gamma$ equals $H$ is usually taken to
mark the onset of equilibrium (see Wagoner 1980). More precisely,
kinetic equilibrium is established by sufficiently rapid elastic
scattering processes, and chemical equilibrium by processes which can
create and destroy particles. Fortunately the particle densities do
not usually become high enough for many-body interactions to be
important and the interaction strengths remain in the perturbative
domain, particularly because of asymptotic freedom for the strong
interactions. Hence the approximation of an ideal gas (see Landau and
Lifshitz 1982) is usually a good one, except near phase
transitions. This vastly simplifies the thermodynamics of the
radiation-dominated (RD) era.

Matters become complicated at temperatures much higher than the masses
of the particles involved, since the cross-section for $2\to2$
processes ultimately decreases ${\propto}T^{-2}$ on dimensional
grounds, hence $\Gamma\,({\propto}T$) then falls behind
$H\,({\propto}T^2)$ at some critical temperature (Ellis and Steigman
1979). Moreover, at temperatures approaching the Planck scale, the
shrinking causal horizon imposes a lower cutoff on the energies of
particles (Ellis and Steigman 1979), while the number of particles in
any locally flat region of space-time becomes negligible (Padmanabhan
and Vasanti 1982). Enqvist and Eskola (1990) have performed a computer
simulation to study the relaxation of a weakly interacting
relativistic gas with an initially non-thermal momentum distribution
towards thermal equilibrium in the early universe. They find that
kinetic equlibrium is achieved after only a few $2\to2$ elastic
collisions, while chemical equilibrium takes rather longer to be
established through $2\to3$ number-{\em changing} processes. In the
extreme case that the universe is created as an initially cold gas of
particles at the Planck scale (e.g. by quantum fluctuations), elastic
scatterings achieve a (maximum) temperature of
$\approx3\times10^{14}\GeV$ while chemical equilibrium is only
established at $\approx10^{12}\GeV$, i.e. well below the grand
unification scale (see also Elmfors \etal 1994). For the QCD gas in
particular, the annihilation rate for quarks to gluons falls behind
$H$ at $\approx3\times10^{14}\GeV$, above which chemical equilibrium
is not achieved (Enqvist and Sirkka 1993).

For an ideal gas, the equilibrium phase space density of particle type
$i$ is 
\begin{equation}
\label{eqdist}
 f_{i}^{\eqm} (q, T) = \left[\exp \left(\frac{E_{i} - \mu_{i}}{T}\right)
  \mp 1 \right]^{-1} ,       
\end{equation} 
where $E_{i}\equiv\sqrt{m_{i}^2+q^2}$, $-/+$ refers to
Bose-Einstein/Fermi-Dirac statistics and $\mu_{i}$ is a possible
chemical potential. The chemical potential is additively conserved in
all reactions. Hence it is zero for particles such as photons and
$Z^0$ bosons which can be emitted or absorbed in any number (at high
enough temperatures)~\footnote{This need not be true
 for $W^{\pm}$ bosons and gluons which carry non-trivial quantum
 numbers. We must {\em assume} that the universe has no net colour or
 hypercharge (see Haber and Weldon 1981).} and consequently equal and
opposite for a particle and its antiparticle, which can annihilate
into such gauge bosons. A finite {\em net} chemical potential for any
species therefore corresponds to a particle-antiparticle asymmetry,
i.e. a non-zero value for any associated conserved quantum
number. Empirically, the net electrical charge of the universe is
consistent with zero and the net baryon number is quite negligible
relative to the number of photons:
$(N_{B}-N_{\bar{B}})/N_{\gamma}\leqsim10^{-9}$ (see Steigman 1976).
Hence for most purposes it is reasonable to set $\mu_e$ and $\mu_B$ to
be zero. The net lepton number is presumably of the same order as the
baryon number so we can consider $\mu_{\nu}$ to be zero for all flavours 
of (massless) neutrinos as well. However if the baryon minus lepton
number $(B-L)$ is not zero, there may well be a large chemical
potential in neutrinos which can influence nucleosynthesis (see
\sref{theoobs}). (Also, even a small asymmetry, comparable
to that observed in baryons, may enable a similarly {\em massive}
particle (see \sref{tc}) to contribute significantly to the
energy density of the universe.)

The thermodynamic observables number density, energy density and
pressure, in equilibrium, are then functions of the temperature alone
(see Harrison 1973):
\begin{equation}
\label{thermdist} 
\eqalign{
 n_{i}^{\eqm} (T) &= g_{i} \int f_{i}^{\eqm} (q, T)\, \frac{\d^3 q}{(2\pi)^3}
  = \frac{g_{i}}{2\pi^2} T^3 I_{i}^{11} (\mp)\ , \\
 \rho_{i}^{\eqm} (T) &= g_{i} \int E_{i} (q)\,f_{i}^{\eqm} (q, T)\, 
  \frac{\d^3 q}{(2\pi)^3} = \frac{g_{i}}{2\pi^2} T^4 I_{i}^{21} (\mp)\ , \\
 p_{i}^{\eqm} (T) &= g_{i} \int \frac{q^2}{3 E_{i} (q)}\,f_{i}^{\eqm} (q, T)\,
  \frac{\d^3 q}{(2\pi)^3} = \frac{g_{i}}{6\pi^2}\,T^4 I_{i}^{03}(\mp)\ ,} 
\end{equation}
where,
\begin{equation} 
\label{integ} 
 I_{i}^{mn}(\mp) \equiv \int_{x_{i}}^{\infty} y^{m} (y^2 - x^2_{i})^{n/2}\ 
  (\e^{y} \mp 1)^{-1} dy\ , \qquad x_{i} \equiv \frac{m_{i}}{T}\ ,
\end{equation} 
$g_{i}$ is the number of internal (spin) degrees of freedom, and $-/+$
refers as before to bosons/fermions. These equations yield the
relation
\begin{equation} 
\label{2law} 
 {\d p^{\eqm} \over \d T} = {(\rho^{\eqm} + p^{\eqm}) \over T}\ ,
\end{equation} 
which is just the second law of thermodynamics (see Weinberg 1972).

For relativistic (R) particles with $x{\ll}1$, the integrals
\eref{integ} are
\begin{equation}
\eqalign{
 {\rm bosons:} &\quad I^{11}_{\R}(-) = 2\zeta(3)\ , \qquad
  I^{21}_{\R}(-) = I^{03}_{\R}(-) = \frac{\pi^4}{15}\ , \\ 
 {\rm fermions:} &\quad I^{11}_{\R}(+) = \frac{3\zeta(3)}{2}\ , \qquad
  I^{21}_{\R}(+) = I^{03}_{\R}(+) = \frac{7\pi^4}{120}\ ,} 
\end{equation} 
where $\zeta$ is the Riemann Zeta function and $\zeta(3)=1.202$;
for example, photons with $g_{\gamma}=2$ have
$n_{\gamma}=\case{2\zeta(3)}{\pi^2} T^3$ and $\rho_{\gamma}=3
p_{\gamma}=\case{\pi^2}{15} T^4$. (Since photons are always in
equilibrium at these epochs, and indeed {\em define} the temperature
$T$, we will not bother with the superscript ${\eqm}$ for $n_{\gamma},
\rho_{\gamma}$ or $p_{\gamma}$.)

For non-relativistic (NR) particles, which have $x{\gg}1$, we recover
the Boltzmann distribution
\begin{equation} 
 n_{\NR}^{\eqm} (T)= \frac{\rho_{\NR}^{\eqm} (T)}{m} =
  \frac{g}{(2\pi)^{3/2}} T^3 x^{3/2} \e^{-x} , \qquad p_{\NR} \simeq 0\ ,
\end{equation} 
independently of whether the particle is a boson or fermion.
Non-relativistic particles, of course, contribute negligibly to the
energy density in the RD era. It should be noted that the Boltzmann
distribution is {\em not} invariant under the cosmic expansion, hence
non-relativistic particles can maintain equilibrium only if they
interact rapidly with a (dominant) population of relativistic
particles (see Bernstein 1988).

It is then convenient to parametrize:
\begin{equation}
\label{energy} 
 \rho_{i}^{\eqm} (T) \equiv \left(\frac{g_{\rho_i}}{2}\right)
  \rho_{\gamma}\ , \qquad {\ie} \quad 
 g_{\rho_i} = \frac{15}{\pi^4}\,g_{i} \,I_{i}^{21} (\mp)\ ,
\end{equation} 
so that $g_{\rho_i}$ equals $g_{i}$ for a relativistic boson,
$\case{7}{8}g_{i}$ for a relativistic fermion, and is negligibly small
($<10\%$ correction) for a non-relativistic particle. When all
particles present are in equilibrium through rapid interactions, the
total number of relativistic degrees of freedom is thus given by
summing over all interacting relativistic bosons (B) and fermions (F):
\begin{equation}
\label{relfree} 
 g_{\R} = \sum_{\B} g_{i} + \frac{7}{8} \sum_{\F} g_{i}\ .
\end{equation} 

At any given time, not all particles will, in fact, be in equilibrium
at a common temperature $T$. A particle will be in kinetic equilibrium
with the background thermal plasma (i.e. $T_{i}=T$) only while it is
interacting, i.e. as long as the scattering rate,
\begin{equation}
 \Gamma_{\scat} = n_{\scat} \langle \sigma_{\scat} v \rangle\ ,
\end{equation} 
exceeds the expansion rate $H$. Here $\langle\sigma_{\scat}v\rangle$
is the (velocity averaged) cross-section for $2\to2$ processes such as
$i\gamma\to{i}\gamma$ and $i\ell^{\pm}\to{i}\ell^{\pm}$ which maintain
good thermal contact between the $i$ particles and the particles (of
density $n_{\scat}$) constituting the background plasma. ($\ell$
refers in particular to electrons which are abundant down to
$T\,\sim\,m_{\el}$ and remain strongly coupled to photons via Compton
scattering through the entire RD era, so that $T_{\el}=T$ always.) The
$i$ particle is said to `decouple' at $T=T_{\D}$ when the condition
\begin{equation}
 \Gamma_{\scat} (T_{\D}) \simeq H (T_{\D})
\end{equation} 
is satisfied. Of course no particle is ever truly decoupled since
there are always {\em some} residual interactions; however such
effects are calculable (e.g. Dodelson and Turner (1992) and are
generally negligible.

If the particle is relativistic at this time (i.e.
$m_{i}\,<\,T_{\D}$), then it will also have been in chemical equilibrium
with the thermal plasma (i.e.
$\mu_{i}+\mu_{\bar{i}}=\mu_{\ell^{+}}+\mu_{\ell^{-}}=\mu_{\gamma}=0$)
through processes such as $i\bar{i}\leftrightarrow\gamma\gamma$ and
$i\bar{i}\leftrightarrow\ell^{+}\ell^{-}$.\footnote{In fact,
 neutrinos, which are both massless and weakly interacting, are the
 only particles in the Standard Model which satisfy this condition.
 The other particles, being both massive and strongly and/or
 electromagnetically interacting, would have self-annihilated when
 they became non-relativistic and would therefore not have survived
 with any appreciable abundance until the epoch of kinetic decoupling
 which generally occurs much later.} Hence its abundance at
decoupling will be just the equilibrium value
\begin{equation}
 n_{i}^{\eqm} (T_{\D}) =  \left(\frac{g_{i}}{2}\right)\,n_{\gamma}
  (T_{\D})\,f_{{\B}, {\F}}\ ,
\end{equation} 
where $f_{\B}=1$ and $f_{\F}=\case{3}{4}$ corresponding to whether
$i$ is a boson or a fermion.  

Subsequently, the decoupled $i$ particles will expand freely without
interactions so that their number in a {\em comoving} volume is
conserved and their pressure and energy density are functions of the
scale-factor $R$ {\em alone}. Although non-interacting, their phase
space distribution will retain the equilibrium form \eref{eqdist},
with $T$ substituted by $T_{i}$, as long as the particles remain {\it
relativistic}, which ensures that both $E_{i}$ and $T_{i}$ scale as
$R^{-1}$. Initially, the temperature $T_{i}$ will continue to track
the photon temperature $T$. Now as the universe cools below various
mass thresholds, the corresponding massive particles will become
non-relativistic and annihilate. (For massive particles in the
Standard Model, such annihilation will be almost total since all such
particles have strong and/or electromagnetic interactions.) This will
heat the photons and other interacting particles, but not the
decoupled $i$ particles, so that $T_{i}$ will now drop below $T$ and,
consequently, $n_{i}/n_{\gamma}$ will decrease below its value at
decoupling.

To calculate this it is convenient, following Alpher \etal (1953), to
divide the total pressure and energy density into interacting (I) and
decoupled (D) parts, which are, respectively, functions of $T$ and $R$
alone:
\begin{equation}
 p = p_{\I} (T) + p_{\D} (R)\ , \qquad \rho = \rho_{\I} (T) + \rho_{\D} (R)\ .
\end{equation} 
The conservation equation \eref{conserv2} written as
\begin{equation}
 R^3 \frac{\d p}{\d T} = \frac{\d}{\d T} \left[ R^3 (\rho + p) \right] 
\end{equation} 
then reduces to
\begin{equation}
 \frac{\d \ln R}{\d \ln T} = - \frac{1}{3} \frac{ (\d \rho_{\I} / 
  \d \ln T)} {(\rho_{\I} + p_{\I}) }\ ,
\end{equation} 
upon requiring the number conservation of decoupled particles
($n_{\D}R^3=\con$) and neglecting the pressure of non-relativistic
decoupled particles. Combining with the second law of thermodynamics
\eref{2law}, we obtain
\begin{equation}
\label{dlnRbydlnT} 
 \frac{\d \ln R}{\d \ln T} = - 1 - \frac{1}{3} 
  \frac{\d \ln \left(\frac{\rho_{\I} + p_{\I}}{T^4}\right)}{\d \ln T}\ ,
\end{equation} 
which integrates to,
\begin{equation}
\label{lnRlnT} 
 \ln R = - \ln T - \frac{1}{3} \ln \left(\frac{\rho_{\I} + 
  p_{\I}}{T^4} \right) + \con .
\end{equation} 
If $(\rho_{\I} + p_{\I})/T^4$ is constant, as for a gas of blackbody
photons, this yields the adiabatic invariant
\begin{equation}
\label{adiabat} 
 R T = \con
\end{equation}  
which we have used earlier to obtain \eref{radera}. The second term on
the RHS of \eref{lnRlnT} is a correction which accounts for departures
from adiabaticity due to changes in the number of interacting species.

(Another possible source of non-adiabaticity is a phase transition
which may release latent heat thus increasing the entropy. The ideal
gas approximation is then no longer applicable and finite temperature
field theory must be used (see Bailin and Love 1986, Kapusta
1988). The standard cosmology assumes parenthetically that such phase
transitions occurred rapidly at their appropriate critical
temperature, generating negligible latent heat, i.e. that they were
second-order.  However, phase transitions associated with spontaneous
symmetry breaking in gauge theories may well be first-order; this
possibility is in fact exploited in the inflationary universe model
(Guth 1981, see Linde 1990) to account for the observed large entropy
content of the universe, as mentioned earlier. We will shortly discuss
the possible generation of entropy during the quark-hadron and
electroweak phase transitions.)

Epochs where the number of interacting species is different can now be
related by noting that \eref{dlnRbydlnT} implies the constancy of the
specific entropy, $S_{\I}$, in a comoving volume:
\begin{equation}
 \frac{\d S_{\I}}{\d T} = 0\ , \qquad S_{\I} \equiv s_{\I} R^3 ,
\end{equation} 
Here, $s_{\I}$, the specific entropy density, sums over all
{\em interacting} species in equilibrium:
\begin{equation} 
 s_{\I} \equiv \frac{\rho_{\I} + p_{\I}}{T} = \sum_{\inter} s_{i}\ ,
\end{equation} 
where, using \eref{thermdist},
\begin{equation} 
 s_{i}(T) = g_{i} \int \frac{3 m_{i}^2 + 4 q^2}{3 E_{i}(q)\,T}\,
  f_{i}^{\eqm}(q, T)\,\frac{\d^3 q}{(2\pi)^3}\ .
\end{equation} 
As with the energy density \eref{energy}, we can conveniently
parametrize the entropy density of particle $i$ in terms of that for
photons:
\begin{equation}
 s_{i} (T) \equiv \left(\frac{g_{s_i}}{2}\right) \left(\frac{4}{3}
  \frac{\rho_{\gamma}}{T}\right)\ ,
\end{equation}
i.e.
\begin{equation}
\label{entrop} 
 g_{s_i} = \frac{45}{4\pi^4} g_{i} \left[I_{i}^{21}(\mp) + \frac{1}{3}
  I_{i}^{03}(\mp)\right] ,
\end{equation} 
so defined that $g_{s_i}$ (like $g_{\rho_i}$) equals $g_i$ for a
relativistic boson, $\case{7}{8} g_{i}$ for a relativistic fermion,
and is negligibly small for a non-relativistic particle. Hence the
number of {\em interacting} degrees of freedom contributing to the
specific entropy density is given by
\begin{equation}
\label{gSI} 
 g_{s_{\I}} \equiv \frac{45}{2\pi^2} \frac{s_{\I}}{T^3} 
  = \sum_{\inter} g_{s_i}\ .
\end{equation} 
This is, of course, the same as $g_{\R}$ \eref{relfree} when all
particles are relativistic. (This parameter has been variously called
$g_{\I}$ (Steigman 1979), $g_{\rm E}$ (Wagoner 1980) and $g^{\prime}$
(Olive \etal 1981a) in the literature.)

It is now simple to calculate how the temperature of a particle $i$
which decoupled at $T_{\D}$ relates to the photon temperature $T$ at a
later epoch. For $T<T_{\D}$, the entropy in the decoupled $i$
particles and the entropy in the still interacting $j$ particles are
{\em separately} conserved:
\begin{equation}
\eqalign{
 S - S_{\I} = s_{i}\,R^3 &= \frac{2\pi^2}{45} g_{s_i}(T)\,(R\,T)_{i}^3 , \\
 S_{\I} = \sum_{j \neq i} s_{j}(T)\,R^3 &= 
  \frac{2\pi^2}{45} g_{s_{\I}}(T)\,(R\,T)^3\ ,}
\end{equation} 
where $S$ is the conserved total entropy at $T>T_{\D}$. Given that
$T_{i}=T$ at decoupling, this then yields for the subsequent ratio of
temperatures (Srednicki \etal 1988, Gondolo and Gelmini 1991):
\begin{equation}
\label{TibyT} 
 \frac{T_{i}}{T} = \left[\frac{g_{s_i} (T_{\D})}{g_{s_i}(T)} 
 \frac{g_{s_{\I}} (T)}{g_{s_{\I}} (T_{\D})}\right]^{1/3} .
\end{equation} 
Note the difference from the expression
$T_{i}/T=[g_{s_{\I}}(T)/g_{s_{\I}}(T_{\D})]^{1/3}$ given by Olive
\etal (1981a), which is not always correct, for example when the
decoupled particles have new interactions which allow them to
subsequently annihilate into other non-interacting particles, thus
changing $g_{s_{i}}$ from its value at decoupling (e.g. Kolb \etal
1986c).

The degrees of freedom specifying the conserved total entropy is then
given, following decoupling, by
\begin{equation}
\label{gs} 
 g_{s} (T) \equiv \frac{45}{2\pi^2} \frac{S}{T^3 R^3} = g_{s_{\I}} (T)
  \left[1 + \frac{g_{s_i}(T_{\D})}{g_{s_{\I}} (T_{\D})}\right] .
\end{equation} 
\noindent 
When the species $i$ becomes non-relativistic and annihilates into the
other relativistic interacting particles before decoupling, the few
remaining decoupled particles have negligible entropy content, hence
$g_{s_i}(T_{\D})\simeq0$. Then $g_{s}$ just counts all {\it
interacting} species at temperature T which have now acquired the
entropy released by the annihilations, i.e. $g_{s}\simeq\,g_{s_{\I}}$
\eref{gSI}. However when the decoupled species is relativistic and
carries off its own entropy which is {\em separately} conserved, then
$g_{s}$ explicitly includes its contribution to the conserved total
entropy, by weighting appropriately by its temperature, which may now
be smaller (according to \eref{TibyT}) than the photon temperature
$T$:
\begin{equation}
\label{gsT} 
\eqalign{
 g_{s} (T) &= \sum_{j \neq i} g_{s_j} (T) + g_{s_i} (T_{i})
  \left(\frac{T_{i}}{T}\right)^3 \\ \nonumber
           &\simeq \sum_{\B} g_{i} \left(\frac{T_{i}}{T}\right)^3 +
            \frac{7}{8} \sum_{\F} g_{i} \left(\frac{T_{i}}{T}\right)^3 .}
\end{equation} 
\noindent 
The last equality follows when all particles are relativistic.  (This
parameter is called $g_{\star{s}}$ by Scherrer and Turner (1986) and
Kolb and Turner (1990), $h$ by Srednicki \etal (1988) and $h_{\eff}$
by Gondolo and Gelmini (1991).) If several different species decouple
while still relativistic, as is possible in extensions of the Standard
Model which contain new weakly interacting massless particles, then
\eref{gs} is easily generalized to (Gondolo and Gelmini 1991)
\begin{equation}
 g_{s} (T) = g_{s_{\I}} (T) \prod_{i{\dec}}
  \left[1 + \frac{g_{s_i}(T_{{\D}_i})}{g_{s_{\I}} (T_{{\D}_i})}\right] \ .
\end{equation} 
We now have an useful fiducial in the total entropy density,
\begin{equation}
 s (T) \equiv \frac{2\pi^2}{45} g_{s}(T) T^3\ ,
\end{equation} 
which {\em always} scales as $R^{-3}$ by appropriately keeping track
of any changes in the number of degrees of freedom. Therefore the
ratio of the decoupled particle density to the blackbody photon
density is subsequently related to its value at decoupling as:
\begin{equation}
\label{nibyngamma} 
\eqalign{
 \frac{(n_{i} / n_{\gamma})_{T}}{(n_{i}^{\eqm} / n_{\gamma})_{T_{\D}}}
  = \frac{g_{s} (T)}{g_{s} (T_{\D})} 
  = \frac{N_{\gamma}(T_{\D})}{N_{\gamma}(T)}\ ,}
\end{equation} 
where $N_{\gamma}=R^3\,n_{\gamma}$ is the total number of blackbody
photons in a comoving volume.

The total energy density may be similarly parametrized as:
\begin{equation}
 \rho (T)  = \sum \rho^{\eqm}_{i} \equiv
              \left(\frac{g_{\rho}}{2}\right) \rho_{\gamma} 
           = \frac{\pi^2}{30} g_{\rho} T^4\ , 
\end{equation}
i.e. 
\begin{equation}
\label{grho}
\eqalign{
 g_{\rho} &= \sum_{j \neq i} g_{\rho_j} (T) + g_{\rho_i} (T_{i})
           \left(\frac{T_{i}}{T}\right)^4 \\ 
          &\simeq \sum_{\B} g_{i} \left(\frac{T_{i}}{T}\right)^4 +
           \frac{7}{8} \sum_{\F} g_{i} \left(\frac{T_{i}}{T}\right)^4,}
\end{equation}
where the last equality follows when all particles are relativistic.
(The parameter $g_{\rho}$ is called $g$ by Steigman (1979), Olive
\etal (1981a) and Srednicki \etal (1988) and $g_{\eff}$ by Gondolo and
Gelmini (1991); more often (e.g. Wagoner 1980, Scherrer and Turner
1986, Kolb and Turner 1990) it is called $g_{\star}$.)

Let us now rewrite \eref{dlnRbydlnT} more compactly as
\begin{equation}
 \frac{\d R}{R} = -\frac{\d T}{T} - \frac{1}{3}
  \frac{\d g_{s_{\I}}}{g_{s_{\I}}} .
\end{equation} 
using \eref{gSI}. (This expression is also given by Srednicki \etal
(1988), however with $g_{s}$ rather than $g_{s_{\I}}$ on the RHS;
admittedly this makes no difference in practice.) Using this, we can
now obtain the relationship between the time $t$ and the temperature
$T$ by integrating the F-L equation \eref{fried1}. Since the
curvature term $k/R^2$ is negligible during the RD era, we have
\begin{equation}
\label{H} 
 H = \sqrt{\frac{8\pi\,\rho}{3\,M_{\Pl}^{2}}} = 1.66\ g_{\rho}^{1/2}
      \frac{T^2}{M_{\Pl}}\ ,
\end{equation} 
and,
\begin{equation}
\label{time} 
\eqalign{
 t &= \int \left(\frac{3\,M_{\Pl}^{2}}{8 \pi\,\rho}\right)^{1/2} 
         \frac{\d R}{R} \\ \nonumber
   &= -\int \left(\frac{45\,M_{\Pl}^{2}}{4 \pi^3}\right)^{1/2}
       g_{\rho}^{-1/2} \left(1 + \frac{1}{3}
       \frac{\d \ln g_{s_{\I}}}{\d \ln T}\right) \frac{\d T}{T^3} .}
\end{equation}
During the periods when ${\d}g_{s_{\I}}/{\d}T\simeq0$, i.e. away from
mass thresholds and phase transitions, this yields the useful commonly
used approximation
\begin{equation}
\label{trad} 
 t = \left(\frac{3\,M_{\Pl}^{2}}{32 \pi\,\rho}\right)^{1/2} 
   = 2.42\ g_{\rho}^{-1/2} \left(\frac{T}{\MeV}\right)^{-2} \sec . 
\end{equation}

The above discussion is usually illustrated by the example of the
decoupling of massless neutrinos in the Standard Model. Taking the
thermally-averaged cross-section to be
$\langle\sigma{v}\rangle\,\sim\,G_{\F}^{2}E^{2}\,\sim\,G_{\F}^{2}T^{2}$,
the interaction rate is
$\Gamma=n\langle\sigma{v}\rangle\,\sim\,G_{\F}^{2}T^{5}$ (since
$n\,\approx\,T^3$). This equals the expansion rate
$H\,\sim\,T^2/M_{\Pl}$ at the decoupling temperature
\begin{equation}
\label{Tdecnu}
 T_{\D} (\nu) \sim (G_{\F}^2 M_{\Pl})^{-1/3} \sim 1 \MeV . 
\end{equation} 
(A more careful estimate of $\langle
\sigma_{\nu\bar{\nu}\to{\el}^+{\el}^-}v\rangle$ (Dicus \etal 1982,
Enqvist \etal 1992a) gives $T_{\D}(\nu_{\mu},\,\nu_{\tau})=3.5$ MeV
for the neutral current interaction and $T_{\D}(\nu_{\el})=2.3$ MeV,
upon adding the charged current interaction.) At this time
$n_{\nu}^{\eqm}=\case{3}{4}n_{\gamma}$ since $T_{\nu}=T$ and
$g_{\nu}=2$. (In the Standard Model, right-handed neutrinos transform
as singlets of $SU(2)_{\L}{\otimes}U(1)_Y$ and have no gauge
interactions, hence these states cannot be excited thermally unless
Dirac masses are introduced (see \sref{numass}).)  Subsequently as $T$
drops below the electron mass $m_{\el}$, the electrons and positrons
annihilate (almost) totally, heating the photons but not the decoupled
neutrinos. From \eref{TibyT} we see that while $g_{\nu}$ does not
change following decoupling, the number of {\em other} interacting
degrees of freedom decreases from 11/2 ($\gamma$ and ${\el}^{\pm}$) to
2 ($\gamma$ only), hence the comoving number of blackbody photons
increases by the factor
\begin{equation}
\label{elecann} 
 \frac{N_{\gamma}\,(T \ll m_{\el})}{N_{\gamma}\,(T = T_{\D} (\nu))} =
  \left[\frac{(RT)_{T \ll m_{\el}}}{(RT)_{T=T_{\D} (\nu)}}\right]^3 = 
  \frac{11}{4}\ ,
\end{equation}
so that subsequently
\begin{equation}
\label{nuabund}
 \left(\frac{n_{\nu}}{n_{\gamma}}\right)_{T \ll m_{\el}} = \frac{4}{11} 
  \left(\frac{n_{\nu}^{\eqm}}{n_{\gamma}}\right)_{T=T_{\D}(\nu)} = 
 \frac{3}{11}\ .
\end{equation} 
The evolution of the neutrino temperature through the period of
${\el}^{\pm}$ annihilation can be computed using \eref{entrop} and
\eref{TibyT} (see Weinberg 1972):
\begin{equation}
\label{TnubyT}
 \frac{T_{\nu}}{T} = \left(\frac{4}{11}\right)^{1/3} 
  \left[1 + \frac{45}{2\pi^4} \left(I^{21}(+) + \frac{1}{3}
  I^{03}(+)\right) \right]^{1/3} .
\end{equation} 
The neutrinos remain relativistic and therefore continue to retain
their equilibrium distribution function hence the degrees of freedom
characterizing the present day entropy and energy densities are :
\begin{eqnarray}
\label{gsgrhonow} 
\eqalign{
 g_{s}\,(T \ll m_{\el}) & = g_{\gamma} + \frac{7}{8}\,N_{\nu}\,g_{\nu}
  \left(\frac{T_{\nu}}{T}\right)^3 = \frac{43}{11}\ , \\
 g_{\rho}\,(T \ll m_{\el}) & = g_{\gamma} + \frac{7}{8}\,N_{\nu}\,g_{\nu}
  \left(\frac{T_{\nu}}{T}\right)^4 = 3.36\ ,}
\end{eqnarray}
for 3 {\em massless} neutrino species ($N_{\nu}$=3). Note that the
increase in the number of comoving photons due to ${\el}^{\pm}$
annihilation \eref{elecann} is indeed given, following
\eref{nibyngamma}, by the ratio
$g_{s}(T_{\D}(\nu))/g_{s}(T_{0})=\case{43}{4}/\case{43}{11}=\case{11}{4}$.

Since neutrino decoupling occurs so close to ${\el}^+ {\el}^-$
annihilation, their residual interactions with the thermal plasma
cause the neutrinos to be slightly heated by the resultant entropy
release (Dicus \etal 1982, Herrera and Hacyan 1989). This effect has
been studied by Dolgov and Fukugita (1992) and, particularly
carefully, by Dodelson and Turner (1992), who solve the governing
Boltzmann equation with both scattering and annihilation processes
included; Hannestad and Madsen (1995) have redone the exercise using
Fermi-Dirac rather than Boltzmann statistics. The asymptotic energy
density in electron neutrinos is found to be raised by $0.8\%$ over
the canonical estimate above, and that for muon and tau neutrinos by
$0.4\%$, while the back reaction due to neutrino heating is found to
suppress the increase in the comoving number of photons by
$0.5\%$. These studies demonstrate that neutrino decoupling is not an
instantaneous process, particularly since the interaction
cross-section increases with the neutrino energy.  Consequently the
spectrum of the decoupled neutrinos deviates slightly from the
Fermi-Dirac form, causing the effective neutrino temperature
($\equiv-q/{\ln}f_{\nu}(q, t)$) to increase with momentum. The
increase is however only by $0.7\%$ even at relatively high momenta,
$q/T\approx10$, justifying the usual approximation of instantaneous
decoupling.

The detailed formalism given above for reconstructing the thermal
history of the RD era is essential for accurately calculating the
abundances of hypothetical massive particles or massless particles
with unusual interactions, which may affect BBN. For the moment we
restrict our attention to the Standard $SU (3)_{\c}{\otimes}SU
(2)_{\L}{\otimes}U (1)_{Y}$ Model and show in \tref{tabthermhist} the
temperature dependence of the number of interacting relativistic
degrees of freedom, $g_{\R}(T)$ \eref{relfree}, as well as the factor
$N_{\gamma}(T_{0})/N_{\gamma}(T)$ \eref{nibyngamma} by which the
comoving blackbody photon number is higher today, at $T=T_{0}$. In
calculating $g_{\R}$ we have assumed that a massive particle remains
relativistic down to $T\,\sim\,m_{i}$ and immediately annihilates
completely into radiation, and that phase transitions happen
instantaneously at the relevant critical temperature with negligible
release of entropy; hence the quoted values are meaningful only when
far away from mass thresholds and phase transitions. Apart from the
massless neutrinos, all particles in the SM are strongly coupled to
the thermal plasma while they are relativistic, hence $g_{s}$
\eref{gsT} equals $g_{\rho}$ \eref{grho} and their common value equals
$g_{\R}$, above the neutrino decoupling temperature $T_{\D}(\nu)$ of a
few MeV, while their low temperature values are given in
\eref{gsgrhonow}. Note that neutrino decoupling has no effect on the
entropy or dynamics, hence $g_{s}$ and $g_{\rho}$ do not change (from
their common value of 43/4 below the muon mass threshold) until
${\el}^{\pm}$ annihilation occurs.


A more careful calculation has been done by Srednicki \etal (1988)
following a similar earlier exercise by Olive \etal (1981a). By
numerical integration over the phase-space density (using
\eref{energy} and \eref{entrop}), these authors obtain $g_{\rho}$ and
$g_{s}$ as a continuous function of $T$ rather than step-wise as in
\tref{tabthermhist}; they also include the (small) contribution to the
energy and entropy density from non-relativistic baryons and
mesons. There is however considerable ambiguity concerning the
thermodynamic history during the quark-hadron phase transition. As the
critical temperature $T_{\c}^{\qh}$ is approached from below, particle
interactions become important and the ideal gas approximation begins
to break down; however at temperatures higher than
$T_{\c}^{*}\approx1\GeV$,\footnote{It is difficult to reliably
calculate $T_{\c}^{*}$ because of non-perturbative effects in the
strongly coupled quark-gluon plasma (see Shuryak 1980, Gross \etal
1981).} the asymptotic freedom of the strong interactions again
permits the system to be decribed as an ideal gas of leptons, quarks
and gauge bosons. Srednicki \etal (1988) present curves for the
behaviour of $g_{\rho}$ and $g_{s}$ in the intervening region
corresponding to two choices (150 and 400 MeV) of $T_{\c}^{\qh}$ and
state that these bound the range of possibilities.\footnote{Srednicki
\etal did not provide any quantitative details as to how these curves
are obtained. It appears (K\,A\,Olive, private communication) that
these authors adopted the na\"{\ii}ve thermodynamic picture (see Olive
1990b) in which a hadron is viewed as a `bag' containing quarks and
gluons so that the pressure and energy density in the region of
interest ($T\sim100-1000$ MeV) are taken to be
$P=\case{\pi^{2}}{90}[2(N_{\c}^{2}-1)+\case{7}{2}N_{\c}N_{\fla}]T^{4}-B,
\rho=3P+4B,$ where $N_{\c}$ (=3) is the number of colours, $N_{\fla}$
(=3) is the number of light quark flavours ($u,\,d,\,s$), and $B$ is
the bag constant representing the vacuum energy difference between the
two phases (which essentially determines $T_{\c}^{\qh}$). In this
picture the pressure in the quark-gluon phase drops steeply with
temperature during `confinement', which occurs at a higher temperature
for a higher adopted value of $B$. The pressure in the hadronic phase
at lower temperatures (calculated assuming non-interacting particles)
is approximately constant hence phase equilibrium is achieved when the
pressure in the two phases become equal at $T \approx 100$ MeV.} They
also show the evolution of $g_{\rho}$ during this epoch for the case
when the phase transition is strongly first-order, a possibility
suggested in the past by lattice gauge calculations which set quark
masses to be zero (see Satz 1985, McLerran 1986).  However recent
lattice computations which have been performed with realistic masses
for the $u$, $d$ and $s$ quarks suggest that there may be a
second-order phase transition or even a `cross-over' (see Toussaint
1992, Smilga 1995). (This is particularly important to keep in mind in
the context of the bound imposed by BBN on superweakly interacting
particles which may have decoupled during this era.) In
\fref{degfreedom} we show both $g_{{\rho}}$ and $g_{s}$ as a function
of temperature as computed by Srednicki \etal (1988). As we have
emphasized, these curves are only meant to indicate the range of
possibilities in this temperature region.


The last three entries in \tref{tabthermhist} are uncertain because of our
ignorance about the mass of the Higgs boson which is responsible for
$SU(2){\otimes}U(1)$ symmetry breaking. It has been assumed here
that the Higgs is sufficiently heavy that the electroweak phase
transition is effectively second-order and occurs at a critical
temperature (see Linde 1979, Weinberg 1980)
\begin{equation}
\label{TcEW} 
 T_{\c}^{\EW} \simeq 300 \GeV \left[ 1 + \left(\frac{m_{H^0}}{150
   \GeV}\right)^{-2} \right]^{-1/2} .
\end{equation} 
Coleman and Weinberg (1973) had studied the possibility that the Higgs
is {\em massless} at tree-level so that the $SU(2){\otimes}U(1)$
symmetry is classically scale-invariant and broken only by radiative
corrections. These corrections generate a small mass,
$m_{H^0}\simeq10\GeV$ (for a light top quark); the critical
temperature is then $T_{\c}^{\EW}\simeq25\GeV$ and the phase
transition is strongly first-order, generating a large and probably
unacceptable amount of entropy (see Sher 1989). However the
Coleman-Weinberg theory is untenable if the top quark mass exceeds
$85\GeV$ as is now established by its recent detection at {\sl
Fermilab} with $m_{t}=180\pm12\GeV$ and, further, such a light Higgs
is now ruled out by \LEP which sets the bound $m_{H^0}>60\GeV$
(Particle Data Group 1996).

Recently, cosmological electroweak symmetry breaking has come under
renewed scrutiny following the realization that fermion-number
violating transitions are unsuppressed at this epoch; the possibility
of generating the baryon asymmetry of the universe then arises if the
neccessary non-equilibrium conditions can be achieved via a
first-order phase transition (see Cohen \etal 1994, Rubakov and
Shaposhnikov 1996). While this topic is outside the scope of the
present review, we note that according to the detailed studies (see
Rom\~ao and Freire 1994, Yaffe 1995), the phase transition is at best
{\it very} weakly first-order. Hence our assumption that any
generation of entropy is insignificant is probably justified although
in extensions of the SM where the Higgs sector is enlarged, e.g. in
supersymmetric models, the phase transition may well be strongly
first-order with substantial entropy generation (e.g. Brignole \etal
1994). These conclusions are however based on perturbation
theory. Recent non-perturbative studies of the $SU(2)$ Higgs model in
three dimensions, using both analytic techniques (e.g. Buchm\"uller
and Philipsen 1995) and lattice simulations (e.g. Kajantie \etal 1996)
show that for large Higgs mass there will be no phase transition but
rather a `cross-over' since there is no gauge-invariant order
parameter. Recent lattice calculations (e.g. Boyd \etal 1995,
Philipsen \etal 1996) also indicate the presence of bound states in
the plasma at high temperatures due to the non-abelian nature of the
colour and electroweak forces; however the consequent departure from
ideal gas behaviour is only of ${\cal O}(10\%)$.

At even higher temperatures, $g_{\R}$ will depend on the adopted
theory. For example, in the minimal $SU(5)$ GUT, with three families
of fermions and a single (complex) {\bf 5} of Higgs plus a {\bf 24}
adjoint of Higgs to break $SU(5)$, the number of degrees of freedom
above the unification scale is given by:
\begin{equation}
\fl
 g_{\R}\ (T \geqsim M_{\GUT})_{SU (5)} = (2\times24 + 24 + 
  2\times5) + \case{7}{8}\ (2\times3\times15) = \case{647}{4}\ .
\end{equation} 
In a supersymmetric model below the SUSY-breaking scale, the degrees
of freedom would at least double overall; in the minimal
supersymmetric Standard Model (MSSM) with three families of fermions
and two (complex) doublets of Higgs plus all the superpartners,
\begin{equation}
\label{gMSSM}
 g_{\R}\ (T \leqsim M_{\SUSY})_{\MSSM} = (24 + 8 + 90)\,
  (1 + \case{7}{8}) = \case{915}{4}\ ,
\end{equation}
when all particles are relativistic. The present experimental limits
(Particle Data Group 1996) allow some supersymmetric particles, if
they exist, to be light enough to possibly affect the last few entries
in \tref{tabthermhist}. Of course given the mass spectrum of any
specific supersymmetric (or other) model, \tref{tabthermhist} can be
recalculated accordingly.

To summarize, although the formulation of kinetic theory in the
expanding universe is far from trivial (see Bernstein 1988), the
thermal history of the universe can be reconstructed fairly reliably
back to the Fermi scale, and, with some caveats, nearly upto the GUT
scale. This is possible primarily because we are dealing with a dilute
radiation dominated plasma in which non-relativistic particles have
negligible abundances and are forced to remain in equilibrium with the
relativistic particles. Interaction rates typically rise faster with
temperature than the expansion rate of the universe, justifying the
usual assumption of equilibrium. The major uncertainties arise where
the ideal gas approximation breaks down, viz. at phase transitions
associated with symmetry breaking, and at very high temperatures when
equilibrium may not be achieved.

\section{Primordial nucleosynthesis\label{primnucl}} 

We now turn to the creation of the light elements towards the end of
the ``first three minutes'' which provides the deepest detailed probe

of the Big Bang.\footnote{The temperature fluctuations in the CMB
 observed (Smoot \etal 1992) by the {\sl Cosmic Background Explorer}
 (\COBE) very probably reflect physical conditions at a much earlier
 epoch, if for example these are due to quantum perturbations
 generated during inflation (see Linde 1990). This interpretation is
 however not sufficiently firmly established as yet to provide a
 reliable ``laboratory'' for particle physics, although this may well
 happen as further observational tests are performed (see Steinhardt
 1995).} The physical processes involved have been well understood for
some time (Hayashi 1950, Alpher \etal 1953, Hoyle and Tayler 1964,
Peebles 1966a,b, Wagoner \etal 1967)~\footnote{Gamow and collaborators
 pioneered such calculations in the 1940s (see Alpher and Herman 1950)
 but did not take into account the crucial role played by the weak
 interactions in maintaining neutron-proton equilibrium; for historical
 accounts see Alpher and Herman (1990) and Wagoner (1990).} and the
final abundances of the synthesized elements are sensitive to a
variety of parameters and physical constants. This enables many
interesting constraints to be derived on the properties of relic
particles or new physics which may influence BBN and alter the
synthesized abundances. It must, of course, first be demonstrated that
the expected elemental abundances in the standard BBN model are
consistent with observations. There is a complication here in that the
light elements are also created and destroyed in astrophysical
environments so their abundances today differ significantly from their
primordial values. The latter can only be inferred after correcting
for the complex effects of galactic chemical evolution (see Tinsley
1980) over several thousand million years and this necessarily
introduces uncertainties in the comparison with theory.

There are many excellent reviews of both theoretical and observational
aspects of BBN (see Peebles 1971, Weinberg 1972, Schramm and Wagoner
1977, Boesgaard and Steigman 1985, Pagel 1992, Reeves 1994). However
these usually quote results obtained by numerical means, while in
order to appreciate the reliability (or otherwise!) of constraints
derived therein one first requires a good analytic understanding of
the physical processes involved. Secondly, as noted above, the
observed elemental abundances have to be corrected for evolutionary
effects and there are differences in the approaches taken by different
authors in inferring the primordial values. It is therefore helpful to
review the essential theory and the actual observational data before
we examine the validity of the standard BBN model and then proceed to
discuss the constraints imposed on new physics.

\subsection{The standard BBN model\label{bbn}} 

It is convenient to consider element synthesis in the early universe
as occuring in two distinct stages: first the decoupling of the weak
interactions which keep neutrons and protons in equilibrium, and
second the onset, a little later, of the nuclear reactions which build
up the light nuclei. It is possible to do this because the very high
value of the entropy per nucleon ($s/n_{\N}\sim10^{11}$) ensures that
the equilibrium abundances of all bound nuclei are quite negligible as
long as free nucleons are in equilibrium. We begin by outlining an
elegant semi-analytic analysis of the first stage by Bernstein \etal
(1989) which follows the evolution of the neutron-to-proton ratio and
allows the yield of $\Hefour$, the primary product of BBN, to be
calculated quite accurately {\em without} any detailed analysis of the
nuclear reaction network. The latter is however neccessary to
calculate the yields of less stable nuclei such as $\Htwo$, $\Hethree$
and $\Liseven$, which are the `left over' products of nuclear
burning. It is seen that the $\Hefour$ abundance depends
sensitively on the Hubble expansion rate at this epoch (and therefore
on the number of neutrino flavours) as well as on the neutron lifetime
(which determines the rate of weak interactions), but only weakly on
the nucleon density. Conversely, the abundances of the other light
elements provide a sensitive probe of the nucleon density.

\subsubsection{Neutron `freeze-out':\label{freezeout}} 

At sufficiently high temperatures (above a few MeV, as we shall see
shortly) neutrons and protons are maintained in both kinetic
equilibrium, i.e. 
\begin{equation}
 T_{\n} = T_{\p} = T_{\el} = T_{\nu_{\el}} = T\ ,
\end{equation}
and chemical equilibrium, i.e. 
\begin{equation}
\mu_{\n} - \mu_{\p} = \mu_{{\el}^{-}} - \mu_{\nu_{\el}}
 = \mu_{\bar{\nu_{\el}}} - \mu_{{\el}^{+}}\ ,
\end{equation}
through the weak processes
\begin{equation}
 {\n} + \nu_{\el} \rightleftharpoons {\p} + {\el}^{-} , \quad 
 {\n} + {\el}^{+} \rightleftharpoons {\p} + \bar{\nu_{\el}} , \quad 
 {\n} \rightleftharpoons {\p} + {\el}^{-} + \bar{\nu_{\el}} \ . 
\end{equation}
Defining $\lambda_{{\n}{\p}}$ as the summed rate of the reactions which
convert neutrons to protons,
\begin{equation}
 \lambda_{{\n}{\p}} = \lambda\ ({\n}\nu_{\el} \to {\p}{\el}^{-}) 
 + \lambda\ ({\n}{\el}^{+} \to {\p}\bar{\nu_{\el}}) 
 + \lambda\ ({\n} \to {\p}{\el}^{-}\bar{\nu_{\el}})\ , 
\end{equation} 
the rate $\lambda_{{\p}{\n}}$ for the reverse reactions which convert
protons to neutrons is given by detailed balance:
\begin{equation}
 \lambda_{{\p}{\n}}  = \lambda_{{\n}{\p}} \ {\e}^{-\Delta m/T(t)} , 
  \qquad \Delta m \equiv m_{\n} - m_{\p} = 1.293 \MeV . 
\label{detbal}
\end{equation} 
For the moment, we ignore the possibility of a large chemical
potential in electron neutrinos which would otherwise appear in the
exponent above (see \eref{detbalchempot}). The chemical potential of
electrons is negligible since any excess of electrons which survives
the annihilation epoch at $T\,\sim\,m_{\el}$ must equal the small
observed excess of protons, given that the universe appears to be
electrically neutral to high accuracy (Lyttleton and Bondi 1959,
Sengupta and Pal 1996), i.e.
\begin{equation}
 \frac{\mu_{\el}}{T} \approx \frac{n_{\el}}{n_{\gamma}} 
  = \frac{n_{\p}}{n_{\gamma}} \sim 10^{-10} .
\end{equation}

The evolution of the fractional neutron abundance $X_{\n}$ is described
by the balance equation 
\begin{equation} 
 \frac{\d X_{\n} (t)}{\d t} = \lambda_{{\p}{\n}} (t) [1 - X_{\n} (t)] -
  \lambda_{{\n}{\p}} (t) X_{\n}(t)\ , \qquad 
 X_{\n} \equiv \frac{n_{\n}}{n_{\N}}\ , 
\end{equation} 
where $n_{\N}$ is the total nucleon density at this time,
$n_{\N}=n_{\n}+n_{\p}$.\footnote{We will make a point of
 referring specifically to {\em nucleons} rather than to baryons as
 other authors do since there may well be other types of stable
 baryons, e.g. `strange quark nuggets' (Witten 1984, see Alcock and 
 Olinto 1988), which do not participate in nucleosynthesis.} The 
equilibrium solution is obtained by setting ${\d}X_{\n}(t)/{\d}t=0$:
\begin{equation} 
\label{nbypeqm}
 X_{\n}^{\eqm} (t) = \frac{\lambda_{{\p}{\n}} (t)}{\Lambda (t)} = 
  \left[1 + {\e}^{\Delta m/T(t)}\right]^{-1} \ , \qquad
  \Lambda \equiv \lambda_{{\p}{\n}}  + \lambda_{{\n}{\p}} \ , 
\end{equation}
while the general solution is
\begin{equation}
\eqalign{
 X_{\n} (t) = \int_{t_{\in}}^{t} \d t'\ I(t, t')\ \lambda_{{\p}{\n}} (t')\ 
  +\ I(t, t_{\in})\ X_{\n} (t_{\in})\ , \\ \nonumber
 I(t,t') \equiv \exp \left[-\int_{t'}^{t} \d t'' \Lambda(t'')\right]\ .} 
\end{equation} 
Since the rates $\lambda_{{\p}{\n}}$ and $\lambda_{{\n}{\p}}$ are very
large at early times, $I(t,t_{\in})$ will be negligible for a suitably
early choice of the initial epoch $t_{\in}$, hence the initial value
of the neutron abundance $X_{\n}(t_{\in})$ plays {\em no} role and
thus does not depend on any particular model of the very early
universe. For the same reason, $t_{\in}$ may be replaced by zero and
the above expression simplifies to (Bernstein \etal 1989)
\begin{equation}
\label{Xn}
\eqalign{
 X_{\n} (t) 
  &= \int_{0}^{t} \d t'\ I(t,t')\ \lambda_{{\p}{\n}} (t')  \\ \nonumber
  &= \frac{\lambda_{{\p}{\n}} (t)}{\Lambda(t)} - \int_{0}^{t} \d t'\ I(t,t')\ 
   {\d \over \d t'}\left[\frac{\lambda_{{\p}{\n}} (t')}{\Lambda(t')}\right] .}
\end{equation}
Since the total reaction rate $\Lambda$ is large compared to the rate
of time variation of the individual rates, this can be written as 
\begin{equation}
\label{Xnapprox}
\eqalign{
 X_{\n} (t) &\simeq \frac{\lambda_{{\p}{\n}}(t)}{\Lambda (t)} - 
  \frac{1}{\Lambda (t)} {\d \over \d t} 
  \left[\frac{\lambda_{{\p}{\n}}  (t)}{\Lambda (t)}\right] \\ \nonumber
 &\simeq X_{\n}^{\eqm} \left[1 + \frac{H}{\Lambda} 
   \frac{\d \ln  X_{\n}^{\eqm}}{\d \ln T}\right] ,}
\end{equation}
using \eref{nbypeqm}. Clearly, the neutron abundance tracks its value
in equilibrium until the inelastic neutron-proton scattering rate
$\Lambda$ decreases sufficiently so as to become comparable to the
Hubble expansion rate $H=\dot{R}/R\simeq-\dot{T}/T$. At this point the
neutrons `freeze-out', i.e. go out of {\em chemical} equilibrium, and
subsequently, as we shall see, $X_{\n}$ relaxes to a constant value
rather than following the exponentially falling value of
$X_{\n}^{\eqm}$. The freeze-out temperature can be approximately
estimated by simply equating the expansion rate,
$H\,\approx\,g_{\rho}^{1/2}T^2/M_{\Pl}$, to the reaction rate per
nucleon,
$\Lambda\,\approx\,n_{\nu}\langle\sigma{v}\rangle\,\sim\,G_{\F}^2T^5$,
where we have used $n_{\nu}\,\sim\,T^3$ and
$\langle\sigma{v}\rangle\,\sim\,G_{\F}^2T^2$ (see discussion following
\eref{lambdanp}). This yields
\begin{equation}
\label{Tfr}
 T_{\fr} \sim \left({g_{\rho}^{1/2} \over {G_{\F}^2
  M_{\Pl}}}\right)^{1/3} \sim 1\ \MeV , 
\end{equation} 
i.e. freeze-out occurs at $t_{\fr}\,\approx\,1\sec$ (using
\eref{trad}). The neutron abundance at this time can be approximated
by its equilibrium value \eref{nbypeqm},
\begin{equation}
\label{Xnfr}
 X_{\n} (T_{\fr}) \simeq X_{\n}^{\eqm} (T_{\fr}) 
  = \left[1 + {\e}^{\Delta m / T_{\fr}}\right]^{-1} . 
\end{equation}
Since the exponent $\Delta m/T_{\fr}$ is of $\Or(1)$, a substantial
fraction of neutrons survive when chemical equilibrium between
neutrons and protons is broken. This results, in turn, in the
synthesis of a significant amount of helium in the early universe. It
is interesting that the individual terms in the exponent above reflect
the widest possible variety of physical interactions which apparently
``conspire'' to make this possible.\footnote{The neutron-proton mass
 difference is determined by the strong and electromagnetic
 interactions, while the freeze-out temperature is fixed by the weak
 and gravitational interactions.} Also, the dependence of $T_{\fr}$
on the energy density driving the expansion makes the helium abundance
sensitive to the number of relativistic particle species (e.g.
massless neutrinos) present, or to any hypothetical non-relativistic
particle which contributes appreciably to the energy density at this
epoch.

Calculation of the asymptotically surviving abundance 
$X_{\n}(t\to\infty)$ requires explicit computation of the reaction
rates (see Weinberg 1972)
\begin{equation}
\label{reacint}
\eqalign{
\fl \lambda ({\n} \nu_{\el} \to {\p} {\el}^{-}) = A \int_{0}^\infty 
    \d q_\nu\ q_\nu^2\ q_{\el}\ E_{\el}\ (1 - f_{\el})\ f_\nu\ , \qquad 
    E_{\el} &= E_{\nu} + \Delta m\ , \\ 
\fl \lambda (n {\el}^+ \to {\p} \bar{\nu_{\el}}) = A \int_{0}^\infty 
    \d q_{\el}\ q_{\el}^2\ q_{\nu}\ E_{\nu}\ (1 - f_{\nu})\ f_{\el}\ , \qquad 
    E_\nu &= E_{\el} + \Delta m\ , \\
\fl \lambda ({\n} \to {\p} {\el}^- \bar{\nu_{\el}}) = 
    A \int_{0}^{q_{0}} \d q_{\el}\ q_{\el}^2\ q_{\nu}\ E_{\nu}\ (1 - f_{\nu})
    (1 - f_{\el})\ , \quad q_{0} &= \sqrt{(\Delta m)^2 - m_{\el}^2}\ .}
\end{equation}
Here $A$ is an effective coupling while $f_{\el}$ and $f_\nu$ are the
distribution functions for electrons and neutrinos. Although the weak
interaction coupling $G_{\F}$ is known quite accurately from muon
decay, the value of $A$, or equivalently, the neutron lifetime, cannot
be directly determined from this alone because neutrons and protons
also interact strongly, hence the ratio of the nucleonic axial vector
($G_{\A}$) and vector ($G_{\V}$) couplings is altered from unity (see
Freedman 1990). Moreover, relating these couplings to the
corresponding experimentally measured couplings for the $u$ and $d$
quarks is complicated by weak isospin violating effects. If we assume
conservation of the weak vector current (CVC), then
$G_{\V}=G_{\F}\cos\theta_{\C}$ where $\theta_{\C}\simeq13^0$ is the
Cabibbo angle which describes the mixing of the quark weak
eigenstates into the mass eigenstates (see Marciano 1991).  However
the weak axial current is {\em not} conserved and $G_{\A}$ for
nucleons differs from that for the first generation quarks. These
non-perturbative effects cannot be reliably calculated, hence $G_{\A}$
(in practice, $G_{\A}/G_{\V}$) must be measured experimentally. The
neutron lifetime is then given by
\begin{equation}
\label{tauninv}
 \tau_{\n}^{-1} = \frac{m_{\el}^5}{2 \pi^3} G_{\V}^2 \left(1 +
  3\,\frac{G_{\A}^2}{G_{\V}^2}\right)\ f\ ,
\end{equation}
where $f=1.715$ is the integral over the final state phase space
(including Columb corrections) and $G_{\V}$ is usually determined
directly from superallowed $0^+\to0^+$ pure Fermi decays of suitable
light nuclei (see Wilkinson 1982). It is thus more reliable to measure
the neutron lifetime directly if possible, and then relate it to the
coupling $A$ in \eref{reacint} in order to obtain the other reaction
rates.

Bernstein \etal (1989) note that the major contribution to the
integrals in \eref{reacint} comes from particles of energy {\it
higher} than the temperature during the BBN era, hence the Fermi-Dirac
distributions may be approximated by their Boltzmann equivalents:
\begin{equation}
 f_{\el} = \left[1 + {\e}^{E_{\el}/T_{\el}}\right]^{-1} \simeq
  {\e}^{-E_{\el}/T_{\el}}\ , \qquad
 f_\nu = \left[1 + {\e}^{E_\nu/t_\nu}\right]^{-1} \simeq 
  {\e}^{-E_\nu/T_\nu}\ .
\end{equation}
Also, since the Boltzmann weights are small in this dilute gas limit,
the Pauli blocking factors in the reaction rates may be neglected:
\begin{equation}
  1 - f_{e,\nu} \simeq 1\ , 
\end{equation}
The electron temperature $T_{\el}$ above equals the photon temperature
$T$ but has been distinguished from the neutrino temperature $T_\nu$
because, as discussed in \sref{thermhist}, the annihilation of
${\el}^{+}{\el}^{-}$ pairs at $T{\leqsim}m_{\el}$ heats the photons
and the (electromagnetically coupled) electrons but not the neutrinos
which have become essentially non-interacting by this time. The
evolution of $T_\nu/T$ is given by entropy conservation \eref{TnubyT};
numerical evaluation of this expression shows that $T_\nu$ remains
within $\approx10\%$ of $T$ until $\approx0.2\MeV$, by which time, as
we shall see below, neutron freeze-out is effectively over. Hence
Bernstein \etal (1989) assume that $T_\nu=T$; the detailed balance
condition \eref{detbal} follows from comparison of the rates
\eref{reacint} to the corresponding rates for the reverse
processes. Their final approximation is to set $m_{\el} = 0$ in
evaluating $\lambda\,({\n}\nu_{\el}\to{\p}{\el}^{-})$ and
$\lambda\,({\n}{\el}^{+}\to{\p}\bar{\nu_{\el}})$ which get most of
their contribution from energies $E_{{\el},\nu}{\gg}m_{\el}$.  These
rates are then equal and given by the formula
\begin{equation} 
\fl  \lambda\ ({\n} \nu_{\el} \to {\p} {\el}^{-}) = 
   \lambda ({\n} {\el}^{+} \to {\p} \bar{\nu_{\el}}) = 
    A\ T^3\ [24\ T^2 + 12\ T\ \Delta m + 2\ (\Delta m)^2]\ , 
\end{equation}
which is accurate to better than $15\%$ until $T$ drops to $m_{\el}$
by which time the rates themselves have become very small. Integration
of the neutron beta decay rate (see \eref{reacint}) now gives the
desired relation between the coupling $A$ and the neutron lifetime
$\tau_{\n}$:
\begin{eqnarray}
\fl \frac{1}{\tau_{\n}} = \frac{A}{5} \sqrt{(\Delta m)^2 - 
  m_{\el}^2} \left[\frac{1}{6}(\Delta m)^4 - 
  \frac{3}{4}(\Delta m)^2 m_{\el}^2 - \frac{2}{3} m_{\el}^4\right] + 
  \frac{A}{4} m_{\el}^4 \Delta m \cosh^{-1}\left(\frac{\Delta m}{m_{\el}}
  \right) \\ \nonumber
\lo = 0.0158\ A\ (\Delta m)^5\ .  
\end{eqnarray}
Hence the total reaction rate can be expressed in terms of the neutron
lifetime as,
\begin{equation}
\label{lambdanp}
\eqalign{
 \lambda_{{\n}{\p}} (t) \simeq 2\ \lambda({\n} \nu_{\el} \to {\p} {\el}^{-}) 
  = \frac{a}{\tau_{\n} y^5} (12 + 6y + y^2) , \\ \nonumber
 y \equiv \frac{\Delta m}{T}, \qquad a = 253 .}
\end{equation}
The contribution of neutron decay itself to $\lambda_{{\n}{\p}}$ has
been neglected here since it is unimportant during the freeze-out
period and becomes comparable to the other terms only for $T\leq0.13$
MeV (corresponding to $y>10$). We see that for $T\gg\Delta{m}$, i.e.
$y{\ll}1$, the reaction rate is $\lambda\approx12a/\tau_{\n}y^5$,
which we have approximated earlier as $\lambda\,\sim\,G_{\F}^2T^5$
(using \eref{tauninv}) in order to estimate $T_{\fr}$ \eref{Tfr}.

The integrating factor in \eref{Xn} can now be calculated:
\begin{equation}
\label{intfact}
\eqalign{
 I(y, y') & = \exp \left[-\int_{y'}^{y} \d y'' {\d t'' \over \d y''}\ 
  \Lambda(y'')\right] \\ \nonumber
 & = \exp\ [K(y) - K(y')]\ ,}
\end{equation}
where,
\begin{equation}
\label{Ky}
\eqalign{
 K(y) & \equiv - b \int_{\infty}^{y} \d y' \left[\frac{12}{y'^4} 
          + \frac{6}{y'^3} + \frac{1}{y'^2} \right] (1 + {\e}^{-y'}) \\ 
        & = b \left[ \left(\frac{4}{y^3} + \frac{3}{y^2} + \frac{1}{y}\right) +
            \left(\frac{4}{y^3} + \frac{1}{y^2}\right) {\e}^{-y} \right] , \\
 {\rm and}, \quad b &= a \left(\frac{45}{4 \pi^3 g_{\rho}}\right)^{1/2}
             \frac{M_{\Pl}}{\tau_{\n}\,(\Delta m)^2} .} 
\end{equation}
The neutron abundance is therefore 
\begin{equation}
\label{Xnfin}
  X_{\n} (y) = X_{\n}^{\eqm} (y) + \int_{0}^y
  \d y'\ {\e}^{y'} [X_{\n}^{\eqm} (y')]^2\ \exp[K(y) - K(y')] .
\end{equation}
The integral can be easily evaluated numerically once the value of $b$
is specified. In the Standard Model, the number of relativistic
degrees of freedom corresponding to photons, electrons and positrons
and 3 species of massless~\footnote{The effects on BBN of a finite
neutrino mass are discussed in \sref{nus}. The results of standard BBN
are however unaffected for $m_{\nu}\leqsim0.1\MeV$ (Kolb and Scherrer
1982). Experimentally, it is only known that $m_{\nu_{\el}}<5.1\eV$,
$m_{\nu_{\mu}}<160\keV$, and $m_{\nu_{\tau}}<24\MeV$ (Particle Data
Group 1996). Hence the $\nu_{\el}$, and probably the $\nu_{\mu}$ too,
are indeed effectively massless but the $\nu_{\tau}$ can, in priniple,
play a more complex role in BBN.} neutrinos ($N_{\nu}=3$) is
$g_{\rho}=43/4$ at this time, hence $b=0.252$, taking
$\tau_{\n}=887\sec$ \eref{taun}. (Subsequently $g_{\rho}$ drops to
3.36 following ${\el}^{+}{\el}^{-}$ annihilation \eref{gsgrhonow};
this raises the total energy density in relativistic particles but the
error incurred by ignoring this is negligible since $X_{\n}$ has
essentially stopped evolving by then; also Bernstein \etal (1989)
actually used $\tau_{\n}=896\sec$ but we have corrected their
numbers.) This yields the asymptotic abundance
\begin{equation} 
\label{Xninfty}
  X_{\n} (y \to \infty) = 0.150\ ,
\end{equation}
which is already achieved by the time $T$ has dropped to about
$0.25\MeV$ ($y\simeq5$), corresponding to $t\simeq20\sec$.

\subsubsection{Element synthesis:\label{synth}} 

Having dealt with the breaking of weak equilibrium between neutrons
and protons, we now consider the onset of nuclear reactions which
build up the light nuclei. This has been traditionally studied by
numerical solution of the complete nuclear reaction network (Peebles
1966b, Wagoner \etal 1967, Wagoner 1969, 1973). More recently the
coupled balance equations for the elemental abundances have been
semi-analytically solved by a novel method of fixed points
(Esmailzadeh \etal 1991) as discussed later. First we outline the
essential physical processes as they pertain to the calculation of the
$\Hefour$ abundance.

Neutrons and protons react with each other to build up light nuclei through 
the following sequence of two-body reactions:
\begin{equation}
\label{reac}
\eqalign{
  {\p} ({\n}, \gamma) {\Htwo}, \\ 
  {\Htwo} ({\p}, \gamma) {\Hethree}, \quad {\Htwo} ({\Htwo}, {\n}) {\Hethree},
   \quad {\Htwo} ({\Htwo}, {\p}) {\Hthree}, \\
  {\Hthree} ({\Htwo}, {\n}) {\Hefour}, \quad {\Hthree} ({\Hefour}, \gamma)
   {\Liseven}, \\
  {\Hethree} ({\n}, {\p}) {\Hthree}, \quad {\Hethree} ({\Htwo},{\p}) {\Hefour},
   \quad {\Hethree} ({\Hefour}, \gamma) {\Beseven}, \\
  {\Liseven} ({\p}, {\Hefour}) {\Hefour}, \quad 
   {\Beseven} ({\n}, {\p}) {\Liseven}  \\
  \vdots }
\end{equation}
The first reaction is the most crucial since deuterium must be formed
in appreciable quantity before the other reactions can proceed at all,
the number densities being in general too low to allow nuclei to be
built up directly by many-body reactions such as $2
{\n}+2{\p}\to{\Hefour}$. The rate (per neutron) of this
reaction (see Weinberg 1972),
\begin{equation}
  \lambda ({\n} {\p} \to {\Htwo} \gamma) = 4.55 \times
   10^{-20}\ n_{\p}\ {\cm}^3 {\sec}^{-1} ,
\end{equation}
is quite large, being determined by the strong interactions, and
exceeds the expansion rate down to quite low temperatures of
$\Or(10^{-3})\MeV$. Hence at the epoch of interest, deuterium will
indeed be present with its equilibrium abundance, given by the Saha equation
\begin{equation}
\label{saha}
  \frac{n_{\Htwo}}{n_{\n} n_{\p}} = \frac{g_{\Htwo}}{g_{\n} g_{\p}} 
   \left(\frac{m_{\Htwo}}{m_{\n} m_{\p}}\right)^{3/2} 
   \left(\frac{T}{2\pi}\right)^{-3/2}
   {\e}^{\Delta_{\Htwo}/T}\ ,
\end{equation}
where $\Delta_{\Htwo}{\equiv}m_{\n}+m_{\p}-m_{\Htwo}=2.23\MeV$ is the
deuteron binding energy, and the $g$'s are statistical
factors. This can be rewritten in terms of the respective mass
fractions as 
\begin{equation}
\label{dsyn}
  \frac{X_{\Htwo}}{X_{\n} X_{\p}} \simeq \frac{24 \zeta(3)}{\sqrt{\pi}} \eta
   \left[\frac{T} {m_{\p}}\right]^{3/2} {\e}^{\Delta_{\Htwo}/T} , 
   \quad X_{i} \equiv \frac{n_{i} A_{i}}{n_{\N}}\ , 
\end{equation}
where,
\begin{equation} 
  \eta \equiv \frac{n_{\N}}{n_{\gamma}} = 2.722 \times 10^{-8}
   \Omega_{\N} h^2 \Theta^{-3}\ , 
\end{equation}
is the ratio of the total number of nucleons (bound or free) to the
number of photons (which remains constant following
${\el}^{+}{\el}^{-}$ annihilation). This quantity is not well known
observationally because it is not clear how much of the dark matter in
the universe is in the form of nucleons. An audit of luminous
material in galaxies and X-ray emitting gas in clusters provides
the lower limit (Persic and Salucci 1992):
\begin{equation}
\label{omeganps}
 \Omega_{{\N}} \equiv \frac{\rho_{\N}}{\rho_{\c}} 
  > 2.2 \times 10^{-3} + 6.1 \times 10^{-4} h^{-1.3}\ .
\end{equation}
(Henceforth, we omit the subscript $_{0}$ on $\Omega$ and
$\Omega_{\N}$.)  A conservative upper limit follows from assuming that
all the gravitating matter permitted by the present age and expansion
rate of the universe is made up of nucleons, i.e.
$\Omega_{{\N}}h^2\leqsim1$ \eref{omegah2}. (Such a high density purely
nucleonic universe cannot create the observed large-scale structure,
given primordial `adiabatic' density fluctuations; however a viable
model {\em can} be constructed assuming primordial isocurvature
fluctuations (Peebles 1987, Cen \etal 1993) which satisfies CMB
anisotropy constraints with $\Omega_{{\N}}\leqsim1$ (e.g. Sugiyama and
Silk 1994).) These considerations require the value of $\eta$ today to
lie in the rather broad range:
\begin{equation}
\label{etarange}
 1.8 \times 10^{-11} \leqsim \eta \leqsim 2.8 \times 10^{-8} , 
\end{equation}
using the limits $0.4{\leqsim}h\leqsim1$ \eref{h} and
$0.993<\Theta<1.007$ \eref{Tcmb}. In the Standard Model, these
constraints also apply during nucleosynthesis since
${\el}^{+}{\el}^{-}$ annihilation is effectively over by this epoch so
the comoving photon number, hence $\eta$, does not change further.

If deuterium synthesis is assumed to begin at a temperature $T_{\ns}$
when $X_{\Htwo}/X_{\n}X_{\p}$ becomes of $\Or(1)$, then for a typical
value $\eta=5\times10^{-10}$, \eref{dsyn} gives
$T_{\ns}\simeq\Delta_{\Htwo}/34$, an estimate which is only
logarithmically sensitive to the adopted nucleon
density.\footnote{Na\"{\ii}vely we would expect deuterium synthesis to
 begin as soon as the average blackbody photon energy of $2.7T$ falls
 below $\Delta_{\Htwo}$ since deuterons would then presumably no
 longer be photodissociated as soon as they are formed. However,
 since there are $\sim10^{10}$ photons per nucleon, there are still
 enough high energy photons in the Wien tail of the Planck
 distribution at this time which are capable of photodissociating
 deuterons, and it takes rather longer for the `deuterium bottleneck'
 to break. There is, in fact, another contributory reason, which we
 will discuss following \eref{QSEasym}.} Bernstein \etal
(1989) obtain a more careful estimate by examination of the rate
equation governing the deuterium abundance. Defining the onset of
nucleosynthesis by the criterion ${\d}X_{\Htwo}/{\d}z=0$ at
$z=z_{\ns}$ (where $z\equiv\Delta_{\Htwo}/T$), they find that the
critical temperature is given by the condition
\begin{equation}
\label{dsynbbf}
 2.9 \times 10^{-6} \left(\frac{\eta}{5 \times 10^{-10}}\right)^2 
  z_{\ns}^{-17/6} \exp(-1.44 z_{\ns}^{1/3})\ {\e}^{z_{\ns}} \simeq 1 .
\end{equation}
Taking $\eta=5\times10^{-10}$, this gives $z_{\ns}\simeq26$, i.e.
\begin{equation}
\label{tns}
  T_{\ns} \simeq \frac{\Delta_{\Htwo}}{26} = 0.086 \MeV .
\end{equation}
At this epoch, $g_{\rho}=3.36$ \eref{gsgrhonow}, hence the
time-temperature relationship \eref{trad} says that nucleosynthesis
begins at
\begin{equation}
\label{3min}
  t_{\ns} \simeq 180 \sec , 
\end{equation}
as widely popularized by Weinberg (1977).

By this time the neutron abundance surviving at freeze-out has been
depleted by $\beta$-decay to 
\begin{equation}
\label{Xndec}
  X_{\n} (t_{\ns}) \simeq X_{\n} (y \to \infty)\ {\e}^{-t_{\ns}/\tau_{\n}} 
   = 0.122\ .
\end{equation}
Nearly {\em all} of these surviving neutrons are captured in $\Hefour$
because of its large binding energy ($\Delta_{\Hefour}=28.3\MeV$) via
the reactions listed in \eref{reac}. Heavier nuclei do not form in any
significant quantity both because of the absence of stable nuclei with
$A$=5 or 8 which impedes nucleosynthesis via
${\n}\,{\Hefour},~{\p}\,{\Hefour}$ or ${\Hefour}\,{\Hefour}$
reactions, and the large Coulomb barrier for reactions such as
${\Hthree}({\Hefour},\gamma){\Liseven}$ and
${\Hethree}({\Hefour},\gamma){\Beseven}$.\footnote{If there are large
 fluctuations in the nucleon density, such as may be induced by a
 first-order quark-hadron phase transition (see Reeves 1991), then
 differential transport of neutrons and protons creates neutron-rich
 regions where heavy elements can indeed be formed through reactions
 such as
 $\H(\n,\gamma)\Htwo(\n,\gamma)\Hthree(\Htwo,\n)\Hefour(\Hthree,\gamma)
 \Liseven(\n,\gamma)\Lieight(\Hefour,\n)\Beleven(\n,\gamma)\Btwelve(\el,
 \nu_{\el})\Ctwelve(\n,\gamma)\Cthirteen(\n,\gamma)\Cfourteen\ldots$
 (see Malaney and Mathews 1993). This will be discussed further in
 \sref{theoobs}.} Hence the resulting {\em mass} fraction of helium,
conventionally referred to as $Y_{\pr}({\Hefour})$, is simply given by
\begin{equation}
\label{Ypapprox}
  Y_{\pr}({\Hefour}) \simeq 2\,X_{\n} (t_{\ns}) = 0.245\ ,
\end{equation}
where the subscript ${\pr}$ denotes primordial. The above calculation
makes transparent how the synthesized helium abundance depends on the
physical parameters. The dominant effect of a smaller neutron lifetime
$\tau_{\n}$ is that freeze-out occurs at a lower temperature with a
smaller neutron fraction (\eref{Ky} and \eref{Xnfin}), hence less
$\Hefour$ is subsequently synthesized; this is only partly negated by
the larger $\beta$-decay factor \eref{Xndec} since only $\approx20\%$
of the neutrons have decayed when nucleosynthesis begins. Increasing
the assumed number of relativistic neutrino species $N_{\nu}$
increases $g_{\rho}$ \eref{gsgrhonow}, speeding up the expansion and
leading to earlier freeze-out and earlier onset of nucleosynthesis,
hence a larger helium abundance. Finally as the nucleon-to-photon
ratio $\eta$ increases, the `deuterium bottleneck' is broken
increasingly earlier (see \eref{dsyn}), allowing a larger fraction of
neutrons to survive $\beta$-decay and be burnt to $\Hefour$, the
abundance of which thus rises approximately logarithmically with
$\eta$.

Bernstein \etal (1989) also consider the effect on neutron freeze-out
of a possible excess of electron neutrinos over antineutrinos,
parametrized by a dimensionless chemical potential,
$\xi_{\nu_{\el}}\equiv\mu_{\nu_{\el}}/T$, which remains constant for
freely expanding neutrinos (see \eref{eqdist}). Anticipating that
$\xi_{\nu_{\el}}$ will be constrained to be sufficiently small, they
neglect the slight increase in expansion rate due to the increased
energy density of the neutrinos and consider only the effect on
neutron-proton interconversions. (They do not consider a chemical
potential for other neutrino types, which would only add to the energy
density without affecting the weak reactions.) The resultant increase
in the rate of $n\nu_{\el}{\to}p{\el}^{-}$ alters the detailed balance
equation \eref{detbal} to
\begin{equation} 
\label{detbalchempot}
  \lambda_{{\p}{\n}}  = \lambda_{{\n}{\p}} \ 
   \exp \left[-\frac{\Delta m}{T(t)} - \xi_{\nu_{\el}}\right] ,
  \qquad \xi_{\nu_{\el}} \equiv \frac{\mu_{\nu_{\el}}}{T}\ ,
\end{equation}
and, hence, lowers the equilibrium neutron abundance to,
\begin{equation} 
\label{Xnchempoteqm}
  X_{\n}^{\eqm} (t, \xi_{\nu_{\el}}) = \left[1 + {\e}^{(y +
   \xi_{\nu_{\el}})}\right]^{-1} , \qquad y \equiv \frac{\Delta m}{T(t)}\ .
\end{equation}
Bernstein \etal (1989) find that this alters the asymptotic neutron abundance
by the {\em same} factor, viz. 
\begin{equation} 
\label{Xnchempot}
 X_{\n} (\xi_{\nu_{\el}}, y \to \infty) = 
  {\e}^{-\xi_{\nu_{\el}}} X_{\n} (y \to \infty)\ . 
\end{equation}

It is now easily shown that the synthesized helium mass fraction
depends on the relevant parameters as
\begin{eqnarray}
\label{Ypbbf}
\fl Y_{\pr}({\Hefour}) = 0.245 + 0.014\,\Delta N_\nu 
                       + 0.0002\,\Delta \tau_{\n} 
                       + 0.009\,\ln \left(\frac{\eta}{5 \times 10^{-10}}\right)
                       - 0.25\,\xi_{\nu_{\el}}\ , \\ \nonumber
 {\where}, \qquad \Delta N_{\nu} \equiv N_{\nu} - 3\ , 
  \qquad \Delta \tau_{\n} \equiv \tau_{\n} - 887 \sec\ .
\end{eqnarray}
For comparison, a recent numerical solution (Walker \etal 1991) of the
nuclear reaction network finds that the helium yield is fitted (to
within $\pm0.001$) in the nucleon density range
$3\times10^{-10}\leqsim\eta\leqsim10^{-9}$ by the formula
\begin{equation} 
\label{Ypwtsso}
\fl Y_{\pr}({\Hefour}) = 0.244 + 0.012\ \Delta N_\nu + 
                  0.00021\,\Delta \tau_{\n} + 
                  0.01\,\ln \left(\frac{\eta}{5 \times 10^{-10}}\right)\ .
\end{equation} 
(There is no term here corresponding to neutrino degeneracy since the
effect of this has not been parametrized by numerical means.) We see
that the semi-analytic result of Bernstein \etal (1989) is impressively
accurate.

Small amounts of deuterium ($X_{\Htwo}\sim10^{-4}$), helium-3
($X_{\Hethree}\sim10^{-5}$) and lithium-7 ($X_{\Liseven}\sim10^{-9}$)
are also left behind when the nuclear reaction rates fall behind the
expansion rate and BBN ends, at $t_{\rm end}\,\approx\,1000\sec$. (The
helium-3 abundance is taken to include that of surviving tritium which
subsequently undergoes beta decay and, similarly, the lithium-7
abundance includes that of beryllium-7.) In contrast to $\Hefour$,
the abundances of these elements are quite sensitive to the nucleon
density since this directly determines the two-body nuclear reaction
rates. The $\Htwo$ and $\Hethree$ abundances drop rapidly with
increasing $\eta$ which ensures more efficient burning to $\Hefour$.
The $\Liseven$ abundance also decreases with increasing $\eta$ in a
regime where its abundance is determined by the competition between
${\Hefour}({\Hthree},\gamma){\Liseven}$ and
${\Liseven}({\p},{\Hefour}){\Hefour}$; however at sufficiently high
$\eta$ ($\geqsim3\times10^{-10}$), its abundance begins increasing
again with $\eta$ due to the increasing production of ${\Beseven}$
through ${\Hefour}({\Hethree},\gamma){\Beseven}$, which makes
$\Liseven$ by electron capture,
${\Beseven}({\el}^{-},\nu_{\el}){\Liseven}$. The reaction rates for
the synthesis of $A>7$ nuclei are not all well known but, even with
extreme values chosen, the mass fraction of elements such as Be and B
does not exceed $10^{-13}$ for any value of $\eta$ (e.g. Thomas \etal
1993).

These results concerning the abundances of $\Htwo$, $\Hethree$ and
$\Liseven$ were originally obtained by numerical solution of the
complete nuclear reaction network (e.g. Wagoner 1969). More recently,
Esmailzadeh \etal (1991) have shown that these abundances are given to
good accuracy by the fixed points of the corresponding rate equations,
as discussed below. The general equation governing the abundance of a
given element is
\begin{equation}
 {\d X \over \d t} = J(t) - \Gamma(t) X\ ,
\end{equation}
where $J(t)$ and $\Gamma(t)$ are the time-dependent source and sink
terms, which, in general, depend on the abundances of the other
elements. The solution to this equation is (Dimopoulos \etal 1988) 
\begin{equation} 
\fl X (t) = \exp \left(-\int_{t_{\in}}^{t} \d t'\ \Gamma (t') \right) \left[ 
 X (t_{\in}) + \int_{t_{\in}}^{t} \d t'\ J(t')\ \exp \left(-\int_{t_{\in}}^{t}
  \d t'' \ \Gamma (t'') \right) \right]\ , 
\end{equation}
where $t_{\in}$, the initial time, may be taken to be zero. These authors
show that if
\begin{equation} 
\label{qse}
 \left|\frac{\dot{J}}{J} - \frac{\dot{\Gamma}}{\Gamma} \right| \ll \Gamma\ ,
\end{equation}
then $X$ approaches its equilibrium value~\footnote{This is distinct
 from the value in nuclear statistical equilibrium (NSE) which
 is given by the Saha equation \eref{dsyn} and
 increases exponentially as the temperature drops, as shown by the
 dashed lines in \fref{abunevol}. Considerations of NSE alone are not
 useful in the present context where the Hubble expansion introduces a
 time-scale into the problem (cf. Kolb and Turner 1990).}
\begin{equation} 
 X^{\eqm} = \frac{J(t)}{\Gamma(t)} 
\end{equation}
on a time scale of $\Or(\Gamma^{-1})$. This state is dubbed
`quasi-static equilibrium' (QSE) since the source and sink terms
nearly cancel each other such that $\dot{X}\simeq0$. (Note
that since $\dot{\Gamma}/\Gamma\approx\dot{J}/J\,\approx\,H$, the
condition \eref{qse} is somewhat more stringent than $\Gamma{\gg}H$
which would be the na\"{\ii}ve criterion for QSE.) As the universe
expands, the nuclear reaction rates slow down rapidly due to the
dilution of particle densities and the increasing importance of
Coulomb barriers; hence $J$ and $\Gamma$ fall rapidly with time. At
some stage $t=t_{\fr}$, $X$ can be said to `freeze-out' if its value
does not change appreciably beyond that point, i.e. if
\begin{equation} 
 \int_{t_{\fr}}^{\infty} \d t\ J(t) \ll X (t_{\fr})\ , \qquad
 \int_{t_{\fr}}^{\infty} \d t\ \Gamma(t) \ll 1\ . 
\end{equation}
Generally freeze-out occurs when $\Gamma{\simeq}H$ and the asymptotic value 
of the elemental abundance is then given by 
\begin{equation} 
\label{QSEasym}
 X (t \to \infty) \simeq X^{\eqm} (t_{\fr}) =
  \frac{J(t_{\fr})}{\Gamma(t_{\fr})}\ .
\end{equation}
It now remains to identify the largest source and sink terms for each
element and calculate the freeze-out temperature and the QSE abundance
at this epoch. This requires careful examination of the reaction
network and details of this procedure are given by Esmailzadeh \etal
(1991). These authors study the time development of the abundances for
a particular choice of the nucleon density and also calculate the
final abundances as a function of the nucleon density. As shown in
\fref{abunevol}, the agreement between their analytic approximations
(dotted lines) and the exact numerical solutions (full lines) is
impressive. The abundances of $\Htwo$, $\Hethree$ and $\Liseven$
are predicted correctly to within a factor of $\approx3$ for the
entire range $\Omega_{\N}\sim0.001-1$, and even the abundance of
$\Hefour$ is obtained to better than $5\%$ for
$\Omega_{\N}\sim0.01-1$. Moreover, this analysis clarifies several
features of the underlying physics. For example, it becomes clear
that in addition to the `deuterium bottleneck' alluded to earlier (see
footnote concerning \eref{dsynbbf}), the synthesis of
$\Hefour$ is additionally delayed until enough tritium has been
synthesized through ${\Htwo}({\Htwo},p){\Hthree}$, since the main
process for making $\Hefour$ is ${\Htwo}({\Hthree},n){\Hefour}$. In
fact, $\Htwo$ and $\Hthree$ are both in QSE when $\Hefour$
forms, hence the former reaction is the {\em only} one whose
cross-section has any perceptible influence on the $\Hefour$
abundance. This behaviour is illustrated in \fref{abunevol} where the
abundance of $\Hefour$ is seen to depart from its NSE curve (dashed
line) at about $0.6\MeV$ and follow the abundances of $\Hthree$ and
$\Hethree$ until these too depart from NSE at about $0.2\MeV$ due to
the `deuterium bottleneck'; subsequently $\Hefour$, $\Hethree$ and
$\Hthree$ all follow the evolution of $\Htwo$ until it finally
deviates from NSE at about $0.07\MeV$ (see Smith \etal 1993).


In figures \ref{abunevol}, \ref{abuneta} and \ref{abunetamc} we show
the elemental yields in the standard Big Bang cosmology (with
$N_{\nu}=3$) obtained using the Wagoner (1969, 1973) computer code,
which has been significantly improved and updated by Kawano (1988,
1992),\footnote{This code has been made publicly available by
 L\,Kawano and has become the {\it de facto} standard tool for BBN
 studies, enabling easy comparison of results obtained by different
 researchers.}  incorporating both new measurements and revised
estimates of the nuclear cross-sections (Fowler \etal 1975, Harris
\etal 1983, Caughlan \etal 1985, Caughlan and Fowler
1988). \Fref{abunevol} shows the evolution of the abundances (by
number) with decreasing temperature for a specific choice of the
nucleon density
($\Omega_{\N}h^2=0.01\Rightarrow\eta=2.81\times10^{-10}$) while
\fref{abuneta} shows the dependence of the final abundances on
$\eta$. In \fref{abunetamc} we show these in more detail, along with
their associated uncertainties calculated by Krauss and Kernan (1995)
as discussed below. This last figure displays the mass fraction
$Y_{\pr}({\Hefour})$ on a {\em linear} scale for clarity.

\subsubsection{Uncertainties:\label{uncert}} 

There have been many studies of the theoretical uncertainties in the
predicted abundances (e.g. Beaudet and Reeves 1983, Yang \etal 1984,
Delbourgo-Salvador \etal 1985, Kajino \etal 1987, Riley and Irvine
1991), in particular that of $\Liseven$ (Kawano \etal 1988, Deliyannis
\etal 1989). Because of the complex interplay between different
nuclear reactions, it is not straightforward to assess the effect on a
particular elemental yield of the uncertainty in some reaction rate.
An illuminating Monte Carlo analysis by Krauss and Romanelli (1990)
exhibited the effect on the abundances corresponding to {\em
simultaneous} variations in all relevant nuclear reaction rates by
sampling them from Gaussian distributions centred on the appropriate
mean values and with widths corresponding to the experimental
uncertainties. This exercise was redone by Smith \etal (1993) using
the latest cross-sections for the eleven most important nuclear
reactions \eref{reac} and their estimated uncertainties (which are
temperature dependent for ${\Hethree}({\Hefour},\gamma){\Beseven}$ and
${\Hthree}({\Hefour},\gamma){\Liseven}$). These authors carefully
discussed the statistical and systematic uncertainties in the
laboratory measurements of relevant cross-sections and emphasized, in
particular, the uncertainties in the `S-factor' which enters in the
extrapolation of a measured cross-section down in energy to obtain its
thermally averaged value at temperatures relevant to
nucleosynthesis. Recently Krauss and Kernan (1995) (see also Kernan
and Krauss 1994) have repeated the exercise with an improved Monte
Carlo procedure, the latest value for $\tau_{\n}$ \eref{taun} and the
new cross-section for the secondary reaction
${\Beseven}({\p},\gamma){\Beight}$ (which however does not affect the
results perceptibly). The dashed lines in \fref{abunetamc} indicate
the region within which $95\%$ of the computed values fall, which thus
correspond to ``$2\sigma$'' bounds on the predicted abundances.

The major uncertainty in the $\Hefour$ abundance is due to the
experimental uncertainty in the neutron lifetime. For many years there
were large discrepancies between different measurements of $\tau_{\n}$
suggestive of unknown systematic errors (see Byrne 1982). Until
recently most BBN calculations adopted a relatively high value, viz.
{\em upwards} of $900\sec$ (Yang \etal 1984, Boesgaard and Steigman
1985, Steigman \etal 1986), although Ellis \etal (1986b) cautioned
that a significantly lower value of $898\pm6\sec$ was indicated (using
\eref{tauninv}) by the precision measurement of
$G_{\A}/G_{\V}=-1.262\pm0.004$ obtained using polarized neutrons (Bopp
\etal 1986). Subsequently several precise direct measurements using
`bottled' neutrons (see Dubbers 1991, Schreckenbach and Mampe 1992)
have shown that the lifetime is indeed lower than was previously
believed. The present weighted average is (Particle Data Group 1994,
1996)~\footnote{The Particle Data Group (1990) had previously quoted
an weighted average $\tau_{\n}=888.6\pm3.5\sec$ and Walker \etal
(1991) adopted the $\95cl$ range 882-896 sec. Smith \etal (1993)
considered only post-1986 experiments which give
$\tau_{\n}=888.5\pm1.9\sec$ but doubled the uncertainty to
$\pm3.8\sec$ in their analysis. Other recent papers (e.g. Kernan and
Krauss 1994, Copi \etal 1995a) use the Particle Data Group (1992)
value of $889.1\pm2.1\sec$.}
\begin{equation}
\label{taun}
  \tau_{\n} = 887 \pm 2\ \sec\ ,
\end{equation} 
as used in our computations and in other recent work (e.g. Krauss and
Kernan 1995, Hata \etal 1995, Olive and Scully 1996, Kernan and Sarkar
1996). As we will see in \sref{rel} the {\em lower} bound is of
particular importance in using BBN to set constraints on new
particles. Variation of the neutron lifetime by $2\sigma$ causes
$Y_{\pr}({\Hefour})$ to change by less than $0.4\%$, while the effect
on the other elemental abundances is comparable, hence negligible in
comparison to their other uncertainties. The uncertainties in the
nuclear cross-sections can alter the calculated yields of $\Htwo$
and $\Hethree$ by upto $\approx15\%$ and $\Liseven$ by upto
$\approx50\%$ but have little effect ($\leqsim0.5\%$) on the
$\Hefour$ abundance. As mentioned earlier, the effect of these
uncertainties on the final abundances are correlated, hence best
studied by Monte Carlo methods.

Finally, there are computational errors associated with the
integration routine in the numerical code for BBN, which can be upto a
few per cent for $\Htwo$, $\Hethree$ and $\Liseven$ but
$\leqsim0.1\%$ for $\Hefour$, with the default settings of the time
steps (Kawano 1992). These have been compensated for by Smith \etal
(1993) but not always taken into account in earlier work. Kernan
(1993) has explored this question in more detail and states that
making the integration time steps short enough that different (order)
Runge-Kutta drivers converge on the same result can produce an
increase in $Y_{\pr}$ of as much as $+0.0017$ relative to results
obtained with the default step size. However direct comparison between
the results of Kernan and Krauss (1994) and those of Walker \etal
(1991) and Smith \etal (1993) does not reveal any difference due to
this reason.

Given the recent sharp fall in the uncertainty in the neutron
lifetime, it is important to include all corrections to the weak
interaction rates which have a comparable effect on the $\Hefour$
abundance. The most detailed such study by Dicus \etal (1982) (see
also Cambier \etal 1982, Baier \etal 1990) finds that including
Coulomb corrections to the weak interaction rates decreases the
calculated value $Y$ by 0.0009, while the corrections due to zero
temperature radiative corrections ($\Delta{Y}=+0.0005$), finite
temperature radiative corrections ($\Delta{Y}=-0.0004$), plasma
effects on the electron mass ($\Delta{Y}=+0.0001$) and, finally, the
slight heating of the neutrinos by ${\el}^{+}{\el}^{-}$ annihilation
($\Delta{Y}=-0.0002$),\footnote{Rana and Mitra (1991) have claimed
 that neutrino heating causes a large change,
 ${\Delta}Y\simeq-0.003$. However by incorporating a careful analysis
 of neutrino heating by Dodelson and Turner (1992) into the BBN code,
 Fields \etal (1993) find ${\Delta}Y=+0.00015$
 for $\eta\sim10^{-10}-10^{-9}$, comparable in magnitude to Dicus
 \etal's estimate although of {\em opposite} sign. Hannestad and
 Madsen (1995) obtain ${\Delta}Y=+0.0001$ from a similar
 analysis incorporating full Fermi-Dirac statistics.} taken together,
change $Y$ by less than 0.0001. Dicus \etal (1982) also state that $Y$
decreases systematically by 0.0013 when the weak rates are computed by
numerical integration rather than being obtained from the approximate
fitting formula given by Wagoner (1973). This amounts to a total
decrease in $Y$ of 0.0022 for the parameter values
($\eta=3\times10^{-10}, N_{\nu}=3, \tau_{\n}=918\sec$) they
adopted; by varying these over a wide range ($\eta=0.3-30\times
10^{-10},\ N_{\nu}=2-10,\ \tau_{\n}=693-961\sec$) Dicus \etal
find an average systematic change of $\Delta{Y}=-0.0025$. Smith \etal
(1993) apply this correction to their results obtained using the
fitted rates while Walker \etal (1991) integrate the rates numerically
with the Coulomb corrections and neutrino heating included, using a
code updated to Caughlan and Fowler's (1988) cross-sections, and state
that the residual uncertainty in $Y_{\pr}$ due to all other effects
does not exceed $\pm0.0002$. (The effect of all these corrections on
the $\Htwo$, $\Hethree$ and $\Liseven$ abundances is only
$1-2\%$, hence negligible in comparison with the other uncertainties
for these elements.)

Subsequently, Kernan (1993) has carefully reexamined the small
corrections discussed by Dicus \etal (1982); although his conclusions
differ in detail, the net correction he finds for Coulomb, radiative
and finite temperature effects is fortuitously the same,
viz. $\Delta{Y}\simeq-0.009$. Further, Seckel (1993) has drawn
attention to the effects of finite nucleon mass which cause a slight
($\approx1\%$) decrease in the weak reaction rates. He finds that
$Y_{\pr}$ has been systematically {\em underestimated} by about 0.0012
in all previous work which ignored such effects.\footnote{Gyuk and Turner
 (1993) have incorporated Seckel's calculations into the BBN code and
 state that the actual correction ranges between 0.0004 and 0.0015 over
 the range $\eta\sim10^{-11}-10^{-8}$, being well approximated by
 $+0.0057\,Y$. However we prefer to follow Seckel's original analysis
 which suggests that the correction is $\eta$-independent.} The
fractional changes in the abundances of the other light elements due
to nucleon mass effects is $\leqsim1\%$. 


The abundances shown in figures\,\ref{abunevol}-\ref{abunetamc} have
been computed by explicit integration of the weak rates using Kawano's
(1992) code, with the lowest possible settings of the time steps in
the (2nd order) Runge-Kutta routine, which allows rapid convergence to
within $0.01\%$ of the true value (Kernan 1993). (We find that doing
so increases $Y$ by 0.0003 on average (for $\eta\sim10^{-10}-10^{-9}$)
relative to the value obtained with the default settings. Also,
explicitly integrating the weak rates reduces $Y$ by 0.0009 on average
relative to using fitting formulae.) To this calculated value $Y$ we
apply a net correction of +0.0003, obtained by adopting the Coulomb,
radiative and finite temperature corrections recalculated by Kernan
(1993) (which includes the new Fields \etal (1993) estimate of
neutrino heating) and the correction for finite nucleon mass given by
Seckel (1993). All this results in an average {\em increase} of
$0.0027$ in $Y_{\pr}$ relative to the values quoted by Smith \etal
(1993), i.e. the true value is fortuitously almost {\em identical} to
that obtained by Wagoner's (1973) procedure of using fitted rates and
a coarse integration mesh and ignoring all corrections!

\subsubsection{Elemental yields:\label{yields}} 

For comparison with the previously given formulae \eref{Ypbbf} and
\eref{Ypwtsso}, our best fit over the range
$\eta\sim3\times10^{-10}-10^{-9}$ is:
\begin{equation}
\label{Ypminefull}
\fl Y_{\pr}({\Hefour}) = 0.2459 + 0.013\ \Delta\ N_\nu 
     + 0.0002\ \Delta \tau_{\n} 
     + 0.01\ \ln \left(\frac{\eta}{5 \times 10^{-10}}\right)\ .
\end{equation}
However, as is evident from \Fref{abunetamc} , any log-linear fit of
this kind overestimates $Y_{\pr}$ for $\eta\leqsim3\times10^{-10}$. A
better fit (to within $\pm0.1\%$) over the entire range
$\eta=10^{-10}-10^{-9}$ is given for the Standard Model ($N_{\nu}=3$)
by
\begin{equation}
\label{Ypmine}
\fl Y_{\pr}({\Hefour}) = 0.2462 
     + 0.01\ \ln \left(\frac{\eta}{5 \times 10^{-10}}\right) 
       \left(\frac{\eta}{5 \times 10^{-10}}\right)^{-0.2} \pm 0.0012\ .
\end{equation}
We have indicated the typical $2\sigma$ error which results, in about
equal parts, from the uncertainty in the neutron lifetime \eref{taun}
and in the nuclear reaction rates. (As shown in \fref{abunetamc}, the
error determined by Monte Carlo actually varies a bit with $\eta$.)
Our values for $Y_{\pr}$ are systematically higher by about $0.0005$
than those shown in \fref{abunetamc} as obtained by Krauss and Kernan
(1995) who use the same neutron lifetime. Such small differences may
arise due to the use of different integration routines, numerical
precision schemes \etc (P\,Kernan, private communication) and provide
an estimate of the systematic computational uncertainty. This should
be borne in mind when discussing the helium abundance to the ``third
decimal place''.\footnote{Note that the $Y_{\pr}$ values in Kernan and
Krauss (1994) are not, as stated therein, higher by 0.003 than those
in Walker \etal (1991) and Smith \etal (1993), all of whom used
$\tau_{\n}=889\sec$; in fact they are higher by only about half that
amount, presumably just due to the incorporation of Seckel's (1993)
nucleon mass correction.}

 
Our best fits for the other elemental abundances over the range $\eta =
10^{-10}-10^{-9}$, together with the typical errors, are: 
\begin{eqnarray}
\label{abundmine}
\eqalign{
\fl \left(\frac{\Htwo}{\H}\right)_{\pr} &= 3.6 \times 10^{-5 \pm 0.06} 
  \left(\frac{\eta}{5 \times 10^{-10}}\right)^{-1.6}\ , \\
\fl \left(\frac{\Hethree}{\H}\right)_{\pr} &= 1.2 \times 10^{-5 \pm 0.06} 
   \left(\frac{\eta}{5 \times 10^{-10}}\right)^{-0.63}\ , \\
\fl  \left(\frac{\Liseven}{\H}\right)_{\pr} &= 1.2 \times 10^{-11 \pm 0.2} 
   \left[\left(\frac{\eta}{5 \times 10^{-10}}\right)^{-2.38} 
     + 21.7 \left(\frac{\eta}{5 \times 10^{-10}}\right)^{2.38}\right]\ .} 
\end{eqnarray}
These are in excellent agreement with the values obtained by Kernan
and Krauss (1994) and Krauss and Kernan (1995). (The error band for
$\Liseven$ obtained by Monte Carlo is actually $\approx10\%$ wider at
$\eta\leqsim2\times10^{-10}$ than is indicated above, as seen in
\fref{abunetamc}.)

We now proceed to discuss the observed elemental abundances and their
inferred primordial values which we can compare with the model
predictions shown in \fref{abunetamc}.

\subsection{Primordial elemental abundances\label{abund}} 

As mentioned earlier, the comparison of the predicted elemental
abundances with observational data is complicated by the fact that the
primordial abundances may have been significantly altered during the
lifetime of the universe by nuclear processing in stars. Moreover this
happens differently for different elements, for example $\Hefour$, a
very stable nucleus, grows in abundance with time as it is synthesized
in stars, while $\Htwo$, which is very weakly bound, is always
destroyed in stars. The history of $\Hethree$ and $\Liseven$ is more
complicated since these elements may be both created and destroyed
through stellar processing. Whenever possible, astronomers endeavour
to measure light element abundances in the most primordial material
available and the recent development of large telescopes and CCD
imaging technology have led to significant advances in the
field. Pagel (1982, 1987, 1992) and Boesgaard and Steigman (1985) have
comprehensively reviewed the observational data and discussed how the
primordial abundances may be inferred by allowing for the effects of
stellar evolution and galactic chemical evolution. The interested
reader is urged to refer to the original literature cited therein to
appreciate the uncertainties involved, both in the measurement of
cosmic abundances today and in the bold astrophysical modelling
necessary to deduce their values over $10\Gyr$ ago
($1\Gyr\equiv10^{9}$\,yr). Subsequently there have been several
observational developments, some of which have been discussed by Pagel
(1993). We review the key results which may be used to confront the
standard BBN model.

\subsubsection{Helium-4:\label{he4}} 

The most important primordially synthesized element, $\Hefour$, has
been detected, mostly through its optical line emission, in a variety
of astrophysical environments, e.g. planetary atmospheres, young
stars, planetary nebulae and emission nebulae --- galactic as well as
extragalactic (see Shaver \etal 1983), as well as in the intergalactic
medium at a redhift of $z\simeq3.2$ (Jakobsen \etal 1994). Hence there
is no doubt about the existence of an ``universal'' helium abundance of
$\approx25\%$ by mass. However, helium is also manufactured in stars,
hence to determine its primordial abundance we must allow for the
stellar helium component through its correlation with some other
element, such as nitrogen or oxygen, which is made only in stars
(Peimbert and Torres-Peimbert 1974). This is best done by studying
recombination lines from HII regions in blue compact galaxies (BCGs)
where relatively little stellar activity has occured, as evidenced by
their low `metal' abundance. (This refers, in astronomical jargon, to
any element heavier than helium!)


In \fref{Ypreg}(a) we show a correlation plot of the observed helium
abundance against that of oxygen and nitrogen as measured in 33 such
selected objects (Pagel \etal 1992). A linear trend is suggested by
the data and extrapolation to zero metal abundance yields the
primordial helium abundance with a small statistical error,
$Y_{\pr}({\Hefour})=0.228\pm0.005$.\footnote{The $\chi^2$ per degree
 of freedom for this fit is only $0.3$, suggesting that the quoted
 statistical measurement errors (typically $\pm 4\%$) may have been
 overestimated (Pagel \etal 1992).} It has however been emphasized,
particularly by Davidson and Kinman (1985), that there may be large
{\em systematic} errors in these abundance determinations, associated
with corrections for (unobservable) neutral helium, underlying stellar
helium absorption lines, collisional excitation \etc; these authors
suggested, in common with Shields (1987), that the systematic error in
$Y_{\pr}$ could be as high as $\pm0.01$. The recent work by Pagel
\etal (1992) has specifically addressed several such sources of error;
for example attention is restricted to objects where the ionizing
stars are so hot that the correction for neutral helium is negligible.
Hence these authors believe that the systematic error has now been
reduced to about the same level as the statistical error, i.e.
\begin{equation} 
\label{Ypptss}
  Y_{\pr}({\Hefour}) = 0.228 \pm 0.005\,(\stat) \pm 0.005\,(\syst)\ .
\end{equation}
Since the (uncertain) systematic error is correlated between the
different data points rather than being random, we cannot assign a
formal confidence level to a departure from the mean value. It is
common practice nonetheless to simply add the errors in
quadrature. Another method is to add the systematic error to the
adopted result and then deduce a $\95cl$ bound from the statistical
error, assuming a Gaussian distribution (Pagel \etal 1992); this gives
the bounds
\begin{equation} 
\label{Yppagellim}
  0.214 < Y_{\pr}({\Hefour}) < 0.242\ (\95cl)\ . 
\end{equation} 
Mathews \etal (1993) have suggested that galactic chemical evolution
causes the correlation between helium and nitrogen to be non-linear at
low metallicity, consequently the extrapolated helium abundance at
zero metallicity is subject to an upward bias. Their own fits to the
data, based on chemical evolution arguments, yield the same value as
Pagel \etal (1992) for the regression with oxygen, but a lower value
$Y_{\pr}({\Hefour})=0.223\pm0.006$ for the regression with
nitrogen.\footnote{These authors state (see also Fuller \etal 1991)
 that this value is ``... $2 \sigma$ below the {\em lower} bound,
 $Y_{\pr}>0.236$, allowed in the standard BBN model with three neutrino
 flavours'' (as quoted by Walker \etal 1991) and suggest various
 modifications to the model to resolve the discrepancy. In fact the
 expected helium abundance in the standard BBN model can be much lower
 (see \fref{abunetamc}) if the uncertain upper bound \eref{He3plusD} on
 ${\Htwo}+{\Hethree}$ is ignored.} However
Pagel and Kazlauskas (1992) argue that the observed constancy of the
N/O ratio at low metallicity favours the linear extrapolation used by
Pagel \etal (1992). Nevertheless, it {\em is} a matter of concern that
the observed slope of the regression against oxygen,
${\d}Y/{\d}Z\approx6$, is several times higher than the value expected
from general chemical evolution arguments (see Pagel 1993). Other
potential problems concern recent observational claims of low
abundances in individual metal-poor galaxies, e.g.
$Y({\Hefour})=0.216\pm0.006$ in {\sl SBS\,0335-052} (Melnick \etal
1992). This particular result is unreliable on technical grounds, viz.
underlying stellar absorption at $\lambda\,4471$ (Pagel 1993); however
it is important that such objects be investigated further. In fact
Skillman and Kennicutt (1993) have recently obtained
\begin{equation}
\label{YZw18}
  Y ({\Hefour}) = 0.231 \pm 0.006\ , 
\end{equation}
in {\sl I\,Zw18}, the most metal-poor galaxy known, in agreement with
the primordial value derived by Pagel \etal (1992). Also, observations
of 11 new metal-poor BCGs by Skillman \etal (1993) yield a similar
result. A combined fit to these data together with that of Pagel \etal gives
\begin{equation}
\label{Ypos}
  Y_{\pr}({\Hefour}) = 0.232 \pm 0.003\ . 
\end{equation} 
after some ``discrepant'' objects are excluded (Olive and Steigman 1995a).  
 
An important question is whether the above analyses have failed to
identify any other systematic errors in the extraction of
$Y_{\pr}$. For example, all the observers cited above use the helium
emissivities given by Brocklehurst (1972). Skillman and Kennicutt
(1993) note that use of the emissivities of Smits (1991) would raise
the derived value \eref{YZw18} in {\sl I\,Zw18} to 0.238 and the
corresponding $2\sigma$ upper bound to 0.25. However the Smits (1991)
emissivities were themselves erroneous and have been corrected by
Smits (1996). Sasselov and Goldwirth (1994) have emphasized that
use of the Smits (1996) emissivities gives a better fit to
detailed line ratios than the Brocklehurst (1972) emissivities which
are known to have problems with the fluxes of the triplet HeI lines
used to extract $Y_{\pr}$. These authors argue that consideration of
additional systematic effects such as inadequacies in the (`Case B')
radiative transfer model used and correction for neutral helium may
raise the upper bound on $Y_{\pr}$ to 0.255 for the data set analysed
by Olive and Steigman 1995a), and as high as 0.258 for the
measurements of {\sl I\,Zw18} by Skillman and Kennicutt (1993).

This question was examined in a study of 10 additional low metallicity
BCGs by Izotov \etal (1994). Whereas use of the emissivities of
Brocklehurst (1972) yields $Y_{\pr}({\Hefour})=0.229\pm0.004$ in
excellent agreement with previous results (e.g. Pagel \etal 1992), use
of the Smits (1996) emissivities raises the value to
$Y_{\pr}({\Hefour})=0.240\pm0.005$. Izotov \etal (1996) have recently
increased their data sample to 27 HII regions in 23 BCGs and
obtain\footnote{The central value is now higher by 0.002 than the result
 $Y_{\pr}({\Hefour})=0.241\pm0.003$ quoted by Thuan \etal (1996)
 because the most metal-deficient BCG ({\sl I\,Zw\,18}) has now been
 excluded from the sample on account of its abnormally low HeI line
 intensities (see discussion in Izotov \etal 1996).}
\begin{equation} 
\label{Ypitl}
  Y_{\pr}({\Hefour}) = 0.243 \pm 0.003\ .
\end{equation}
It is seen from \fref{Ypreg}(b) that use of the new emissivities
(together with the new correction factors (Kingdon and Ferland 1995)
for the collisional enhancement of He\,I emission lines) also
decreases the dispersion of the data points in the regression plots
and the derived slope, ${\d}Y/{\d}Z\approx1.7\pm0.9$, is now smaller
than found before (e.g. by Pagel \etal 1992) and in agreement with
general expectations in chemical evolution models. Systematic effects
due to deviations from Case B recombination theory, temperature
fluctuations, Wolf-Rayet stellar winds and supernova shock waves are
demonstrated to be negligible while different corrections for
underlying stellar absorption and fluorescent enhancement in the HeI
lines alter $Y_{\pr}$ by at most $\pm 0.001$. (For example, use of the
older collisional enhancement correction factors (Clegg 1983) gives
$Y_{\pr}=0.242\pm0.004$.) Thus, while the widely adopted bound
$Y_{\pr}({\Hefour})<0.24$ (e.g. Walker \etal 1991, Smith \etal 1993,
Kernan and Krauss 1994) based on \eref{Yppagellim} may be
``reasonable'', a more ``reliable'' upper bound to the primordial
helium abundance is (Kernan and Sarkar 1996)
\begin{equation} 
\label{Ypitllim}
  Y_{\pr}({\Hefour}) < 0.25\ .
\end{equation} 
Krauss and Kernan (1995) also consider a value of $Y_{\pr}$ as large
as 0.25. Copi \etal (1995a) favour a ``reasonable'' bound of 0.243 and
an ``extreme'' bound of 0.254.

\subsubsection{Deuterium:\label{d}} 

The primordial abundance of deuterium is even harder to pin down since
it is easily destroyed in stars at temperatures exceeding
$\approx6\times10^5\dK$); in fact, its spectral lines have not been
detected in any star, implying ${\Htwo}/{\H}<10^{-6}$ in stellar
atmospheres. It is seen in the giant planets, which reflect the
composition of the pre-Solar nebula, with an abundance
${\Htwo}/{\H}\approx(1-4)\times10^{-5}$. It is also detected in the
local interstellar medium (ISM) through its ultraviolet absorption
lines in stellar spectra but its abundance shows a large scatter,
${\Htwo}/{\H}\approx(0.2-4)\times10^{-5}$, suggesting localized
abundance fluctuations and/or systematic errors. Even among the
cleanest lines of sight (towards hot stars within about 1 kpc) the
abundance as measured by the {\sl Copernicus} and {\sl IUE} satellites
varies in the range ${\Htwo}/{\H}\approx(0.8-2)\times10^{-5}$ (Laurent
1983, Vidal-Madjar 1986). From a careful analysis of the available
data, McCullough (1992) finds that after discarding some unreliable
measurements, the remaining 7 {\sl IUE} and 14 {\sl Copernicus}
measurements are all consistent with an interstellar abundance of
\begin{equation}
\label{Dism}
 \left(\frac{\Htwo}{\H}\right)_{\ISM} = 1.5 \pm 0.2 \times 10^{-5}\ .
\end{equation}
Linsky \etal (1993, 1995) have measured
${\Htwo}/{\H}=1.60\pm0.09\,(\stat)\,^{+0.05}_{-0.10}\,(\syst)\times10^{-5}$
towards the star {\sl Capella} at $12.5\kpc$. using the {\sl Hubble
Space Telescope} (see \fref{Dmeas}). However since the
Lyman-$\alpha$ line (of hydrogen) is severely saturated even towards
such a nearby star, such observations, although very accurate, cannot
test whether there are real spatial variations in the interstellar
deuterium abundance (Pagel 1993). Also the entire data set is still
too limited to reveal any correlation of the $\Htwo$ abundance with
the metallicity (Pasachoff and Vidal-Madjar 1989).


It has been argued that there are no important astrophysical sources
of deuterium (Epstein \etal 1976) and ongoing observational attempts
to detect signs of deuterium synthesis in the Galaxy have so far not
contradicted this belief (see Pasachoff and Vidal-Madjar 1989). If
this is indeed so, then the lowest $\Htwo$ abundance observed today
should provide a {\em lower} bound to the primordial abundance.
McCullough's (1992) analysis of the observations discussed above then
implies:
\begin{equation} 
\label{Dismlim}
 \left(\frac{\Htwo}{\H}\right)_{\pr} > 1.1 \times 10^{-5}\ (\95cl).
\end{equation}
which we consider a ``reliable'' bound (Kernan and Sarkar 1996). 
 
To obtain an upper bound to the primordial abundance, it has been
traditional to resort to models of galactic chemical evolution which
indicate that primordial $\Htwo$ has been depleted due to cycling
through stars (`astration') by a factor of about $2-10$ (e.g. Audouze
and Tinsley 1976, Clayton 1985, Delbourgo-Salvador \etal 1985,
Vangioni-Flam \etal 1994). The depletion factor may be, moreover,
variable within the Galaxy (e.g. Delbourgo-Salvador \etal 1987),
leading to large fluctuations in the observed interstellar abundance
today; unfortunately, as mentioned above, this hypothesis cannot be
observationally tested. Hence by these arguments the primordial
abundance of deuterium is very approximately bounded to be less than a
few times $10^{-4}$. It is obviously crucial to detect deuterium
outside the Solar system and the nearby interstellar medium in order
to get at its primordial abundance and also, of course, to establish
its cosmological origin. Adams (1976) had proposed searching for
Lyman-series absorption lines of deuterium in the spectra of distant
quasars, due to foreground intergalactic clouds made of primordial
unprocessed material. However problems arise in studying such quasar
absorption systems (QAS) because of possible confusion with
neighbouring absorption lines of hydrogen and multi-component velocity
structure in the clouds (Webb \etal 1991). The recent availability of
large aperture ground-based telescopes, e.g. the 10-mt {\sl Keck
Telescope}, has provided the required sensitivity and spectral
resolution, leading to several detections.  Songaila \etal (1994) find
\begin{equation}
\label{Dlya1}
 \left(\frac{\Htwo}{\H}\right)_{\rm QAS (1)} \approx (1.9-2.5)\times 10^{-4}\ ,
\end{equation}
in a chemically unevolved cloud at $z=3.32$ along the line of sight to
the quasar {\sl Q0014+813}, and note that there is a $3\%$ probability
of the absorption feature being a misidentified Ly-$\alpha$ line of
hydrogen. Carswell \etal (1994) obtain ${\Htwo}/{\H}=10^{-3.6\pm0.3}$
in the same cloud but the confusion probability in their data is said
to be as high as $15\%$. However, further observations have resolved
$\Htwo$ lines at $z=3.320482$ and $z=3.320790$, thus eliminating the
possibility of such confusion; the measured abundances in the two
clouds are, respectively, ${\Htwo}/{\H}=10^{-3.73\pm0.12}$ and
$10^{-3.72\pm0.09}$ (where the errors are {\em not} gaussian) (Rugers
and Hogan 1996a). These authors also set an independent lower limit of
${\Htwo}/{\H}\geq1.3\times10^{-4} (\95cl)$ on their sum from the Lyman
limit opacity. Recently, they have detected
${\Htwo}/{\H}=1.9^{+0.6}_{-0.9}\times10^{-4}$ in another QAS at
$z=2.797957$ towards the same quasar; The errors are higher because
the ${\rm D}$ feature is saturated, even so a $\95cl$ lower limit of
${\Htwo}/{\H}>0.7\times10^{-4}$ is obtained (Rugers and Hogan 1996b).
There have been other, less definitive, observations of QAS consistent
with this abundance, e.g. ${\Htwo}/{\H}\approx10^{-3.95\pm0.54}$ at
$z=2.89040$ towards {\sl GC0636+68} (Hogan 1995a),
${\Htwo}/{\H}\leqsim1.5\times10^{-4}$ at $z=4.672$ towards {\sl
BR1202-0725} (Wampler \etal 1995) and
${\Htwo}/{\H}\leqsim10^{-3.9\pm0.4}$ at $z=3.08$ towards {\sl
Q0420-388} (Carswell \etal 1996).\footnote{The metallicity in the last
 two QAS are, respectively, about $1/10$ of Solar and 2 times Solar,
 showing that significant stellar processing has already occured even
 at such high redshifts!} However, very recently, other observers have
found much lower values in QAS at $z=3.572$ towards {\sl Q1937-1009}
(Tytler \etal 1996) and at $z=2.504$ towards {\sl Q1009+2956} (Burles
and Tytler 1996); their average abundance is
\begin{equation}
\label{Dlya2}
 \left(\frac{\Htwo}{\H}\right)_{\rm QAS (2)} = 2.4 \pm 0.3\ (\stat) \pm 0.3\ 
                                               (\syst) \times 10^{-5}\ .
\end{equation}
Unlike the cloud in which the abundance \eref{Dlya1} was measured,
these QAS also exhibit absorption due to C and Si, whose synthesis in
stars would have been accompanied by destruction of $\Htwo$. Tytler
\etal (1996) argue that this must have been negligible since the
metallicity is very low. Although this is true averaged over the
cloud, large fluctuations in the observed ${\Htwo}$ abundance are
possible since the mass of absorbing gas covering the QSO image is
very small; thus ${\Htwo}$ may well have been significantly depleted
in it by a star which was not massive enough to eject `metals' (Rugers
and Hogan 1996b). Tytler and Burles (1996) point out in response that
the line of sight through the QAS is $\sim10$\,Kpc long so it would be
difficult for metals to be removed from two independent lines of
sight, leaving the same $\Htwo$ abundance in each. They also note that
their data quality is superior to the other detections and upper
limits. In view of this confusing observational situation and keeping
in mind that $\Htwo$ is {\em always} destroyed by stellar processing,
we adopt the high $\Htwo$ measurement \eref{Dlya1} in the chemically
unevolved cloud as a conservative upper limit on its primordial
abundance.
\begin{equation}
\label{Dlyalim}
 \left(\frac{\Htwo}{\H}\right)_{\pr} \leqsim 2.5 \times 10^{-4}\ ,
\end{equation}    
although one cannot attach a confidence level to this number. Given
the contradictory observations, it is probably premature to interpret
\eref{Dlya1} as providing a {\em lower} bound on the primordial
deuterium abundance (cf. Krauss and Kernan 1995).

Edmunds (1994) has argued that a primordial abundance as large as in
\eref{Dlya1} cannot be reduced to the present ISM value \eref{Dism} in
a simple `closed box' chemical evolution model. If so, the lower QAS
measurement \eref{Dlya2} would be favoured as the primordial
one. However this chemical evolution model is well known to be
inadequate in many respects (see Rana 1991) and cannot yield any
reliable conclusions. A definitive resolution of the discrepancy
between the two values can only come from further observations which
are in progress

\subsubsection{Helium-3:\label{he3}} 

The abundance of $\Hethree$ is similarly uncertain, with the
additional complication that it is capable of being both produced and
destroyed in stars. It has been detected through its radio
recombination line in galactic HII regions although initial attempts
to measure its abundance gave rather widely varying results in the
range $\approx(1-15)\times10^{-5}$ along with some upper bounds at the
$\approx10^{-5}$ level (Bania \etal 1987). Balser \etal (1994) have
recently made considerable progress in overcoming the observational
problems involved in determining the rather weak line parameters and
in modelling the HII regions; they now obtain more stable abundances
in the range
\begin{equation}
\label{He3HII}
 \left(\frac{\Hethree}{\H}\right)_{\rm HII} \approx (1 - 4) \times 10^{-5}\ , 
\end{equation}
in a dozen selected regions. Also Rood \etal (1992) have detected a
large abundance (${\Hethree}/{\H}\approx10^{-3}$) in the planetary
nebula {\sl NGC\,3242}, in accord with the expectation (Rood \etal
1976, Iben and Truran 1978) that stars comparable in mass to the Sun
create $\Hethree$. However massive stars destroy $\Hethree$, hence to
determine the overall consequence of astration requires detailed
modelling of stellar evolution and averaging over some assumed initial
mass function (IMF) of stars. A detailed study by Dearborn \etal
(1986) considered several possibilities for the initial helium and
`metal' abundances and averaged over a Salpeter type IMF:
${\d}N/{\d}M\,\propto\,M^{-1.35}$ (see Scalo 1986). These authors
found that the net fraction, $g_{3}$, of $\Hethree$ which survives
stellar processing is quite sensitive to the assumed initial
abundances. When averaged over stars of mass $3\,M_{\odot}$ and above,
$g_{3}$ is as large as $0.47$ for a standard Pop\,I composition
($28\%$ helium, $2\%$ metals) but as small as $0.04$ for an extreme
low metal model ($25\%$ helium, $0.04 \%$ metals).  However if stars
of mass down to $0.8\,M_{\odot}$ are included, then $g_{3}$ exceeds
$0.5$ for any assumed composition.\footnote{This is presumably because
stars in the mass range $0.8-3\,M_{\odot}$, which contribute
dominantly to the average over the assumed power law mass function,
were assumed not to destroy {\em any} $\Hethree$.} If however there
has indeed been {\em net} creation of $\Hethree$ in stars, it is
puzzling that the galactic observations find the highest $\Hethree$
abundances in the outer Galaxy where stellar activity is {\em less}
than in the inner Galaxy. While regions with high abundances do lie
preferentially in the Perseus spiral arm, there are large
source-to-source variations which do not correlate with stellar
activity (Balser \etal 1994). Secondly, in this picture the present
day interstellar $\Hethree$ abundance should be significantly higher
than its proto-Solar abundance as measured in meteorites
\eref{Solarsys}; however several of the ISM abundances are {\em less}
than the Solar system value.

To reconcile these discrepancies, Hogan (1995b) has suggested that
there may in fact be net {\em destruction} of $\Hethree$ in
$\approx\,1-2\,M_{\odot}$ stars through the same mixing process which
appears to be needed to explain other observations, e.g. the
$^{12}$C/$^{13}$C ratio (Wasserburg \etal 1995). A plausible mechanism
for this has also been suggested (Charbonnel 1995). This is indeed
essential if the primordial deuterium abundance is as high as is
indicated by the recent measurement \eref{Dlya1} in a QAS. Although
the $\Htwo$ can be astrated down to its much lower abundance in the
ISM, the $\Hethree$ thus produced would exceed observational bounds
unless it too is destroyed to a large extent. These considerations
have prompted reexamination of the usual assumptions about the
chemical evolution of $\Hethree$ (e.g. Vangioni-Flam and Cass\'e 1995,
Copi \etal 1995c, Galli \etal 1995, Olive \etal 1995, Palla \etal
1995, Scully \etal 1996). It is clear that until this is is better
understood, one cannot use the $\Hethree$ abundance to sensibly
constrain BBN (cf. Wilson and Rood 1994).

\subsubsection{Deuterium + Helium-3:\label{d+he3}} 

Yang \etal (1984) had suggested that the uncertainties in determining
the primordial abundances of $\Htwo$ and $\Hethree$ may be
circumvented by considering their {\em sum}. Since $\Htwo$ is burnt
in stars to $\Hethree$, a fraction $g_{3}$ of which survives stellar
processing, the primordial abundances may be related to the abundances
later in time in a simple `one-cycle' approximation to galactic
chemical evolution. Neglecting the possible production of $\Hethree$
in light stars yields the inequality
\begin{equation}
\label{yangevol}
  \left(\frac{{\Htwo} + {\Hethree}}{\H}\right)_{\pr} < 
   \left(\frac{{\Htwo} + {\Hethree}}{\H}\right) +
   \left(\frac{1}{g_{3}} - 1\right) \left(\frac{\Hethree}{\H}\right)\ .
\end{equation}
The terms on the RHS may be estimated at the time of formation of
the Solar system, about $4.5\Gyr$ ago, as follows. The abundance of
$\Hethree$ in the Solar wind, deduced from studies of gas-rich
meteorites, lunar rocks and metal foils exposed on lunar missions, may
be identified with the sum of the pre-Solar abundances of $\Hethree$
and $\Htwo$ (which was burnt to $\Hethree$ in the Sun), while the
smallest $\Hethree$ abundance found in carbonaceous chondrites,
which are believed to reflect the composition of the pre-Solar nebula,
may be identified with the pre-Solar abundance of $\Hethree$ alone
(Black 1971, Geiss and Reeves 1972). Such abundance determinations are
actually made in ratio to $\Hefour$; combining these data with the
standard Solar model estimate $({\Hefour}/{\H})_{\odot}=0.10\pm0.01$
(see Bahcall and Ulrich 1988), Walker \etal (1991) obtain:
\begin{equation}
\label{Solarsys}
\fl 1.3 \times 10^{-5} \leqsim \left(\frac{\Hethree}{\H}\right)_{\odot} 
     \leqsim 1.8 \times 10^{-5}\ , \qquad
 3.3 \times 10^{-5} \leqsim \left(\frac{{\Htwo}+{\Hethree}}{\H}\right)_{\odot}
  \leqsim 4.9 \times 10^{-5} .
\end{equation}
Although these are quoted as ``$2\sigma$'' bounds, we have chosen to
view these as approximate inequalities since these authors have not
estimated the systematic uncertainties. Walker \etal also interpret
the work by Dearborn \etal (1986) on the survival of $\Hethree$ in
stars to imply the lower limit
\begin{equation}
\label{g3}
 g_{3} \geqsim 0.25\ .
\end{equation}
Using these values in \eref{yangevol} then bounds the primordial sum
of $\Htwo$ and $\Hethree$ as (Yang \etal 1984, Walker \etal 1991)
\begin{equation}
\label{He3plusD}
 \left(\frac{{\Htwo}+{\Hethree}}{\H}\right)_{\pr} \leqsim 10^{-4}\ .
\end{equation}
(When confronting low nucleon density models, with
$\eta\leqsim2\times10^{-10}$, we may view this as essentially an upper
bound on primordial $\Htwo$ since the relative abundance of $\Hethree$
is then over a factor of 10 smaller.) Olive \etal (1990) obtained a
similar bound in the `instantaneous recycling' approximation, i.e.
assuming that some fraction of gas undergoes several cycles of stellar
processing instantaneously. Steigman and Tosi (1992, 1995) obtain even
stronger bounds in more elaborate models of galactic chemical
evolution, for particular choices of the initial mass function, star
formation rate, matter infall rate \etc. Combining these results with
the Solar system abundances discussed above, Hata \etal (1996a) find
the ``primordial'' abundances
\begin{equation}
\label{He3D}
 1.5\times10^{-5} \leq \left(\frac{\Htwo}{\H}\right)_{\pr} 
  \leq 10^{-4}, \qquad
  \left(\frac{\Hethree}{\H}\right)_{\pr} \leq 2.6 \times 10^{-5} .
\end{equation}
In our opinion, all these bounds should be viewed with caution given
the many astrophysical uncertainties in their derivation. Indeed there
are now several observational indications that the
bound \eref{He3plusD} (or its minor variants) used in many recent
analyses (e.g. Walker \etal 1991, Smith \etal 1993, Kernan and Krauss
1994, Copi \etal 1995a, Krauss and Kernan 1995) is overly restrictive,
and the bounds \eref{He3D} used by Hata \etal (1995) even more
so. Geiss (1993) has recently reassessed the Solar system data and
quotes more generous errors in the derived abundances:
\begin{equation}
\label{Solarsysgeiss}
\fl \left(\frac{\Hethree}{\H}\right)_{\odot} = 1.5\pm1.0 \times 10^{-5}\ ,
     \qquad \left(\frac{{\Htwo}+{\Hethree}}{\H}\right)_{\odot} = 
     4.1 \pm 1.0 \times 10^{-5}\ .
\end{equation}
If we use these numbers to estimate ``$2\sigma$'' limits, the bound
\eref{He3plusD} is immediately relaxed by a factor of 2! In fact it is
not even clear if the Solar system abundances provide a representative
measure at all, given that observations of $\Hethree$ elsewhere in the
Galaxy reveal large (and unexplained) source-to-source variations
(Balser \etal 1994). Further, the survival fraction of $\Hethree$ may
have been overestimated (Hogan 1995b). Note that if the primordial
$\Htwo$ abundance is as indeed as high as $2.5\times10^{-4}$
\eref{Dlyalim}, then not more than $10\%$ of the $\Hethree$ into which
it was burnt could have survived stellar processing, in conflict with
the ``theoretical'' lower limit of $25\%$ \eref{g3}. Indeed, Gloeckler
and Geiss (1996) have recently found, using the {\sl Ulysses}
spacecraft, that
${\Hethree}/{\Hefour}=2.2^{+0.7}_{-0.6}\,(\stat)\,\pm0.2\,(\syst)\times
10^{-4}$ in the local interstellar cloud. This is close to the value
of $1.5\pm0.3\times10^{-4}$ in the pre-solar nebula (see
\eref{Solarsysgeiss}), demonstrating that the $\Hethree$ abundance has
hardly increased since the formation of the solar system.
 
The Solar system abundances \eref{Solarsys} imply
$1.8\times10^{-5}\leqsim({\Htwo}/{\H})_{\odot}\leqsim3.3\times10^{-5}$,
hence Walker \etal (1991) as well as Smith \etal (1993) adopt for the
primordial value $({\Htwo}/{\H})_{\pr}\geqsim1.8\times10^{-5}$. We
consider this to be less reliable than the lower bound \eref{Dismlim}
from direct observations of interstellar deuterium, particularly since
Geiss (1993) quotes large errors from a reassessment of the Solar
system data: $({\Htwo}/{\H})_{\odot}=(2.6\pm1.0)\times10^{-5}$. Copi
\etal (1995a) adopt a ``sensible'' lower bound of $1.6\times10^{-5}$
while Krauss and Kernan (1995) opt for a lower bound of
$2\times10^{-5}$.

\subsubsection{Lithium-7:\label{li-7}} 

Lithium, whose common isotope is $\Liseven$, is observed in the
atmospheres of both halo (Population II) and disk (Population I)
stars, with widely differing abundances (see Michaud and Charbonneau
1991). This is not unexpected since $\Liseven$, like $\Htwo$, is
easily destroyed (above $2\times10^{6}\dK$) hence only the lithium on
the stellar surface survives, to an extent dependent on the amount of
mixing with the stellar interior, which in turn depends on the stellar
temperature, rotation \etc. For Pop\,I stars in open clusters with
ages in the range $\sim0.1-10\Gyr$, the observed abundances range upto
${\Liseven}/{\H}\approx10^{-9}$ (e.g. Hobbs and Pilachowski 1988).
However in the somewhat older Pop\,II halo dwarfs, the abundance is
observed to be about $10$ times lower and, for high temperatures and
low metallicity, nearly independent of the stellar temperature and the
metal abundance (Spite and Spite 1982, Spite \etal 1987, Rebolo \etal
1988, Hobbs and Thorburn 1991). For a sample of 35
such stars with [Fe/H]$<-1.3$,\footnote{Square brackets indicate the
 logarithmic abundance relative to the Solar value, i.e. [Fe/H]$<-1.3$
 means Fe/H$<5\times10^{-2}$(Fe/H)$_{\odot}$.} and $T\geqsim5500\dK$
and the weighted average of the lithium-7 abundance is (Walker \etal
1991)
\begin{equation}
\label{Li7wtsso}
 \left(\frac{\Liseven}{\H}\right)^{\I\I} = 10^{-9.92 \pm 0.07}\ 
  (\95cl)\ . 
\end{equation}
This has been used to argue that the Pop\,II abundance reflects the
primordial value in the gas from which the stars formed, with the
higher abundance in the younger Pop\,I stars created subsequently, for
example by supernovae (Dearborn \etal 1989). Indeed evolutionary
modelling (ignoring rotation) of halo stars indicate that they are
essentially undepleted in lithium (Deliyannis \etal 1990). Taking both
observational errors and theoretical uncertainties (mostly the effects
of diffusion) into account, these authors find the fitted initial
abundance to be:
\begin{equation} 
\label{Li7ddk}
 \left(\frac{\Liseven}{\H}\right)_{\pr}^{\I\I} = 10^{-9.80 \pm 0.16}\ 
  (\95cl)\ . 
\end{equation}
This is assumed to be the primordial abundance by Walker \etal (1991)
and Smith \etal (1993) since any production of lithium after the
Big Bang, but before halo star formation is presumed to be unlikely
(see Boesgaard and Steigman 1985). Kernan and Krauss (1994) also adopt 
this value.

Recently the situation has taken a new turn with the discovery that
there are several extremely metal-poor Pop\,II halo dwarfs with {\it
no} detectable lithium. Thorburn (1994) has determined accurate
abundances for 80 stars with [Fe/H]$<-1.9$ and $T>5600\dK$, of which 3
are lithium deficient with respect to the others by a factor of over
10 (see \fref{Li7meas}). Ignoring these reveals a weak, though
statistically significant, trend of increasing $\Liseven$ abundance
with both increasing temperature and increasing metallicity which had
not been apparent in older data (cf. Olive and Schramm 1992) but is
also seen for a smaller sample of extreme halo dwarfs by Norris \etal
(1994). Thorburn (1994) interprets this as indicating lithium
production by galactic cosmic ray spallation processes. Indeed
beryllium (Gilmore \etal 1992) and boron (Duncan \etal 1992) have also
been seen in several metal-poor halo stars with abundances {\em
proportional} to the metallicity and in the ratio B/Be\,$\approx10$,
which does point at such a production mechanism rather than a
primordial origin. This should also have created lithium at the level
of about $35\%$ of its Pop\,II abundance; much of the observed
dispersion about the Pop\,II `plateau' would then be due to the
$\approx2$\,Gyr range in age of these stars. Thorburn (1994) therefore
identifies the primordial abundance with the observed average value in
the hottest, most metal-poor stars, viz.
\begin{equation}
\label{Li7thorburn}
  \left(\frac{\Liseven}{\H}\right)_{\pr}^{\I\I} = 10^{-9.78 \pm 0.20} 
   \ (\95cl)\ ,
\end{equation}  
which agrees very well with the value \eref{Li7ddk} inferred by
Deliyannis \etal (1990). Subsequently, Thorburn's results have been
questioned by Molaro \etal (1995) who find {\em no} significant
correlation of the lithium abundance, in a sample of 24 halo dwarfs,
with either the temperature (when this is determined by a
spectroscopic method rather than by broad-band photometry) or the
metallicity (determined using an updated stellar atmosphere model).
Thus they reaffirm the purely primordial origin of the Pop\,II
$\Liseven$ `plateau' and argue that the observed dispersion is
entirely due to measurement errors alone. However, Ryan \etal (1996),
who include data on 7 new halo dwarfs, have confirmed the original
finding of Thorburn (1994); they note that Molaro \etal (1995) did not
test whether $\Liseven$ is {\em simultaneously} correlated with $T$
and [Fe/H]. In any case, the abundance Molaro \etal (1995) derive for
24 stars with $T>5700\dK$ and [Fe/H]$<-1.4$ is fortuitously identical
to that given by Thorburn (1994) so \eref{Li7thorburn} remains the
best estimate of the primordial Pop\,II $\Liseven$ abundance. Even so
this leaves open the question of why several stars which are in all
respects similar to the other stars which define the Pop\,II
`plateau', are so lithium deficient. Until this is clarified, it may
be premature to assert that the Pop\,II abundance of lithium reflects
its primordial value.


Further, the observation that ${\Liseven}/{\H}\approx10^{-9}$ in
Pop\,I stars as old as $\approx10\Gyr$ (in {\sl NGC\,188}) then
requires the galactic ${\Liseven}/{\H}$ ratio to rise by a factor of
about 10 in the first $\approx2-5\Gyr$ and then remain constant for
nearly $10\Gyr$ (Hobbs and Pilchaowski 1988). This encourages the {\em
opposite} point of view, viz. that the (highest) Pop\,I abundance is
that of primordially synthesized lithium, which has been (even more)
depleted in the older Pop\,II stars, for example through turbulent
mixing driven by stellar rotation (Vauclair 1988). The observational
evidence for a $\pm25\%$ dispersion in the Pop\,II $\Liseven$
`plateau' is consistent with this hypothesis (Deliyannis \etal
1993). Rotational depletion was studied in detail by Pinsonneault
\etal (1992) who note that the depletion factor could have been as
large as $\approx10$. Chaboyer and Demarque (1994) have demonstrated
that models incorporating rotation provide a good match to the
observed $\Liseven$ depletion with decreasing temperature in Pop\,II
stars and imply a primordial abundance
\begin{equation}
\label{Li7popI}
  \left(\frac{\Liseven}{\H}\right)_{\pr}^{\I} = 10^{-8.92 \pm 0.1} ,
\end{equation}
corresponding to the highest observed Pop\,I value. (However the trend
of increasing $\Liseven$ abundance with increasing metallicity seen
by Thorburn (1994) cannot be reproduced by these models.) Studies of
galactic chemical evolution (Mathews \etal 1990a, Brown 1992) show that
both possibilities can be accomodated by the observational data, {\em
including} the bound ${\Liseven}/{\H}<10^{-10}$ on the interstellar
$\Liseven$ abundance in the {\sl Large Magellanic Cloud} along the
line of sight to {\sl Supernova\,1987A} (e.g. Baade \etal
1991). Although this apparently supports the Pop\,II abundance, the
bound is considerably weakened by the uncertain correction for the
depletion of lithium onto interstellar grains.

Recently Smith \etal (1992) have detected ${\Lisix}$ with an abundance
\begin{equation}
\label{Lisix}
  \left(\frac{\Lisix}{\Liseven}\right)^{\I\I} = 0.05 \pm 0.02\ ,
\end{equation}
in {\sl HD\,84937}, one of the hottest known Pop\,II
stars. (Interestingly enough, to fit the observed spectrum requires
line broadening of $\approx5\km{\sec}^{-1}$, suggestive of rotation.)
Since ${\Lisix}$ is much more fragile than $\Liseven$, this has been
interpreted (e.g. Steigman \etal 1993) as arguing against significant
rotational depletion of primordially synthesized lithium since this
would require the undepleted star to have formed with comparable
amounts of ${\Lisix}$ and $\Liseven$, whereas
${\Lisix}/{\Liseven}\approx10^{-4}$ in standard BBN. The simplest
interpretation is that the ${\Lisix}$ (and some fraction of the
$\Liseven$) was created by cosmic ray spallation processes. However
this argument does not hold if there is some primordial source of
${\Lisix}$, as may happen in non-standard models.\footnote{In this
 scenario protogalactic matter has been astrated by a large factor
 (Audouze and Silk 1989) implying that the primordial abundance of
 deuterium, another fragile isotope, should also be quite large,
 viz. ${\Htwo}/{\H}=(7\pm3)\times10^{-4}$ (Steigman \etal 1993). This
 is however not inconsistent with the recent direct observations of
 primordial deuterium \eref{Dlya1}, if the $\Hethree$ created
 by the astration of deuterium is also destroyed.} Moreover Hobbs and
Thorburn (1994) have found the same relative abundance of ${\Lisix}$
in the cooler evolved subgiant {\sl HD 201891}, which however has
${\Liseven}/{\H}=7.9\times10^{-11}$, a factor of 2 below the
Pop\,II plateau, indicating that some depletion {\em has}
occured. Vauclair and Charbonnel (1995) have pointed out that a mass
loss of order $\sim10^{-13}-10^{-12}\,M_{\odot}\,{\yr}^{-1}$ through
stellar winds can deplete $\Liseven$ without depleting $\Lisix$; their
preferred primordial abundance would then be the upper envelope of the
Pop\,II value i.e. $({\Liseven}/{\H})_{\pr}\approx10^{-9.5\pm0.1}$.

Given these considerations, we believe the Pop\,I value \eref{Li7popI}
to be ``reliable'' and the new Pop\,II value \eref{Li7thorburn} to be
``reasonable'' (Kernan and Sarkar 1996). Copi \etal (1995a) consider
an upper bound of $({\Liseven}/{\H})_{\pr}\leq3.5\times10^{-10}$ on
the basis of the Pop\,II value, allowing for depletion by a factor of
2. Krauss and Kernan (1995) take the (older) Pop\,II value
\eref{Li7ddk} to be primordial but also consider an upper bound as
high as $5\times10^{-10}$ to allow for some depletion.

\subsection{Theory versus observations\label{theoobs}} 

We now determine the restrictions imposed on the nucleon-to-photon
ratio by comparing the {\em inferred} bounds on the abundances of
light elements with the $\95cl$ limits on their computed values.  To
begin with, we consider each element separately, as in previous work,
although this procedure is, strictly speaking, statistically incorrect
since the different elemental yields are correlated. Nevertheless it
is an useful exercise to establish the approximate range of $\eta$ for
which there is concordance between the various abundances. First,
consider the ``reliable'' abundance bounds \eref{Ypitllim},
\eref{Dismlim}, \eref{Li7popI}:
\begin{equation}
\label{reliable}
\eqalign{
 Y_{\pr}({\Hefour}) < 0.25 \quad
  &\Rightarrow \quad \eta < 9.1 \times 10^{-10}\ , \\
 \left(\frac{\Htwo}{\H}\right)_{\pr} > 1.1 \times 10^{-5} \quad 
  &\Rightarrow \quad \eta < 1.1 \times 10^{-9}\ , \\
 \left(\frac{\Liseven}{\H}\right)_{\pr}^{\I} < 1.5 \times
  10^{-9} \quad &\Rightarrow \quad 4.1 \times 10^{-11} < \eta < 1.4
  \times 10^{-9}\ .} 
\end{equation}
The ``reasonable'' abundance bounds \eref{Yppagellim}, \eref{Dlyalim}
and \eref{Li7thorburn} yield:
\begin{equation}
\label{reasonable}
\eqalign{
 Y_{\pr}({\Hefour}) < 0.24 \quad &\Rightarrow \quad
  \eta < 3.4 \times 10^{-10}\ , \\
 \left(\frac{\Htwo}{\H}\right)_{\pr} \leqsim 2.5 \times 10^{-4} \quad 
  &\Rightarrow \quad \eta \geqsim 1.3 \times 10^{-10}\ , \\
 \left(\frac{\Liseven}{\H}\right)_{\pr}^{\I\I} <
  2.6 \times 10^{-10} \quad &\Rightarrow \quad 1.0 \times 10^{-10} < 
  \eta < 5.9 \times 10^{-10}\ .}
\end{equation}
Finally, the indirect bound \eref{He3plusD}
\begin{equation}
\label{indirect}
 \left(\frac{{\Htwo}+{\Hethree}}{\H}\right)_{\pr} \leqsim 10^{-4}
  \quad \Rightarrow \quad \eta \geqsim 2.6 \times 10^{-10}\ ,
\end{equation}
provides a restrictive, albeit rather uncertain, lower limit on
$\eta$. The situation is illustrated in \fref{concord} where we
illustrate the consistency of standard BBN with the observations,
viz. the primordial $\Hefour$ abundance \eref{Ypitl} inferred from
BCGs, the ISM $\Htwo$ limit \eref{Dismlim} and the two conflicting
measurements \eref{Dlya1} and \eref{Dlya2} in QAS, and, finally, the
$\Liseven$ abundances in Pop\,II \eref{Li7thorburn} and Pop\,I
\eref{Li7popI} stars.


\subsubsection{Standard nucleosynthesis:\label{stanbbn}} 

Adopting the ``reliable'' bounds on extragalactic $\Hefour$,
interstellar $\Htwo$ and Pop\,I $\Liseven$, we see from
\eref{reliable} that BBN can {\em conservatively} limit $\eta$ to only
within a factor of about 20:
\begin{equation}
\label{etarellim}
 4.1 \times 10^{-11} < \eta < 9.1 \times 10^{-10} \quad
  \Rightarrow \quad 0.0015 < \Omega_{\N} h^2 < 0.033\ .
\end{equation}
However this still improves on the observational uncertainty in $\eta$
\eref{etarange} by a factor of about 70. Note that the upper limit to
$\eta$ comes from $\Hefour$, the element whose abundance is the {\em
least} sensitive to the nucleon density. The one from interstellar
$\Htwo$, which was historically crucial in establishing the
consistency of BBN (Reeves \etal 1973), is slightly less restrictive
although arguably more robust and therefore still valuable. The Pop\,I
$\Liseven$ abundance provides a weak lower limit.

On the basis of the ``reasonable'' bounds quoted in \eref{reasonable},
$\eta$ can be pinned down to within a factor of about 3:
\begin{equation}
\label{etareaslim}
 1.3 \times 10^{-10} < \eta < 3.4 \times 10^{-10} \quad
  \Rightarrow \quad 0.0048 < \Omega_{\N} h^2 < 0.013\ ,
\end{equation}
i.e. assuming that the recent high $\Htwo$ abundance measurement in a
Lyman-$\alpha$ cloud bounds its primordial value and that the
systematic error does not exceed the statistical error in the
$\Hefour$ abundance determination. The Pop\,II $\Liseven$ abundance
provides a slightly less restrictive lower limit to $\eta$.

Finally, if we accept the upper bound on the sum of primordial $\Htwo$
and $\Hethree$ inferred {\em indirectly} from Solar system abundances
and stellar evolution arguments, then $\eta$ is known \eref{indirect}
to within about $15\%$ when combined with the ``reasonable'' upper
bound on $\Hefour$:
\begin{equation} 
\label{etaindlim}
 \eta \simeq (2.6-3.4) \times 10^{-10} \quad
  \Rightarrow \quad \Omega_{\N} h^2 \simeq 0.011 \pm 0.0015\ .
\end{equation}
We emphasize that only this last constraint has been highlighted in
the literature; for example, Smith \etal (1993) quote
$2.9\leq(\eta/10^{10})\leq3.8$, corresponding to their adopted bounds
$([{\Htwo}+{\Hethree}]/{\H})_{\pr}\leq9\times10^{-5}$ and
$Y_{\pr}({\Hefour})\leq0.24$. Other groups have relied on $\Liseven$
rather than $\Hefour$ to provide the upper bound to $\eta$,
e.g. Walker \etal (1991) adopt
$({\Liseven}/{\H})_{\pr}^{\I\I}\leq1.4\times10^{-10}$ and quote
$2.8\leq(\eta/10^{-10})\leq4.0$, while Copi \etal (1995a) adopt the
more generous bounds
$([{\Htwo}+{\Hethree}]/{\H})_{\pr}\leq1.1\times10^{-4}$ and
$({\Liseven}/{\H})_{\pr}^{\I\I}\leq3.5\times10^{-10}$ to derive
$2.5\leq(\eta/10^{-10})\leq6.0$. In contrast, if primordial deuterium
has indeed been detected with an abundance \eref{Dlya1} of
${\Htwo}/{\H}\simeq(1.9-2.5)\times10^{-4}$, then the implied nucleon 
density is about a factor of 2 smaller:\footnote{Dar (1995) finds that
 a value of $\eta\simeq1.6\times10^{-10}$ then provides the best fit
 (with a confidence level exceeding $70\%$) to the $\Htwo$, $\Hefour$
 and $\Liseven$ abundances, assuming
 $Y_{\pr}({\Hefour})=0.228\pm0.005$ and
 $({\Liseven}/{\H})_{\pr}^{\I\I}=1.7\pm0.4\times10^{-10}$. (Note
 however that his calculated abundances are systematically lower than
 in \eref{Ypmine} and \eref{abundmine} and his adopted
 abundances differ from those given here.)}
\begin{equation}
\label{etaDlya1}
 \eta \simeq (1.3-1.9) \times 10^{-10} \quad \Rightarrow \quad 
  \Omega_{\N} h^2 \simeq 0.0059 \pm 0.0011\ ,
\end{equation}
while if the primordial abundance is instead in the $\95cl$ range
${\Htwo}/{\H}\simeq(1.5-3.3)\times10^{-5}$ \eref{Dlya2},
the implied nucleon density is about a factor of 2 bigger:
\begin{equation}
\label{etaDlya2}
 \eta = (4.6-8.1) \times 10^{-10} \quad \Rightarrow \quad 
  \Omega_{\N} h^2 = 0.023 \pm 0.0064\ (\95cl).
\end{equation}
Of course both possiblities are consistent with the ``reliable''
bounds \eref{etarellim}.


The above procedure of deriving limits on $\eta$ using one element at
a time ignores the fact that the different elemental yields are
correlated. Taking this into account in a statistically consistent
manner would lead to more stringent constraints than those obtained
above using the symmetric $\95cl$ limits from the Monte Carlo
procedure (Kernan and Krauss 1994). As seen in \fref{He4vsHe3Dmc}, the
$\Htwo$ abundance is strongly anti-correlated with the $\Hefour$
abundance; hence those Monte Carlo runs in which the predicted
$\Hefour$ is lower than the mean, and which therefore may be allowed
by some adopted observational upper bound, will also generally predict
a higher than average $\Htwo$ abundance, which may be in conflict with
the corresponding observational upper bound. Krauss and Kernan (1995)
determine the number of runs (as $\eta$ is varied) which result in
abundances simultaneously satisfying the upper bounds on $\Hefour$,
$\Liseven$ and ${\Htwo}+{\Hethree}$ and the lower bound on
$\Htwo$. The maximum value of $\eta$ is then found by requiring that
50 runs out of 1000 (upto $\sqrt{N}$ statistical fluctuations) satisfy
all the constraints. Using their method and adopting the ISM bound
\eref{Dismlim} ${\Htwo}/{\H}>1.1\times10^{-5}$ and the Pop\,II bound
\eref{Li7thorburn} ${\Liseven}/{\H}<2.6\times10^{-10}$, Kernan and
Sarkar (1996) find that the maximum allowed value of $\eta$ varies
linearly with the adopted upper bound to the $\Hefour$ as
\begin{equation}
\label{etalimI}
 \eta_{\max} = [3.19 + 375.7\ (Y_{\pr}^{\max} - 0.240)] \times 10^{-10}\ ,
\end{equation}
upto $Y_{\pr}^{\max}=0.247$; for higher $Y_{\pr}$, the Pop\,II
$\Liseven$ bound \eref{Li7thorburn} does not permit $\eta$ to exceed
$5.7\times10^{-10}$, as shown in \fref{etalim}\,(a). If we choose
instead to use the more conservative Pop\,I bound \eref{Li7popI}
${\Liseven}/{\H}<1.5\times10^{-9}$, the constraint is further relaxed
to
\begin{equation}
\label{etalimII}
\fl \eta_{\max} = [3.28 + 216.4\,(Y_{\pr}^{\max} - 0.240) + 
               34521\,(Y_{\pr}^{\max} - 0.240)^2] \times 10^{-10}\ , 
\end{equation}
for $Y_{\pr}<0.252$ and saturates at $1.06\times10^{-9}$ for higher
values, essentially due to the ISM $\Htwo$ bound, as shown in
\fref{etalim}\,(b). Thus for $Y_{\pr}^{\max}=0.25$
\eref{Ypitllim}, we find
\begin{equation}
\label{etamax}
  \eta \leq 8.9 \times 10^{-10} \quad \Rightarrow \quad 
  \Omega_{\N} h^2 \leq 0.033\ ,
\end{equation}
which is slightly more stringent than the value \eref{reliable}
determined using the (symmetric) $\95cl$ bounds on the $\Hefour$
abundance alone.


The above bounds on $\eta$ were obtained assuming the validity of
standard BBN amd may be altered in variant models as reviewed by
Malaney and Mathews (1993). Below, we briefly discuss deviations
which are permitted within the context of the Standard Model of
particle physics~\footnote{As an exotic example of possibilities (far)
 beyond the SM, Bartlett and Hall (1991) have speculated
 that the comoving number of photons may {\em decrease} after the
 nuclosynthesis epoch if they become coupled to a cold `hidden sector'
 via some mixing mechanism at a temperature of $\Or(10)\keV$. Then the
 universe may indeed have the critical density in nucleons without
 violating the upper bound on the nucleon-to-photon ratio from
 BBN!} since we will discuss the effect of new physics,
in \sref{cons}. We will also not consider the effect of gross
departures from the standard cosmology, e.g. alternate theories of
gravity and anisotropic world-models. In general, such deviations tend
to speed up the expansion rate and increase the synthesized helium
abundance, thus further tightening the constraints derived from
standard BBN.

\subsubsection{Inhomogeneous nucleosynthesis:\label{inhombbn}} 

The most well motivated departure from standard BBN is the possibility
that nucleosynthesis occurs in an inhomogeneous medium, e.g. due to
fluctuations generated by a first-order quark-hadron phase transition
at $T_{\c}^{\qh}\approx150-400\MeV$, a possibility emphasized by
Witten (1984). As noted earlier (see footnote just before
\eref{Ypapprox}) the signature for this would be the synthesis of
significant amounts of elements beyond helium, although there is
continuing controversy about the extent to which this would happen,
due to the difficulty of adequately modelling the problem
(e.g. Applegate \etal 1988, Malaney and Fowler 1988, Terasawa and Sato
1991, Jedamzik \etal 1994b). Observationally, there are no indications
for an universal `floor' in the abundances of such elements,
particularly beryllium and boron, which would suggest a primordial
origin (see Pagel 1993) and indeed their abundances are reasonably
well understoood in terms of cosmic ray spallation processes
(e.g. Prantzos \etal 1993, Steigman \etal 1993). Furthermore, recent
theoretical developments suggest that the quark-hadron phase
transition is effectively second-order (see Bonometto and Pantano
1993) and does not generate significant fluctuations in the nucleon
distribution (e.g. Banerjee and Gavai 1992). Nevertheless it is
interesting to study the effect of hypothetical fluctuations on
nucleosynthesis to see to what extent the standard picture may be
altered. Such models (e.g. Kurki-Suonio \etal 1990, Mathews \etal
1990b, Jedamzik \etal 1994a) can satisfy the conservative
observational bounds \eref{reliable} on $\Hefour$, $\Htwo$ and
$\Liseven$ (Pop\,I) with a higher nucleon density than in standard
BBN; the upper limit to $\eta$ is raised to
\begin{equation}
\label{etainhom}
 \eta \leqsim 2 \times 10^{-9} \quad \Rightarrow \quad 
  \Omega_{\N} h^2 \leqsim 0.073\ .
\end{equation}
However the less reliable bound \eref{Li7thorburn} on $\Liseven$
(Pop\,II) and the indirect bound \eref{He3plusD} on
$({\Htwo}+{\Hethree})/{\H}$ {\em are} violated unless $\eta$ remains
in about the same range \eref{etaindlim} as is required by homogeneous
nucleosynthesis on the basis of the same bounds. Thus even allowing
for hypothetical and rather fine-tuned inhomogeneities, an
Einstein-DeSitter universe with $\Omega_{\N}=1$ is
disfavoured. Although a nucleon-dominated universe which is open,
e.g. having $\Omega_{\N}\approx0.15$, is still allowed if one invokes
inhomogeneous nucleosynthesis, there is no clear test of such a
scenario. In particular the expected yields of `r'-process elements
(heavier than Si) is over a factor of $10^{5}$ below presently
observable bounds (Rauscher \etal 1994).

\subsubsection{`Cascade' nucleosynthesis:\label{casbbn}} 
 
In models with evaporating primordial black holes (Carlson \etal 1990)
(as also relic massive decaying particles (Dimopoulos \etal 1988)),
the nucleon density can be much higher with $\Omega_{\N}\approx1$,
since the photon and hadronic cascades triggered by the decay products
(see \sref{nonrel}) can reprocess the excess $\Hefour$ and
$\Liseven$ and create acceptable amounts of $\Htwo$ and
$\Hethree$ for decay lifetimes in the range
$\approx(2-9)\times10^{5}\sec$.\footnote{Another way in which a
 decaying particle can allow a large nucleon density is if it creates
 non-thermal electron antineutrinos during nucleosynthesis with
 energies of $\Or(\MeV)$ which can convert protons into neutrons at
 late times. Thus enough $\Htwo$ can be created (and not burnt
 further due to the low prevailing densities) while ${\Beseven}$ (which
 would have subsequently decayed to overproduce $\Liseven$) is
 destroyed (Scherrer 1984, Terasawa and Sato 1987). However the
 increased expansion rate due to the decaying particle also boosts the
 neutron fraction at freeze-out, hence the final $\Hefour$
 abundance. To allow $\Omega_{\N}h^{2}\approx0.2$ subject to the
 constraint $Y_{\pr}\leq0.25$ requires rather fine-tuned parameters
 e.g. a tau neutrino with $m_{\nu_{\tau}}\approx20-30\MeV$,
 $\tau\approx200-1000\sec$ and $m_{\nu_{\tau}}
 n_{\nu_{\tau}}/n_{\nu_{\el}}\approx0.03-0.1\MeV$ (Gyuk and Turner
 1994). Of course this possibility does not exist within the Standard
 Model and appears contrived even in extensions thereof since the
 decays must not create any visible energy (see \sref{nudecays}).}
The abundance of each element is determined by the fixed point of the
balance equation incorporating its production by hadronic showers and
destruction by photodissociation. The final $\Hefour$ abundance
depends only on the product of the decaying particle abundance and
baryonic branching ratio, while the other abundances are determined by
the ratio of the particle mass to the baryonic branching ratio. The
final abundance of $\Liseven$ can be made consistent with either the
Pop\,I or Pop\,II value but a large amount of ${\Lisix}$ is also
produced with ${\Lisix}/{\Liseven}\approx3-10$, in apparent conflict
with the observed bound of ${\Lisix}/{\Liseven}\leqsim0.1$ in Pop\,II
stars. Dimopoulos \etal (1988) have argued that since ${\Lisix}$ is
much more fragile than $\Liseven$, it may have been adequately
depleted through rotational mixing (see Deliyannis \etal 1990). Indeed
${\Lisix}$ has been recently detected in two Pop\,II stars with an
abundance \eref{Lisix} consistent with a primordial source,
although admittedly there are difficulties in reconciling such a
scenario with our present understanding of galactic chemical evolution
(e.g. Audouze and Silk 1989, Steigman \etal 1993).

The more modest aim of having a purely nucleonic universe with
$\Omega_{\N}\approx0.15$ can be achieved without (over)producing
${\Lisix}$ in the scenario of Gnedin and Ostriker (1992) wherein an
early generation of massive stars collapse to form black holes with
accretion disks which emit high energy photons capable of
photodissociating the overproduced helium and lithium. These authors
confirm, as was noted earlier by Dimopoulos \etal (1988), that
reprocessing by photodisintegration alone cannot allow values higher
than $\Omega_{\N}\approx0.2$, contrary to the results of Audouze \etal 
(1985).

\subsubsection{Neutrino degeneracy:\label{nudegen}} 

Finally we consider the possible role of neutrino degeneracy, which
was first studied by Wagoner \etal (1967). As mentioned earlier a
chemical potential in electron neutrinos can alter neutron-proton
equilibrium \eref{Xnchempot}, as well as increase the expansion rate,
the latter effect being less important. Consequently only the
abundance of $\Hefour$ is significantly affected and the allowed range
of $\eta$ is still determined by the adopted primordial abundances of
the other elements. For example, imposing $0.21{\leq}Y_{\pr}\leq0.25$
then requires (e.g. Yahil and Beaudet 1976, David and Reeves 1980,
Scherrer 1983, Terasawa and Sato 1988, Kang and Steigman 1992)
\begin{equation}
\label{chempotelim}
 - 0.06 \leqsim \xi_{\nu_{\el}} \leqsim 0.14\ ,
\end{equation}
assuming the chemical potential in other neutrino types, which can
only increase the expansion rate, to be negligible. (For orientation,
a value of $\xi_\nu\equiv\mu_\nu/T\approx\sqrt{2}$ is equivalent to
adding an additional neutrino flavour (with $\xi_\nu=0$) which we
consider in \sref{rel}.) Now, if both $\xi_{\nu_{\el}}$ and
$\xi_{\nu_{\mu,\tau}}$ are non-zero, then the lowering of the $n/p$
ratio at freeze-out (due to $\xi_{\nu_{\el}}$) may be compensated for
by the net speed-up of the expansion rate (due to
$\xi_{\nu_{{\el},\mu,\tau}}$), thus enabling the $\Hefour$ and the
${\Htwo}, {\Hethree}$ abundances to be all within observational bounds
even for large values of the nucleon density which are normally
disallowed (Yahil and Beaudet 1976). Even the surviving $\Liseven$
abundance, which is determined by a more complex interplay between
reactions with different $\eta$ dependence, may be made to match
either its Pop\,I value (David and Reeves 1980) or its Pop\,II value
(Olive \etal 1991, Starkman 1992, Kang and Steigman 1992). For
example, with $\xi_{\nu_{\el}}\approx1.4-1.6$ {\em and}
$\xi_{\nu_{\mu},\nu_{\tau}}\approx25-30$, one can have $\Omega_{\N}
h^{2}$ as high as $\approx1$ (e.g. Starkman 1992). However such an
universe would have been radiation dominated until well after the
(re)combination epoch, making it difficult to create the observed
large-scale structure (Freese \etal 1983). Taking such constraints
into account, Kang and Steigman (1992) quote the limits
\begin{equation}
\label{chempotmutau}
 - 0.06 \leqsim \xi_{\nu_{\el}} \leqsim 1.1\ , \quad 
 |\xi_{\nu_{\mu}, \nu_{\tau}}| \leqsim 6.9\ , \quad
 \eta \leqsim 1.9 \times 10^{-10}\ ,
\end{equation}
i.e. a critical density nucleonic universe is not permitted. In any
case, earlier theoretical studies which allowed the possibility of
generating such large lepton numbers (e.g. Langacker \etal 1982) need
to be reconsidered since we now know that ($B-L$ conserving) fermion
number violation is unsuppressed even in the Standard Model at
temperatures above the electroweak phase transition (see Shaposhnikov
1991, 1992). This would have converted part of any primordial lepton
asymmetry into a baryon asymmetry, hence one cannot plausibly have
$\xi_{\nu}\gg\eta$ without considerable fine-tuning (e.g. arranging
for large cancellations between lepton asymmetries of opposite signs
in different flavour channels). Therefore unless the lepton asymmetry
is somehow generated {\em after} the electroweak phase transition, or
unless the asymmetry is so large as to prevent the phase transition
itself (see Linde 1979), it is reasonable to conclude that neutrino
degeneracy cannot significantly affect the standard BBN model.

\section{Constraints on new physics\label{cons}} 

Having established the consistency of standard BBN, we will now use it
to constrain new physics beyond the Standard
$SU(3)_{\c}{\otimes}SU(2)_{\L}{\otimes}U(1)_{Y}$ Model. This has
usually been done for specific models but one can identify two general
classes of constraints, viz. those pertaining to stable particles
(e.g. new massless neutrinos, goldstone bosons) which are relativistic
during nucleosynthesis, and those concerned with massive, decaying
particles (e.g. massive neutrinos, gravitinos) which are
non-relativistic at this time. The extent to which we need to be
cautious in this enterprise depends on the sensitivity of the physics
under consideration to the light element abundances. For example, in
constraining massive decaying particles (\sref{nonrel}), we can use
together ``reliable'' as well as ``indirect'' bounds on elemental
abundances, because the constraints can be simply scaled for different
choices of the bounds. However in constraining the number of neutrino
species or other light stable particles we must be more careful since
the result is sensitive to the lower limit to $\eta$ following from
the abundance bounds on elements other than $\Hefour$ and cannot be
simply scaled for different choices of such bounds.

\subsection{Bounds on relativistic relics\label{rel}} 

The Standard Model contains only $N_{\nu}=3$ weakly interacting
massless neutrinos but in extensions beyond the SM there are often new
superweakly interacting massless (or very light) particles. Since
these do not usually couple to the $Z^0$ vector boson, there is no
constraint on them from the precision studies at \LEP of the
`invisible width' of $Z^0$ decays which establish the number of
$SU(2)_{\L}$ doublet neutrino species to be (LEP Elcetroweak Working
Group 1995)
\begin{equation}
\label{NnuLEP}
 N_{\nu} = 2.991 \pm 0.016\ ,
\end{equation}
and rule out any new particle with full strength weak interactions
which has mass smaller than $\approx\,m_{Z^0}/2$. (Previous
experiments which had set somewhat weaker bounds on $N_{\nu}$ are
reviewed by Denegri \etal (1990).)
 
Peebles (1966a) (see also Hoyle and Tayler 1964) had emphasized some
time ago that new types of neutrinos (beyond the $\nu_{\el}$ and
$\nu_{\mu}$ then known) would increase the relativistic energy
density, hence the expansion rate, during primordial nucleosynthesis,
thus increasing the yield of $\Hefour$. Subsequently Shvartsman
(1969) pointed out that new superweakly interacting particles would
have a similar effect and could therefore be constrained through the
observational bound on the helium abundance. This argument was later
quantified by Steigman \etal (1977) for new types of neutrinos and by
Steigman \etal (1979) for new superweakly interacting particles. While
the number of (doublet) neutrinos is now well known from laboratory
experiments, the constraint on superweakly interacting particles from
nucleosynthesis continues to be an unique probe of physics beyond the
Standard Model and is therefore particularly valuable.  

As we have seen earlier, increasing the assumed number of relativistic
neutrino species $N_{\nu}$ increases $g_{\rho}$ \eref{gsgrhonow},
hence the expansion rate \eref{H}, causing both earlier freeze-out
with a larger neutron fraction (see \eref{Xnfin}) and earlier onset of
nucleosynthesis (see \eref{3min}), all together resulting in a larger
value of $Y_{\pr}({\Hefour})$ (see \eref{Ypminefull}).\footnote{For a
very large speed-up
 factor, there is no time for $\Htwo$ and $\Hethree$ to be burnt
 to $\Hefour$, hence $Y_{\pr}$ begins to decrease with $N_{\nu}$
 and drops below $25\%$ for $N_{\nu}\geqsim6600$ (Ellis and Olive
 1983). However this would overproduce $\Htwo$ and $\Hethree$ by
 several orders of magnitude (Peebles 1971, Barrow 1983).} One can
parametrize the energy density of new relativistic particles in terms
of the equivalent number $N_{\nu}$ of doublet neutrinos so that the
limit on $\Delta N_{\nu}\,({\equiv}N_{\nu}-3)$ obtained by comparing
the expected $\Hefour$ yield with its observational upper bound
constrains the physics which determines the relic abundance of the new
particles. (A complication arises if the tau neutrino is sufficiently
massive so as to be non-relativistic during nucleosynthesis, as we
will consider later; then the number of relativistic (doublet) neutrinos
during nucleosynthesis would be 2 rather than 3.) The interaction rate
keeping a superweakly interacting particle in thermal equilibrium will
typically fall behind the Hubble expansion rate at a much higher
`decoupling' temperature than the value of a few MeV for (doublet)
neutrinos. As discussed in \sref{thermhist}, if the comoving
specific entropy increases afterwards, e.g. due to annihilations of
massive particles in the Standard Model, the abundance of the new
particles will be diluted relative to that of neutrinos, or
equivalently their temperature will be lowered relatively
(see \eref{TibyT}) since neutrinos in the SM have the
same temperature as that of photons down to $T\,\sim\,m_{\el}$. Hence
the energy density during nucleosynthesis of new massless particles
$i$ is equivalent to an effective number $\Delta N_{\nu}$ of
additional doublet neutrinos:
\begin{equation}
\label{Nnuequiv}
 \Delta N_{\nu} = f_{\B,\F} \sum_{i} \frac{g_{i}}{2} 
  \left(\frac{T_{i}}{T_{\nu}}\right)^4\ ,
\end{equation}
where $f_{\B}=8/7$ (bosons) and $f_{\F}=1$ (fermions), and
$T_{i}/T_{\nu}$ is given by \eref{TibyT}. Thus the number of
such particles allowed by a given bound on $\Delta N_{\nu}$ depends on
how small $T_{i}/T_{\nu}$ is, i.e. on the ratio of the number of
interacting relativistic degrees of freedom at $T_{\D}$ (when $i$
decouples) to its value at a few MeV (when neutrinos decouple). Using
\tref{tabthermhist} we see that $T_{\D}>m_{\mu}$ implies
$T_{i}/T_{\nu}<0.910$ while $T_{\D}>T_{\c}^{\qh}$ bounds
$T_{i}/T_{\nu}<0.594$; the smallest possible value of $T_{i}/T_{\nu}$
in the Standard Model is 0.465, for $T_{\D}>T_{\c}^{\EW}$ (Olive \etal
1981a).

For example, consider a new fermion $\F$, e.g. a singlet (sterile)
neutrino. Its decoupling temperature $T_{\D}$ can be approximately
calculated, as for a doublet neutrino \eref{Tdecnu}, by equating the
interaction rate to the Hubble rate. For definiteness, consider
annihilation to leptons and parametrize the cross-section as
\begin{equation}
\label{cross}
 \langle \sigma v \rangle_{{\ell} \bar{\ell} \to \F \bar{\F}} = 
  \left(\frac{n_{\F}^{\eqm}}{n_{\ell}^{\eqm}}\right)^2 
  \langle \sigma v \rangle_{\F \bar{\F} \to {\ell} \bar{\ell}} 
  \equiv \alpha T^2 .
\end{equation}  
Then equating the annihilation rate $\Gamma_{\ann}^{\eqm}=n_{\F}
\langle\sigma{v}\rangle_{F\bar{F}\to{\ell}\bar{\ell}}$ to the
expansion rate $H$ \eref{H} gives,
\begin{equation}
\label{Tdecnew}
 T_{\D} \simeq 7.2 \times 10^{-7}\ \GeV g_{\rho}^{1/6} \alpha^{-1/3} .
\end{equation}
Thus the smaller the coupling, the earlier the particle decouples,
e.g. $T_{\D}>m_{\mu}$ if
$\alpha<1.4\times10^{-15}{\GeV}^{-4}$,\footnote{A more careful
 analysis of decoupling actually yields a less stringent bound
 \eref{alphabound}.} and $T_{\D}>T_{\c}^{\qh}$ if
$\alpha<2.1\times10^{-16}{\GeV}^{-4}(T_{\c}^{\qh}/0.3\GeV)^{-3}$. The
energy density of the new particle during nucleosynthesis is then,
respectively, equivalent to $\Delta N_{\nu}=0.69$ and $\Delta
N_{\nu}=0.12$. Thus if the observationally inferred bound was,
say $\Delta N_{\nu}<1$, then one such singlet neutrino would be
allowed per fermion generation only if they decoupled above
$T_{\c}^{\qh}$. This requirement would impose an interesting
constraint on the particle physics model in which such neutrinos
appear.

We can now appreciate the significance of the precise bound on
$N_{\nu}$ from nucleosynthesis. This depends on the adopted elemental
abundances as well as uncertainties in the predicted values. Taking
into account experimental uncertainties in the neutron lifetime and in
nuclear reaction rates, the $\Hefour$ abundance can be calculated to
within $\pm0.5\%$ (see \eref{Ypmine}). In contrast, the
observationally inferred upper bound to $Y_{\pr}$ is uncertain by as
much as $\approx4\%$ (compare \eref{Yppagellim} and
\eref{Ypitllim}). More importantly, the bound on $N_{\nu}$ can only be
derived if the nucleon-to-photon ratio $\eta$ (or at least a lower
bound to it) is known (see \eref{Ypminefull}). This involves
comparison of the expected and observed abundances of other elements
such as $\Htwo$, $\Hethree$ and $\Liseven$ which are much more poorly
determined, both observationally and theoretically. It is a
wide-spread misconception that the $\Hefour$ abundance {\em alone}
constrains $N_{\nu}$; in fact the effect of a faster expansion rate
can be balanced by the effect of a lower nucleon density so that
$N_{\nu}$ is {\em not at all constrained} for
$\eta\leqsim5\times10^{-11}$ which is quite consistent with the direct
observational limit \eref{etarange}, as well as the reliable upper
bound \eref{reliable} to the $\Liseven$ abundance! Of course with such
a low nucleon density large amounts of $\Htwo$, $\Hethree$ and
$\Liseven$ would be created, in conflict with the ``reasonable''
observational bounds \eref{reasonable}. Hence one can derive a lower
limit to $\eta$ from the abundances of these elements and then
constrain $N_{\nu}$ given an observational upper bound on
$Y_{\pr}$. Thus the reliability of the BBN constraint on $N_{\nu}$ is
essentially determined by the reliability of the lower limit to
$\eta$, a fact that has perhaps not been always appreciated by
particle physicists who use it to constrain various interesting
extensions of the Standard Model. To emphasize this point, we briefly
review the history of this constraint and comment in particular on
those mentioned by the Particle Data Group (1992, 1994, 1996).
 
Steigman \etal (1977) originally quoted the constraint
$N_{\nu}\leqsim7$ following from their assumption that
$\Omega_{\N}h^{2}\geqsim0.01$ (i.e. $\eta>2.8\times10^{-10}$) and the
conservative bound $Y_{\pr}\leq0.29$. Yang \etal (1979) argued that a
more restrictive bound $Y_{\pr}\leq0.25$ was indicated by observations
and concluded that in this case {\em no} new neutrinos beyond
$\nu_{\el}$, $\nu_{\mu}$ and $\nu_{\tau}$ were allowed. Their adopted
limit $\Omega_{\N} h^{2}\geqsim0.01$ was based on the assumption
(following Gott \etal 1974) that the dynamics of galaxies is governed
by nucleonic matter. Following the growing realization that the dark
matter in galaxies could in fact be non-baryonic, Olive \etal (1981b)
proposed a much weaker limit of $\eta>2.9\times10^{-11}$ following
from just the observed {\em luminous} matter in galaxies, and noted
that {\em no} constraint on $N_{\nu}$ could then be derived for any
reasonable bound on $Y_{\pr}$. These authors presented a systematic
analysis of how the inferred constraint on $N_{\nu}$ varies with the
assumed nucleon density, neutron lifetime and $\Hefour$ abundance and
emphasized the need for a detailed investigation of the other
elemental abundances to better constrain $\eta$ and $N_{\nu}$.

This was done by Yang \etal (1984) who proposed that the sum of
primordial $\Htwo$ and $\Hethree$ could be bounded by considerations
of galactic chemical evolution; using Solar system data
\eref{Solarsys} they inferred
$[({\Htwo}+{\Hethree})/{\H}]_{\pr}\leqsim10^{-4}$ and from this
concluded $\eta \geq 3\times10^{-10}$ (see \eref{indirect}). This
yielded the often quoted constraint
\begin{equation}
\label{Nnu4}
 N_{\nu} \leq 4\ ,
\end{equation}
assuming that $Y_{\pr}\leq0.25$ and $\tau_{\n}>900\sec$. Ellis \etal
(1986b) pointed out that this constraint could be relaxed since (a)
laboratory experiments allowed for the neutron lifetime to be as low
as $883\sec$ , (b) Possible systematic observational errors allowed
for $Y_{\pr}({\Hefour})$ to be as high as 0.26, and (c) the
observational indication that there is net destruction of $\Hethree$
in stars (see discussion before \eref{He3HII}) allowed for $[({\Htwo}+
{\Hethree})/{\H}]_{\pr}$ to be as high as $5\times10^{-4}$. Thus {\em
conservatively} BBN allows upto
\begin{equation}
\label{Nnu5.5}
 N_{\nu} \leqsim 5.5\ ,
\end{equation}
neutrinos. However Steigman \etal (1986) reasserted the constraint in
\eref{Nnu4}.

Subsequently precision measurements of the neutron lifetime
(e.g. Mampe \etal 1989) confirmed that it was lower than had been
previously assumed. Moreover, Krauss and Romanelli (1990) quantified
the uncertainties in the theoretical predictions by a Monte Carlo
method taking all experimental uncertainties in input reaction rates
into account. Combining these results with a detailed study of the
$\Liseven$ abundance evolution in halo stars, Deliyannis \etal (1989)
presented a new lower limit of $\eta>1.2\times10^{-10}$ on the basis
of the Pop\,II $\Liseven$ observations. They noted that this would
allow upto
\begin{equation}
\label{Nnu5}
 N_{\nu} \leq 5
\end{equation}
neutrino species to be consistent with a primordial $\Hefour$ mass
fraction less than $25\%$. Olive \etal (1990) continued to adopt the
indirect bound $[({\Htwo}+{\Hethree})/{\H}]_{\pr}\leq1.1\times10^{-4}$
and assumed a more stringent upper bound $Y_{\pr}\leq0.24$
\eref{Yppagellim}, so that their derived constraint on $N_{\nu}$
became even more restrictive
\begin{equation}
\label{Nnu3.4}
 N_{\nu} \leq 3.4\ .
\end{equation}
Making nearly identical assumptions, viz. $\tau_{\n}>882\sec$,
$[({\Htwo}+{\Hethree})/{\H}]_{\pr}\leq10^{-4}$,
$Y_{\pr}({\Hefour})<0.24$, Walker \etal (1991) quoted an even tighter
limit 
\begin{equation}
\label{Nnu3.3}
 N_{\nu} \leq 3.3\ .
\end{equation}
Subsequently, the possible detection of a large primordial deuterium
abundance in a Lyman-$\alpha$ cloud (Songaila \etal 1994, Rugers and
Hogan 1996a,b), as well as observational indications that the helium-4
abundance may have been systematically underestimated (Sasselov and
Goldwirth 1994, Izotov \etal 1994, 1996) have further justified the
caution advocated by Ellis \etal (1986b) in deriving this important
constraint. Nevertheless Copi \etal (1995a) have recently reasserted
the upper limit of 3.4 neutrino species, continuing to adopt similar
bounds as before, viz. $\tau_{\n}>885\sec$,
$[({\Htwo}+{\Hethree})/{\H}]_{\pr}\leq1.1\times10^{-4}$ and
$Y_{\pr}({\Hefour})<0.243$.


Kernan and Krauss (1994) have emphasized that the procedure used by
all the above authors is statistically inconsistent since the
abundances of the different elements are {\em correlated} and the use
of symmetric confidence limits on the theoretical abundances is overly
conservative. (Moreover, only Deliyanis \etal (1989) allowed for
errors in the expected yields due to reaction rate uncertainties.)
Just as in the case of the derived limits on $\eta$ (see discussion
before \eref{etalimI}), a correct analysis allowing for correlations
would yield a tighter constraint on $N_{\nu}$. Kernan and Krauss
(1994) illustrate this by considering the abundance bounds
$[({\Htwo}+{\Hethree})/{\H}]_{\pr}\leq10^{-4}$ and
$Y_{\pr}({\Hefour})\leq0.24$ advocated by Walker \etal (1991) and
determining the $\95cl$ limits on $\eta$ and $N_{\nu}$ by requiring
that at least 50 out of 1000 Monte Carlo runs lie within the {\em
joint} range bounded by both ${\Htwo}+{\Hethree}$ and $\Hefour$. As
shown in \fref{He4vsHe3Dmc}\,(a), this imposes tighter constraints
than simply requiring that 50 runs lie, either to the left of the
$({\Htwo}+{\Hethree})$ bound (for low $\eta$), or below the $\Hefour$
bound (for high $\eta$). Moreover, the procedure of simply checking
whether the symmetric $\95cl$ limit for an individual elemental
abundance is within the observational bound gives an even looser
constraint. \Fref{He4vsHe3DNnu}(a) plots the number of Monte Carlo
runs (out of 1000) which satisfy the joint observational bounds as a
function of $\eta$ for different values of $N_{\nu}$; it is seen that
the $\95cl$ limit is
\begin{equation}
\label{Nnu3.04}
 N_{\nu} < 3.04\ ,
\end{equation}
rather than 3.3 as quoted by Walker \etal (1991).\footnote{If
 correlations had not been included, the limit would have been 3.15,
 the difference from Walker \etal being mainly due to the $\approx1\%$
 increase in the (more carefully) calculated $\Hefour$ abundance.} As
emphasized by Kernan and Krauss (1994), this is an extremely stringent
constraint, if indeed true, on physics beyond the Standard Model. For
example even a singlet neutrino which decouples above $T_{\c}^{\EW}$
will be equivalent to 0.047 extra neutrino species, and is therefore
{\em excluded}. More crucially, the helium mass fraction (for
$N_{\nu}$=3) is now required to exceed $0.239$ for consistency with
the assumed bound on ${\Htwo}+{\Hethree}$, so a measurement below this
value would rule out standard BBN altogether!  Hence these authors
draw attention again to the possibility that the systematic
uncertainty in the usually quoted value \eref{Ypptss} of
$Y_{\pr}({\Hefour})$ has been underestimated. The $N_{\nu}$ limit can
be simply parametrized as
\begin{equation}
\label{Nnu3.81}
  N_{\nu} \leq 3.07 + 74.1\ (Y_{\pr}^{\max} - 0.24)\ ,
\end{equation}
where $\tau_{\n}=887\pm2\sec$ has been used (Krauss and Kernan 1995),
so $N_{\nu}\leq3.8$ for $Y_{\pr}^{\max}=0.25$ \eref{Ypitllim}. Olive
and Steigman (1995b) assign a low systematic error of $\pm0.005$ to
their extrapolated primordial helium abundance
$Y_{\pr}({\Hefour})=0.232\pm0.003$ \eref{Ypos} and thus obtain a best
fit value of $N_{\nu}=2.17\pm0.27\,(\stat)\pm0.42\,(\syst)$. This is
unphysical if there are indeed at least 3 massless neutrinos so they
compute the upper limit on $N_{\nu}$ restricting attention to the {\it
physical} region alone (see Particle data Group 1996), obtaining
\begin{equation}
\label{Nnu3.6}
 N_{\nu}<3.6\ .
\end{equation}
They also impose the weaker condition $N_{\nu}\geq2$ (as would be
appropriate if the $\nu_\tau$ was massive and decayed before
nucleosynthesis) to obtain the bound $N_{\nu}<3.2$.


As we have discussed earlier, the indirect bound $[({\Htwo}+
{\Hethree})/{\H}]_{\pr}\leq10^{-4}$ \eref{He3plusD} used above is
rather suspect and it would be more conservative to use the
``reasonable'' observational bounds
${\Htwo}/{\H}\leqsim2.5\times10^{-4}$ \eref{Dlyalim} and
$({\Liseven}/{\H})_{\pr}^{\I\I}\leq2.6\times10^{-10}$
\eref{Li7thorburn} to constrain $\eta$. A Monte Carlo exercise has
been carried out for this case (Kernan and Sarkar 1996) and yields the
constraint,
\begin{equation}
\label{Nnu4.53}
 N_{\nu} \leq 3.75 + 78\ (Y_{\pr}^{\max} - 0.24)\ ,
\end{equation}
if we require all constraints to be {\em simultaneously}
satisfied. Thus, as shown in \fref{He4vsHe3DNnu}(b), the conservative
limit is $N_{\nu}\leq4.53$, if the $\Hefour$ mass fraction is as high
as $25\%$ \eref{Ypitllim}. Equivalently, we can derive a limit on the
`speedup rate' of the Hubble expansion due to the presence of the
additional neutrinos which contribute 7/4 each to $g_{\rho}$, the
number of relativistic degrees of freedom \eref{grho}, increasing it
above its canonical value of 43/4 at this epoch. Then the
time-temperature relationship \eref{trad} becomes modified as
$t\to{t'}=\xi^{-1}t$, where
\begin{equation}
\label{speedup}
 \xi \equiv \left[1 + \frac{7}{43} (N_{\nu} -3)\right]^{1/2}\ .
\end{equation}
Since $\xi$ cannot far exceed unity, we obtain using \eref{Nnu4.53}, 
\begin{equation}
\label{speeduplim}
 \xi - 1 \leqsim 0.061 + 6.3\,(Y_{\pr}^{\max} - 0.24)\ .
\end{equation}

In contrast to our conservative approach, Hata \etal (1995)
deduce the even more restrictive values
$({\Htwo}/{\H})_{\pr}=3.5^{+2.7}_{-1.8}\times10^{-5}$ and
$({\Hethree}/{\H})_{\pr}=1.2\pm0.3\times10^{-5}$ (``$\95cl$'') using a
chemical evolution model normalized to Solar system abundances and
convolving with BBN predictions (Hata \etal 1996a). Combining this with
the estimate
$Y_{\pr}({\Hefour})=0.232\pm0.003\,(\stat)\pm0.005\,(\syst)$ by Olive
and Steigman (1995a), and adopting
$({\Liseven}/{\H})_{\pr}=1.2^{+4.0}_{-0.5}\times10^{-10}$
(``$\95cl$'') these authors obtain $N_{\nu}=2.0\pm0.3$. Thus they are
led to conclude that $N_{\nu}<2.6\,(\95cl)$ so the Standard Model
($N_{\nu}=3$) is ruled out at the $98.6\%$ c.l.! However the
confidence levels quoted on their adopted elemental abundances are
unreliable for the detailed reasons given in \sref{abund}, hence
we believe this conclusion is not tenable.

The BBN bound on $N_{\nu}$ is under renewed discussion (e.g. Fields
and Olive 1996, Fields \etal 1996, Cardall and Fuller 1996a, Copi
\etal 1996, Hata \etal 1996b) following the first direct measurements
of deuterium at high redshifts which have called into question the
chemical evolution models employed in earlier work (e.g. Yang \etal
1984, Steigman \etal 1986, Walker \etal 1991). We consider the bounds
given in \eref{Nnu4.53} and \eref{speeduplim} to be conservative and
advocate their use by particle physicists seeking to constrain models
of new physics.

\subsection{Bounds on non-relativistic relics\label{nonrel}} 

The presence during nucleosynthesis of a {\em non-relativistic}
particle, e.g. a massive neutrino, would also increase the energy
density, hence the rate of expansion, and thus increase the
synthesized abundances. This effect is however different from that due
to the addition of a new relativistic particle, since the energy
density of a non-relativistic particle decreases $\propto\,T^3$
(rather than $\propto\,T^4$) hence the speed-up rate due to the
non-relativistic particle is not constant but increases steadily with
time. If the particle thus comes to matter-dominate the universe much
earlier than the canonical epoch \eref{matdom}, then it must
subsequently decay (dominantly) into relativistic particles so that
its energy density can be adequately redshifted, otherwise the bounds
on the age and expansion rate of the universe today would be violated
(Sato and Kobayashi 1977, Dicus \etal 1978a). If such decays are into
{\em interacting} particles such as photons or
electromagnetically/strongly interacting particles which increase the
entropy, the nucleon-to-photon ratio will decrease (Miyama and Sato
1978, Dicus \etal 1978b).\footnote{This assumes thermalization of the
released energy which is very efficient for decay lifetimes
$\leqsim{10^5}\sec$ (Illarianov and Sunyaev 1975, Sarkar and Cooper
1984). For longer lifetimes thermalization is incomplete, but then the
absence of a spectral distortion in the CMBR sets equally restrictive
constraints on the decaying particle abundance (e.g. Ellis \etal 1992,
Hu and Silk 1994).} As we have seen, the observationally inferred
upper bound on the synthesized $\Hefour$ abundance implies an {\it
upper} limit to $\eta_{\ns}$, the nucleon-to-photon ratio during the
nucleosynthesis epoch, while observations of luminous matter in the
universe set a {\em lower} limit \eref{etarange} to the same ratio
today. Hence we can require particle decays after nucleosyntheis to
not have decreased $\eta$ by more than a factor $\eta_{\ns}/\eta_{0}$,
having calculated the elemental yields (and $\eta_{\ns}$) taking into
account the increased expansion rate due to the decaying
particle. However if the particle decays into {\it non-interacting}
particles, e.g. neutrinos or hypothetical goldstone bosons which do
not contribute to the entropy, then the only constraint comes from the
increased expansion rate during nucleosynthesis. There may be
additional effects in both cases if the decays create electron
neutrinos/antineutrinos which can alter the chemical balance between
neutrons and protons (Dicus \etal 1978a) and thus affect the yields of
$\Htwo$ and $\Hethree$ (Scherrer 1984).

First we must calculate how the dynamics of the expansion are altered
from the usual radiation-dominated case. Given the thermally-averaged
self-annihilation cross-section of the $x$ particle, one can obtain
the relic abundance in ratio to photons the $x$ particle would have at
$T\,\ll\,m_{\el}$, assuming it is stable, using the methods outlined
by Srednicki \etal (1988) and Gondolo and Gelmini (1991). An
approximate estimate may be obtained from the simple `freeze-out'
approximation (see Kolb and Turner 1990) of determining the
temperature at which the self-annihilation rate falls behind the
Hubble expansion rate. The surviving relic abundance is then given by
the equilibrium abundance at this temperature:
\begin{equation}
\label{freezeoutabun}
 \left(\frac{m_{x}}{\GeV}\right) \left(\frac{n_{x}}{n_{\gamma}}\right) \approx
 \left(\frac{\langle\sigma v\rangle}{8\times10^{-18}{\GeV}^{-2}}\right)^{-1}\ .
\end{equation}
Since we will generally be concerned with decay lifetimes much longer
than $\approx1\sec$, this can be taken to be the initial value of the
decaying particle abundance. We can now identify the temperature
$T_{\m}$ at which the particle energy density $\rho_{x}\,(\simeq
m_{x}n_{x}$) would equal the radiation energy density
$\rho_{\R}\,(=\pi^2g_{\rho}/30T^4)$, viz.
\begin{equation}
\label{Tm}
 T_{\m} \equiv \frac{60\,\zeta (3)}{g_{\rho}\,\pi^4} \frac{m_{x}
  n_{x}}{n_{\gamma}}\ .
\end{equation}  
If the particle decays at a temperature below $T_{\m}$, then it would
have matter-dominated the universe before decaying and thus
significantly speeded up the expansion. The usual time-temperature
relationship \eref{trad} is thus altered to
\begin{equation}
\label{tradm}
\fl t \simeq -\left[\frac{3 M_{\Pl}^2}{8 \pi (\rho_{x} + 
  \rho_{\R})}\right]^{1/2}
  \int \frac{dT}{T} = \left(\frac{5}{\pi^3 g_{\rho}}\right)^{1/2}
  \frac{M_{\Pl}}{T_{\m}^2} \left[\left(\frac{T_{\m}}{T} - 
  2 \right)\left(\frac{T_{\m}}{T} + 1 \right)^{1/2} + 2 \right] . 
\end{equation}
This reduces in the appropriate limit ($T_{\m}\ll\,T$) to the
radiation-dominated case. After the particles decay, the universe
reverts to being radiation-dominated if the decay products are
massless. If we assume that all the $x$ particles decay {\it
simultaneously} when the age of the universe equals the particle
lifetime, then the temperature at decay, $T_{\d}$, is given by the
above relationship setting $t=\tau_{x}$.

\subsubsection{Entropy producing decays:\label{visdec}} 

First let us consider the case when the decays create
electromagnetically interacting particles.  Following Dicus \etal
(1978a,b) we assume that the effect of the decays is to cause a `jump' in
the temperature, which we obtain from energy conservation to be
\begin{equation}
\label{Tjump}
 T\,(t > \tau_{x}) = [T\,(t < \tau_{x})]^{3/4} [T\,(t < \tau_{x}) + 
  f_{\gamma} g_{\rho} T_{\m}]^{1/4}\ , 
\end{equation}
where $f_{\gamma}$ is the fraction of $\rho_{x}$ which is ultimately
converted into photons. The resulting change in $\eta$ is 
\begin{equation}
\label{etachange}
 \frac{\eta\,(t < \tau_{x})}{\eta\,(t > \tau_{x})} =
  \left(1 + \frac{f_{\gamma}\,g_{\rho}\,T_{\m}}{2\,T_{\d}}\right)^{3/4} \leq
  \frac{\eta_{\ns}}{\eta_{0}}\ . 
\end{equation}
(In fact, radiative particle decays which follow the usual exponential
decay law cannot {\em raise} the photon temperature in an
adiabatically cooling universe (cf. Weinberg 1982), but only slow down
the rate of decrease, as noted by Scherrer and Turner (1985). However
their numerical calculation shows that this does not significantly
affect the change in $\eta$, which turns out to be only $\approx10\%$
larger than the estimate above.) From \eref{tradm} we obtain
$\tau_{x}\,\propto\,T_{\m}^{-1/2}T_{\d}^{3/2}$ for
$T_{\m}\,\gg\,T_{\d}$, i.e. if the $x$ particles decay well after the
universe has become dominated by their energy density.  In this
approximation, the constraint on the decay lifetime is (Ellis \etal
1985b)
\begin{equation}
\label{tauxlim}
 \left(\frac{\tau_{x}}{\sec}\right) \leqsim 0.8\ f_{\gamma}^{3/2}
  \left(\frac{T_{\m}}{\MeV}\right)^{-2}
  \left(\frac{\eta_{\ns}}{\eta_{0}}\right)^2 ,
\end{equation}
if we take $g_{\rho}=3.36$, i.e. for $T_{\d}{\ll}m_{\el}$. 

As mentioned above, the upper limit to $\eta_{\ns}$ corresponding to
the conservative requirement $Y_{\pr}({\Hefour})<0.25$ \eref{Ypitllim}
depends on the extent to which the expansion rate during
nucleosynthesis is influenced by the massive particle.  Such limits
were obtained by Kolb and Scherrer (1982) (following Dicus \etal
1978b) who modified the standard BBN code to include a massive
neutrino (with the appropriate energy density) and examined its effect
on the elemental yields. The effect should be proportional to the
neutrino energy density, which rises $\propto\,m_{\nu}$ as long as the
neutrinos remain relativistic at decoupling, i.e. for $m_{\nu}$ less
than a few MeV, and falls thereafter $\propto\,m_{\nu}^{-2}$ (e.g. Lee
and Weinberg 1977). Indeed the synthesized abundances are seen to
increase with increasing neutrino mass upto $m_{\nu}\approx5\MeV$, and
fall thereafter as $m_{\nu}$ increases further. Kolb and Scherrer
found that a neutrino of mass $m_{\nu}\sim0.1-10\MeV$ alters the
$\Hefour$ abundance {\em more} than a massless neutrino and that the
neutrino mass has to exceed 20 MeV before the change in the abundance
becomes acceptably small, while for $m_{\nu}\,\geqsim\,25\MeV$ there
is negligible effect on nucleosynthesis. (In fact the abundances of
$\Htwo$ and $\Hethree$ are also increased, and by a factor which may
even exceed that for the $\Hefour$ abundance. This is because these
abundances are sensitive to the expansion rate at
$T\approx0.04-0.08\MeV$ when the strong interactions which burn
deuterium freeze-out, and the massive particle may come to
matter-dominate the expansion precisely at this time.) From the
results of Kolb and Scherrer (1982) it can be inferred that
$\eta_{\ns}\,\propto\,T_{\m}^{-1/2}$ for $T_{\m}\geqsim10^{-2}\MeV$
and $\eta_{\ns}\approx\con$ for $T_{\m}\leqsim10^{-2}\MeV$, hence it
is easy to generalize the constraints obtained for neutrinos to any
other particle which is non-relativistic during nucleosynthesis,
i.e. with a mass larger than a few MeV. Ellis \etal (1985b) used these
values of $\eta_{\ns}$ to obtain the following restrictions on the
energy density of the decaying particle as a function of the lifetime:
\begin{equation}
\label{nxlimvis}
\eqalign{
  \left(\frac{m_{x}}{\GeV}\right) \left(\frac{n_{x}}{n_{\gamma}}\right) 
  &\leqsim 6.0 \times 10^{-3}\left(\frac{\tau_{x}}{\sec}\right)^{-1/3}
  f_{\gamma}^{-1/2} \left(\frac{\eta_{0}}{1.8 \times 10^{-11}}\right)^{-2/3} 
   \\ \nonumber
  &{\for} \quad t_{\ns} \leqsim \tau_{x} \leqsim 3.8 \times 10^5
   f_{\gamma}^{3/2} {\sec}\ ,\\
  &\leqsim 5.1 \times 10^{-2} \left(\frac{\tau_{x}}{\sec}\right)^{-1/2} 
   f_{\gamma}^{3/4} \left(\frac{\eta_{0}}{1.8 \times 10^{-11}}\right)^{-1} 
   \\ \nonumber
  &{\for} \quad \tau_{x} \geqsim 3.8 \times 10^5 f_{\gamma}^{3/2} {\sec}\ .} 
\end{equation}
These constraints should be valid for radiative decays occuring after
the begining of nucleosynthesis at $t_{\ns}\simeq180\sec$ as indicated
by the dotted line in \fref{nxlimentexp}(a). Scherrer and Turner
(1988a) have performed a numerical study in which the cosmological
evolution is computed taking into account the exponentially decreasing
energy density of the massive particle and the correspondingly
increasing energy density of its massless decay products, without
making any approximations (c.f. the assumption above that
$T_{\m}\,\gg\,T_{\d}$). These authors were thus able to study how the
constraint weakens as $\tau_{x}$ decreases below $t_{\ns}$, as shown
by the full line in \fref{nxlimentexp}(a). The reason the curve turns
up sharply is that helium synthesis is unaffected by decays which
occur prior to the epoch ($T\approx0.25\MeV$) when the ${\n}/{\p}$
ratio freezes out (see \eref{Xninfty}).\footnote{We have
 corrected for the fact that these authors referred to the value of
 $n_{x}/n_{\gamma}$ at $T\approx100\MeV$, i.e. before
 ${\el}^{+}{\el}^{-}$ annihilation, while we always quote the value at
 $T{\ll}m_{\el}$, i.e. the abundance the particle would have today if
 it had not decayed. Also, they adopted a slightly different bound:
 $\eta_{0}\,\geqsim\,3\times10^{-11}$.} In addition, Scherrer and
Turner studied the effect on the $\Htwo$ and $\Hethree$ abundances and
imposed the indirect bound
$[({\Htwo}+{\Hethree})/{\H}]_{\pr}\leqsim10^{-4}$ \eref{He3plusD} to
obtain a more restrictive constraint shown as the dashed line in
\fref{nxlimentexp}(a). All these curves are drawn assuming
$f_{\gamma}=1$ and can be scaled for other values of $f_{\gamma}$ (or
$\eta_{0}$) following \eref{nxlimvis}.

When the decays occur {\em before} the nucleosynthesis era, the
generation of entropy can only be constrained by requiring that the
baryon asymmetry generated at earlier epochs should not have been
excessively diluted, as was noted by Harvey \etal (1981). Scherrer
and Turner (1988a) assumed that the initial nucleon-to-photon ratio is
limited by $\eta_{\in}<10^{-4}$, as was believed to be true for GUT
baryogenesis (see Kolb and Turner 1983), and combined it with the lower
limit on the value of $\eta$ today, to obtain the constraints shown as
dot-dashed lines in \fref{nxlimentexp}(a). Obviously these are very
model dependent since the initial value of $\eta$ may be higher by
several orders of magnitude, as is indeed the case in various non-GUT 
models of baryogenesis (see Dolgov 1992).


\subsubsection{`Invisible' decays:\label{invisdec}} 

We should also consider the possibility that $f_{\gamma}=0$, i.e.  the
decays occur into massless particles such as neutrinos or hypothetical
goldstone bosons. In this case there is no change in the entropy,
hence the constraints discussed above do not apply. However we can
still require that the speed-up of the expansion rate during
nucleosynthesis not increase the synthesized abundances
excessively. As mentioned earlier, Kolb and Scherrer (1982) found that
when a massive neutrino was incorporated into the standard BBN code,
the observational bound
$Y_{\pr}({\Hefour})<0.25$ \eref{Ypitllim} is respected when
the neutrino mass exceeds 20 MeV. We can generalize their result to
any particle which is non-relativistic at nucleosynthesis by demanding
that it should not matter-dominate the expansion any earlier than a 20
MeV neutrino. This implies the constraint (Ellis \etal 1985b)
\begin{equation}
\label{nxlimstable}
 \left(\frac{m_{x}}{\GeV}\right) \left(\frac{n_{x}}{n_{\gamma}}\right)
  \leqsim 1.6 \times 10^{-4}\ ,
\end{equation}
which is valid for particles which decay after nucleosynthesis,
i.e. for $\tau_{x}{\geqsim}t_{\ns}$ as indicated by the dotted line in
\fref{nxlimentexp}(b). Scherrer and Turner (1988b) obtain a similar
requirement from a detailed numerical calculation, as shown by the
full line in the same figure. Again, they were able to study how the
constraint relaxes as $\tau_{x}$ becomes smaller than $t_{\ns}$. Since
the decay products are massless, the effect is then the same as the
addition of new relativistic degrees of freedom. Imposing the
additional indirect bound
$[({\Htwo}+{\Hethree})/{\H}]_{\pr}\leqsim10^{-4}$ \eref{He3plusD},
equivalent in this context to allowing one new neutrino species (see
\eref{Nnu4}), then yields the constraint
\begin{equation}
\label{nxliminvis}
 \left(\frac{m_{x}}{\GeV}\right) \left(\frac{n_{x}}{n_{\gamma}}\right)
  \leqsim 9.8 \times 10^{-4}\ \left(\frac{\tau_{x}}{\sec}\right)^{-1/2}\ ,
\end{equation}
valid for $\tau_{x}\ll\tau_{\ns}$ as shown by the dashed line in
\fref{nxlimentexp}(b). This requirement is {\em more} restrictive than
the corresponding one \eref{nxlimvis} for decays which create entropy,
hence the two constraints should be weighted with the appropriate
branching ratios in order to obtain the correct constraint for a
particle whose decays produce both non-interacting particles and
photons. Neutrino decay products actually present a special case since
these are not entirely non-interacting. Indeed if decaying particles
create a (non-thermal) population of electron (anti)neutrinos, these
will bias the chemical balance between protons and neutrons towards
the latter through the reaction ${\n}{\el}^{+}\to{\p}\bar{\nu_{\el}}$;
the reverse reaction ${\n}\nu_{\el}\to{\p}{\el}^{-}$ is negligible by
comparison since protons always outnumber neutrons by a large factor
(Scherrer 1984). This effect is important only when the mass of the
decaying particle is of $\Or(10)\MeV$ as will be discussed later in
the context of a massive unstable tau neutrino (see \sref{nus}).

Note also that if the particle lifetime exceeds the age of the
universe then the only constraint comes from requiring that it
respects the bound \eref{omegah2} on the present energy density.
Using \eref{freezeoutabun} and
$\Omega_{x}h^2=3.9\times10^{7}(m_x/\GeV)(n_x/n_\gamma)$, this requires
\begin{equation} 
\label{nxlimtoday}
 \left(\frac{m_x}{\GeV}\right) \left(\frac{n_x}{n_\gamma}\right) 
  \leqsim 2.6 \times 10^{-8}\ .
\end{equation}
A particle which saturates this bound would of course be the
(dominant) constituent of the dark matter; however from the preceeding
discussion it is clear that such an abundance is still too small to
have affected nucleosynthesis.
 
Far more stringent constraints than those discussed above come from
consideration of the direct effects of the decay products on the
synthesized elemental abundances. High energy photons or leptons from
the decaying particles can initiate electromagnetic cascades in the
radiation-dominated thermal plasma, thus creating many low energy
photons with $E_{\gamma}\sim\Or(10)\MeV$ which are capable of
photodissociating the light elements (Lindley 1979). Such
photofissions can occur only for $t\,\geqsim\,10^4\sec$, i.e. after
nucleosynthesis is {\em over}, since at earlier epochs the blackbody
photons are energetic enough and numerous enough that photon-photon
interactions are far more probable than photon-nucleus interactions
(Lindley 1985). When the $x$ particle decays into energetic quarks or
gluons, these fragment into hadronic showers which interact with the
ambient nucleons thus changing their relative abundances. (The
alteration of elemental abundances by direct annihilation with
antinucleons has also been considered (e.g Khlopov and Linde 1984,
Ellis \etal 1985b, Halm 1987, Dominguez-Tenreiro 1987); however
Dimopoulos \etal (1988) have shown that the effect of the hadronic
showers is far more important.) If such hadronic decays occur {\em
  during} nucleosyntheis, the neutron-to-proton ratio is increased
resulting in the production of more $\Htwo$ and $\Hefour$ (Reno
and Seckel 1988). However when hadronic decays occur {\em after}
nucleosynthesis, the result is destruction of $\Hefour$ and creation
of $\Htwo$ and $\Hethree$, as well as both ${\Lisix}$ and
$\Liseven$ (Dimopoulos \etal 1988).

When the $x$ particle has both radiative and hadronic decay modes, the
situation is then simplified by noting that for
$\tau_{x}\sim10^{-1}-10^4\sec$, radiative decays do not play a
significant role while hadronic decays are constrained by the
concommitant overproduction of $\Htwo$ and $\Hefour$ by the
hadronic showers (Reno and Seckel 1988). For longer lifetimes, the
situation is more complicated since elements may be simultaneously
both created and destroyed by photo- and hadro- processes. It has been
argued that for $\tau_{x}\geqsim10^5\sec$, the most stringent
constraint on radiative decays comes from constraining the possible
overproduction of $\Htwo$ and $\Hethree$ through photofission of
$\Hefour$, since the simultaneous destruction of the former by
photofission is negligible by comparison (Ellis \etal 1985b,
Juszkiewicz \etal 1985). A somewhat weaker constraint obtains from
constraining the depletion of the $\Hefour$ abundance itself (Ellis
\etal 1985b, Dimopoulos \etal 1989). These constraints are strengthened
if hadronic decays also occur since these too destroy $\Hefour$ and
create $\Htwo$ and $\Hethree$. However all these constraints are
found to be modified when the development of the electromagnetic
cascades is studied taking $\gamma-\gamma$ (M\"oller) scattering into
account; this reveals that $\Hefour$ destruction is significant only
for $\tau_{x}\geqsim5\times10^6\sec$ (Ellis \etal 1992). It has also
been argued that in the interval $\tau_{x}\sim10^3-10^5\sec$,
$\Htwo$ is photodissociated but not $\Hefour$, so that the
strongest constraint on radiative decays now comes from requiring that
$\Htwo$ should not be excessively depleted (Juszkiewicz \etal 1985,
Dimopoulos \etal 1989). Again, reexamination of the cascade process
indicates that the appropriate interval is shifted to
$\tau_{x}\sim5\times10^4-2\times10^6\sec$ (Ellis \etal 1992). This
particular constraint may appear to be circumvented if hadronic decay
channels are also open since hadronic showers {\em create} $\Htwo$;
however such showers also create the rare isotopes ${\Lisix}$ and
$\Liseven$ and are thus severely constrained by observational limits
on their abundance (Dimopoulos \etal 1989). This ensures that the
$\Htwo$ photofission constraint is not affected by such
hadronic decays.

\subsubsection{Electromagnetic showers:\label{emshower}} 

Let us begin by examining the manner in which a massive particle
decaying into photons or leptons generates electromagnetic cascades in
the radiation-dominated thermal plasma of the early universe. The
dominant mode of energy loss of a high energy photon (of energy
$E_{\gamma}$) is ${\el}^{+}{\el}^{-}$ pair-production on the low
energy blackbody photons (of energy $\epsilon_{\gamma}$) while the
produced electrons and positrons (of energy $E_{\el}$) lose energy by
inverse-Compton scattering the blackbody photons to high
energies. Pair-production requires $E_{\gamma} \epsilon_{\gamma}\geq
m_{\el}^2$, while $E_{\el} \epsilon_{\gamma}{\geq}m_{\el}^2$ implies
that scattering occurs in the Klein-Nishina regime in which the
electron loses a large fraction of its energy to the scattered
photon. Thus a primary photon or lepton triggers a cascade which
develops until the photon energies have fallen below the
pair-production threshold,
$E_{\max}=m_{\el}^2/\epsilon_{\gamma}$. Subsequently the photons
undergo Compton scattering on the electrons and pair-production on the
ions of the thermal plasma. If the density of the blackbody photons is
large enough, the cascade is termed `saturated' implying that nearly
all of the primary paricle energy is converted into photons with
energy below $E_{\max}$. Note that such an electromagnetic cascade can
be initiated even when the decay particle is a neutrino since it can
initiate pair-production, $\nu\bar{\nu}\to{\el}^{+}{\el}^{-}$, off the
(anti)neutrinos of the thermal background, if its energy is
sufficiently high (e.g. Gondolo \etal 1993), or even off the decay
(anti)neutrinos (Frieman and Giudice 1989). Once high energy electrons
have been thus created, the subsequent development of the shower
proceeds as before.

The spectrum of the `breakout' photons below the pair-production
threshold was originally found by Monte Carlo simulations of the
cascade process to be (Aharonian \etal 1985, Dimopoulos \etal 1988) 
\begin{equation}
\label{casspec1}
\eqalign{
 \frac{\d N}{\d E_{\gamma}} &\propto E_{\gamma}^{-3/2} \quad {\for} \quad 
 0 \leq E_{\gamma} \leq E_{\max} \equiv \frac{m_{\el}^2}{\epsilon_{\gamma}},\\ 
 &\propto 0 \qquad {\for} \quad E_{\gamma} > E_{\max}\ ,}
\end{equation}
when the background photons are assumed to be monoenergetic.
Subsequently an analytic study of the kinetic equation for the cascade
process showed that the spectrum actually steepens further to
$E_{\gamma}^{-1.8}$ for $E_{\gamma}\geqsim0.3E_{\max}$ (Zdziarski and
Svensson 1989). This feature had not been recognized in the Monte
Carlo simulations due to insufficient statistics. In the cosmological
context, the background photons are not monoenergetic but have a
Planck distribution at temperature $T$. Na\"{\ii}vely we would expect
that the pair-production threshold is then
$E_{\max}\,\approx\,m_{\el}^2/T$. However the prim{\ae}val plasma is
radiation-dominated, i.e. the number density of photons is very large
compared to the number density of electrons and nuclei. Hence even
when the temperature is too low for a high energy photon to
pair-produce on the {\em bulk} of the blackbody photons,
pair-production may nevertheless occur on the energetic photons in the
Wien tail of the Planck distribution (Lindley 1985). Although the
spectrum here is falling exponentially with energy, the number of
photons with $\epsilon_{\gamma}\geqsim25T$ is still comparable to the
number of thermal electrons since $n_{\gamma}/n_{\el}\geqsim10^9$.
Therefore pair-production on such photons is as important as Compton
scattering on electrons or pair-production on ions, the respective
cross-sections being all comparable. Hence the value of $E_{\max}$ is
significantly lowered below the above estimate, as seen by equating
the mean free paths against pair-production on photons and Compton
scattering on electrons (Zdziarski and Svensson 1989):
\begin{equation}
 E_{\max} \simeq \frac{m_{\el}^2}{20.4 T\ [1 + 0.5\, 
  \ln (\eta/7\times10^{-10})^2 + 0.5\,\ln (E_{\max}/m_{\el})^2]}\ .
\end{equation}
(Note that at the energies relevant to photofission processes
($E_{\gamma}<100\MeV$), pair-production on ions is unimportant by
comparison with Compton scattering.)  Although various authors have
noted this effect, they have used quite different estimates of
$E_{\max}$, viz. $m_{\el}^2/12T$ (Lindley 1985, Juszkiewicz \etal
1985), $2m_{\el}^2/25T$ (Salati \etal 1987), $m_{\el}^2/18T$ (Kawasaki
and Sato 1987), $m_{\el}^2/25T$ (Dimopoulos \etal 1988, 1989) and
$m_{\el}^2/32T$ (Dominguez-Tenreiro 1987). Moreover all these authors
assumed the spectrum to be of the form \eref{casspec1} whereas for a
blackbody target photon distribution it actually steepens to
$E_{\gamma}^{-1.8}$ for $E_{\gamma}{\geqsim}0.03E_{\max}$ (Zdziarski
1988).

Subsequently it was noted that $\gamma-\gamma$ elastic
scattering is the {\em dominant} process in a radiation-dominated
plasma for photons just below the pair-production threshold (Zdziarski
and Svensson 1989), hence $E_{\max}$ really corresponds to the energy 
for which the mean free paths against $\gamma-\gamma$
scattering and $\gamma-\gamma$ pair-production are equal. For a Planck
distribution of background photons this is (Ellis \etal 1992)
\begin{equation}
\label{Emax}
 E_{\max} \simeq \frac{m_{\el}^2}{22T} \ ;
\end{equation}
photons pair-produce above this energy and scatter elastically below
it.  Another effect of $\gamma-\gamma$ scattering is reprocessing of
the cascade spectrum leading to further reduction in the number of
high energy photons. The spectrum now falls like $E_{\gamma}^{-1.5}$
upto the energy $E_{\crit}$ where $\gamma-\gamma$ scattering and
Compton scattering are equally probable and then steepens to
$E_{\gamma}^{-5}$ before being cutoff at $E_{\max}$ by the onset of
pair-production (Zdziarski 1988). The value of $E_{\crit}$ depends
weakly on the photon energy; at the energies of $\sim2.5-25\MeV$
relevant to the photofission of light nuclei, it is
\begin{equation}
\label{Ecrit}
 E_{\crit} \simeq \left(\frac{m_{\el}^2}{44 T}\right) 
  \left(\frac{\eta}{7\times10^{-10}}\right)^{1/3} ,
\end{equation}
i.e. effectively $E_{\crit}{\simeq}E_{\max}/2$.

We can now study how the yields in the standard BBN model are
altered due to photofission by the cascade photons from a hypothetical
decaying particle. Let ${\d}N_{x}/{\d}E$ denote the spectrum of high
energy photons (or electrons) from massive particle decay, normalized
as
\begin{equation}
\label{norm}
 \int_{0}^{\infty} E \frac{\d N_{x}}{\d E} \d E = f_{\gamma}m_{x} \ ,
\end{equation}
where $f_{\gamma}$ is the fraction of the $x$ particle mass released in
the form of electromagnetically interacting particles (easily
calculable once the decay modes and branching ratios are specified). A
decay photon (or electron) of energy $E$ initiates a cascade with the
spectrum 
\begin{equation}
\label{casspec2}
\eqalign{
 \frac{\d n_{E}}{\d E_{\gamma}}  
  &= \frac{24 \sqrt{2}}{55} \frac{E}{E_{\max}^{1/2}} E_{\gamma}^{-3/2}
   \quad {\for} \quad 0 \leq E_{\gamma} \leq E_{\max}/2, \\ 
  &= \frac{3}{55} E E_{\max}^{3} E_{\gamma}^{-5}
   \quad {\for} \quad E_{\max}/2 \leq E_{\gamma} \leq E_{\max}, \\
  &= 0 \quad \for \quad E_{\gamma} > E_{\max},}
\end{equation}
where we have normalized the cascade spectrum as
\begin{equation}
\label{casnorm}
 \int_{0}^{E_{\max}} E_{\gamma} {\d n_{E} \over \d E_{\gamma}}
  \d E_{\gamma} = E \ , \qquad E_{\max} = {m_{\el}^2 \over 22 T} \ . 
\end{equation}
Recently, Kawasaki and Moroi (1995a,b) have claimed that numerical
solution of the governing Boltzman equations yields a different
cascade spectrum which has significant power beyond the cutoff
$E_{\max}$ and is also less steep below $E_{\max}/2$. We note that
Protheroe \etal (1995) obtain results in agreement with those above
from a detailed Monte Carlo simulation of the cascade process.

To write the balance equation for the change in the abundance of
element $i$ with total photofission cross-section $\sigma_i$ (above
threshold $Q_i$), we note that recombination of the dissociated
nuclei, in particular $\Htwo$, is negligible for $t\geqsim10^{4}\sec$,
hence (Ellis \etal 1985b)
\begin{equation}
\label{photorate}
\fl \frac{\d X_{\in}}{\d t}\vert_{\photo} = - \frac{\d n_{x}}{\d t} 
   \int_{0}^{\infty}
  \frac{\d N_{x}}{\d E} \d E \left(\int_{Q_i}^{E} 
   \frac{\d n_{E}}{\d E_{\gamma}}
  \frac{X_{\in} \sigma_i}{n_{\el} \sigma_{\C}} \d E_{\gamma} - \sum_{j \neq i} 
  \int_{Q_i}^{E} \frac{\d n_{E}}{\d E_{\gamma}} 
  \frac{X_j \sigma_{j \rightarrow i}}{n_{\el} \sigma_{\C}} \d E_{\gamma} 
   \right), 
\end{equation}
where $\sigma_{j{\to}i}$ is the partial cross-section for
photofission of element $i$ to element $j$ and $\sigma_{\C}$ is
the cross-section for Compton scattering on the thermal electrons of
density
\begin{equation}
\label{ne}
 n_{\el} \simeq \left(1 - \frac{Y}{2}\right)\ n_{\N} = 
  \frac{7}{8} \eta n_{\gamma} \ ,
\end{equation}
for a H+He plasma, taking $Y({\Hefour})=0.25$. (Anticipating the
stringent constraints on the particle abundance to be derived shortly,
we assume that $\eta$ is not altered significantly by the entropy
released in particle decays.) Since the number density of $x$
particles decreases from its initial value $n_{x}^{\in}$ as
\begin{equation}
 \frac{\d n_{x}}{\d t} = - \frac{n_{x}^{\in}}{\tau_{x}} \exp
  \left(-\frac{t}{\tau_{x}}\right), 
\end{equation}
the time-integrated change in the elemental abundance (in a comoving
volume) is given by (Ellis \etal 1992)
\begin{equation}
\label{photoint}
\eqalign{
\fl \int_{t_i^{\min}}^{\infty} \frac{\d X_{\in}}{\d t}\vert_{\photo} \ \d t 
  &\simeq \left(m_{x} \frac{n_{x}}{n_{\gamma}}\right) {f_{\gamma} \over \eta}
   \left(1 - \frac{Y}{2}\right)^{-1}
   \left[ - X_{\in} \beta_i(\tau_{x}) + \sum_{j \neq i} X_j \beta_{j 
   \to i} (\tau_{x}) \right], \\ 
  \beta_i(\tau_{x}) 
   &\equiv \int_{t_i^{\min}}^{\infty} \frac{\d t}{\tau_{x}} \exp
    \left(-\frac{t}{\tau_{x}}\right) 
    \int_{Q_i}^{E_{\max}(t)} \left(\frac{1}{E} \frac{\d n_{E}}{\d E_{\gamma}}
    \right) \frac{\sigma_i (E_{\gamma})}{\sigma_{\C} (E_{\gamma})} 
    \d E_{\gamma}\ , \\
  \beta_{j \to i}(\tau_{x}) 
   &\equiv \int_{t_j^{\min}}^{\infty} \frac{\d t}{\tau_{x}} \exp
   \left(-\frac{t}{\tau_{x}}\right) \int_{Q_j}^{E_{\max}(t)} \left(\frac{1}{E}
   \frac{\d n_{E}}{\d E_{\gamma}}\right) \frac{\sigma_{j \to
   i}(E_{\gamma})}{\sigma_{\C} (E_{\gamma})} \d E_{\gamma} \ ,}
\end{equation}
(We have dropped the superscript ${\in}$ on $n_{x}$ above and
hereafter since, as before, we will be concerned with particles which
decay late, long after they fall out of chemical equilibrium. Hence
the usual freeze-out abundance (e.g. \eref{freezeoutabun}), can be
sensibly taken to be the initial abundance, with due allowance made
for whether the particle decays occur before or after ${\el}^+{\el}^-$
annihilation.) The time $t_{i}^{\min}$ at which photofission of
element $i$ starts can be computed (from the time-temperature
relationship for a radiation-dominated universe \eref{trad} with
$g_{\rho}=3.36$ for $T{\ll}m_{\el}$) corresponding to the critical
temperature at which the cascade cutoff energy $E_{\max}$ \eref{Emax}
equals the threshold $Q_i$ for the most important photofission
reactions:
\begin{equation}
\label{photothresh}
\fl
\eqalign{
 Q_{\gamma{\Htwo}\to{\p}{\n}} = 2.23\ \MeV, \\
 Q_{\gamma{\Hethree}\to{\p}{\Htwo}} = 5.49\ \MeV, \quad 
 Q_{\gamma{\Hethree}\to{2\p}{\n}} = 7.72\ \MeV, \\
 Q_{\gamma{\Hefour}\to{\p}{\Hthree}} = 19.8\ \MeV, \quad 
 Q_{\gamma{\Hefour}\to{\n}{\Hethree}} = 20.6\ \MeV, \quad 
 Q_{\gamma{\Hefour}\to{\p}{\n}{\Htwo}} = 26.1\ \MeV, \\
 \vdots}
\end{equation}
As we shall see, the decaying particle abundance is constrained to be
sufficiently small that the effect on the dynamics of the expansion is
negligible. Hence it is consistent to take the input abundances to
be those obtained in the standard BBN model and study how these may be
altered by photofission processes.


The integrals $\beta_i$ and $\beta_{j{\to}i}$ have been computed
numerically for various values of $\tau_{x}$ using the cascade
spectrum \eref{casspec2} and the known cross-sections for photofission
processes (see Faul \etal 1981, Gari and Hebach 1981, Govaerts \etal
1981). As seen in \fref{photorates}, $\beta_i$ rises sharply from zero
above a critical value of $\tau_{x}$ (which increases as the square of
the photofission threshold $Q_i$), peaks at a value which is nearly
the same ($\approx1\GeV^{-1}$) for all light elements, and
subsequently falls off rather slowly with increasing $\tau_{x}$. This
reflects the fact that the relevant photofission cross-sections are
all of order a few millibarns above threshold and fall rapidly
thereafter with increasing energy. Photofission begins when the
cascade cutoff energy just crosses the photofission threshold and the
dominant effect is that of photons with energies just over this
threshold. This implies that when photofission of $\Hefour$ occurs,
the resultant production of $\Htwo$ and $\Hethree$
($\gamma{\Hefour}\to{\p}{\Hthree},\,{\n}{\Hethree},\,{\p}{\n}{\Htwo};~
{\Hthree}\to{\Hethree}\,{\el}^{-}\bar{\nu_{\el}}$) far dominates their
destruction since the abundance of $\Hefour$ is $\sim10^4$ times
greater (Ellis \etal 1985b). As seen in \fref{photorates},
photofission of $\Htwo$ begins at $\tau_{x}\approx10^4\sec$ and
becomes significant at $\tau_{x}\approx10^5\sec$ while photofission of
$\Hefour$ begins at $\tau_{x}\approx10^6\sec$ and begins significant
at $\tau_{x}\approx10^7\sec$. Therefore for $\tau_{x}\geqsim10^6\sec$,
\eref{photoint} can be simplified to read, for the
difference between the initial and final mass fractions of
${\Htwo}+{\Hethree}$:
\begin{equation}
\label{He3plusDprod}
\fl X_{\fin} ({\Htwo} + {\Hethree}) - X_{\in}({\Htwo} + {\Hethree}) \simeq 
  Y_{\in} ({\Hefour}) \left(m_{x}
  \frac{n_{x}}{n_{\gamma}}\right) \frac{f_{\gamma}}{\eta} 
  \left(1 - \frac{Y}{2}\right)^{-1} r\ \beta_{{\Hefour}}(\tau_{x}), 
\end{equation}
where, $r\equiv[(\frac{3}{4}\sigma_{\gamma{\Hefour}\to{\n}{\Hethree},
  ~{\p}{\Hthree}}+\frac{1}{2}\sigma_{\gamma{\Hefour}\to{\p}{\n}{\Htwo}})/
\sigma_{\gamma{\Hefour}\to{\rm all}}]\simeq0.5$, and the subscripts
$\in$ and $\fin$ refer to the initial and final values. To obtain the
most conservative constraint on the abundance of the decaying particle
we must consider the maximum value allowed for
$[X_{\fin}({\Htwo}+{\Hethree})-X_{\in}({\Htwo}+{\Hethree})]
\eta/Y_{\in}({\Hefour})$. Since $\Hefour$ cannot have been destroyed
significantly (without overproducing $\Htwo$ and $\Hethree$) we
take its initial abundance to be the maximum permitted value, i.e.
$Y_{\in}({\Hefour})<0.25$, which implies that $\eta<9.2\times10^{-10}$
\eref{reliable}. Hence a minimum mass fraction
$X_{\in}({\Htwo}+{\Hethree})>3.8\times10^{-5}$ would have been
primordially synthesized (see \fref{abunetamc}). The maximum
final abundance after photoproduction consistent with `standard'
galactic chemical evolution is bounded by
$X_{\fin}({\Htwo}+{\Hethree})\leqsim2.5\times10^{-4}$, using
\eref{He3plusD} and taking into account that comparable
numbers of $\Hethree$ and $\Htwo$ nuclei are photoproduced. Using
these numbers and taking $f_{\gamma}=1$ yields the upper limit (full
line) on $m_{x}n_{x}/n_{\gamma}$ shown in \fref{nxlimcascade}
above which ${\Htwo}+{\Hethree}$ is overproduced. For reference, the
dashed line indicates the constraint obtained earlier by Ellis \etal
(1985b) using the same argument but with a less sophisticated treatment
of the cascade process. A similar constraint was obtained by
Juszkiewicz \etal (1985). Recently, Protheroe \etal
(1995) have performed a Monte Carlo simulation of the cascade process
and quoted bounds on $\Omega_{x}/\Omega_{\N}$ for three choices of
$(\eta/10^{-10})=2.7,\,3.3,\,5.4$. We have rescaled their bound taking
$\eta$ to be $9.2\times10^{-10}$ for fair comparison with Ellis \etal
(1992) and plotted this as the dotted line in
\fref{nxlimcascade}. The two results are seen to be in
good agreement. We cannot however reproduce either the {\em
  less} stringent constraint quoted by Kawasaki and Sato (1987) or the
{\em more} stringent constraint given by Kawasaki and Moroi
(1995a);\footnote{Kawasaki and Moroi (1995b) claim that the
 discrepancy arises because Ellis \etal (1992) and Protheroe \etal
 (1995) did not allow for the standard synthesis of
 ${\Htwo}+{\Hethree}$. In fact these authors did so, albeit in a more
 conservative (and consistent) manner.} 
since both these results were obtained entirely by numerical
integration of the governing equations, we cannot easily identify the
reasons for the discrepancy.  Dimopoulos \etal (1989) did not
consider the constraint on the decaying particle abundance from
photoproduction of ${\Htwo}+{\Hethree}$. These authors criticized
Ellis \etal (1985b) for having neglected the photofission of $\Htwo$
by comparison, but as shown above this is quite justified since the
correction is only of $\Or(10^{-4})$.


For $\tau\leqsim10^6\sec$, photofission of $\Hefour$ is not
significant so $\Htwo$ and $\Hethree$ are not produced but only
destroyed. Assuming that hadronic decay channels are not open,
\eref{photorate} now reads for the change in the $\Htwo$ mass fraction
alone
\begin{equation}
\label{photoD}
 \frac{X_{\in} ({\Htwo})}{X_{\fin} ({\Htwo})} \simeq \exp \left[ 
  \left(\frac{m_{x} n_{x}}{n_{\gamma}}\right) \frac{f_{\gamma}}{\eta} 
  \left(1 - \frac{Y}{2}\right)^{-1} \beta_{\Htwo} (\tau_{x}) \right]  
\end{equation}
Again, to obtain the most conservative constraint on the particle
abundance, we must maximize the quantity
$\eta\ln[X_{\in}({\Htwo})/X_{\fin}({\Htwo})]$ subject to the
observational constraint that the $\Htwo$ abundance after photofission
must exceed the observational bound \eref{Dismlim}. Using
\eref{abundmine} we see that this quantity peaks at
$\approx6.5\times10^{-10}$ for $\eta\approx4\times10^{-10}$. The
corresponding upper limit on the decaying particle abundance is
indicated in \fref{nxlimcascade} above which $\Htwo$ is excessively
depleted. The dot-dashed line alongside is the upper limit obtained by
Dimopoulos \etal (1989) from similar considerations but ignoring
$\gamma-\gamma$ scattering. All the above constraints apply to any
decaying particle which can generate electromagnetic cascades above
the photofission threholds; this requires
$m_{x}\approx2E_{\gamma}\geqsim5-50\MeV$ depending on which element is
being considered (see \eref{photothresh}).

\subsubsection{Hadronic showers:\label{hadshower}} 

As mentioned earlier, when hadronic decay channels are open, $\Htwo$
is {\em produced} by hadronic showers and this requires
reconsideration of the constraint derived above. In fact, even if the
particle decays exclusively into photons, the resulting
electromagnetic cascades will be effectively hadronic for
$E_{\gamma}\epsilon_{\gamma}>\Or(1){\GeV}^2$; furthermore there is
always a $\approx1\%$ probability for the (virtual) decay photon to
convert into a $q \bar{q}$ pair over threshold. Hence hadronic showers
will be generated if the particle is heavier than $\approx1\GeV$ even
if it has no specific hadronic decay channels (Reno and Seckel 88). As
discussed in detail by Dimopoulos \etal (1988, 1989), the main effect
of hadronic showers is the destruction of the ambient $\Hefour$
nuclei and the creation of ${\Htwo}, {\Hethree}, {\Lisix}$ and
$\Liseven$. The average number of $i$ nuclei created per $x$
particle decay, $\xi_i$, can be computed by modelling the hadronic
shower development using ${\el}^{+}{\el}^{-}$ jet data. The balance
equation for an elemental abundance now reads
\begin{equation}
\label{photohadrorate}
 \frac{\d X_{i}}{\d t} = \frac{\d X_{i}}{\d t}\vert_{\photo} + \frac{\d 
  X_{i}}{\d t}\vert_{\hadro}\ ,
\end{equation}
where the first term on the RHS is given by \eref{photorate} and the
second term is (Dimopoulos \etal 1988, 1989)
\begin{equation}
\label{hadrorate}
 \frac{\d X_{i}}{\d t}\vert_{\hadro} = r_{\B}^{\star} \xi_i 
  \frac{\d n_{x}}{\d t}\ , \qquad r_{\B}^{\star} \equiv
  \left(\frac{\nu_{\B}}{5} \right) r_{\B} {\cal F}\ .
\end{equation}
Here $r_{\B}^{\star}$ is an `effective' baryonic branching ratio
defined in terms of the true baryonic branching ratio $r_{\B}$, the
baryonic multiplicity $\nu_{\B}$, and a factor ${\cal F}$ representing
the dependence of the yields $\xi_i$ on the energy of the primary
shower baryons. The ${\el}^{+}{\el}^{-}$ jet production data suggests
that for $m_{x}=1$\,TeV, there are $\approx5$ nucleon-antinucleon pairs
produced with $\approx5\GeV$ energy/nucleon. For other values of
$m_{x}$, $\nu_{\B}$ depends logarithmically on the energy, except near
the baryon production threshold where the dependence is somewhat
stronger. 

Considering the effects of hadroproduction alone, \eref{hadrorate}
integrates to read (Dimopoulos \etal 1989)
\begin{equation}
 \frac{n_{x}^{\in}}{n_{\gamma}} < \frac{({\cal N}_i^{\max} - {\cal
   N}_i^{\min})\,\eta}{r_{\B}^{\star}\,\xi_i}\ , 
\end{equation}
where ${\cal N}_i^{\max}$ and ${\cal N}_i^{\min}$ are, respectively, the
maximum (observed) and minimum (synthesized) abundance of element $i$
by number, relative to hydrogen. This constraint can be imposed
on the hadroproduction of $\Htwo$, $\Hethree$, ${\Lisix}$ and
$\Liseven$. Taking ${\cal N}_{\Liseven}^{\max}\approx2\times10^{-10}$
and ${\cal N}_{\Liseven}^{\min}\approx5\times10^{-11}$, this gives
(Dimopoulos \etal 1989)
\begin{equation}
 \frac{n_{x}^{\in}}{n_{\gamma}} < 1.5 \times 10^{-5} 
  \frac{\eta}{r_{\B}^{\star}}\ .
\end{equation}
A similar constraint follows from requiring ${\cal N}_{({\Htwo}+
{\Hethree})}^{\max}\approx10^{-4}$. An even stricter constraint can be
obtained if we assume that the primordial abundance of ${\Lisix}$ did
not exceed its presently observed value, i.e. ${\cal
N}_{\Lisix}^{\max} \approx10^{-11}$. This yields (Dimopoulos \etal
1989)
\begin{equation}
 \frac{n_{x}^{\in}}{n_{\gamma}} < 3 \times 10^{-7} 
  \frac{\eta}{r_{\B}^{\star}}\ .
\end{equation}
The fact that these constraints are so very restrictive considerably
simplifies the situation when {\em both} electromagnetic and hadronic
showers occur. For a given $x$ particle abundance, the hadronic
branching ratio must be very small in order not to overproduce
lithium.  This ensures that the production of $\Htwo$ by hadronic
showers is quite negligible relative to its production by
electromagnetic showers. This is borne out by numerical solution of
\eref{photohadrorate} taking both kinds of showers into account
(Dimopoulos \etal 1989). It is true that in a small region of
parameter space ($\tau_{x}\sim10^{5}-10^{6}\sec$,
$m_{x}n_{x}/n_{\N}\sim10^{1}-10^{3}\GeV$,
$r_{\B}^{\star}n_{x}/n_{\N}\sim10^{-4}-10^{-2}$), the photodestruction
of $\Htwo$ {\em is} compensated for by hadroproduction of $\Htwo$ but
this also results in the production of an excessive amounts of
${\Lisix}$. If this is indeed inconsistent with observations
(e.g. Steigman \etal 1993), the constraints derived from consideration
of photofission processes are {\em not} evaded even if hadronic decay
channels are also open.

For $\tau_{x}<10^4\sec$, photofission does not occur for any element
and standard nucleosynthesis is unaffected by electromagnetic showers.
However hadronic showers can induce interconversions between the
ambient protons and neutrons thus changing the equilibrium $n/p$
ratio. This has been studied in detail by Reno and Seckel (1988) as
discussed below. The transition rate for a thermal nucleon to convert
to another nucleon is the usual weak interaction rate plus the rate due
to hadronic showers, given by
\begin{equation}
 \Gamma_{{\p} \to {\n}} = \frac{\Gamma_{x} n_{x}}{X ({\p})\,
  n_{\N}} \sum {\cal P}_{x_{i}} f_{{\p}{\n}}^i\ , \qquad \Gamma_{{\n}
  \to {\p}} = \frac{\Gamma_{x} n_{x}}{X ({\n})\,n_{\N}} \sum 
  {\cal P}_{x_{i}} f_{{\n}{\p}}^i\ ,
\end{equation}
where, $\Gamma_{x}\equiv\tau_{x}^{-1}$, $X({\p})$ and $X({\n})$ are the
proton and neutron fractions, ${\cal P}_{x_i}$ is the average number of
hadronic species $i$ per $x$ particle decay and $f_{{\p}{\n}}^i$,
$f_{{\n}{\p}}^i$ are the probabilities for $i$ to induce the respective
transitions. The fragmentation process can be modelled using data on
jet multiplicities from ${\el}^{+}{\el}^{-}$ annihilation experiments
(Reno and Seckel 1988): 
\begin{equation}
 {\cal P}_{x_i} \simeq N_{\jet}\,\langle n_{\ch}(E_{\jet}) \rangle\,B _{\h}
  \left(\frac{n_i}{n_{\ch}}\right) .
\end{equation}
Here $B_{\h}$ is the hadronic branching ratio for $x$ decay, $N_{\jet}$
is the number of jets, and $n_i$, the charge multiplicity of species
$i$, has been expressed as a fraction of the average charge
multiplicity ${\langle}n_{\ch}(E_{\jet})\rangle$ at a given energy
$E_{\jet}$. The transition probability is computed as the ratio of the
strong interaction rate to the sum of the decay and absorption rates
for the injected hadrons:
\begin{equation}
 f_{{\p}{\n}}^i = \frac{\Gamma_{{\p}{\n}}^i}{\Gamma_D^i + \Gamma_A^i}\ ,
   \qquad 
 f_{{\n}{\p}}^i = \frac{\Gamma_{{\n}{\p}}^i}{\Gamma_D^i + \Gamma_A^i}\ .
\end{equation}
When the decaying particle carries no baryon number, the decay hadrons
can be thought of as being injected in pairs so that $i$ can refer to
mesons as well as to baryon-antibaryon pairs
(i.e. $i=n\bar{n},p\bar{p},\ldots$).  The injected hadrons (except
$K_L$) are stopped before they interact with the ambient neutrons so
that threhold values of cross-sections can be used (Reno and Seckel
1988). The variable quantifying the effect of hadronic decays is then
the $x$ particle abundance multiplied by a parameter $F$ defined as 
\begin{equation}
 F \equiv \frac{N_{\jet}B _{\h}}{2} \frac{\langle n (E_{\jet}) \rangle}
  {\langle n (E_{33 {\GeV}}) \rangle}\ ,
\end{equation}
so that $F\simeq1$ for $m_{x}=100\GeV$, if we take
$E_{\jet}{\simeq}m_{x}/3$, $n_{\jet}B_{\h}=2$, i.e. assuming that $x$
decays into 3 particles at the parton level and $N_{\jet}$ equals the
number of (non-spectator) quarks at the parton level.

The neutron fraction in the thermal plasma is always less than 0.5,
being $X_{\n}\simeq0.2$ at $T\simeq1\MeV$ where the weak interaction
rate freezes-out \eref{Xnfr}, and decreasing by beta decay to 0.12 at
0.09 MeV \eref{Xndec} when nuclear reactions begin. Since there are
always more protons than neutrons, the overall effect of hadronic
decays in the interval $\tau_{x}\sim1-200\sec$ is to convert protons
into neutrons. (For injection of $p\bar{p}$ pairs, the neutron
fraction is actually reduced but this is compensated for by the
effects of mesons and $n\bar{n}$ injection.) The additional neutrons
thus produced are all synthesized into $\Hefour$ and hence hadronic
decays in this lifetime interval are constrained by the observational
upper bound to the helium abundance. Reno and Seckel (1988) obtain
upper limits on $Fn_{x}/n_{\gamma}$ as a function of $\tau_{x}$ for
$\eta=3\times10^{-10}$ and $10^{-9}$, adopting the bound
$Y_{\pr}({\Hefour})<0.26$. We have rescaled their results (for the
case when $x$ does not itself carry baryon number) to the more
stringent constraint $Y_{\pr}({\Hefour})<0.25$ \eref{Ypitllim}; this
requires that we restrict ourselves to the case $\eta=3\times10^{-10}$
since for $\eta=10^{-9}$, $Y_{\pr}$ already exceeds 0.25 even in the
absence of $x$ decays (see also Lazarides \etal 1990). We calculate
the resulting upper limit on $m_{x}n_{x}/n_{\gamma}$ taking $B_{\h}=1$
(the limit scales inversely as $B_{\h}$) and show this in
\fref{nxlimcascade}. This constraint gets more stringent as $\tau_{x}$
increases from $0.1\sec$ to $100\sec$ since the neutron fraction is
dropping in this time interval. At later times the neutron fraction is
effectively zero (since all neutrons are now bound in nuclei) and the
only free neutrons are those created by $x$ decay. These can bind into
$\Htwo$ but $\Htwo$ cannot burn further to $\Hefour$ since the
corresponding reaction rate is now too low due to the small
densities. For $\tau_{x}\sim100-1000\sec$, the $\Hethree$ abundance is
also increased by ${\Htwo}-{\Htwo}$ burning. For
$\tau_{x}\geqsim10^4\sec$ the released neutrons decay before forming
$\Htwo$. Hence in the interval $10^2-10^4\sec$, the appropriate
constraint on hadronic decays is the indirect bound
$({\Htwo}+{\Hethree})/{\H}<10^{-4}$ \eref{He3plusD}. The corresponding
upper limit on $m_{x}n_{x}/n_{\gamma}$, extracted from Reno and Seckel
(1988), is also shown in \fref{nxlimcascade}.

The above arguments apply to {\em any} form of energy release in the
early universe, for example to annihilations of massive particles as
they turn non-relativistic and freeze-out. This has been considered by
Hagelin and Parker (1990) and by Frieman \etal (1990), however their
modelling of the cascade process was not accurate
(e.g. $\gamma-\gamma$ scattering was not included). To obtain the
correct constraints, the bounds shown in figures \ref{nxlimentexp} and
\ref{nxlimcascade} should be imposed on the energy released in
such annihilations.

The constraints derived from considerations of entropy generation
(\sref{visdec}) and speedup of the expansion rate
(\sref{invisdec}) apply to any particle which is non-relativistic
during nucleosynthesis, i.e. heavier than $\sim0.1\MeV$, and are
independent of the mass (except insofar as the mass may determine the
relic abundance). However the bounds based on the development of
electromagnetic (\sref{emshower}) and hadronic
(\sref{hadshower}) cascades require the mass of the decaying (or
annihilating) particle to be significantly higher. For example, to
generate an electromagnetic shower capable of efficiently
photodissociating $\Htwo$, the initiating photon/electron must have an
energy exceeding twice the relevant threshold \eref{photothresh},
i.e. about $5\MeV$ and this would require the decaying particle to be
at least $10\MeV$ in mass; for $\Hefour$, the required mass is closer
to $100\MeV$. To generate a hadronic cascade, the mass would have to
be even higher, typically in excess of a few hundred MeV. Thus when
dealing with a mass of $\Or(\MeV)$, e.g. a massive $\nu_\tau$, it is
more reliable to just calculate the spectrum of the photons scattered
by the decay ${\el}^{\pm}$ and then evaluate the extent to which say
deuterium is photodissociated (Sarkar and Cooper 1984, Scherrer
1984).\footnote{The inverse-Compton energy
 loss rate $\langle\dot{E}\rangle$ is high in a radiation-dominated
 universe, so that $-\int\d\,E'/\langle\dot{E'}\rangle\ll\,t$ at the
 epochs of interest. Hence the relevant transport equation for the
 electrons (see Blumenthal and Gould 1970) can be simply solved even in
 the relativistic Klein-Nishina limit to obtain the electron energy
 spectrum (modified by inverse-Compton scattering) given the source
 spectrum from neutrino decays. The spectrum of the Compton scattered
 high energy photons is then obtained by appropriate integration over
 the blackbody source distribution.}

\section{Applications\label{appl}} 

So far we have attempted to keep the discussion of constraints as
`model-independent' as possible in order that the results may be
applied to any type of particle, including those which have not yet
been thought of! Of course most discussions in the literature refer to
specific particles, whether known or hypothetical. Of the known
particles, the most interesting are neutrinos since laboratory
experiments have set only weak limits on their properties. Turning to
hypothetical particles, the most interesting are those predicted by
suggested solutions to the naturalness problems of the Standard Model,
e.g. technicolour and supersymmetry. Constraints on both categories
have important implications for the nature of the dark matter in the
universe.

\subsection{Neutrinos\label{nus}} 

The oldest and most popular application of cosmological constraints
has been to neutrinos. Although neutrinos are massless in the Standard
Model and interact weakly, their large relic abundance ensures that
even a small neutrino mass or magnetic moment would have observable
consequences in cosmology. These properties arise in (unified)
theories incorporating new physics, e.g.  lepton number violation at
high energies, hence the cosmological arguments provide a sensitive
probe of such physics. We discuss below only those constraints which
arise from nucleosynthesis; other cosmological constraints, e.g. from
stellar evolution, have been discussed in a number of recent reviews
(see Kolb \etal 1989, Fukugita and Yanagida 1994, Gelmini and Roulet
1995).

\subsubsection{Neutrino masses:\label{numass}} 

Combining the relic neutrino abundance \eref{nuabund} with the
observational bound \eref{omegah2} on the present energy density
imposed by the age and expansion rate of the universe, one obtains the
well-known upper limit (Gershte\v{\ii}n and Zeldovich 1966, Cowsik and
McCleland 1972, Szalay and Marx 1976)
\begin{equation}
\label{nuuplim}
 \sum_{i} m_{\nu_i} \left(\frac{g_{\nu_i}}{2}\right) \leqsim 94\ \eV\ ,
\end{equation}
where the sum is over all species which were relativistic at
decoupling, i.e. with $m_{\nu_i}\leqsim1\MeV$. Alternatively, if the
neutrino is more massive and falls out of chemical equilibrium before
kinetic decoupling with the relic abundance \eref{freezeoutabun}, then
one obtains the {\em lower} limit (Lee and Weinberg 1977, Dicus \etal
1977, Vysotski\v{\ii} \etal 1977)
\begin{equation}
\label{nulolim}
 m_{\nu_i} \geqsim 2 \GeV\ .
\end{equation}
Thus no stable neutrino with Standard Model weak interactions can have
a mass in the range $\sim100\eV-2\GeV$. The experimental mass
limits are (Particle Data Group 1996)
\begin{equation}
\label{numasses}
 ``{m_{\nu_{\el}}}" < 5.1\ \eV, \quad 
 ``{m_{\nu_{\mu}}}" < 160\ \keV, \quad 
 ``{m_{\nu_{\tau}}}" < 24\ \MeV,
\end{equation}
all at $\95cl$. (The quotes are to remind us that these are really
bounds on the mass eigenstates coupled dominantly to the respective
charged leptons (Schrock 1981) as discussed below.)  Thus only the
electron neutrino mass is experimentally known to be below the
cosmological upper bound. The muon and tau neutrinos are then required
by cosmology to also be lighter than $100\eV$, or have their relic
abundance suppressed by some means, e.g. decays or enhanced
self-annihilations.

The Standard Model contains only massless neutrinos in left-handed (LH)
doublets but its successful phenomenology would be unaffected by the
addition of right-handed (RH) neutrinos as isosinglets and/or additional
Higgs bosons to violate lepton number conservation, thus allowing a
Dirac and/or Majorana mass (see Langacker 1988, Mohapatra and Pal
1991, Valle 1991, Gelmini and Roulet 1995). A Majorana mass term is
naturally generated by a dimension-5 operator in extensions of the
SM with intrinsic left-right symmetry and lepton-quark
symmetry, such as $SO(10)$ and its subgroup
$SU(2)_{\L}{\otimes}SU(2)_{\R}{\otimes}U(1)_{B-L}$. A Dirac neutrino
can be viewed as a mass-degenerate pair of Majorana neutrinos with
opposite $CP$ eigenvalues (see Kayser \etal 1989) and arises, for
example, in extended $SO(10)$ models. In general, one should consider
a mass matrix mixing $p$ two-component neutrino spinors belonging to
$SU(2)_{\L}$ doublets with $q$ two-component neutrino spinors which
are singlets under $SU(2)_{\L}\otimes\,U(1)_{Y}$. The \LEP
measurement of the $Z^0$ decay width fixes $p$ to be 3 but allows any
number of singlets (more precisely, any number of light neutral states
in representations with zero third component of weak isospin). The
mass terms in the Lagrangian have the following form in terms of the
spinors $\rho$ describing the current eigenstates
\begin{equation}
 {\cal L}^{\nu}_{\rm mass} = -\frac{1}{2} \rho^{\T} \sigma_{2} M \rho + \hc
  \ , \qquad M \equiv \left[\begin{array}{clcr}
  M_{\L} & D \\ D^{T} & M_{\R}\\ \end{array} \right] ,
\end{equation}
where $\sigma_{2}$ is the Pauli matrix, $M$ is a complex symmetric
matrix in which the ($3\times3$) submatrix $M_{\L}$ describes the
masses arising in the doublet sector from (combinations of) VEVs of
Higgs fields transforming as weak isotriplets, $D$ is the
($3\times\,q$) Dirac mass matrix coming from the VEVs of doublet Higgs
fields, and $M_{\R}$ is a ($q\times\,q$) matrix describing the masses
in the singlet sector which are already $SU(2)_{\L}\otimes\,U(1)_{Y}$
invariant. The {\em physical} neutrino mass eigenstates, $\rho_{m}$,
are then given by
\begin{equation}
 \rho = U \rho_{m}\ ,
\end{equation}
where $U$ is a unitary matrix which diagonalizes $M$ in terms of the
$3+q$ physical masses. Written out explicitly in the notation used by
experimentalists, this says that the (LH) weak flavour
eigenstates $\nu_{\alpha{\L}}$ ($\alpha={\el},\mu,\tau$) which appear
in the weak interaction coupled via $W$ to $e,\mu,\tau$, are in
general related to the mass eigenstates $\nu_{i{\L}}$
($i=1,2,{\ldots}3+q$) through a leptonic Cabibbo-Kobayashi-Masakawa
(CKM) mixing
\begin{equation}
\label{CKMnu}
 \nu_{\alpha{\L}} = \Sigma_{i} U_{\alpha{i}} \nu_{i{\L}}\ .
\end{equation}
(Henceforth the subscript $_{\L}$ will be implied unless otherwise
specified.)  This allows flavour-changing processes such as neutrino
oscillations, when the neutrino mass differences are very small
relative to the momenta so they propagate coherently, and neutrino
decays, when the mass differences are sufficiently large that the
propagation is incoherent (see Bilenky and Petcov 1987, Oberauer and
von Feilitzsch 1992).

Before proceeding to study the effects of neutrino oscillations and
decays, we review recent studies of the effect of a large
neutrino mass on BBN (Kolb \etal 1991, Dolgov and Rothstein 1993,
Dodelson \etal 1994, Kawasaki \etal 1994). In
\eref{nuuplim} we have set $g_{\nu}=2$ since only LH
neutrinos (and RH antineutrinos) have full-strength weak
interactions. If neutrinos have Dirac masses, then the non-interacting
RH neutrino (and LH antineutrino) states can also
come into thermal equilibrium through spin-flip scattering at
sufficiently high temperatures, thus doubling
 $g_{\nu}$.\footnote{Actually in most extensions of the Standard Model
 wherein neutrinos have masses, these are associated with the
 violation of global lepton number and are Majorana in nature, so this
 process is irrelevant. Even for Dirac neutrinos, the RH
 states are not populated at decoupling for a mass of $\Or(100)\eV$,
 hence it is always valid to take $g_{\nu}=2$ in \eref{nuuplim}.} The
rate at which the RH states are populated is
$\propto\,(m_{\nu}/T)^2$, hence Shapiro \etal (1980) had concluded that
equilibrium would {\em not} be achieved for a mass of
$\approx30\eV$ which was indicated at that time for the $\nu_{\el}$ (and
which is of order the cosmological bound \eref{nuuplim} for stable
neutrinos). However given the weaker mass limits for the other
neutrinos, as also the possibility that these may be {\em unstable},
one must consider whether their RH states may have been populated
during nucleosynthesis. Obviously this will further speed up the
expansion and be in conflict with the bound on $N_{\nu}$, hence an
upper limit on the Dirac mass can be derived by requiring that the
spin-flip scattering rate fall behind the Hubble expansion rate {\em
before} the quark-hadron phase transition. Then the RH states do not
share in the entropy release and are diluted adequately (Fuller and
Malaney 1991, Enqvist and Uibo 1993). Dolgov \etal (1995) also include
the production of wrong-helicity states through
$\gamma\gamma\to\pi^0\to\nu\bar{\nu}$ (Lam and Ng 1991) as well as
$\pi^\pm\to\mu\nu_\mu$, which are insensitive to $T_{\c}^{\qh}$. An
updated and corrected calculation (Fields \etal 1996) yields the
constraints:
\begin{equation}
\label{Dirbound}
 m_{\nu_{\mu}} < 310 \keV\ , \qquad 
 m_{\nu_{\tau}} < 370 \keV\ ,
\end{equation}
imposing $N_{\nu}<4$ and taking $T_{\c}^{\qh}>100\MeV$. These bounds
improve (degrade) by a factor of about 2, if the BBN limit on
$N_{\nu}$ is tightened (weakened) to 3.3 (4.5). 

When the neutrino mass exceeds a few MeV, they are non-relativistic at
decoupling so their relic energy density \eref{freezeoutabun} falls
inversely as the self-annihilation cross-section, hence decreases with
increasing mass (for $m_\nu\ll\,m_{W}$). The constraint on the
speed-up rate (or on $N_\nu$) during BBN then implies a {\em lower}
bound on $m_\nu$.\footnote{This implicitly assumes that the
 nucleon-to-photon ratio is unaffected by the massive neutrino, so that
 the $N_\nu$ bound is directly applicable. However the neutrino {\em
 must} decay subsequently and the entropy thus released will in fact
 lower $\eta$ (see \sref{visdec}). Therefore these analyses are only
 valid if the decays create {\em no} entropy at all, which is rather 
 unnatural.} Initial calculations suggested a lower limit {\em above} the 
present experimental bound on $m_{\nu_\tau}$, e.g. Kolb \etal (1991)
quoted $m_{\nu}>25\MeV$ corresponding to the constraint $N_\nu<3.4$
while Dolgov and Rothstein (1993) obtained $m_{\nu}>35\MeV$ for
$N_\nu<3.6$. Recently Hannestad and Madsen (1996) claimed that careful
solution of the Boltzmann equation (including scattering reactions)
lowers the relic abundance substantially, thus opening an allowed
region for a Majorana $\nu_\tau$ above 16 MeV (adopting various
abundance bounds which effectively imply $N_\nu<3.4$). As we have
discussed in \sref{rel}, such restrictive bounds are no
longer justified by the data, hence this conclusion is unreliable.
Moreover for such massive neutrinos, there are additional effects on
the abundances since their annihilations create electron neutrinos,
viz. $\nu_\tau\bar{\nu_\tau}\to\nu_{\el}\bar{\nu_{\el}}$, which can
bias neutron-proton interconversions. Taking such effects into account
excludes the high mass window, even for a relaxed upper bound on
$N_\nu$ (Dolgov \etal 1996).

For any of the above bounds to be valid, the neutrino must be present
at the time of nucleosynthesis, i.e. its lifetime must exceed
$\sim1-1000\sec$. However the present (redshifted) energy density of
the decay products may be excessive unless the decays occur early
enough (Dicus \etal 1977). Making the conservative assumptions that
the decay products are all massless and that their energy-density has
always dominated the universe, we obtain the bound\footnote{This
 assumes that all neutrinos decay instantaneously at $t=\tau_{\nu}$;
 numerical integration over an exponential distribution of decay times
 relaxes the bound on the lifetime by about $50\%$ (Dicus \etal 1978a,
 Mass\'o and Pomarol 1989). Note that the bound shown by Kolb and Turner 
 (1990) is incorrect.}
\begin{equation}
\label{nulifebound}
 \tau_{\nu} \leqsim \left(\frac{m_{\nu}}{94\eV} 
  \frac{g_{\nu}}{2}\right)^{-2} (\Omega h^2)^2 t_{0}
 \leqsim 3\times10^{12} \sec 
  \left(\frac{m_{\nu}}{10\keV} \frac{g_{\nu}}{2}\right)^{-2},
\end{equation}
using the inequality $\Omega\,h^2<1/3$ for $t_{0}>10^{10}\yr$, $h>0.4$
which holds for a radiation-dominated universe (Pal 1983). We will see
below that this lifetime bound {\em cannot} be satisfied by neutrinos
allowed by the mass bound \eref{Dirbound}, unless they have
`invisible' decays into hypothetical Goldstone bosons. However a
strict upper bound on the mass of the $\nu_{\tau}$
\eref{nutaumass} can still be obtained (irrespective of
whether it is Dirac or Majorana) by considering its decays into
Standard Model particles, as we discuss below.

\subsubsection{Neutrino decays:\label{nudecays}}

The most studied decay mode for neutrinos has been the radiative
process $\nu_{i}\to\nu_{j}\gamma$ and the consequences of such decays,
both in astrophysical sites of neutrino production and in the early
universe, have been widely investigated (see Maalampi and Roos 1990,
Sciama 1993). In the Standard Model extended to allow for Dirac
neutrino masses (without unpaired singlets) this decay mode is
severely supppressed by a leptonic Glashow-Iliopoulos-Maiani (GIM)
mechanism and consequently has a rather long lifetime (Marciano and
Sanda 1977, Petcov 1977, see Pal and Wolfenstein 1982)
\begin{equation}
\label{taunugamma}
 \fl \tau_{\nu_{i}\to\nu_{j}\gamma} 
  \simeq \frac{2048\pi^4}{9\alpha G_{\F}^2 m_{\nu_{i}}^5}
   \frac{1}{|\sum_{\alpha} U_{\alpha{j}}^* U_{\alpha{i}} F (r_{\alpha})|^{2}}
  > 2.4\times10^{14} \sec \left(\frac{m_{\nu_{i}}}{\MeV}\right)^{-5},
\end{equation}
where $r_{\alpha}\equiv(m_{{\ell}_\alpha}/M_{W})^2$,
$F(r_{\alpha})\simeq-\case{3}{2}+\case{3}{4}\,r_{\alpha}$ for
$r_{\alpha}\ll\,1$ (i.e. for $\ell_{\alpha}=\tau$ , and we have
assumed $m_{\nu_{i}}\,\gg\,m_{\nu_{j}}$). Nieves (1983) has noted that
the next-order process $\nu_{i}\to\nu_{j}\gamma\gamma$ is not GIM
suppressed and may therefore possibly dominate over single photon
decay. For $m_{\nu_{j}}\,\ll\,m_{\nu_{i}}\,\ll\,m_{\el}$,
\begin{equation}
\label{taunugamma2}
 \fl \tau_{\nu_{i}\to\nu_{j}\gamma\gamma} 
  = \frac{552960\pi^5 m_{\el}^4}{\alpha^2 G_{\F}^2 m_{\nu_{i}}^9} 
   \frac{1}{|U_{{\el}j}^*U_{{\el}i}|^{2}} 
  > 1.1\times10^{12} \sec |U_{{\el}i}|^{-2}
     \left(\frac{m_{\nu_{i}}}{\MeV}\right)^{-9}. 
\end{equation}
However, when $m_{\nu_{i}}\,>\,m_{\el}$, the lifetime increases to 
(Sarkar and Cooper 1984):
\begin{equation}
\label{taunugamma2p}
 \fl \tau_{\nu_{i}\to\nu_{j}\gamma\gamma} 
  = \frac{384\pi^5}{\alpha^2 G_{\F}^2 m_{\nu_{i}}^5} 
   \frac{1}{|U_{{\el}j}^*U_{{\el}i}|^{2}} 
  > 1.1\times10^{10} \sec |U_{{\el}i}|^{-2}
     \left(\frac{m_{\nu_{i}}}{\MeV}\right)^{-5}. 
\end{equation}
In any case, for $m_{\nu_{i}}\geqsim2m_{\el}$, the tree-level charged current
decay $\nu_i\to{\el}^-{\el}^+\nu_{\el}$ takes over, with the
lifetime 
\begin{equation}
\label{taueenu}
 \fl \tau_{\nu_i\to{\el}^-{\el}^+\nu_{\el}} 
  = \frac{192\pi^3}{G_{\F}^2 m_{\nu_{i}}^5} 
    \frac{1}{|U_{{\el}i}|^{2} f(m_{\el}/m_{\nu_i}) }
  = 2.4\times10^{4} \sec |U_{{\el}i}|^{-2} 
          \left(\frac{m_{\nu_i}}{\MeV}\right)^{-5},
\end{equation}
for $m_{\nu_{i}}\gg\,m_{\el}$ where $f(x)$ is a phase-space factor
($\simeq1$ for $x\,\ll\,1$).

Experiments at PSI and TRIUMF have set upper limits on the mixing
$|U_{{\el}i}|^2$ of any $\nu_{i}$ with mass in the range $4-54\MeV$
which can be emitted along with an electron in pion decay (see Bryman
1993), by measuring the branching ratio
$R_{\pi}=(\pi\to{\el}\nu)/(\pi\to\mu\nu)$ and/or searching for
additional peaks in the energy spectrum of $\pi\to{\el}\nu$ decays
(e.g. Bryman \etal 1983, Azeluos \etal 1986, De Leener-Rosier \etal
1991, Britton \etal 1992, 1994). Similar, although less stringent
bounds are obtained from studies of kaon decay (Yamazaki \etal
1984). Direct searches have been also carried out for decays of
heavier neutrinos with masses upto a few GeV produced through mixing
in accelerator beams of muon and electron neutrinos as well as for
unstable tau neutrinos produced through decays of $D_{s}$ charmed
mesons in `beam-dump' experiments (Bergsma \etal 1983, Cooper-Sarkar
\etal 1985, Bernardi \etal 1986). Searches have also been carried out
for radiative decays of electron and muon neutrinos (e.g. Oberauer
\etal 1987, Krakauer \etal 1991) in low energy reactor and accelerator
beams.

As we have discussed in \sref{nonrel}, primordial nucleosynthesis
restricts such decays in several distinct ways. The most general is
the constraint \eref{nxlimvis} on entropy generation
subsequent to nucleosynthesis which imposes an upper bound on the
lifetime given the relic energy density of the decaying neutrino as a
function of its mass (Sato and Kobayashi 1977, Miyama and Sato
1978). Using \eref{taueenu} this can be converted into a {\em
lower} limit on the mixing $|U_{{\el}i}|^2$ of a massive
neutrino. Kolb and Goldman (1979) noted that the limit thus extracted
from the lifetime bound obtained from consideration of the $\Htwo$
abundance by Dicus \etal (1978b) is {\em higher} than the upper limit
on this mixing as deduced from $\pi$ and $K$ decays, if
$m_{\nu}\leqsim9\MeV$. This conclusion was shown (Sarkar and Cooper
1984) to hold even using the less restrictive lifetime
bound \eref{nxlimvis} following from the more reliable constraint
$Y_{\pr}({\Hefour})<0.25$, but using improved experimental limits on
the mixing (Bryman \etal 1983). For Majorana neutrinos which have a
higher relic abundance, Krauss (1983b) found (using the Dicus \etal
(1978b) lifetime bound) that any neutrino mass below $23\MeV$ was
ruled out on the basis of this argument. A similar conclusion was
arrived at by Terasawa \etal (1988) who adopted the even more generous
bound $Y_{\pr}({\Hefour})<0.26$. The present situation is illustrated
in \fref{nutaudecay} which shows the upper bound on $\tau_{\nu}$
inferred from \fref{nxlimentexp}(a) (corresponding to the requirement
$Y_{\pr}({\Hefour})<0.25$), where we have calculated the relic
neutrino abundance assuming it is a Dirac particle. For $m_{\nu}$ less
than about $15\MeV$ these are {\em below} the lower bound to the
lifetime calculated from the best current limits on the mixing
$|U_{{\el}i}|^2$ (Britton \etal 1992, 1994, De Leener-Rosier \etal
1991). Recently Dodelson \etal (1994) have made a comprehensive study
of the lifetime bounds on an unstable neutrino, taking into account
many (small) effects ignored in previous calculations. They adopt the
more restrictive bound $Y_{\pr}({\Hefour})<0.24$, which leads to more
stringent constraints than those obtained previously, extending down
to a lifetime of $\Or(10^2)\sec$. However, as discussed in
\sref{he4}, this bound can no longer be considered reliable.  


It was believed (Lindley 1979, Cowsik 1981) that the radiative decays
of neutrinos heavier than about $5\MeV$ can be restricted further by
constraining the photofission of deuterium by the decay
photons. However Kolb and Scherrer (1982) argued that this constraint
does not apply to the {\em dominant} decay mode
$\nu_i\to{\el}^-{\el}^+\nu_{\el}$ since the rapid thermalization of
the decay ${\el}^{\pm}$ by scattering against the background photons
severely suppresses $\Htwo$ photofission. This is erroneous since
the background photons are themselves energetic enough during the BBN
era to be Compton scattered by the decay ${\el}^{\pm}$ to energies
above the threshold for $\Htwo$ photofission. Sarkar and Cooper
(1984) calculated the spectrum of the scattered photons and concluded
that the $\Htwo$ abundance would be depleted by a factor exceeding
100 unless the neutrino decay lifetime is less than about $20-100\sec$
for $m_{\nu}\sim5-100\MeV$. A similar constraint was obtained by
Krauss (1984). Subsequently Lindley (1985) pointed out that the
scattered photons were much more likely to undergo $\gamma-\gamma$
scattering on the energetic photons in the Wien tail of the thermal
background than photodissociate deuterium (see also Scherrer
1984). Taking this into account relaxes the upper bound on the
lifetime by a factor of about 100 (Lindley 1985) as shown in
\fref{nutaudecay}.\footnote{Kawasaki \etal (1986) and Terasawa \etal
 (1988) also studied this constraint using numerical methods but found
 it to be weaker by a factor of about 2 than the semi-analytic result
 of Lindley.} Thus Krauss (1985) and Sarkar (1986) concluded that
cosmological and laboratory limits appeared to allow an unstable
$\nu_{\tau}$ with a lifetime of $\Or(10^3)\sec$ and a mass between
$20\MeV$ and its (then) upper limit of $70\MeV$. Subsequently the
experimental limit on the $\nu_{\tau}$ mass has come down to $24\MeV$
while the laboratory limits on the mixing angle $|U_{{\el}i}|^2$ have
improved further. As shown in \fref{nutaudecay}, the experimental
lower bound on $\tau_{\nu_{i}\to{\el}^-{\el}^+\nu_{\el}}$ now exceeds
the cosmological upper bounds from $\Htwo$ photofission and entropy
generation, for a neutrino mass in the range $1-25\MeV$. Thus the
conclusion of Sarkar and Cooper (1984), viz. that
\begin{equation}
\label{nutaumass}
 m_{\nu_{\tau}} < 2 m_{\el}\ ,
\end{equation}
is reinstated. We emphasize that this bound is more general than
the similar one \eref{Dirbound} which only applies to Dirac neutrinos
which decay `invisibly' after nucleosynthesis.

Sarkar and Cooper (1984) had argued that a $\nu_{\tau}$ lighter than
$1\MeV$ has no decay modes which are fast enough to satisfy the
constraint \eref{nulifebound} from the energy density. The radiative
decay $\nu_{\tau}\to\nu_{\el}\gamma$ is generally too slow and in any
case is observationally required to have a lifetime greater than the
age of the universe (see Sciama 1993). The `invisible' decay
$\nu_{\tau}\to\nu_{\el}\bar{\nu_{\el}}\nu_{\el}$ is also GIM
suppressed but may be mediated sufficiently rapidly by Higgs scalars
in the left-right symmetric model
$SU(2)_{\L}{\otimes}SU(2)_{\R}{\otimes}U(1)_{B-L}$ (e.g. Roncadelli
and Senjanovi\'c 1981) or through GIM-violation by
flavour-changing-neutral-currents (FCNC) in other extensions of the
Standard Model (e.g. De R\'ujula and Glashow 1980, Hosotani
1981). However, this neccessarily enhances the radiative decay mode as
well (McKellar and Pakvasa 1983, Gronau and Yahalom 1984) and is thus
ruled out observationally. It is thus neccessary to invent a new
massless (or very light) particle for the neutrino to decay
into. Since giving neutrinos masses usually involves the spontaneous
violation of lepton number, a candidate particle is the associated
Goldstone boson, the Majoron (Chikashige \etal 1980, Gelmini and
Roncadelli 1981). The neutrino decay lifetime in such models is
usually too long (e.g. Schechter and Valle 1982) but can be made
sufficiently short if the model is made contrived enough. Although
Majorons which have couplings to the $Z^0$ are now ruled out by \LEP,
there still remain some viable models with, e.g. singlet Majorons (see
Gelmini and Roulet 1995). If neutrinos can indeed have fast decays
into Majorons then the BBN bounds on visible decays discussed above
are not relevant. Nevertheless the constraints based on the expansion
rate are still valid and additional constraints obtain if the final
state includes electron (anti)neutrinos (e.g. Terasawa and Sato 1987,
Kawasaki \etal 1994, Dodelson \etal 1994). Interestingly enough, such
decays can slightly reduce the $\Hefour$ abundance for a mass of
$\Or(1)\MeV$ and a lifetime of $\Or(1)\sec$, thus {\em weakening} the
bound on $N_{\nu}$.

\subsubsection{Neutrino oscillations:\label{nuosc}}

Neutrino {\em flavour} oscillations are not relevant in the early
universe since the number densities of all flavours are equal in
thermal equilibrium and all three species decouple at about the same
temperature.  However if there is mixing between the left-handed
(active) and right-handed (sterile) neutrinos, then oscillations may
bring these into thermal equilibrium boosting the expansion rate while
depleting the population of active (electron) neutrinos which
participate in nuclear reactions. Thus powerful bounds on such mixing
can be deduced from consideration of the effects on
nucleosynthesis. These considerations are particularly relevant to
reports of experimental anomalies attributed to the existence of
sterile neutrinos, e.g. the $17\keV$ anomaly in $\beta$-decay (Hime
and Jelly 1991, see Hime \etal 1991) and the recently reported
33.9\,MeV anomaly in $\pi$ decay (Armbruster \etal 1995, see Barger
\etal 1995)

The first estimates of these effects (e.g. Khlopov and Petcov 1981,
Fargion and Shepkin 1981, Langacker \etal 1986, Manohar 1987) did not
take into account the coherent forward scattering of the active
species (N\"otzold and Raffelt 1988) which provides a correction to
the average momentum ($\langle{p_{\el}}\rangle=3.15T$) of $\nu_{\el}$
in the thermal plasma. For $1\MeV\leqsim\,T\ll\,100\MeV$, this is
\begin{equation}
\label{effenergy}
 V_{\el} = \sqrt{2} G_{\F} n_{\gamma} \left(L - A \frac{T^2}{M_{W}^2}\right)\ ,
\end{equation}
where $A=4(1+0.5\cos^2\theta_{\W})(7\zeta(4)/2\zeta(3))^2=55$ and
$L$ is a sum of terms proportional to the lepton and baryon
asymmetries in the plasma (and therefore appears with opposite sign in
the corresponding effective energy for $\bar{\nu_{\el}}$). In the
presence of neutrino oscillations, the lepton asymmetry, if not too
large, is dynamically driven to zero on a time-scale large compared to
the oscillation time (Enqvist \etal 1990a, 1991); thus the average
self-energy correction is
$\langle{V_{\el}}\rangle=V_{\el}(L=0)$.\footnote{However for
 large neutrino degeneracy the first term in \eref{effenergy}
 dominates and oscillations between different active flavours becomes 
 the important process (Savage \etal 1991). If the degeneracy is as 
 large as $L\geqsim10^{-5}$, the constraints on active-sterile mixing 
 are evaded (Foot and Volkas 1995a); such an asymmetry can arise due 
 to the oscillations themselves (Foot \etal 1996).} Hence a
Mikheyev-Smirnov-Wolfenstein (MSW) resonant transition of $\nu_{\el}$
into $\nu_{\s}$ (and {\em simultaneously} $\bar{\nu_{\el}}$ into
$\bar{\nu_{\s}}$) can occur satisfying
$V_{\el}=\Delta\,m^2\cos2\theta_{\vac}/\langle{p_{\el}}\rangle$, only
if the mass-difference squared,
$\Delta\,m^2\equiv\,m_{\nu_{\s}}^2-m_{\nu_{\el}}^2$, is {\em
negative}, i.e.  opposite to the case in the Sun. If this occurs after
electron neutrino decoupling (at about $2\MeV$) but before neutron
freeze-out is complete (at about $0.2\MeV$), then the surviving
neutron fraction is larger, leading to increased helium
production.\footnote{In fact,
 resonant transitions of $\nu_{\el}$ to the (very slightly cooler)
 $\nu_{\mu}$ or $\nu_{\tau}$ in this temperature interval can have a
 similar but smaller effect; $Y_{\pr}$ is increased by at most $0.0013$
 (Langacker \etal 1987).}
Enqvist \etal (1990b) have examined these effects using the
semi-analytic formulation of Bernstein \etal (1989) and conclude that
the survival probability $P(\theta_{\vac})$ of
$\nu_{\el}/\bar{\nu_{\el}}$ must exceed 0.84 in order not to alter
$Y_{\pr}({\Hefour})$ by more than $4\%$. Using the Landau-Zener
formula for the probability of transition between adiabatic states,
they derived a severe bound on the vacuum mixing angle
$\theta_{\vac}$.

The case when $\Delta\,m^2$ is positive is more interesting, having
been proposed in the context of solutions to the Solar neutrino
problem (see Bahcall 1989). Here the major effect is that the sterile
neutrinos, which nominally decouple at a very high temperature and
thus have a small abundance relative to active neutrinos,
$n_{\nu_{\s}}/n_{\nu_{\a}}\approx0.1$ (Olive and Turner 1982), can be
brought back into thermal equilibrium through $\nu_{\a}-\nu_{\s}$
oscillations. The production rate (through incoherent scattering) is
\begin{equation}
 \Gamma_{\nu_{\s}} \simeq \frac{1}{2} (\sin^{2}2\theta_{\m})\
  \Gamma_{\nu_{\a}}\ ,
\end{equation}
where $\Gamma_{\nu_{\a}}$ is the total interaction rate of the active
species ($\Gamma_{\nu_{\el}}\simeq4G_{\F}^2T^5,
\Gamma_{\nu_{\mu,\tau}}=2.9G_{\F}^2T^5$) and the mixing angle in
matter is related to its vacuum value as
\begin{equation}
 \sin^{2} 2\theta_{\m} = (1 - 2x\cos 2\theta_{\vac} + x^2)^{-1} 
  \sin^{2} 2\theta_{\vac}\ ,
\end{equation}
where $x\equiv2\langle{p}\rangle\langle{V}\rangle/\Delta\,m^2$. (One
requires $\sin^{2}2\theta_{\m}<0.15$ to be able to ignore non-linear
feedback processes which would reduce $\Gamma_{\nu_{\s}}$.) The ratio
of the sterile neutrino production rate to the Hubble rate
\eref{H}, $\Gamma_{\nu_{\s}}/H$, thus has a maximum at
\begin{equation}
 T_{\max} = B_{\nu_{\a}} (\Delta\,m^2)^{1/6}\ ,
\end{equation}
where $B_{\nu_{\el}}=10.8$ and $B_{\nu_{\mu,\tau}}=13.3$. If
$\Gamma_{\nu_{\s}}/H>1$ at this point, then the sterile neutrinos will
be brought into equilibrium thus boosting the expansion rate, hence the
synthesized helium abundance. The BBN bound on $N_{\nu}$ can now be
translated into an upper bound on the mixing. For example, Barbieri
and Dolgov (1990, 1991) used $N_{\nu}<3.8$ (which requires
$T_{\max}>m_{\mu}$ (see \eref{Tdecnew}) to obtain
$(\sin\theta_{\vac})^{4}\Delta\,m^2<6\times10^{-3}{\eV}^2$ while
Kainulainen (1990) used $N_{\nu}<3.4$ (which requires
$T_{\max}>T_{\c}^{\qh}$) to obtain
$(\sin\theta_{\vac})^{4}\Delta\,m^2<3.6\times10^{-4}{\eV}^2$. (Somewhat
weaker bounds obtain for $\nu_{\mu}/\nu_{\tau}$ oscillations into
singlets.) Barbieri and Dolgov (1990) also noted that if
$\nu_{\el}-\nu_{\s}$ oscillations occur {\em after} electron neutrinos
decouple, then their number density is depleted, giving rise to a
(negative) neutrino chemical potential which increases the helium
abundance (see \eref{Ypbbf}). Then the bound $N_{\nu}<3.8$
corresponds to the excluded region $\sin^{2}2\theta_{\vac}\geqsim0.4$ and
$\Delta\,m^2\geqsim2\times10^{-7}{\eV}^2$.

Recently Enqvist \etal (1992b) have performed a thorough examination
of both cases, improving on approximations made in the earlier
estimates of the collision rates through detailed calculations. (In
fact we have quoted above their values for $\Gamma_{\nu_{\a}}$ and
$B_{\nu_{\a}}$.) They consider several possible constraints from
nucleosynthesis, viz. $N_{\nu}<3.1,3.4,3.8$ and perform numerical
calculations to determine the allowed parameters in the
$\Delta\,m^2-\sin^{2}2\theta_{\vac}$ plane. In contrast to the
previous results, these authors find that BBN considerations {\em rule
out} the large mixing-angle MSW solution to the Solar neutrino
problem. They also consider $\nu_{\mu}-\nu_{\s}$ mixing and show that
this cannot be a solution to the atmospheric neutrino anomaly (see
Beier \etal 1992, Perkins 1993). Their results are confirmed by the
similar calculations of Shi \etal (1993). Of course all these results
are {\em invalidated} if the bound on $N_{\nu}$ is relaxed to exceed
4, which we have argued (\sref{rel}) is allowed by the
observational data (Cardall and Fuller 1996b, Kernan and Sarkar 1996b).

These bounds have been discussed (e.g. Dixon and Nir 1991, Babu and
Rothstein 1991, Enqvist \etal 1992c,d, Cline 1992, 
Cline and Walker 1992) in connection with the $17\keV$ neutrino which
was seen to be emitted in $\beta$-decay with a mixing of about $1\%$
with the electron neutrino (Simpson 1985, see Hime 1992). The most
likely interpretation of this state was that it was either a singlet
neutrino or else a member of a pseudo-Dirac pair. The theoretical
possibilities as well as the constraints from nucleosynthesis (and
other cosmological/astrophysical arguments) have been comprehensively
reviewed by Gelmini \etal (1992). However the experimental evidence
now disfavours the existence of this particle (see Particle Data Group
1994), the signal for which was faked by a conspiracy of systematic
errors (Bowler and Jelley 1994). With regard to the KARMEN anomaly
(Armbruster \etal 1994) which has been interpreted as due a singlet
neutrino of mass $33.9\MeV$ mixing with all three doublet neutrinos
(Barger \etal 1995), the mixing angles are not known but are
restricted within certain limits
e.g. $|U_{{\el}x}|^2<8.5\times10^{-7}$,
$|U_{{\mu}x}|^2<2\times10^{-3}$, $|U_{{\tau}x}|^2<1$. If the mixing is
sufficiently large, the $x$ particle would be brought into equilibrium
at a temperature of a few GeV. Although its abundance would thus be
suppressed relative to doublet neutrinos by the entropy production in
the quark-hadron transition, it would still have a large energy
density during BBN since it would have become {\em non-relativistic}
by then. The $x$ decays can cause photofission of the synthesized
abundances (Langacker \etal 1986) so the mixing angles are required to
be large enough that such decays occur sufficiently early, obeying the
cosmological mass-lifetime constraints shown in
\fref{nutaudecay}. Such constraints have also been discussed in
connection with hypothetical sterile neutrinos having masses larger
than a GeV (Bamert \etal 1995).

\subsubsection{Neutrino magnetic moments:\label{numagmom}}

A massless neutrino has no electromagnetic properties but when the
Standard Model is extended to include a Dirac neutrino mass, this
generates a magnetic dipole moment (e.g. Lee and Schrock 1977)
\begin{equation}
\label{SMmu}
 \mu_{\nu} = \frac{3 e G_{\F}}{8\sqrt{2}\pi^2} 
           = 3.2\times10^{-19} \left(\frac{m_{\nu}}{\eV}\right)\mu_{\B}\ ,
\end{equation}
A Majorana neutrino, being its own antiparticle, has zero magnetic
(and electric) dipole moments by $CPT$ invariance. This refers to the
{\em diagonal} moments; in general flavour-changing transition
magnetic (and electric) moments exist for both Dirac and Majorana
neutrinos. The neutrino magnetic moment may be significantly enhanced
over the above estimate in extensions of the SM (see Pal 1992).

A Dirac magnetic moment allows the inert RH states to be produced in
the early universe through $\nu_{\L}{\el}\to\nu_{\R}{\el}$ scattering
(Morgan 1981a) with cross-section
\begin{equation}
 \sigma_{\nu_{\L}{\el}\to\nu_{\R}{\el}} 
 = \pi \left(\frac{\alpha}{m_{\el}}\right)^2 
    \left(\frac{\mu_{\nu}}{\mu_{\B}}\right)^2 
    \ln \left(\frac{q_{\max}^2}{q_{\min}^2}\right) \ ,
\end{equation}
Fukugita and Yazaki (1987) noted that
$q_{\max}=3.15T$ whereas $q_{\min}\sim2\pi/l_{\D}$ where
$l_{\D}=(T/4\pi\,n_{\el}\alpha)^{1/2}$ is the Debye length in the
plasma of electron density $n_{\el}$. By the arguments of
\sref{rel}, the $\nu_{\R}$ should go out of equilibrium early enough
that its abundance is adequately diluted by subsequent entropy
generation. This requires $\mu_{\nu}\leqsim1.5\times 10^{-11}\mu_{\B}$
according to the approximate calculation of Morgan (1981a). A more
careful analysis (Fukugita and Yazaki 1987) gives
\begin{equation} 
\label{muupbound}
 \mu_{\nu} < 5 \times 10^{-11} \mu_{\B} 
              \left(\frac{T_{\c}^{\qh}}{200\MeV}\right)^2 ,
\end{equation}
corresponding to the usual constraint $N_{\nu}<4$. (The conservative
constraint $N_{\nu}<4.5$ would not change this significantly.) This is
more stringent than direct experimental bounds,
e.g. $\mu_{\nu_{\el}}<1.8\times10^{-10}\mu_{\B}$ (Derbin \etal 1994)
and $\mu_{\nu_{\mu}}<1.7\times10^{-9}\mu_{\B}$ (Krakauer \etal
1990). If there is a primordial magnetic field $B$ then
spin-procession can further populate the RH states
(e.g. Shapiro and Wasserman 1981), leading to the correlated bound
$\mu_{\nu}<10^{-16}\mu_{\B}(B/10^{-9}{\rm G})^{-1}$ (Fukugita \etal
1988).

Giudice (1990) pointed out that the bound \eref{muupbound} applies only
to neutrinos which are relativistic at nucleosynthesis and can
therefore be evaded by the tau neutrino which is experimentally
allowed to have a mass upto 24 MeV. Indeed whereas such a massive
$\nu_{\tau}$ would nominally have too high a relic abundance, a
magnetic moment of $\Or(10^{-6})\mu_{\B}$ would enable it to
self-annihilate rather efficiently (through $\gamma$ rather than $Z^0$
exchange) so as to make $\Omega_{\nu_\tau}\sim1$ today. There is no
conflict with nucleosynthesis since the $\nu_{\tau}$ energy density
during BBN can be much less than that of a relativistic
neutrino. In fact Kawano \etal (1992) calculate that if the magnetic
moment of a MeV mass $\nu_{\tau}$ exceeds 
\begin{equation}
\label{mulobound}
 \mu_{\nu_{\tau}} \geqsim 7\times10^{-9} \mu_{\B}\ ,
\end{equation}
its energy density during BBN is sufficiently reduced that it
satisfies $Y_{\pr}({\Hefour})<0.24$ (see also Grasso and Kolb
1996). (For a much larger $\mu_{\nu_{\tau}}$, the self-annihilation of
$\nu_{\tau}$s is so efficient that effectively $N_{\nu}\simeq2$ rather
than 3, hence the $\Hefour$ abundance is actually reduced, rather than
increased, relative to standard BBN!) However the direct experimental
bound $\mu_{\nu_{\tau}}<5.4\times10^{-7}\mu_{\B}$ rules out the
possibility that a MeV mass tau neutrino can constitute the dark
matter (Cooper-Sarkar \etal 1992).

\subsubsection{New neutrino interactions:\label{nunewint}}

Apart from a magnetic moment, neutrinos may have additional
interactions in extensions of the Standard Model and this can be
constrained by BBN in a similar manner (Hecht 1971, Morgan 1981b). For
example, Grifols and Mass\'o (1987) have calculated a bound on the
neutrino charge-radius\footnote{However, this is better interpreted as
 a bound on the scattering cross-section (mediated through any process)
 since the neutrino charge-radius is {\em not} a gauge-invariant
 quantity (Lee and Shrock 1977).} defined through the expression
\begin{equation}
\label{chargeradbound}
 \langle r^2 \rangle \equiv 
 \left(\frac{\sigma_{{\el}^+{\el}^-\to\nu_{\R}\nu_{\R}}}
            {\pi\alpha^2q^2/54}\right)^{1/2} 
  < 7 \times 10^{-33}\ {\cm}^2 \ ,
\end{equation}
corresponding to the constraint $N_{\nu}<4$. Mass\'o and Toldr\`a
(1994a) have also considered a hypothetical vector-type interaction
{\em between} neutrinos, ${\cal
H}=F_{\V}(\bar{\nu_i}\gamma^{\mu}\nu_{i})(\bar{\nu_j}\gamma^{\mu}\nu_{j})$,
which can bring RH states into equilibrium. Requiring as
before that this does not happen below the quark-hadron transition
implies the limit
\begin{equation}
 F_{\V} < 3 \times 10^{-3} G_{\F}\ .
\end{equation}  
Another BBN constraint on non-standard interactions was derived by
Babu \etal (1991).

Kolb \etal (1986c) have studied hypothetical `generic' interacting
species, viz. particles which maintain good thermal contact with
neutrinos (or photons) throughout the BBN epoch. They show that the
effect on BBN depends on the particle mass and cannot be simply
parametrized in terms of $\Delta\,N_{\nu}$. An example is a massive
neutrino in the triplet-Majoron model (Gelmini and Roncadelli 1981)
which maintains equilibrium with light neutrinos through exchange of
Majorons --- the Goldstone boson associated with global lepton number
violation. Stringent bounds are then imposed on the Majoron couplings;
however this model has in any case been experimentally ruled out by
\LEP.

Interactions mediated by new gauge bosons will be considered in the
context of extended technicolour (\sref{tc}) and
superstring-motivated models (\S\,\ref{ss}).

\subsection{Technicolour\label{tc}} 

This is an attractive mechanism for spontaneously breaking the
electroweak $SU(2)_{\L}\otimes\,U(1)_{Y}$ symmetry {\em
non-perturbatively}, without introducing fundamental Higgs bosons.  It
does so in a manner akin to the breaking of the
$SU(2)_{\L}\otimes\,SU(2)_{\R}$ chiral symmetry of the (nearly)
massless $u$ and $d$ quarks by the formation of a $q\bar{q}$
condensate at $\Lambda_{\QCD}$ when the $SU(3)_{\c}$ colour force
becomes strong (see Farhi and Susskind 1981). A generic technicolour
scenario thus invokes new hyper-strong interactions with an intrinsic
scale of $\Lambda_{\TC}\approx0.5\TeV$, due to gauge interactions
with $N_{\TC}\geq3$ unbroken technicolours. These interactions bind
techniquarks $Q_{\T}$ in the fundamental $N_{\TC}$ representation of
$SU(N_{\TC})$, forming $Q_{\T}\bar{Q_{\T}}$ `technimeson' and
$Q_{\T}^{N_{\TC}}$ `technibaryon' bound states. The latter will have
integer spin if $N_{\TC}$ is even, and the choice often favoured is
$N_{\TC}=4$. In this case the lightest technimeson would be expected
to be short-lived with $\tau\,\ll\,1$ sec, thus evading BBN
constraints, but the lightest technibaryon, which has a mass
\begin{equation}
 m_{\TB} \simeq m_{\p} \left(\Lambda_{\TC} \over \Lambda_{\QCD}\right)  
  = m_{\p} \left(v \over f_\pi\right) = 2.4\,{\TeV}\ ,
\end{equation} 
is likely to be metastable, by analogy with the proton of QCD. Indeed
as in QCD, there is no renormalizable interaction that can cause
technibaryon decay. However, the minimal technicolour model must in
any case be extended to incorporate quark and lepton masses, and one
might anticipate that it is unified in some kind of techni-GUT.
Therefore one expects, in general, higher-order effective
non-renormalizable interactions which cause technibaryon decay, of the
form
\begin{equation}
 {\cal L}_{\ETC} = \frac{Q_{\T}^{N_{\TC}} f^n}{\Lambda_{\ETC}^{3/2 
  (N_{\TC} + n) - 4}}\ ,
\end{equation}
where $f$ is a quark or lepton field and $\Lambda_{\ETC}$ is some mass
scale $\gg\,\Lambda_{\TC}$ at which the effective interaction is
generated.  These would imply a technibaryon lifetime
\begin{equation}
\label{tauTB}
 \tau_{\TB} \simeq \frac{1}{\Lambda_{\TC}} 
  \left(\frac{\Lambda_{\ETC}}{\Lambda_{\TC}}\right)^{3(N_{\TC}+n) - 8}\ , 
\end{equation} 
i.e. $\sim10^{-27}(\Lambda_{\ETC}/\Lambda_{\TC})^4\sec$, for the
favoured case $N_{\TC}=4$ with the minimal choice $n=0$.

Estimating the self-annihilation cross-section of technibaryons to be
(e.g. Chivukula and Walker 1990)
\begin{equation}
\label{TBann}
 {\langle \sigma v \rangle}_{\TB} \simeq 
  {\langle\sigma v \rangle}_{{\p}\bar{\p}} 
  \left(\frac{m_{\p}}{m_{\TB}}\right)^2 \simeq 3 \times 10^{-5} {\GeV}^2 ,
\end{equation}
the {\em minimum} expected relic abundance is
$m_{\TB}\,n_{\TB}/n_\gamma\,=\,3\times10^{-13}\GeV$, where we have
taken account of entropy generation following freeze-out at
$\approx70\GeV$. Unstable technibaryons with such a small abundance
are not constrained by nucleosynthesis (Dodelson 1989). However
technibaryons may have a much higher relic density if they possess an
asymmetry of the same order as the baryon asymmetry (Nussinov
1985). If the latter is due to a net $B-L$ generated at some high
energy scale, then this would be subsequently distributed among {\em
all} electroweak doublets by fermion-number violating processes in the
Standard Model at temperatures above the electroweak scale (see
Shaposhnikov 1991, 1992), thus naturally generating a technibaryon
asymmetry as well. If such $B+L$ violating processes cease being
important below a temperature
$T_{*}\,\simeq\,T_{c}^{\EW}\,\approx\,300\GeV$, then the
technibaryon-to-baryon ratio, which is suppressed by a factor
$[m_{\TB}(T_{*})/T_{*}]^{3/2}\e^{-m_{\TB}(T_{*})/T_{*}}$, is just
right to give $\Omega_{\TB}\,\simeq\,1$ (Barr \etal 1990),
i.e. $m_{\TB}\,n_{\TB}/n_\gamma\,=\,3\times10^{-8}\GeV$. As can be
seen from figures \ref{nxlimentexp} and \ref{nxlimcascade}, lifetimes
between $\approx1\sec$ and $\approx10^{13}\sec$ are forbidden for
particles with such an abundance. If technibaryons decay with a longer
lifetime, i.e. after (re)combination, their decay products would be
directly observable today. Constraints from the diffuse gamma ray
background (Dodelson 1989) as well as the diffuse high energy neutrino
background (Gondolo \etal 1993) then extend the lower lifetime bound
all the way to $\approx3\times10^{17}\yr$ (Ellis \etal 1992),
i.e. such particles should be essentially stable and an important
component of the dark matter.\footnote{Although technibaryons with
masses upto a few TeV are experimentally ruled out as constituents of
the Galactic dark matter if they have {\em coherent} weak interactions
(e.g. Ahlen \etal 1987, Caldwell \etal 1988, Boehm \etal 1991), the
{\em lightest} technibaryon may well be an electroweak singlet (as
well as charge and colour neutral), thus unconstrained by such direct
searches (e.g. Chivukula \etal 1993).}  These lifetimes bounds imply
that
\begin{equation}
 \Lambda_{\ETC} \leqsim 6 \times 10^{9}\ {\GeV} \quad {\rm or} \quad 
  \Lambda_{\ETC} \geqsim 10^{16}\ {\GeV}\ .
\end{equation}
The former case would be applicable to any extended technicolour (ETC)
model containing interactions that violate technibaryon number, while
the latter case could accomodate a techni-GUT at the usual grand
unification scale.

Another constraint on ETC models follows from the BBN bound on
$N_{\nu}$. Krauss \etal (1993) note that such models typically contain
right-handed neutrinos (see King 1995) which can be produced through
$\nu_{\L}\bar{\nu_{\L}}\to\nu_{\R}\bar{\nu_{\R}}$ which proceeeds
through the exchange of ETC gauge bosons of mass $M_{\ETC}$. Thus one
should require this process to go out of equilibrium before
 the quark-hadron
phase transition (\S\,\ref{rel}). Krauss \etal (1995) adopt the bound
$N_{\nu}\,\leqsim\,3.5$ to derive
\begin{equation}
 \frac{M_{\ETC}}{g_{\ETC}} \geqsim 2 \times 10^{4} \GeV\ ,
\end{equation}
where $g_{\ETC}$ is the relevant ETC gauge coupling. While the above
cosmological constraints on technicolour are not particularly
restrictive, the basic idea has in any case fallen into disfavour
because of the difficulties in constructing realistic phenomenological
models which are consistent with experimental limits on
flavour-changing neutral currents and light technipion states (see
King 1995). Also the radiative corrections to SM parameters are
generally expected to be large (see Lane 1993), in conflict with the
experimental data (see Langacker 1994).

\subsection{Supersymmetry and supergravity\label{susy}} 

Because of such experimental difficulties with dynamical electroweak
symmetry breaking, it is now generally accepted that the problems
associated with a fundamental Higgs boson are better cured by
supersymmetry, in a manner consistent with all such experimental
constraints (see Adriani \etal 1993, Baer \etal 1995). As noted
earlier, the quadratically divergent radiative corrections to the mass
of a fundamental Higgs scalar can be cancelled by postulating that for
every known fermion (boson), there is a boson (fermion) with the {\em
same} interactions. Thus each particle of the Standard Model must be
accompanied by its superpartner --- spin-$\half$ partners for the
gauge and Higgs bosons and spin-0 partners for the leptons and quarks
--- in the minimal supersymmetric Standard Model (see Nilles 1984,
Haber and Kane 1985). The Lagrangian consists of a supersymmetric part
with gauge interactions as in the SM while the Yukawa interactions are
derived from the `superpotential'
\begin{equation}
\label{mssmsp}
 P_{\rm MSSM} = h_u Q H_2 u^c + h_d Q H_1 d^c + h_e L H_1 e^c
              + \mu H_1 H_2\ . 
\end{equation}
The chiral superfields $Q$ contain the LH quark doublets, $L$ the LH
lepton doublets, and $u^c$, $d^c$, $e^c$ the charge conjugates of the
RH up--type quarks, RH down--type quarks and RH electron--type leptons
respectively. Two Higgs fields are required to give masses separately
to the up-type charge 2/3 quarks, and to the down-type charge-1/3
quarks and leptons; the last term is a mixing between them which is
permitted by both gauge symmetry and supersymmetry (see
\S\,\ref{nmssm}).

Supersymmetry must neccessarily be broken in the low energy world
since the superpartners of the known particles (with the same mass)
have not been observed. However if supersymmetry is to provide a
solution to the hierarchy problem, the mass-splitting $\tilde{m}$
between ordinary particles and their superpartners cannot be
significantly higher than the electroweak scale. Although
superparticles particles have not yet been directly produced at
accelerators, they would influence through their virtual effects, the
evolution with energy of the gauge couplings in the Standard
Model. Interestingly enough, the precision data from \LEP demonstrate
that only in this case would there be the desired unification of all
three couplings (e.g Ellis \etal 1991, Amaldi \etal 1991, Langacker
and Luo 1991, see de Boer 1994); further this happens at a
sufficiently high energy ($\approx2\times10^{16}\GeV$), so as to
account for the failure to detect proton decay upto a lifetime of
$\sim10^{32}\yr$ (see Perkins 1984, Particle Data Group 1996) which
rules out most non-SUSY GUTs (see Langacker 1981, Enqvist and
Nanopoulos 1986). Reversing the argument, a SUSY-GUT can then predict,
say the weak mixing angle, to a precision better than $0.1\%$, in
excellent agreement with experiment (see Dimopoulos 1995, Ellis
1995). Another attraction of supersymmetry is that it provides a
natural mechanism for breaking of the electroweak symmetry at the
correct energy scale through radiative corrections to the Higgs mass,
if the top quark is sufficiently heavy (see Ib\'a\~nez and Ross 1993);
the recently discovered top quark does indeed have the required mass
(Ross and Roberts 1992).

The first phenomenological models to be constructed attempted to
incorporate global supersymmetry down to the electroweak scale (see
Fayet and Ferrara 1975). Such models therefore contain a massless
goldstino ($\tilde{G}$), the spin-$\half$ fermion associated with the
spontaneous breaking of supersymmetry at a scale of $\Or(\TeV)$, as
well as a new light spin-$\half$ fermion, the photino
($\tilde{\gamma}$). Both the goldstino and photino couple to matter
with strength comparable to a doublet neutrino (see Fayet 1979). Thus,
having two degrees of freedom each, they count as two extra neutrino
species during nucleosynthesis and thus are in conflict with the often
used bound $N_{\nu}<4$ (Dimopoulos and Turner 1982), or even our
conservative bound $N_{\nu}<4.5$. Therefore it is neccessary to make
these particles decouple earlier than the quark-hadron phase
transition in order to dilute their abundance. To weaken their
interactions adequately then requires that the supersymmetry breaking
scale be raised above $\approx10\TeV$ (Sciama 1982)

However models with global supersymmetry have severe difficulties,
e.g. in generating the neccessary mass-splitting of ${\cal O}(m_W)$
between ordinary particles and their superpartners, and because they
possess a large cosmological constant which cannot even be fine tuned
to zero (see Fayet 1984). Thus it is neccessary to consider {\em
local} supersymmetry, i.e. supergravity (see Van Nieuwenhuizen 1981),
which provides an implicit link with gravity. (Indeed superstring
theories (see Green \etal 1993) which unify gravity with the other
interactions, albeit in a higher-dimensional space, yield supergravity
as the effective field theory in four-dimensions at energies small
compared to the compactification scale which is of order $M_{\Pl}$.)
In supergravity models, the goldstino is eliminated by the super-Higgs
mechanism which gives a mass to the gravitino, spin-$\3half$
superpartner of the graviton (Deser and Zumino 1977). The
helicity-$\3half$ components interact only gravitationally; however if
the gravitino mass is very small then its interactions are governed by
its helicity-$\half$ component which is just the goldstino associated
with global SUSY breaking (Fayet 1979). At energies much smaller than
the superpartner masses, the typical scattering cross-section is
$\sigma_{\tilde{G}{\el}\to\tilde\gamma{\el}}
\approx0.4\sigma_{\nu{\el}\to\nu{\el}}(m_{3/2}/10^{-5}\eV)^{-2}$
(Fayet 1979), assuming that the photino is also light. Hence a
sufficiently light gravitino would have the same cosmological
abundance as a massless two-component neutrino. Requiring that the
gravitino decouple at a temperature $T>m_\mu$ (see \eref{Tdecnew})
then implies (Fayet 1982)
\begin{equation}
\label{fayetbound}
 m_{3/2} \geqsim 10^{-2}\ \eV\ , 
\end{equation}
which is rather more restrictive than the lower limit of
$\sim10^{-6}\eV$ deduced from laboratory experiments (see Fayet 1987).

In fact the gravitino is expected to be much heavier in the class of
supergravity models which have been phenomenologically most successful
(see Nath \etal 1984, Nilles 1984). Here supersymmetry is broken by
non-perturbative dynamics at a scale $\Lambda$ in a `hidden sector'
which interacts with the visible sector only through gravitational
interactions (Witten 1981b, see Nilles 1990).\footnote{Alternatively,
the `messenger sector' can have gauge interactions so that the soft
masses are generated by radiative corrections while the gravitino,
which interacts only gravitationally, remains light:
$m_{3/2}\sim\,m_{W}^2/M_{\Pl}\sim10^{-6}\eV$ (e.g. Dine and Nelson
1993, Dine \etal 1996).} Supersymmetry breaking is then communicated
to the low energy world only through `soft' supersymmetry breaking
masses for the sfermions and gauginos (superpartners of the fermions
and gauge bosons) and a mass for the gravitino, all of which are of
order the effective supersymmetry breaking scale in the visible
sector, i.e. close to the electroweak scale (e.g. Barbieri \etal 1982,
Chamseddine \etal 1982, Nilles \etal 1983, Alvarez-Gaume \etal
1983). Thus if supersymmetry is to solve the gauge hierarchy problem,
the gravitino mass must be no higher than $\sim1\TeV$. This however
poses a serious cosmological problem as we discuss below.

\subsubsection{The gravitino problem, baryogenesis and 
 inflation:\label{gravprob}} 

At high energies, the dominant interactions of the gravitino with
other particles and their superpartners at high energies come from its
helicity-$\3half$ component (rather than its helicity-$\half$
goldstino component). For example it can decay into a gauge boson
$A_{\mu}$ and its gaugino partner $\lambda$ through a dimension-5
operator, with lifetime
$\tau_{3/2\,\to\,A_{\mu}\lambda}\approx\,4M_{\Pl}^2/N_{\c}m_{3/2}^3$
where $N_{\c}$ is the number of available channels, e.g.
\begin{eqnarray}
\label{gravtau}
 \tau_{3/2 \to \tilde{\gamma} \gamma} &= 3.9\times10^{5} \sec
  &\left(\frac{m_{3/2}}{\TeV}\right)^{-3}, \nonumber\\
 \tau_{3/2 \to \tilde{g} g} &= 4.4\times10^{4} \sec
  &\left(\frac {m_{3/2}}{\TeV}\right)^{-3}\ ,
\end{eqnarray}
assuming $m_{A_{\mu},\,\lambda}\,\ll\,m_{3/2}$.

It was first noted by Weinberg (1982) that notwithstanding their very
weak interactions, massive gravitinos would have been abundantly
produced in the early universe at temperatures close to the Planck
scale and would thus come to matter-dominate the universe when the
temperature dropped below their mass. Their subsequent decays would
then completely disrupt primordial nucleosynthesis, thus creating a
cosmological crisis for supergravity. It was suggested (Ellis \etal
1983, Krauss 1983a) that this problem could be solved by invoking an
inflationary phase, just as for GUT monopoles.  However, unlike the
latter, gravitinos can be recreated by scattering processes during the
inevitable reheating phase following inflation as well as (in a
model-dependent manner) through direct decays of the scalar field
driving inflation (Nanopoulos \etal 1983). The gravitino abundance
produced by $2\to2$ processes involving gauge bosons and gauginos
during reheating was computed by Ellis \etal (1984b) to be,
\begin{equation}
\label{ekn}
 \frac{n_{3/2}}{n_{\gamma}} = 2.4\times10^{-13} 
  \left(\frac{T_{\R}}{10^9\GeV}\right) 
  \left[1 - 0.018 \ln \left(\frac{T_{\R}}{10^9\GeV}\right)\right]\ ,
\end{equation}
at $T\,\ll\,m_{\el}$, where $T_{\R}$ is the maximum temperature
reached during reheating. This is a conservative lower bound to the
true abundance, for example Kawasaki and Moroi (1995a) estimate an
abundance higher by a factor of 4 after including interaction terms
between the gravitino and chiral multiplets. Recently Fischler (1994)
has claimed that gravitinos can be brought into thermal equilibrium at
temperatures well below the Planck scale via interactions of their
goldstino component with a cross-section which increases as $T^2$ due
to the breaking of supersymmetry by finite temperature effects. If so,
their production during reheating would be far more efficient and
yield a relic abundance
\begin{eqnarray}
\label{fischlerabun}
 \frac{n_{3/2}}{n_{\gamma}} &\approx
  &\frac{g_*^{1/2}\alpha_{\s}^3 T^3}{m_{3/2}^3 M_{\Pl}^2} \nonumber\\
  &\sim & 3\times10^{-13} \left(\frac{T_{\R}}{10^5\GeV}\right)^3
            \left(\frac{m_{3/2}}{\TeV}\right)^{-2} .
\end{eqnarray}
However Leigh and Rattazzi (1995) argue on general grounds that there
can be no such enhancement of gravitino production. Ellis \etal (1996)
perform an explicit calculation of finite-temperature effects for
$m_{3/2},\tilde{m}\ll\,T\ll\Lambda$ and demonstrate that these do not
alter the estimate in \eref{ekn}.

The BBN constraints on massive decaying particles shown in figures
\ref{nxlimentexp} and \ref{nxlimcascade} then provide a restrictive
upper limit to the reheating temperature after inflation, dependent on
the gravitino lifetime. For example, Ellis \etal (1985b) quoted
\begin{equation} 
\label{TRlimens}
 T_{\R} \leqsim 2.5 \times 10^8 \GeV 
                 \left(\frac{m_{3/2}}{100\GeV}\right)^{-1}, 
 \quad \for \quad m_{3/2} \leqsim 1.6 \TeV\ ,
\end{equation} 
(taking $f_{\gamma}=0.5$), from simple
considerations of ${\Htwo}+{\Hethree}$ overproduction due to
$\Hefour$ photofission which gave the constraint
\begin{equation} 
 m_{3/2} \frac{n_{3/2}}{n_{\gamma}} \leqsim 3\times10^{-12} \GeV
                                             f_{\gamma}^{-1} , 
\end{equation}
shown as a dashed line in \fref{nxlimcascade}. However, as is seen
from the figure, a more detailed calculation of this process (Ellis
\etal 1992) actually yields a more restrictive constraint for a
radiative lifetime greater than $\sim2\times10^7\sec$, corresponding
to $m_{3/2}\leqsim300\GeV$, but a less stringent constraint for
shorter lifetimes.\footnote{Kawasaki and Moroi (1995a) quote a limit
 more stringent by a factor of about 100, of which, a factor of 4 comes
 from their more generous estimate of the relic gravitino
 abundance. The remaining discrepancy is because they obtain (by
 numerical integration of the governing equations) a significantly more
 stringent constraint on ${\Htwo}+{\Hethree}$ overproduction, which, as
 noted earlier, disagrees with both the analytic estimate of Ellis
 \etal (1992) as well as the Monte Carlo calculation of Protheroe \etal
 (1995).}. Hence the true bound is (taking $f_{\gamma}=0.5$),
\begin{equation} 
\label{TRlimHe4phot}
 T_{\R} \leqsim 10^8 \GeV \quad 
  \for \quad m_{3/2} = 100\ \GeV.
\end{equation} 
Using the results of Ellis \etal (1992) we have obtained the upper
bound on $T_{\R}$ implied by the relic abundance \eref{ekn} as a
function of the gravitino mass (calculated using the
lifetime \eref{gravtau}) and show this in \fref{TR}. For a gravitino
mass of $1\TeV$, the radiative lifetime is about $4\times10^5\sec$ and
the best constraint now comes from requiring that the photofission of
deuterium not reduce its abundance below the observational lower limit
(Juszkiewicz \etal 1985, Dimopoulos \etal 1989). The improved
calculation of Ellis \etal (1992) gives for this bound,
\begin{equation} 
\label{TRlimDphot}
 T_{\R} \leqsim 2.5 \times 10^9 \GeV \quad 
  \for \quad m_{3/2} = 1\ \TeV,
\end{equation}
where we have taken $f_{\gamma}\simeq0.8$ as is appropriate for such a
massive gravitino.  Photofission processes become ineffective for
$\tau\leqsim10^4\sec$ but now there are new constraints from the
effect of hadrons in the showers on the $^4$He abundance (Reno and
Seckel 1989, Dimopoulos \etal 1989). If the gravitino mass is $10\TeV$
with a corresponding lifetime of
$\tau_{{3/2}\to\tilde{g}g}\sim50\sec$, this bound is
\begin{equation} 
\label{TRlimhad} 
T_{\R} \leqsim 6 \times 10^9 \GeV \quad
  \for \quad m_{3/2} = 10\ \TeV.
\end{equation}
Weinberg (1982) had suggested that the entropy release in the decays
of a gravitino of mass exceeding $\approx10\TeV$ would reheat the
universe to a temperature high enough to restart nucleosynthesis, thus
evading the cosmological problem. However, as noted earlier, particle
decays following an exponential decay law cannot actually raise the
temperature but only slow down its rate of decrease (Scherrer and
Turner 1985), hence one should really require the gravitino to be
massive enough that it decays before the begining of nucleosynthesis.
A careful calculation by Scherrer \etal (1991) taking into account the
effects of hadronic decays shows that the lower bound on the mass is
then
\begin{equation}
\label{m3/2lolim}
 m_{3/2} \geqsim 53\ \TeV.
\end{equation}
However such a large gravitino mass cannot be accomodated in (minimal)
supergravity models without destabilizing the hierarchy.


Other constraints on the gravitino abundance follow from examination
of the effects of the annihilation of antiprotons produced in the
decay chain
${3/2}\to\tilde{g}g,\,\tilde{g}\,\to\,q\bar{q}\tilde{\gamma}$ (Khlopov
and Linde 1984, Ellis \etal 1985b, Halm 1987, Dominguez-Tenreiro 1987)
but these are not as restrictive as those given above. The effects of
the decay ${3/2}\to\nu\tilde{\nu}$ have been studied by de Laix and
Scherrer (1993). Correcting earlier estimates by Frieman and Giudice
(1989) and Gratsias \etal (1991), the tightest bound they obtain is
$T_{\R}\leqsim2\times10^{10}\GeV$ for $m_{3/2}=10\TeV$. (Rather
different bounds are obtained by Kawasaki and Moroi (1995c) by
numerical solution of the governing equations but, as noted earlier,
their cascade spectrum disagrees with that obtained by other workers.)

If the gravitino is in fact the LSP, then we can demand that its relic
abundance respect the cosmological bound \eref{nxlimtoday} on the
present energy density in massive stable particles.  Using \eref{ekn},
this requires (Ellis \etal 1985b):
\begin{equation} 
\label{TRlimstable}
 T_{\R} \leqsim 10^{12} \left(\frac{m_{3/2}}{100\GeV}\right)^{-1}\ \GeV\ ,
\end{equation}
while the decays of the next-to-lightest supersymmetric particle
(NLSP), typically the neutralino $\chi^0$, can presumably be made
consistent with the BBN constraints since such particles can usually
self-annihilate sufficiently strongly to reduce their relic abundance
to an acceptable level. Moroi \etal (1993) have reexamined this
question and taken into account the (small) additional gravitino
production from the NLSP decays. They note that the relic $\chi^0$
abundance according to recent calculations (e.g. Drees and Nojiri
1993) is in fact sufficiently high (essentially due to improved lower
limits on sparticle masses) that the ${\Htwo}+{\Hethree}$
photoproduction constraint calculated by Ellis \etal (1992) requires
$\tau_{\chi^0}\leqsim5\times 10^{6}\sec$. For this to be so, the
neutralino has to be sufficiently heavy relative to the gravitino, e.g
for $m_{\chi^0}=50\GeV$, the gravitino mass must be less than
$3.4\GeV$, but for $m_{\chi^0}=1\TeV$ the gravitino mass can be as
high as $772\GeV$.

Moroi \etal (1993) also reevaluate the bound \eref{TRlimstable} on
$T_{\R}$ following from the relic energy density argument for a stable
gravitino which is much lighter than the sfermions. They consider the
regime $m_{3/2}\ll\tilde{m}\ll\,T$ for which they find that the
gravitino (goldstino) annihilation cross-section is enhanced
proportional to $m_{3/2}^{-2}$ so that the bound on $T_{\R}$ from
consideration of the relic energy density decreases proportionally to
$m_{3/2}$. Of course this bound evaporates when the gravitino becomes
sufficiently light that its relic abundance from thermal equilibrium
(at the Planck scale) comes within observational constraints. This
limiting mass is (Pagels and Primack 1982)
\begin{equation}
\label{m3/2uplim}
 m_{3/2} \leqsim 1\ \keV;
\end{equation}
the increase by a factor of $\approx10$ over the corresponding bound
\eref{nuuplim} for neutrinos is because of the dilution by a factor of
$\approx10$ of gravitinos (which decouple at $T\,\gg\,T_{\c}^{\EW}$)
relative to neutrinos which decouple at a few MeV (see
\tref{tabthermhist}). Moroi \etal also consider the effect of a light
gravitino on the speed-up rate during BBN following Fayet (1982) (see
\eref{fayetbound}). They find that the relevant dominant
thermalization process is gravitino (goldstino) annihilation to light
lepton pairs and obtain the bound
$m_{3/2}>10^{-4}\eV(m_{\tilde\ell}/100\GeV)$ by requiring decoupling
at $T>T_{\c}^{\qh}$ (corresponding to the constraint
$N_\nu<3.3$). However Gherghetta (1996) notes that in the more
appropriate regime $m_{3/2}\ll\,T\ll\tilde{m}$, the dominant
equilibrating process is actually gravitino (goldstino) annihilation
into photons, the cross-section for which is considerably smaller.
Thus the mass bound is weakened to
\begin{equation}
 m_{3/2} \geqsim 10^{-6}\eV \left(\frac{m_{\tilde\gamma}}{100\GeV}\right)^{1/2}\ ,
\end{equation}
if the gravitinos are required to decouple at $T>m_\mu$, corresponding
to the constraint $N_\nu<3.7$ (compare with the bound
\eref{fayetbound} obtained assuming a light photino). Note that such a
reduction of the annihilation cross-section would also degrade the
bounds on $T_{\R}$ quoted by Moroi \etal (1993).

The realization that the F-R-W universe we inhabit cannot have
achieved a temperature higher than $\sim10^{9}\GeV$, if the production
of electroweak scale relic gravitinos is to be adequately suppressed,
had a big impact on phenomenological models of baryogenesis.  A decade
ago when these bounds were first presented (e.g. Ellis \etal 1985b)
baryogenesis was generally believed to be due to the
out-of-equilibrium $B$-violating decays of heavy bosons with masses of
order the unification scale (see Kolb and Turner 1983). In
supersymmetric models, protons can decay efficiently through
dimension-5 operators (see Enqvist and Nanopoulos 1986), hence the
experimental lower bounds
$\tau_{\p\to{\mu^+}{K^0},\nu{K^+}}>10^{32}\yr$ (Particle Data Goup
1996) then implies that the mass of the relevant (Higgs triplet)
bosons is rather large, viz. $m_{\tilde{H_3}}\geqsim10^{16}\GeV$
(e.g. Ellis \etal 1982). It may be possible to suppress the dangerous
dimension-5 operators (e.g. Coughlan \etal 1985) but even so one has a
lower limit $m_{\tilde{H_3}}\geqsim10^{11}\GeV$ from consideration of
the (unavoidable) dimension-6 operators . Such heavy particles {\em
cannot} be thermally generated after inflation given the associated
bounds on gravitino production, hence the standard scenario for
baryogenesis is no longer viable!

This conflict motivated studies of various alternative
possibilities. One suggestion was that the Higgs triplets could be
created directly through the decays of the scalar field driving
inflation (Coughlan \etal 1985, Mahajan 1986). For this to be
possible, the inflaton field should of course be significantly heavier
than $10^{11}\GeV$. However the amplitude of the CMB temperature
fluctuations observed by \COBE, interpreted as due to quantum
fluctuations of the inflaton field (see Liddle and Lyth 1993, Turner
1993), suggest an inflaton mass which is comparable in value
(e.g. Ross and Sarkar 1996), thus leaving little room for manouver. A
similar idea is to invoke decays of the inflaton into squarks, which
decay in turn creating a baryon asymmetry if the superpotential
includes a term which violates $R$-parity (Dimopoulos and Hall
1987). Here, the reheat temperature is required to be very low, less
than a few GeV, in order that the generated asymmetry is not washed
out by scattering processes and inverse decays. Another suggestion due
to Fukugita and Yanagida (1986) which has received much attention is
that the out-of-equilibrium $L$-violating decays of heavy right-handed
neutrinos generate a {\em lepton} asymmetry, which is subsequently
reprocessed by $B-L$ conserving fermion number violating interactions
in the Standard Model (see below) into a baryon asymmetry. The
$\nu_{\R}$ should thus have a mass less that $\sim10^{11}\GeV$, which
in turn imposes an interesting {\em lower} limit on the masses of the
light (left-handed) neutrinos generated by the `see-saw' mechanism,
$m_{\nu_{\L}}\sim\,m_{q,l}^2/m_{\nu_{\R}}$ (see Pl\"umacher 1996).

A different approach is to try and evade the gravitino problem
altogether by decoupling the gravitino mass from the electroweak
scale, i.e. making it lighter than $1\keV$ \eref{m3/2uplim}), or
heavier than $50\TeV$ \eref{m3/2lolim}. As mentioned
earlier, this is not possible in minimal supergravity but can be
achieved in `no-scale' supergravity models which are based on
non-compact K\"ahler manifolds (see Lahanas and Nanopoulos
1987).\footnote{Here the scale of supersymmetry breaking (hence the
gravitino mass) is classically undetermined and can be suitably fixed
by radiative corrections. However, the construction of an acceptable
cosmology is then beset by the `Polonyi problem', viz. the late
release of entropy stored in very weakly coupled fields in such models
(e.g. German and Ross 1986, Ellis \etal 1986c, Bertolami 1988).} Also
the gravitino is naturally light in gauge-mediated supersymmetry
breaking models (Dine and Nelson 1993, Dine \etal 1996) which are
consequently free of cosmological problems.

Thus the identification of the gravitino problem stimulated the study
of mechanisms for low temperature baryogenesis. Coincidentally two
such possibilities were proposed at that time and these have since
come under intense scrutiny (see Dolgov 1992). The first followed the
realization that since fermion number is violated (although $B-L$ is
conserved) even in the Standard Model at temperatures above the
electroweak scale (Kuzmin \etal 1985), a baryon asymmetry may be
generated utilising the (small) CP violation in the CKM mixing of the
quarks if the neccessary out-of-equilibrium conditions can be achieved
during the electroweak phase transition. (These detailed studies
indicate however that successful baryogenesis probably requires
extension of the SM to include, e.g. supersymmetry, in order to
increase the sources of $CP$ violation as well as make the electroweak
phase transition sufficiently first-order (see Krauss and Rey 1992,
Cohen \etal 1994, Rubakov and Shaposhnikov 1996). The second
mechanism, specific to supergravity models, is based on the
observation that sfermion fields will develop large expectation values
along `flat' directions in the scalar potential during the
inflationary phase (Affleck and Dine 1984). A baryon asymmetry can
then be generated at a much lower temperature, when the Hubble
parameter becomes less than the effective mass, as the coherent
oscillations in the fields decay (e.g. Ellis \etal 1987). Recently,
Dine \etal (1995, 1996) have emphasized the role of non-renormalizable
operators in the superpotential in stabilizing the flat direction and
shown that a baryon asymmetry of the required magnitude can indeed
arise.

In inflationary models, the reheat temperature $T_{\R}$ is determined
by the couplings to matter fields of the scalar `inflaton' field
$\phi$, which drives inflation at an energy scale $\Delta$. Inflation
ends when $\phi$ evolves to the minimum of its potential and begins to
oscillate about it until it decays, converting its vacuum energy,
$\Delta^4$, into radiation. The inflaton is required to be a gauge
singlet in order that its quantum fluctuations during the vacuum
energy dominated `De Sitter phase' of expansion do not generate
temperature fluctuations in the CMB in excess of those observed by
\COBE (see Olive 1990a). Thus its couplings to matter fields are
neccessarily very weak and reheating is a slow process. The inflaton
mass is $m_{\phi}\sim\Delta^2/M_{\Pl}$ and its decay width is
\begin{equation} 
\label{gammaphi}
 \Gamma_{\phi} \sim \frac{m_{\phi}^3}{M_{\Pl}^2}\ ,
\end{equation}
so the maximum temperature reached during reheating
is\footnote{Recently, several authors (Kofman \etal 1996, see Boyanovsky
 \etal 1996) have reexamined the dynamics of reheating and identified
 mechanisms, e.g. parametric resonance, which can drastically change
 the simple picture discussed here, for specific couplings of the
 inflaton to matter fields.}  
\begin{equation}
\label{Tr}
 T_{\R} \sim (\Gamma_{\phi} M_{\Pl})^{1/2} \sim \frac{\Delta^3}{M_{\Pl}^2}\ .
\end{equation}
The requirement that this be less than the phenomenological bound
imposed by the gravitino problem poses a serious challenge for
inflationary models (see Bin\'etruy 1985, Enqvist 1986, Olive
1990a). As mentioned earlier the quantum fluctuations of the inflaton
field generates density perturbations as it `rolls' down its potential
(see Brandenberger 1985) and these induce temperature fluctuations in
the CMB. The \COBE measurement of the CMB quadrupole anisotropy then
fixes the amplitude of these perturbations on spatial scales of order
the present Hubble radius and imposes an independent constraint on the
ratio of the vacuum energy to the slope of the inflaton scalar
potential. Since the vacuum energy is already bounded by the reheat
constraint, this translates into an upper bound on the slope of the
potential at the point where these fluctuations are generated, which
can be identified by solving the equation of motion for the
inflaton. Ross and Sarkar (1996) have shown that when the scalar
potential is given by minimal $N=1$ supergravity (Cremmer \etal 1979,
1983), this constraint is {\em not} satisfied by generic `chaotic'
inflationary models (see Linde 1990) wherein the inflaton evolves
towards its global minimum at the origin from an initial VEV beyond
the Planck scale. However it is easy to satisfy this constraint in a
`new' inflationary model (Holman \etal 1984) where the inflaton
evolves from an initial value at the origin towards its minimum at the
Planck scale. The scalar potential is then as flat as is required,
with $\Delta\sim3\times10^{14}\GeV$, i.e. $m_{\phi}$ of
$\Or(10^{11})\GeV$ as mentioned earlier. Two observationally testable
consequences of this model are that gravitational waves contribute
negligibly to the CMB anisotropy and the spectrum of scalar density
perturbations is `tilted' with a slope of about 0.9, which improves
the fit to the observed clustering and motions of galaxies in an
universe dominated by cold dark matter (Sarkar 1996, Adams \etal
1996).

A related issue is the aforementioned Polonyi problem (Coughlan \etal
1983, 1984, Dine \etal 1984, Goncharov \etal 1984) associated with
very weakly coupled light scalar fields which are driven out to large
VEVs along flat directions during the De Sitter phase (if the Hubble
parameter exceeds their mass). Subsequently the field evolves towards
its minimum just like the inflaton field and eventually reaches the
minimum and oscillates about it converting the stored vacuum energy
into radiation. However this happens very late for a light singlet
field (see \eref{gammaphi}) hence the universe reheats to a
temperature {\em below} the value of $\Or(10)\MeV$ required for
starting off successful nucleosynthesis. Hence it is essential to
address this problem which is particularly acute in models derived
from (compactified) string theories, because of the associated
`moduli' fields which do exhibit such flat directions and have masses
at most of order the electroweak scale, giving a reheat temperature of
\begin{equation} 
\label{Trmoduli}
 T_{\R} \sim m_{\phi}^{3/2} M_{\Pl}^{-1/2} \sim 10^{-6} \GeV 
\end{equation}
(e.g. de Carlos \etal 1993, Banks \etal 1994). One way to evade this
is to invoke a second phase of inflation with a Hubble parameter {\em
smaller} than the moduli mass, with a reheat temperature high enough
not to disturb nucleosynthesis (Randall and Thomas 1995); the second
epoch of inflation must of course be short enough not to erase the
density perturbations produced in the initial inflationary era but
long enough to solve the moduli problem. A natural mechanism for such
`thermal inflation' thas recently been identified in models with an
intermediate symetry breaking scale (Lyth and Stewart 1995,
1996). There are other possible solutions to the moduli problem
(e.g. Dine \etal 1995, 1996, Thomas 1995, Banks \etal 1995a,b, Ross
and Sarkar 1996, Linde 1996) --- for example all moduli may have VEVs
fixed by a stage of symmetry breaking {\em before} inflation, or the
moduli minima may be the {\em same} during and after inflation,
corresponding to a point of enhanced symmetry. In the first case, the
`dilaton' field, which determines the string coupling, should also
acquire a (non-perturbative) mass much higher than the electroweak
scale since otherwise the curvature of the potential in the dilaton
direction is too great to allow for a period of inflation (Brustein
and Steinhardt 1993). In both cases the implication is that the moduli
{\em cannot} be treated (cf. Kounnas \etal 1994, Bin\'etruy and Dudas
1995, Dimopoulos \etal 1995) as dynamical variables at the electroweak
scale, determining the couplings in the low energy theory. All this is
a direct consequence of the requirement that standard BBN not be
disrupted by the late release of entropy!

\subsubsection{The `$\mu$--problem' and the NMSSM:\label{nmssm}} 

As we have seen, the major motivation for adding (softly broken)
supersymmetry to the Standard Model is to bring under control the
quadratic divergences associated with a fundamental Higgs boson and
make it `natural' for its mass to be at the electroweak scale. However
the minimal version of the supersymmetric Standard Model has a new
naturalness problem associated with the mixing term $\mu\,H_{1}H_{2}$
between the two Higgs doublets (see \eref{mssmsp}). For successful
phenomenology $\mu$ should also be of order the electroweak scale but
there is no symmetry which ensures this --- the `$\mu$--problem'
(e.g. Hall \etal 1983, Kim and Nilles 1984). To address this, the MSSM
is extended by the addition of a singlet Higgs superfield $N$ in the
next-to-minimal supersymmetric Standard Model (NMSSM) (e.g. Nilles
\etal 1983, Derendinger and Savoy 1984). By invoking a $Z_{3}$
symmetry under which every chiral superfield $\Phi$ transforms as
$\Phi\to\e^{2\pi{i}/3}\Phi$, the allowed terms in the superpotential
are
\begin{equation}
\label{nmssmsuperpot}
 P_{\rm NMSSM} = h_u Q H_2 u^c + h_d Q H_1 d^c + h_e L H_1 e^c + \lambda N
              H_1 H_2 - \frac{1}{3} k N^3\ ,
\end{equation}
while the Higgs part of the soft supersymmetry breaking potential is
extended by the inclusion of two additional trilinear soft terms
$A_{\lambda}$ and $A_{k}$ to
\begin{eqnarray}
 V_{\rm soft}^{\rm Higgs} &=
  &- \lambda A_{\lambda}(NH_{1}H_{2} + \hc)
  - \frac{1}{3} k\,A_k (N^3 + \hc) \nonumber\\
  &&+ m^2_{H_{1}} \vert H_{1} \vert^2
  + m^2_{H_{2}} \vert H_{2} \vert^2
  + m^2_{N} \vert N \vert^2, 
\end{eqnarray} 
where $H_{1}H_{2}=H_{1}^0H_{2}^0-H^-H^{+}$. The $\mu$--term can now be
simply set to zero by invoking the $Z_{3}$ symmetry. An effective
$\mu$--term of the form $\lambda\langle{N}\rangle$ will still be
generated during $SU(2)_{\L}\,\otimes\,U(1)_{Y}$ breaking but it is
straightforward to arrange that $\langle{N}\rangle$ is of order a soft
supersymmetry breaking mass. Apart from solving the `$\mu$--problem'
the NMSSM has interesting implications for supersymmetric Higgs
phenomenology (e.g. Ellis \etal 1989, Elliot \etal 1994) and dark
matter (e.g. Olive and Thomas 1991, Abel \etal 1993).

Unfortunately, the NMSSM has a cosmological problem. The $Z_{3}$ of
the model is broken during the phase transition associated with
electroweak symmetry breaking in the early universe. Due to the
existence of causal horizons in an evolving universe, such
spontaneously broken discrete symmetries lead to the formation of
domains of different degenerate vacua separated by domain walls
(Zel'dovich, Kobzarev and Okun 1975, Kibble 1976). These have a
surface energy density $\sigma\sim\nu^3$, where $\nu$ is a typical VEV
of the fields, here of order the electroweak scale. Such walls would
come to dominate the energy density of the universe and create
unacceptably large anisotropies in the CMB unless their energy scale
is less than a few MeV (see Vilenkin 1985). Therefore cosmology
requires electroweak scale walls to disappear well before the present
era. Following the original suggestion by Zel'dovich \etal (1975),
this may be achieved by breaking the degeneracy of the vacua,
eventually leading to the dominance of the true vacuum. This happens
when the pressure $\varepsilon$, i.e. the difference in energy density
between the distinct vacua, begins to exceed the tension $\sigma/R$,
where $\sigma$ is the surface energy density of the walls and $R$ the
scale of their curvature. When $R$ becomes large enough for the
pressure term to dominate, the domain corresponding to the true vacuum
begins to expand into the domains of false vacuum and eventually fills
all of space. It has been argued (Ellis \etal 1986a, Rai and
Senjanovi\'c 1994) that strong gravitational interactions at the
Planck scale, which are expected to explicitly violate any discrete
symmetry, would cause just such a non-degeneracy in the minima of
$\Or(\nu^5/M_{\Pl})$ where $\nu$ is a generic VEV, of $\Or(m_{W})$ in
the present case.

Abel \etal (1995) have tested whether the above solution is indeed
viable by studying the cosmological evolution of the walls under the
influence of the tension, the pressure due to the small explicit
$Z_{3}$ breaking and the friction due to particle reflections (see
Vilenkin and Shellard 1994). They find that in order to prevent wall
domination of the energy density of the universe one requires
$\varepsilon>\sigma^2/M_{\Pl}^2$, a pressure which can be produced by
dimension-6 operators in the potential. A much tighter constraint
however comes from requiring that primordial nucleosynthesis not be
disrupted by the decays of the walls into quarks and leptons. The
energy density released in such collisions at time $t$ is
\begin{equation}
 \frac{\rho_{\rm walls}}{n_{\gamma}} \sim \frac{\sigma}{t n_{\gamma}}
  \approx 7\times10^{-11}\GeV \left(\frac{\sigma}{m_{W}^3}\right) 
   \left(\frac{t}{\sec}\right)^{1/2} .
\end{equation}
The walls must disappear before $t\approx0.1\sec$ to ensure that
the hadrons produced in their decays do not alter the
neutron-to-proton ratio, resulting in a $^4$He mass fraction in excess
of $25\%$ (see \fref{nxlimcascade}). This requires the
magnitude of explicit $Z_{3}$ breaking to be
\begin{equation}
\label{wallpressure}
 \varepsilon \geqsim \lambda^{\prime} \frac{\sigma m_{W}^2}{M_{\Pl}}\ ,
\end{equation}
with $\lambda^{\prime}\sim10^{-7}$. Thus addition of a dimension-5
non-renormalizable operator to the superpotential is sufficient to
evade the cosmological constraints. However this creates a naturalness
problem since introduction of non-renormalizable terms together with
soft supersymmetry breaking produces corrections to the potential
which are quadratically divergent and thus proportional to powers of
the cut-off $\Lambda$ in the effective supergravity theory
(e.g. Ellwanger 1983, Bagger and Poppitz 1993). Since the natural
scale for this cut-off is $M_{\Pl}$, these can destabilize the
hierarchy, forcing the singlet VEV (and hence the scale of electroweak
symmetry breaking) upto, at least, the hidden sector scale of
$\approx(m_{3/2}M_{\Pl})^{1/2}\sim10^{11}\GeV$. By examining the
possible dimension-5 $Z_{3}$ breaking terms, Abel \etal (1995)
demonstrate that this can be averted only if the coefficient
$\lambda^{\prime}$ in \eref{wallpressure} is smaller than
$3\times10^{-11}$. Thus the NMSSM has either a cosmological domain
wall problem or a hierarchy problem.

A possible solution is to reintroduce the $\mu$ term in the
superpotential in such a way as to avoid the introduction of the
dangerous non-renormalizable operators. By allowing specific couplings
of the hidden sector fields to the visible sector (Giudice and Masiero
1988), one can retain $Z_{3}$ symmetry in the full theory but break it
spontaneously when supersymmetry is broken; then the hierarchy is not
destabilized by tadpole diagrams. Nevetheless allowed operators which
would give $N$ a mass of order the SUSY breaking scale still have to
be set to zero by hand and this constitutes a naturalness problem of
at least one part in $10^{9}$ (Abel \etal 1995).

\subsubsection{$R$-parity breaking and LSP decays:\label{Rpbreak}} 

Apart from the terms in the MSSM superpotential shown in
\eref{mssmsp}, one can in general add other gauge-invariant terms such
as (Hall and Suzuki 1984)
\begin{equation}
\label{Rbreak}
 P_{\not R} = \lambda_{ijk} L_i L_j e^c_k
                       +  \lambda'_{ijk} L_i Q_j d^c_k
                       +  \lambda''_{ijk} u^c_i d^c_j d^c_k \ ,
\end{equation}
where $L_i$ and $Q_i $ are the $SU(2)$-doublet lepton and quark
superfields and $e^c_i, u^c_i, d^c_i$ are the singlet
superfields. These are phenomenologically dangerous since the
$\lambda$ and $\lambda'$ couplings violate lepton number, while
$\lambda''$ couplings violate baryon number. Hence such couplings are
usually eliminated by enforcing a discrete symmetry termed
$R$--parity, $R\equiv(-1)^{3B+L+2S}$ (Farrar and Fayet 1978). This has
the additional important consequence that the lightest superpartner
(LSP) is absolutely stable and therefore a good dark matter candidate.
However, an exact $R$-parity is not essential from a theoretical point
of view, since rapid proton decay can be prevented by simply requiring
that all the $\lambda''_{ijk}$ in \eref{Rbreak} be zero as a
consequence of an underlying symmetry.\footnote{$R$-parity may also be
 broken spontaneously if a neutral scalar field, viz. the sneutrino,
 gets a VEV (Aulakh and Mohapatra 1982) but this gives rise to a
 problematical massless Goldstone boson, the Majoron, unless explicit
 $\Rp$ is also introduced in some way. Moreover, this induces a
 neutrino mass and the scenario is thus constrained by the BBN bounds
 (\S\,\ref{nudecays}) on an unstable $\nu_{\tau}$ (e.g. Ellis \etal
 1985a).} It has been argued (Campbell \etal 1991, Fischler \etal 1991)
that the other terms must also be zero or very small in order for a
primordial baryon asymmetry to survive since this requires that $B$
and/or $L$-violating interactions should not have come into thermal
equilibrium above the electroweak scale when ($B-L$ conserving)
fermion number violation is already unsuppressed in the Standard Model
(Kuzmin \etal 1985). However this can be evaded, for example, through
lepton mass effects which allow a baryon asymmetry to be regenerated
at the electroweak scale through sphaleron processes if there is a
primordial flavour-dependent lepton asymmetry (Kuzmin \etal 1987,
Dreiner and Ross 1993).\footnote{There is however no such loophole for
 $\Delta\,B=2,\,\Delta\,L=0$ interactions such as the dim-9 operator
 $(qqqqqq)/M^5$ (usually heavy Higgs exchange in unified theories)
 which mediate neutron-antineutron oscillations (Mohapatra and Marshak
 1980). Such processes involve only quark fields, hence no flavour
 symmetry can be separately conserved because of CKM mixing. The
 experimental lower limit $\tau_{n-\bar{n}}\geqsim10^{8}\sec$ (Particle
 Data Group 1996) implies that such oscillations can have no influence
 on nucleosynthesis (Sarkar 1988). Demanding that such processes do not
 come into thermal equilibrium in the early universe requires
 $M\geqsim10^{14}\GeV$ (Campbell \etal 1991, 1992b).} There are also
other possibilities for protecting a baryon asymmetry (e.g. Campbell
\etal 1992a, Cline \etal 1994) so this is not a firm constraint.

The cosmological consequences of $R$-parity violation have been
examined by Bouquet and Salati (1987). The LSP, which is usually the
neutralino, is now unstable against tree-level decays (similar to a
heavy neutrino) with lifetime
\begin{equation}
 \tau \approx 
  \frac{10^{-16}\sec}{\lambda^2} 
  \left(\frac{m_{\chi^0}}{10\GeV}\right)^{-5}
  \left(\frac{m_{\tilde{f}}}{100\GeV}\right)^{4}\ .
\end{equation}
where $m_{\tilde{f}}$ is the mass of the squark/slepton as appropriate
to the $\Rp$ coupling $\lambda$ under consideration. Such decays must
of course occur early enough so as not to disturb BBN, hence the
arguments of \S\,\ref{nonrel} impose a {\em lower} bound on the $\Rp$
coupling. The precise bound on the lifetime depends on the decay mode,
e.g. whether or not the final state includes hadrons. The relic
abundance of a LSP heavier than a few MeV freezes-out before BBN,
therefore the usual calculation (e.g. Ellis \etal 1984a) can be used,
ignoring the effect of LSP decays. For example, a neutralino of mass
$10\GeV$ has a relic abundance of
$n_{\chi^0}/n_\gamma\sim3\times10^{-8}\GeV\,(m_{\tilde{f}}/100\GeV)^4$;
if it decays through the operators $\lambda'_{LQd}$ or
$\lambda''_{udd}$ creating hadronic showers, then \fref{nxlimcascade}
shows that we can require $\tau_{\chi^0}\leqsim1\sec$, i.e.
\begin{equation}
 \lambda \geqsim 10^{-8} .
\end{equation}
(Laboratory experiments looking for $\Rp$ effects are sensitive to a
{\em maximum} lifetime of $\Or(10^{-6})\sec$ so can only probe
$\lambda\geqsim10^{-5}$.) Of course $\lambda$ may be very small (or
indeed zero!) making the neutralino lifetime longer than the age of
the universe. Then the BBN constraints do not apply but arguments
based on the absence of a high energy neutrino background require the
lifetime to be higher than $\approx3\times10^{17}\yr$ (Gondolo \etal
1993), thus implying $\lambda\leqsim10^{-21}$ (e.g. Campbell \etal
1992b). Similar arguments have been used to constrain the
destabilization of the LSP through $\Rp$ in the singlet sector of the
NMSSM (Allahverdi \etal 1994).

\subsubsection{Superstring models and new gauge bosons:\label{ss}} 

Phenomenological models ``motivated'' by the superstring often contain
additional neutral particles in each fermion generation, notably
right-handed neutrinos which are singlets of the Standard Model (see
Ellis 1987, Hewett and Rizzo 1989). These are often massless or very
light and thus relativistic at the time of nucleosynthesis. However
they couple to matter not through the $Z^0$ but through a hypothetical
new neutral gauge boson $Z'$ which is experimentally required to be
heavier than the $Z^0$. Thus the $\nu_{\R}$ energy density at the time
of nucleosynthesis will be suppressed relative to the conventional
$\nu_{\L}$ if its interactions are sufficiently weak (i.e. the $Z'$ is
sufficiently heavy) to move the $\nu_{\R}$ decoupling back earlier
than some epoch of entropy generation, e.g. $\mu^{+}\mu^{-}$
annihilation or the quark-hadron phase transition. Then each
$\nu_{\R}$ will only count as a fraction of a $\nu_{\L}$ and possibly
satisfy the BBN bound on $N_{\nu}$, which thus translates into a lower
bound on the mass of the $Z'$. Ellis \etal (1986b) argued for the
conservative constraint $N_{\nu}\leqsim5.5$ \eref{Nnu5.5}
and noted that this would permit one additional $\nu_{\R}$ per
generation if these decoupled before $\mu^{+}\mu^{-}$ annihilation
thus ensuring $T_{\nu_{\R}}/T_{\nu_{\L}}<0.59$. As discussed in
\S\,\ref{rel}, this requires $\alpha\equiv
\langle\sigma{v}\rangle_{{\ell}^+{\ell}^-\to\nu_{\R}\bar{\nu_{\R}}}
<1.4\times10^{-15}{\GeV}^{-4}$, but a more careful analysis of
decoupling yields a bound {\em less} stringent than this na\"{\ii}ve
estimate (Ellis \etal 1986b)
\begin{equation} 
\label{alphabound}
 \langle\sigma{v}\rangle_{{\ell}^+{\ell}^-\to\nu_{\R}\bar{\nu_{\R}}}
 < 7 \times 10^{-15} T^2\ {\GeV}^4\ .
\end{equation}
For the
$SU(3)_{\c}{\otimes}SU(2)_{\L}{\otimes}U(1)_{Y}{\otimes}U(1)_{\rm E}$
model obtained from Calabi-Yau compactification (Witten 1985b, Dine
\etal 1985), the couplings of the new gauge boson $Z_{\eta}$ are
specified, e.g. $g_{\eta}=e/\cos\theta_{\W}$, enabling the
annihilation cross-section above to be computed (Ellis \etal
1986b). The above argument then gives $m_{Z'}\geqsim330\GeV$, which
has only recently been matched by direct experimental bounds on such a
new gauge boson (Particle Data Group 1996). Of course the cosmological
bound would be more stringent still if one were to adopt a more
restrictive constraint, e.g. taking $N_{\nu}\leqsim4$ would require
$m_{Z'}\geqsim780\GeV$ (Ellis \etal 1986b). Even more restrictive
bounds have been quoted (Steigman \etal 1986, Gonzalez-Garcia and
Valle 1990, Lopez and Nanopoulos 1990, Faraggi and Nanopoulos 1991)
but these were based on an approximate treatment of decoupling and
adopted more restrictive, but less reliable, constraints from BBN.

\subsubsection{Supersymmetry breaking:\label{ssbreak}} 

Perhaps the most crucial issue in phenomenological supergravity
concerns how supersymmetry is actually broken. As mentioned earlier,
the most popular option is to break supersymmetry non-perturbatively
in a hidden sector which interacts with the visible sector only
through gravitational interactions (see Amati \etal 1988, Nilles 1990)
or through gauge interactions (Dine and Nelson 1993). A spontaneously
broken $R$-symmetry is neccessary and sufficient for such dynamical
supersymmetry breaking (Affleck \etal 1985, Nelson and Seiberg 1994)
and implies the existence of a pseudo Nambu-Goldstone boson, the
$R$-axion. In renormalizable hidden sector models, the axion has a
decay constant of $\Or(M_{\SUSY})$ and a mass of
$\Or(m_{3/2}^{1/2}M_{\SUSY}^{1/2})\sim10^{7}\GeV$, where
\begin{equation} 
 M_{\SUSY} \sim (m_{3/2} M_{\Pl})^{1/2} \sim 10^{11} \GeV
\end{equation} 
is the effective scale of SUSY breaking in the hidden sector. The
axion field is set oscillating during inflation and the energy density
contained in such oscillations after reheating is released as the
axions decay into both visible particles and gravitinos. As discussed
by Bagger \etal (1994) this is constrained by the bounds on massive
decaying particles discussed in \S\,\ref{nonrel}. The implied upper
limit on the reheat temperature is found to be competetive with that
obtained (\S\,\ref{gravprob}) from considerations of thermal gravitino
generation. (Bagger \etal (1994) also note that in visible sector
models wherein SUSY breaking is communicated through gauge
interactions, it is neccessary to have $M_{\SUSY}$ higher than
$\sim10^{5}\GeV$ in order to make the axion heavier than $\sim100\MeV$
so as to evade astrophysical bounds (see Raffelt 1990).)

In non-renormalizable hidden sector models, the scale of gaugino
condensation in the hidden sector due to Planck scale interactions is
\begin{equation}
 \Lambda \sim M_{\SUSY}^{2/3} M_{\Pl}^{1/3}\ , 
\end{equation} 
and the $R$-axion mass is of
$\Or(M_{\SUSY}^{2}/M_{\Pl})\sim10^{3}\GeV$ while its decay constant is
of $\Or(M_{\Pl})$. Banks \etal (1994) note that this will give rise to
a Polonyi problem and conclude that all such models are thus ruled
out. However Rangarajan (1995a) has specifically considered the
$E_{8}\otimes\,E'_{8}$ superstring model compactified on a Calabi-Yau
manifold (Gross \etal 1985) and calculated the energy density in the
coherent oscillations of the axions given their low temperature
potential (e.g. Choi and Kim 1985). He finds that the axions decay
before nucleosynthesis as required (\S\,\ref{nonrel}) if
\begin{equation}
 \Lambda \geqsim 10^{13} \GeV\ ,
\end{equation}
consistent with the expected value of
$\Lambda\simeq5\times10^{13}\GeV$ (Derendinger \etal 1985). The decay
of the axion oscillations does increase the comoving entropy by a
factor of $\sim10^7$, but this is deemed acceptable if baryogenesis
occurs by the Affleck-Dine mechanism (Rangarajan 1995b).

Yet another application of BBN bounds has been to orbifold
compactifications of the superstring wherein supersymmetry is broken
{\em perturbatively} by the Scherk-Schwarz mechanism at the
electroweak scale (e.g. Rohm 1984, Antoniadis \etal 1988, Ferrara \etal
1989) by postulating the existence of a large internal dimension (see
Antoniadis 1991). Thus, in addition to the MSSM, these models contain
a repeating spectrum of Kaluza-Klein (KK) modes all the way up to the
Planck scale, whose spacing ($\epsilon\,\approx\,1/2R$, where $R$ is
the radius of compactification) is comparable to the supersymmetry
breaking scale of $\Or(\TeV)$. Such modes can be directly excited at
forthcoming accelerators such as the LHC (Antoniadis \etal 1994),
hence this possibility is of great experimental interest. These modes
will also be excited in the early universe and this radically alters
the thermal history (Abel and Sarkar 1995). The KK modes are labelled
by quantum numbers of internal momenta/charges which are of the form
\begin{equation} 
 P_{\rm LR}=\frac{n}{R}\pm \frac{mR}{2}\ ,
\end{equation}
where $R$ represents some internal radius of compactification. The
winding modes ($m\neq0$) have masses of $\Or(M_{\Pl})$ and need not be
considered further, while the particles in the $n$th KK mode have
masses $m_n\,\sim\,n\epsilon$. Thus above the compactification scale,
when the temperature rises by $\epsilon$, two new levels of (gauge
interacting) KK excitations become relativistic, so that the number of
relativistic degrees of freedom increases {\it linearly} with
temperature. For example, the number of entropic degrees of freedom
rises above the limiting value $\hat{g}_s=915/4$ \eref{gMSSM} in the
MSSM according to
\begin{equation}                                                     
\label{KKdof}
 g_s (T) = \hat{g}_s + \frac{T}{\epsilon} g_{s\KK}\ ,
\end{equation}
where the constant $g_{s\KK}$ is determined by evaluating the entropy
density of the plasma. (In the spontaneously broken string theories,
each KK level comes in $N=4$ multiplets, so that KK gauge bosons
contribute 8 bosonic and 8 fermionic degrees of freedom in the vector
and fermionic representations of SO(8) respectively; in the minimal
case in which the KK excitations are in $SU(3){\otimes}SU(3)_{\c}$
multiplets, this gives e.g. $g_{s\KK}=1400$.) Thus at a temperature
much higher than the KK level-spacing ($T\,\gg\,\epsilon$), nearly all
the entropy is contained in the KK modes and almost none in the matter
multiplets. By entropy conservation, the Hubble expansion rate is then
altered from its usual form \eref{H} as
\begin{equation}
\label{HKK}
 H = - \frac{4}{3} \frac{\dot{T}}{T} 
  = 1.66\ \sqrt{\frac{g_{s\KK}}{\epsilon}}\ \frac{T^{5/2}}{M_{\Pl}}\ .
\end{equation}
Now consider the history of the universe starting from the maximum
temperature it reached, viz. the reheating temperature $T_{\R}$
($\gg\,\epsilon$) at the end of inflation. The entropy is initially
evenly spread out amongst the strongly (as opposed to gravitationally)
interacting KK modes and the massless matter multiplets. Until the
temperature drops below the first KK level, the evolution of the
universe is therefore governed by the KK modes, whose contribution to
the entropy is continually decreasing as the temperature drops. During
this period there is production of massive gravitons and gravitinos
which can only decay to the massless (twisted) particles since their
decays to the (untwisted) KK modes is kinematically suppressed. The
effect of the decaying particles on the abundances of the light
elements then imposes a severe bound on $T_{\R}$ (Abel and Sarkar
1995). For (hadronic) decays occuring before the begining of
nucleosynthesis, the requirement that $Y_{\pr}({\Hefour})$ not be
increased above $25\%$ translates into the bound
\begin{equation}
 T_{\R} \leqsim 2\times10^{4} \GeV \left(\frac{\epsilon}{\TeV}\right)^{1/3} ,
\end{equation}                                   
while for later decays, consideration of $^{2}$H photofission imposes
an even stricter bound. However the reheat temperature expected in
these models is expected to be significantly larger than the usual
value \eref{Tr}, viz.
\begin{equation}
\label{Tr4d}
  T_{\R} \sim \left(\frac{g_{s\KK}}{\epsilon}\right)^{-1/4}
   (\Gamma_{\phi} M_{\Pl})^{1/2}
  \sim 10^6\ \GeV\ \left(\frac{\epsilon}{\TeV}\right)^{1/4} ,
\end{equation}
for an inflaton mass $m_{\phi}\sim10^{11}\GeV$ as is required to
reproduce the \COBE measurement of CMB fluctuations. A possible
solution would appear to be a second phase of inflation with
$m_{\phi}\sim\epsilon$ to dilute the KK states but the reheat
temperature is then of $\Or(10^{-6})\GeV$ i.e. too low for
nucleosynthesis to occur. Thus it appears to be difficult to construct
a consistent cosmological history for four-dimensional superstring
models with tree-level supersymmetry breaking, notwithstanding their
many theoretical attractions.

\subsection{Grand unification and cosmic strings\label{gutstring}} 

Phase transitions associated with the spontaneous breaking of a
symmetry in the early universe can create stable topological defects
in the associated Higgs field, viz. domain walls, strings and
monopoles (Kibble 1976, see Vilenkin and Shellard 1994). Stable domain
walls are cosmologically unacceptable, as we have seen earlier
(\S\,\ref{nmssm}), and so are monopoles, which are neccessarily created
during GUT symmetry breaking (see Preskill 1984). Such monopoles would be
expected to have a relic abundance comparable to that of baryons, but
are $\sim10^{16}$ times heavier, so would clearly lead to cosmological
disaster. Further, direct searches have failed to find any monopoles
(see Particle Data Group 1996). The most attractive mechanism for
getting rid of then is to invoke an inflationary phase, with reheating
to a temperature well below the GUT scale, as is also required
independently from consideration of the gravitino problem
(\S\,\ref{gravprob}).

By contrast, cosmic strings have an acceptably small relic
energy density and have been studied in great detail because they
provide an alternative to inflationary scalar field fluctuations as
the source of the perturbations which seed the growth of large-scale
structure (see Brandenberger 1991, 1994). Detailed numerical studies
(e.g. Albrecht and Stebbins 1992) find that this requires the string
tension $\mu$ to be in the range
\begin{equation}
\label{galaxystring}
 \mu \approx 1-4 \times 10^{-6} M_{\Pl}^{2}\ ,
\end{equation}
interestingly close to the GUT scale, and in agreement with the value
$\mu=2\pm0.5\times10^{-6}M_{\Pl}^{2}$ obtained (Coulson \etal 1994,
see also Bennett \etal 1992) by normalizing the associated CMB
fluctuations to the \COBE data. An interesting constraint on such GUT
scale strings follows from Big Bang nucleosynthesis (e.g. Hogan and
Rees 1984, Brandenberger \etal 1986, Bennett 1986, Quir\`os 1991). The
evolving network of cosmic strings generates gravitational radiation
which contributes to the total relativistic energy density, thus the
bound on the speed-up rate translates into an upper bound on the
string tension. From a detailed numerical study, Caldwell and Allen
(1992) find that the bound $N_{\nu}\leq3.4$ \eref{Nnu3.4} implies
\begin{equation}
\label{bbnstring}
 \mu < 7 \times 10^{-6} M_{\Pl}^{2}\ .
\end{equation}
These authors illustrate how the bound is weakened if one adopts a
more conservative bound, e.g. $N_{\nu}\leq4$ implies
$\mu<1.6\times10^{-5}M_{\Pl}^{2}$. In general, consideration of the
effect of the gravitational wave background on pulsar timing
observations gives more stringent bounds (see Hindmarsh and Kibble
1995) but these too are consistent with the value \eref{galaxystring}
required for structure formation.

It has been noted that cosmic strings are likely to be superconducting
so that large currents can be induced in them by a primordial magnetic
field (Witten 1985a). In addition to gravitational waves, such a
string also radiates electromagnetic radiation at an ever-increasing
rate as its motion is damped thus increasing the current. Thus the end
point is expected to be the catastrophic release of the entire energy
content into high energy particles; such explosions will send shock
waves into the surrounding intergalactic medium and galaxy formation
may take place in the dense shells of swept-up matter (Ostriker,
Thomson and Witten 1986). However this is constrained by the stringent
bounds on such energy release during the nucleosynthesis era (Hodges
and Turner 1988, Sigl \etal 1995) and the idea is essentially ruled
out by other constraints from the thermalization of the blackbody
background (e.g. Wright \etal 1994).

\subsection{Miscellaneous bounds\label{misc}} 

There have been other applications of BBN constraints to hypothetical
particles which do not fit into the categories considered above. For
example bounds on scalars and pseudo-scalars (e.g. Bertolini and
Steigman 1992) have been applied to hadronic axions (Chang and Choi
1993), to a particle which couples to two photons but not to leptons
or quarks (Mass\'o and Toldr\`a 1994b) and to Majoron emission in
$\beta\beta$-decay (Chang and Choi 1994). Bounds have been derived on
`shadow matter' in superstring theories (Kolb \etal 1985, Krauss \etal
1986), on the time-evolution of possible new dimensions (Kolb \etal
1986a, Barrow 1987) and on `mirror fermions' (e.g. Fargion and Roos
1984, Senjanovi\'c 1986, Berezhiani and Mohapatra 1995, Foot and
Volkas 1995b). Carlson and Glashow (1987) have ruled out a suggested
solution to the orthopositronium decay puzzle involving
`milli-charged' particles (also discussed by Davidson and Peskin 1994)
while Escribano \etal (1995) have ruled out another solution involving
exotic particle emission. For lack of space, we do not discuss these
results except to caution that many of them assume an overly
restrictive limit on $N_{\nu}$ and should be suitably rescaled to the
conservative bound \eref{Nnu4.53}.

\subsection{Implications for the dark matter\label{dm}} 

The nature of the dark matter which is observed to dominate the
dynamics of individual galaxies as well as groups and clusters of
galaxies (see Faber and Gallagher 1979, Trimble 1987, Ashman 1992) is
one of the key problems at the interface of particle physics and
cosmology. It may just be ordinary matter in some non-luminous form,
e.g. planets, white dwarfs, black holes \etc (see Lynden-Bell and
Gilmore 1990, Carr 1994). However the BBN bound on the abundance of
nucleons in {\em any} form constrains this possibility and implies
that most of the dark matter is in fact non-nucleonic.

\subsubsection{`Baryonic' dark matter:\label{bdm}} 

The usually quoted BBN value of $\Omega_{\N}\approx0.011h^{-2}$
\eref{etaindlim} is significantly higher than the value
\eref{omeganps} obtained from direct observations of luminous matter
\eref{omeganps}, as shown in \fref{omegan}. This suggests that most
nucleons are dark and, in particular, that much of the dark matter in
galactic halos, which contribute $\Omega\approx0.05h^{-1}$ (see Binney
and Tremaine 1987), may be nucleonic. However if the indications of a
high primordial $\Htwo$ abundance (Songaila \etal 1994, Rugers and
Hogan 1996a,b) are correct, then the implied lower value of
$\Omega_{\N}\approx0.0058\,h^{-2}$ \eref{etaDlya1} is close to its
observational lower limit (for high values of $h$), leaving little
room for nucleonic dark matter. Conversely, if the primordial $\Htwo$
abundance is as low as found by Tytler \etal 1996, the corresponding
value of $\Omega_{\N}\approx0.023\,h^{-2}$ \eref{etaDlya2} would
suggest that much of the halo dark matter is nucleonic. Either
possibility is consistent with searches to date for gravitational
microlensing events expected (Paczy\'{n}ski 1986) for a halo dominated
by dark compact objects. The 8 candidate events detected by the {\sl
MACHO} collaboration imply that about half of the halo mass can be in
the form of such objects having a most probable mass of
$0.5^{+0.3}_{-0.2}M_{\odot}$ (Alcock \etal 1995, 1996).


Even more interesting is the comparison with clusters of galaxies
which clearly have a large nucleonic content, particularly in the form
of X-ray emitting intracluster gas. White \etal (1993) have reviewed
the data on the well-studied {\sl Coma} cluster, for which they find
the nucleonic mass fraction
\begin{equation}
\label{comafrac}
 f_{\N} \equiv \frac{M_{\N}}{M_{\tot}} \geq 0.009 + 0.05\,h^{-3/2}\ , 
\end{equation}
where the first term corresponds to the luminous matter in the cluster
galaxies (within the Abell radius, $r_{\A}\approx1.5\,h^{-1}$ Mpc) and
the second to the intracluster X-ray emitting gas. Using results from
hydrodynamical simulations of cluster formation these authors show
that cooling and other dissipative effects could have enhanced
$f_{\N}$ within $r_{\A}$ by a factor of at most $\Upsilon\approx1.4$
over the global average. Similarly large nucleonic fractions (between
$10\%$ and 22$\%$) have also been found in a sample of 13 other
clusters (White and Fabian 1995). If such structures are indeed fair
tracers of the universal mass distribution, then $f_{\N}$ is related
to the global density parameters as
\begin{equation}
  f_{\N} = \Upsilon \frac{\Omega_{\N}}{\Omega}\ .
\end{equation}
Thus for $\Omega=1$ as expected from inflation, the {\sl Coma}
observations can be consistent with standard BBN only for a low
deuterium abundance (and low values of $h$) as shown in
\fref{omegan}. Observations of large-scale structure and CMB
anisotropy do favour high $\Omega_{\N}$ and low $h$ for a critical
density universe dominated by cold dark matter (e.g. White \etal 1996,
Adams \etal 1996). Conversely if the deuterium abundance is indeed
high, then to achieve consistency would require $\Omega\approx0.1$
(for which there is, admittedly, independent observational evidence;
see e.g. Coles and Ellis 1994). The dark matter in {\sl Coma} and
other clusters would then be comparable to that in the individual
galactic halos. It is presently controversial whether this is indeed
evidence for $\Omega<1$ or whether the nucleonic enhancement factor
$\Upsilon$ and/or the total cluster mass have been underestimated;
also there may be sources of non-thermal pressure in clusters
(viz. magnetic fields, cosmic rays) which would lower the inferred
pressure of the X-ray emitting plasma, hence the value of $f_{\N}$
(see Felten and Steigman 1995).

In principle there may exist baryonic matter which does not
participate in nuclear reactions and is utherefore unconstrained by the
above arguments. Two examples are planetary mass black holes (Crawford
and Schramm 1982, Hall and Hsu 1990) and strange quark nuggets (Witten
1984, see Alcock and Olinto 1988) which, it has been speculated, can
be formed in cosmologically interesting amounts during a strongly
first-order quark-hadron phase transition. As discussed earlier
(\S\,\ref{inhombbn}), the fluctuations induced by such a violent phase
transition should have resulted in the synthesis of observable
(although rather uncertain) amounts of heavy elements (see Malaney and
Mathews 1993). Also, according to our present theoretical
understanding, this phase transition is relatively smooth (see
Bonometto and Pantano 1993, Smilga 1995).

\subsubsection{`Non-Baryonic' dark matter:\label{nbdm}} 

Given that the dark matter, at least in galactic halos, is unlikely to
be baryonic, it is interesting to consider whether it may be composed
of relic particles. This is well motivated (see Hall 1988, Sarkar
1991, Ellis 1994) since extensions of the Standard Model often contain
new massive particles which are stable due to some new conserved
quantum number. Alternatively, known particles which are
cosmologically abundant, i.e. neutrinos, can constitute the dark
matter if they acquire a small mass, e.g. through violation of global
lepton number. Thus there are many particle candidates for the dark
matter corresponding to many possible extensions of the SM (see
Srednicki 1990). In order to optimize experimental strategies for
their detection (see Primack \etal 1988, Smith and Lewin 1990), it is
important to narrow the field and it is here that constraints from
cosmological nucleosynthesis play an important role.

It is generally assumed that dark matter particles must be weakly
interacting since dark matter halos appear to be
non-dissipative. However since there is as yet no `standard' model of
galaxy formation (see White 1994), it is legitimate to ask whether
dark matter particles may have electromagnetic or strong interactions,
given that their interaction lifetime exceeds the age of the Galaxy
due to the low density of interstellar space (Goldberg and Hall
1986). This possibility has been studied in detail and various
constraints identified (De R\'ujula, Glashow and Sarid 1990,
Dimopoulos \etal 1990, Chivukula \etal 1990). According to the
standard relic abundance calculation, such particles would have
survived freeze-out with a {\em minimum} relic abundance of
$\sim10^{-12}-10^{-10}$ per nucleon (Dover \etal 1979, Wolfram
1979). These would have then bound with ordinary nuclei during
nucleosynthesis, creating anomalously heavy isotopes of the light
elements (Dicus and Teplitz 1980, Cahn and Glashow 1981). Sensitive
searches for such isotopes have been carried out in a variety of
terrestrial sites, all with negative results (see Rich \etal 1987,
Smith 1988). The best limits on the concentration of such particles
are $\leqsim10^{-29}-10^{-28}$ per nucleon in the mass range
$\sim10-10^{3}\GeV$ (Smith \etal 1982), $\leqsim10^{-24}-10^{-20}$ per
nucleon in the mass range $\sim10^{2}-10^{4}\GeV$ (Hemmick \etal 1990)
and $\leqsim6\times10^{-15}$ in the mass range $\sim10^{4}-10^{8}\GeV$
(Verkerk \etal 1992). Thus it is reasonable to infer that dark matter
particles are electrically neutral and weakly interacting.\footnote{In
 principle, dark matter particles may be strongly {\em
 self}-interacting (Carlson \etal 1992); this possibility is mildly
 constrained by the bound \eref{speeduplim} on the speed-up
 rate. Nucleons themselves are strongly interacting so would be
 expected to have a negigibly small relic abundance from a state of
 thermal equilibrium. Their observed abundance then requires a primordial
 matter-antimatter {\em asymmetry}, which is $\sim10^9$ times greater than
 their freeze-out abundance.}

Apart from the above general constraint, BBN would not appear to be
relevant to individual particle candidates for the dark matter, since
by definition their energy density is negligible relative to that of
radiation during nucleosynthesis and furthermore, they are required to
be stable or at least very long-lived. Nevertheless, BBN does provide
another important constraint since it implies that the comoving
entropy cannot have changed significantly (barring very exotic
possibilities) since the MeV era. Thus the relic abundance of, for
example, a `cold dark matter' particle is unlikely to have been much
altered from its value \eref{freezeoutabun} at freeze-out, which we can
rewrite as
\begin{equation}
\label{cdm}
 \Omega_x h^2 \simeq 
 \left(\frac{\langle\sigma v\rangle}{3\times10^{-10}{\GeV}^{-2}}\right)^{-1}.
\end{equation}
Thus it would be natural for a massive particle to constitute the dark
matter if it is weakly interacting, i.e. if its interactions are fixed
by physics above the Fermi scale. This is arguably the most direct
hint we have today for an intimate connection between particle physics
and cosmology, beyond their respective standard models.

\section{Conclusions\label{concl}} 

In the words of Ya'B Zeldovich, cosmology has long provided the ``poor
man's accelerator'' for particle physics. As terrestrial accelerators
come closer to the ultimate limits of technology and resources, it is
imperative that our understanding of the cosmological laboratory be
developed further, in particular since it offers probes of phenomena
which can {\em never} be recreated in laboratories on Earth, however
powerful our machines become. (This is not just to do with the
energies available but because the early universe provides an {\em
equilibrium} thermal environment, in contrast to the non-equilibrium
environment of particle collisions in an accelerator.) There is an
understandable reluctance, at least among experimentalists, to treat
cosmological constraints on the same footing as as the results of
repeatable and controlled laboratory experiments. However many
theorists are already guided almost exclusively by cosmological
considerations since there is simply no other experimental data
available at the energies they are interested in. We therefore close
with the following plea concerning the improvement of constraints from
Big Bang nucleosynthesis.

In the comparison of the abundance data with the theoretical
expectations, we have noted the rather unsatisfactory state of the
observational situation today. Whereas there has been some concerted
effort in recent years towards precise abundance determinations, the
quoted numbers are still plagued by uncertain systematic errors and
workers in this field use rather subjective criteria,
e.g. ``reasonable'' and ``sensible'', to determine abundance
bounds. In this regard, a comparison with the experimental style in
high energy physics is illuminating. Thousands of person-years of
effort have been invested in obtaining the precise parameters of the
$Z^{0}$ resonance in ${\el}^+{\el}^-$ collisions, which measures the
number of light neutrino species (and other particles) which couple to
the $Z^{0}$. In comparison, relatively little work has been done by a
few small teams on measuring the primordial light element abundances,
which provide a complementary check of this number as well as a probe
of new superweakly interacting particles which do not couple to the
$Z^{0}$. In our view, such measurements ought to constitute a {\em
key} programme for cosmology, with the same priority as, say, the
measurement of the Hubble constant or of the cosmological density
parameter. Our understanding of the $2.73\dK$ cosmic microwave
background has been revolutionized by the accurate and consistent
database provided by the \COBE mission. A similar revolution is
overdue for primordial nucleosynthesis.  \vfill \ack

I am indebted to many colleagues with whom I have enjoyed discussions
and collaborations on cosmological constraints, in particular Steven
Abel, Jeremy Bernstein, Ramanath Cowsik, John Ellis, Kari Enqvist,
Graciela Gelmini, Peter Kernan, Roger Phillips, Graham Ross, Dennis
Sciama and David Seckel. I am grateful to Bernard Pagel and Julie
Thorburn for their critical remarks concerning elemental abundances,
to Larry Kawano for the upgraded Wagoner computer code, and to all the
authors who provided figures from their publicataions. I thank Michael
Birkel, Herbi Dreiner and Cedric Lacey for carefully reading parts of
the manuscript and all the people who provided valuable comments on
the preprint version. Finally I would like to thank the editor,
Richard Palmer, for his courtesey and patience.

This work was supported by an Advanced Fellowship awarded by the
Particle Physics \& Astronomy Research Council, UK.

\References


\item[] Abel S A and Sarkar S {1995} {\PL} B{\bf 342} {40} 
\item[] Abel S A, Sarkar S and Whittingham I B {1993} {\NP} B{\bf 392} {83}
\item[] Abel S A, Sarkar S and White P L {1995} {\NP} B{\bf 454} {663}
\item[] Adams J A, Ross G G and Sarkar S {1996} {\preprint} 
         {OUTP-96-48P (hep-ph/9608336)} [submitted to {\PL} B]
\item[] Adams, T F {1976} {\AA} {\bf 50} {461}
\item[] Adriani O \etal (L3 collab.) {1993} {\PRep} {\bf 236} {1}
\item[] Affleck I and Dine M {1985} {\NP} B{\bf 249} {361}
\item[] Affleck I, Dine M and Seiberg N {1985} {\NP} B{\bf 256} {557}
\item[] Aharonian F A, Kirillov-Ugryumov V G and Vardanian V V
         {1985} {\ApSS} {\bf 115} {201}
\item[] Ahlen S P \etal {1987} {\PL} B{\bf 195} {603}
\item[] Akerlof C W and Srednicki M (ed) {1993} {\it Relativistic
         Astrophysics and Particle Cosmology: Proc. TEXAS/PASCOS Symp.}
         {(New York: New York Academy of Sciences)}
\item[] Albrecht A and Stebbins A {1992} {\PRL} {\bf 68} {2121}
\item[] Alcock C and Olinto A {1988} {\ARNPS} {\bf 38} {161} 
\item[] Alcock C \etal (MACHO collab.) {1995} {\PRL} {\bf 74} {2867}
\item[]  \dash {1996} {\preprint} {astro-ph/9606165} [submitted to {\ApJ}]
\item[] Allahverdi R, Campbell B A and Olive K A {1994} {\PL} 
         B{\bf 341} {166}
\item[] Allen C W {1973} {\it Astrophysical Quantities} {(London: 
         Athalone Press)}
\item[] Alpher R A and Herman R C {1950} {\RMP} {\bf 22} {153}
\item[]  \dash {1990} {\it Modern Cosmology in Retrospect} {ed R~Bertotti 
          \etal (Cambridge: Cambridge University Press) p~129}
\item[] Alpher R A, Follin J W and Herman R C {1953} {\PR} {\bf 92} {1347}
\item[] Alvarez E, Dominguez-Tenreiro R D, Ib\'a\~nez Cabanell J M and
         Quir\'os M (ed) {1987} {\it Cosmology and Particle Physics:
         Proc. 17th GIFT Seminar on Theoretical Physics}
         {(Singapore: World Scientific)}
\item[] Alvarez-Gaume L, Polchinski J and Wise M {1983} {\NP} B{\bf 221} {495}
\item[] Amaldi U, de Boer W and Furstenau H {1991} {\PL} B{\bf 260} {447}
\item[] Amati D, Konishi K, Meurice Y, Rossi G C and Veneziano G
         {1988} {\PRep} {\bf 162} {169}
\item[] Antoniadis I {1991} {\it Particles, Strings and Cosmology: PASCOS-91} 
         {(Singapore: World Scientific) p~718}
\item[] Antoniadis A, Bachas C, Lewellen D and Tomaras T {1988} {\PL} 
         B{\bf 207} {441}
\item[] Antoniadis A, Benakli K and Quir\`os M {1994} {\PL} B{\bf 331} {313}
\item[] Applegate J H, Hogan C J and Scherrer R J {1988} {\ApJ} {\bf 329} {572}
\item[] Armbruster B \etal (KARMEN collab.) {1995} {\PL} B{\bf 348} {19}
\item[] Ashman K M {1992} {\PASP} {\bf 104} {1109}
\item[] Astbury A, Campbell B A, Israel W, Khanna F C, Pinfold J L and
         Page D (ed) {1995} {\it Particle Physics and Cosmology: Proc. 9th
         Lake Louise Winter Institute} {(Singapore: World Scientific)}
\item[] Audouze J, Lindley D and Silk J {1985} {\ApJ} {\bf 293} {L53}
\item[] Audouze J and Silk J {1989} {\ApJ} {\bf 342} {L5}
\item[] Audouze J and Tinsley B M {1976} {\ARAA} {\bf 14} {43}
\item[] Aulakh C S and Mohapatra R N {1982} {\PL} {\bf 119B} {136}
\item[] Azeluos G \etal {1986} {\PRL} {\bf 56} {2241}


\item[] Baade D, Cristiani S, Lanz T, Malaney R A, Sahu K C and Vladilo G 
         {1991} {\AA} {\bf 251} {253}
\item[] Babu K S and Rothstein I Z {1992} {\PL} B{\bf 275} {112}
\item[] Babu K S, Mohapatra R N and Rothstein I Z {1991} {\PRL} {\bf 67} {545}
\item[] Baer H \etal {1995} {\preprint} {LBL-37016 (hep-ph/9503479)} 
\item[] Bagger J and Poppitz E {1993} {\PR} {\bf 71} {2380}
\item[] Bagger J, Poppitz E and Randall L {1994} {\NP} B{\bf 426} {3}
\item[] Bahcall J N {1989} {\it Neutrino Astrophysics} {(Cambridge: 
         Cambridge University Press)}
\item[] Bahcall J N and Ulrich R K {1988} {\RMP} {\bf 60} {297}
\item[] Baier R, Pilon E, Pire B and Schiff D {1990} {\NP} B{\bf 336} {157}
\item[] Bailin D and Love A {1986} {\it Introduction to Gauge Field 
         Theory} {(Bristol: Adam Hilger)}
\item[] Balser D S, Bania T M, Brockway C J, Rood R T and Wilson T L 
         {1994} {\ApJ} {\bf 430} {667}
\item[] Bamert P, Burgess C P and Mohapatra R N {1995} {\NP} B{\bf 438} {3} 
\item[] Banerjee B and Gavai R V {1992} {\PL} B{\bf 293} {157}
\item[] Bania T M, Rood R T and Wilson T L {1987} {\ApJ} {\bf 323} {30}
\item[] Banks T, Kaplan D and Nelson A E {1994} {\PR} D{\bf 49} {779}
\item[] Banks T, Berkooz M and Steinhardt P J {1995a} {\PR} D{\bf 52} {705}
\item[] Banks T, Berkooz M, Shenker S H, Moore G and Steinhardt P J {1995b}
         {\PR} D{\bf 52} {3548}
\item[] Barbieri R and Dolgov A D {1990} {\PL} B{\bf 237} {440}
\item[]  \dash {1991} {\NP} B{\bf 349} {742}
\item[] Barbieri R, Ferrara S and Savoy C A {1982} {\PL} {\bf 119B} {343}
\item[] Barger V, Phillips R J N and Sarkar S {1995} {\PL} B{\bf 352}
         {365} (erratum B{\bf 356} {617})
\item[] Barr S M, Chivukula R S and Farhi E {1990} {\PL} B{\bf 241} {387} 
\item[] Barrow J D {1983} {\PL} {\bf 125B} {377}
\item[]  \dash {1987} {\PR} D{\bf 35} {1805}
\item[] Bartlett J G and Hall L J {1991} {\PRL} {\bf 66} {541}
\item[] Beaudet G and Reeves H {1983} {\AA} {\bf 134} {240}
\item[] Beier E W \etal {1992} {\PL} B{\bf 283} {446}
\item[] Bennett D P {1986} {\PR} D{\bf 34} {3492}
\item[] Bennett D P, Stebbins and A Bouchet F R {1992} {\ApJ} {\bf 399} {L5}
\item[] Berezhiani Z G and Mohapatra R N {1993} {\PR} D{\bf 52} {6607} 
\item[] Bergsma F \etal (CHARM collab.) {1983} {\PL} {\bf 128B} {361}
\item[] Bernardi G \etal (PS191 collab.) {1986} {\PL} {\bf 166B} {479}
\item[] Bernstein J {1988} {\it Kinetic Theory in the Expanding 
         Universe} {(Cambridge: Cambridge University Press)}
\item[] Bernstein J, Brown L M and Feinberg G {1989} {\RMP} {\bf 61} {25}
\item[] Bertolami O {1988} {\PL} B{\bf 209} {277}
\item[] Bertolini S and Steigman G {1992} {\NP} B{\bf 387} {193}
\item[] Bethke S and Pilcher J E {1992} {\ARNPS} {\bf 42} {251}
\item[] Bilenky S and Petcov S T {1987} {\RMP} {\bf 59} {671}
\item[] Bin\'etruy P {1985} {\it Proc. 6th Workshop on Grand Unification}
         {ed S~Rudaz and T~Walsh (Singapore: World Scientific) p~403}
\item[] Bin\'etruy P and Dudas E {1995} {\NP} B{\bf 442} {21}
\item[] Binney J and Tremaine S {1987} {\it Galactic Dynamics} {(Princeton:
         Princeton University Press)}
\item[] Birkinshaw M and Hughes J P {1994} {\ApJ} {\bf 420} {33}
\item[] Black D C {1971} {\Nature} {\bf 231} {1480}
\item[] Blandford R D and Narayan R {1992} {\ARAA} {\bf 30} {311}
\item[] Blumenthal G R and Gould R J {1970} {\RMP} {\bf 42} {237} 
\item[] Boehm F \etal {1991} {\PL} B{\bf 255} {143}
\item[] Boesgaard A M and Steigman G {1985} {\ARAA} {\bf 23} {919}
\item[] Bonometto S A and Pantano O {1993} {\PRep} {\bf 228} {175}
\item[] Bopp P \etal {1986} {\PRL} {\bf 56} {919}
\item[] B\"orner G {1988} {\it The Early Universe} {(Berlin: Springer-Verlag)}
\item[] Bouquet A and Salati P {1987} {\NP} B{\bf 284} {557}
\item[] Bowler M and Jelley N A {1994} {\PL} B{\bf 331} {193}
\item[] Boyanovsky D, de Vega H J, Holman R and Salgado J F J {1996}
         {\preprint} {astro-ph/9609007}
\item[] Boyd G \etal {1995} {\PRL} {\bf 75} {4169}
\item[] Brandenberger R {1985} {\RMP} {\bf 57} {1}
\item[]  \dash {1991} {\PS} T{\bf 36} {114}
\item[]  \dash {1994} {\IJMP} A{\bf 9} {2117}
\item[] Brandenberger R, Albrecht A and Turok N {1986} {\NP} B{\bf 277} {655}
\item[] Brignole A, Espinosa J R, Quir\'os M and Zwirner F {1994} {\PL}
         B{\bf 324} {181}
\item[] Britton D I \etal {1992} {\PR} D{\bf 46} {885}
\item[]  \dash {1994} {\PR} D{\bf 49} {28}
\item[] Brocklehurst M {1972} {\MN} {\bf 157} {211}
\item[] Brown L {1992} {\ApJ} {\bf 389} {251}
\item[] Brustein R and Steinhardt P J {1993} {\PL} B{\bf 302} {196}
\item[] Bryman D A {1993} {\CNPP} {\bf 21} {101}
\item[] Bryman D A \etal {1983} {\PRL} {\bf 50} {1546}
\item[] Buchm\"uller W and Philipsen O {1995} {\NP} B{\bf 443} {47}
\item[] Burkhardt H and Steinberger J {1991} {\ARNPS} {\bf 41} {55}
\item[] Burles S and Tytler D {1996} {\preprint} {astro-ph/9603070} 
         [submitted to {\Science}]
\item[] Busculic D \etal (ALEPH collab.) {1995} {\PL} B{\bf 349} {585}
\item[] Byrne J {1982} {\RPP} {\bf 45} {115}


\item[] Cahn R and Glashow S {1981} {\Science} {\bf 213} {607}
\item[] Caldwell D O \etal {1988} {\PRL} {\bf 61} {510}
\item[] Caldwell R R and Allen B {1992} {\PR} D{\bf 45} {3447}
\item[] Cambier J-L, Primack J and Sher M {1982} {\NP} B{\bf 209} {372}
\item[] Campbell B A, Davidson S, Ellis J and Olive K A 
         {1991} {\PL} {\bf B256} {457}
\item[]  \dash {1992a} {\PL} {\bf B297} {118}
\item[]  \dash {1992b} {\AP} {\bf 1} {77}
\item[] Cardall C Y and Fuller G M {1996a} {\preprint} {astro-ph/9603071}
\item[]  \dash {1996b} {\PR} D{\bf 54} {1260}
\item[] Carlson E D and Glashow S L {1987} {\PL} B{\bf 193} {168}
\item[] Carlson E D, Esmailzadeh R, Hall L J and Hsu S D H
         {1990} {\PRL} {\bf 65} {2225}
\item[] Carlson E D, Machacek M and Hall L J {1992} {\ApJ} {\bf 398} {43}
\item[] Carr B J {1994} {\ARAA} {\bf 32} {531}
\item[] Carroll S M, Press W H and Turner E L {1992} {\ARAA} {\bf 30} {499} 
\item[] Carswell R F, Rauch M, Weymann R J, Cooke A J and Webb J K 
         {1994} {\MN} {\bf 268} {L1}
\item[] Carswell R F \etal {1996} {\MN} {\bf 278} {506} 
\item[] Caughlan G R and Fowler W A {1988} {\ADNDT} {\bf 40} {283}
\item[] Caughlan G R, Fowler W A, Harris M J and Zimmerman B A 
         {1985} {\ADNDT} {\bf 32} {197}
\item[] Cen R, Ostriker J P and Peebles J {1993} {\ApJ} {\bf 415} {423}
\item[] Chaboyer B and Demarque P {1994} {\ApJ} {\bf 433} {510}
\item[] Chamseddine A, Arnowitt R and Nath P {1982} {\PRL} {\bf 49} {970}
\item[] Chang S and Choi K {1993} {\PL} B{\bf 316} {51}
\item[]  \dash {1994} {\PR} D{\bf 49} {12}
\item[] Charbonnel C {1995} {\ApJ} {\bf 453} {L41}
\item[] Cheng H-Y {1988} {\PRep} {\bf 158} {1}
\item[] Cheng T-P and Li L-F {1984} {\it Gauge Theory of Elementary Particle
         Physics} {(Oxford: Oxford University Press)}
\item[] Chikashige Y, Mohapatra R N and Peccei R {1980} {\PRL} {\bf 45} {1926}
\item[] Chivukula R S and Walker T P {1990} {\NP} B{\bf 329} {445}
\item[] Chivukula R S, Cohen A G, Dimopoulos S and Walker T P 
         {1990} {\PRL} {\bf 65} {957}
\item[] Chivukula R S, Cohen A G, Luke M and Savage M J 
         {1993} {\PL} B{\bf 298} {380}
\item[] Chivukula R S, Dugan M J, Golden M and Simmons E H 
         {1995} {\ARNPS} {\bf 45} {255}
\item[] Choi K and Kim J E {1985} {\PL} {\bf 165B} {71}
\item[] Clayton D D {1985} {\ApJ} {\bf 290} {428}
\item[] Clegg R E S {1983} {\MN} {\bf 229} {31P}
\item[] Cline J M {1992} {\PRL} {\bf 68} {3137}
\item[] Cline J M and Walker T P {1992} {\PRL} {\bf 68} {270}
\item[] Cline J M, Kainulainen K and Olive K A {1994} {\PR} {\bf D49} {6394}
\item[] Cohen A G, Kaplan D B and Nelson A E {1994} {\ARNPS} {\bf 47} {27}
\item[] Coleman S and Weinberg E {1973} {\PR} D{\bf 7} {1888}
\item[] Coles P and Ellis G F R {1994} {\Nature} {\bf 370} {609}
\item[] Collins P D B, Martin A D and Squires E J 
         {1989} {\it Particle Physics and Cosmology} {(New York: Wiley)}
\item[] Cooper-Sarkar A M \etal (BEBC WA66 collab.) 
         {1985} {\PL} {\bf 160B} {207}
\item[] Cooper-Sarkar A M, Sarkar S, Guy J, Venus W, Hulth P O and Hultqvist K 
         {1992} {\PL} B{\bf 280} {153}
\item[] Copi C J, Schramm D N and Turner M S {1995a} {\Science} {\bf 267} {192}
\item[]  \dash {1995b} {\PRL} {\bf 75} {3981}
\item[]  \dash {1995c} {\ApJ} {\bf 455} {L95}
\item[]  \dash {1996} {\preprint} {FERMILAB-Pub-96/122-A [astro-ph/9606059}
                       [submitted to {\PRL}]
\item[] Costa G and Zwirner F {1986} {\RNC} {\bf 9} {1}
\item[] Coughlan G, Fischler W, Kolb E W, Raby S and Ross G G 
         {1983} {\PL} {\bf 131B} {59}
\item[] Coughlan G, Holman R, Ramond P and Ross G G {\PL} {\bf 140B} {44}
\item[] Coughlan G, Ross G G, Holman R, Ramond P, Ruiz-Altaba M and Valle J W F
         {1985} {\PL} {\bf 158B} {401}
\item[] Coulson D, Ferreira P, Graham P and Turok N 
         {1994} {\Nature} {\bf 368} {27}
\item[] Cowsik R {1981} {\it Cosmology and Particles} 
         {ed J~Audouze \etal (Gif Sur Yvette: Editions Fronti\'eres) p~157}
\item[] Cowsik R and McCleland J {1972} {\PRL} {\bf 29} {669}
\item[] Crawford M and Schramm D N {1982} {\Nature} {\bf 298} {538}
\item[] Cremmer E, Julia B, Scherk J, Ferrara S, Girardello L and van 
         Nieuwenhuizen P {1979} {\NP} B{\bf 147} {105}
\item[] Cremmer E, Ferrara S, Girardello L and van Proyen A 
         {1983} {\NP} B{\bf 212} {413}


\item[] Dar A {1995} {\ApJ} {\bf 449} {550}
\item[] David Y and Reeves H {1980} {\PTRS} A{\bf 296} {415}
\item[] Davidson K and Kinman T D {1985} {\ApJS} {\bf 58} {321}
\item[] Davidson S and Peskin M {1994} {\PR} D{\bf 49} {2114}
\item[] Dearborn D S P, Schramm D N and Steigman G {1986} {\ApJ} {\bf 302} {35}
\item[] Dearborn D S P, Schramm D N, Steigman G and Truran J 
         {1989} {\ApJ} {\bf 347} {455}
\item[] de Boer W {1994} {\PPNP} {\bf 33} {201}
\item[] de Carlos  B, Casas J A, Quevedo F and Roulet E 
         {1993} {\PL} B{\bf 318} {447}
\item[] Dekel A {1994} {\ARAA} {\bf 32} {371}
\item[] de Laix A A and Scherrer R J {1993} {\PR} D{\bf 48} {562}
\item[] Delbourgo-Salvador P, Gry C, Malinie G and Audouze J,
         {1985} {\AA} {\bf 150} {53} 
\item[] Delbourgo-Salvador P, Audouze J and Vidal-Madjar A 
         {1987} {\AA} {\bf 174} {365} 
\item[] De Leener-Rosier N \etal {1991} {\PR} D{\bf 43} {3611}
\item[] Deliyannis C P, Demarque P, Kawaler S D, Krauss, L M and Romanelli P 
         {1989} {\PRL} {\bf 62} {1583}
\item[] Deliyannis C P, Demarque P and Kawaler S D {1990} {\ApJS} {\bf 73} {21}
\item[] Deliyannis C P, Pinsonneault M H and Duncan D K 
         {1993} {\ApJ} {\bf 414} {740}
\item[] Denegri D, Sadoulet B and Spiro M {1990} {\RMP} {\bf 62} {1}
\item[] Derbin A V \etal {1994} {\it Yad. Phys.} {\bf 57} {236}
\item[] Derendinger J-P and Savoy C A {1984} {\NP} B{\bf 237} {307}
\item[] Derendinger J-P, Ib\'a\~nez L and Nilles H P 
         {1985} {\PL} {\bf 155B} {65}
\item[] De R\'ujula A and Glashow S {1981} {\PRL} {\bf 45} {942}
\item[] De R\'ujula A, Nanopoulos D V and Shaver P A (ed) 
         {1987} {\it An Unified View of the Macro- and Micro-Cosmos} 
         {(Singapore: World Scientific)}
\item[] De R\'ujula A, Glashow S and Sarid U {1990} {\NP} B{\bf 333} {173}
\item[] Deser S and Zumino B {1977} {\PRL} {\bf 38} {1433}
\item[] Dicke R H and and Peebles P J E {1979} {\it General Relativity: An
         Einstein Centenary Survey} {ed S~W~Hawking and W~Israel (Cambridge:
         Cambridge University Press) p~504}
\item[] Dicus D A and Teplitz V L {1980} {\PRL} {\bf 44} {218}
\item[] Dicus D A, Kolb E W and Teplitz V L {1977} {\PRL} {\bf 39} {168}
\item[]  \dash {1978a} {\ApJ} {\bf 221} {327}
\item[] Dicus D A, Kolb E W, Teplitz V L and Wagoner R V 
         {1978b} {\PR} D{\bf 17} {1529} 
\item[] Dicus D A, Kolb E W, Gleeson A M, Sudarshan E C G, Teplitz V L and 
         Turner M S {1982} {\PR} D{\bf 26} {2694}
\item[] Dimopoulos S {1995} {\it Proc XXVII Conf. on High Energy
         Physics, Glasgow} {ed P~J~Bussey and I~G~Knowles (Bristol: 
         IOP Publishing) Vol~I, p~93}
\item[] Dimopoulos S and Hall L J {1987} {\PL} {\bf 196B} {135}
\item[] Dimopoulos S and Turner M S {1982} {\it The Birth of the Universe}
         {ed J~Audouze and T~Tran~Thanh~Van (Gif Sur Yvette: Editions 
         Fronti\'eres) p~113}
\item[] Dimopoulos S, Esmailzedeh R, Hall L J and Starkman G D
         {1988} {\ApJ} {\bf 330} {545} 
\item[]  \dash {1989} {\NP} B{\bf 311} {699}
\item[] Dimopoulos S, Eichler D, Esmailzadeh R and Starkman G 
         {1990} {\PR} D{\bf 41} {2388}
\item[] Dimopoulos S, Giudice G F and Tetradis N {1995} {\NP} B{\bf 454} {59}
\item[] Dine M (ed) {1988} {\it String Theory in Four Dimensions} 
         {(Amsterdam: North-Holland)}
\item[]  \dash {1990} {\ARNPS} {\bf 40} {145}
\item[] Dine M, Fischler W and Nemechansky D {1984} {\PL} {\bf 136B} {169}
\item[] Dine M, Kaplunovsky V, Mangano M, Nappi C and Seiberg N 
         {1985} {\NP} {\bf B259} {519}
\item[] Dine M and Nelson A E {1993} {\PR} D{\bf 48} {1277}
\item[] Dine M, Randall L and Thomas S {1995} {\PRL} {\bf 75} {398}
\item[]  \dash {1996} {\NP} B{\bf 458} {291}
\item[] Dine M, Nelson A E, Nir Y and Shirman Y {1996} {\PR} D{\bf 53} {2658}
\item[] Dixon L J and Nir, Y {1991} {\PL} B{\bf 266} {425}
\item[] Dodelson S {1989} {\PR} D{\bf 40} {3252}
\item[] Dodelson S and Turner M S {1992} {\PR} D{\bf 46} {3372}
\item[] Dodelson S, Guyk G and Turner M S {1994} {\PR} D{\bf 49} {5068}
\item[] Dolgov A D {1992} {\PRep} {\bf 222} {309}
\item[] Dolgov A D and Fukugita M {1992} {\PR} D{\bf 46} {5378}
\item[] Dolgov A D and Rothstein I Z {1993} {\PRL} {\bf 71} {476}
\item[] Dolgov A D and Zel'dovich Ya B {1981} {\RMP} {\bf 53} {1}
\item[] Dolgov A D, Kainulainen K and Rothstein I Z 
         {1995} {\PR} D{\bf 51} {4129}
\item[] Dolgov A D, Pastor S and Valle J W F {1996} {\PL} B{\bf 383} {193}
\item[] Dominguez-Tenreiro R {1987} {\ApJ} {\bf 313} {523}
\item[] Dover C B, Gaisser T K and Steigman G {1979} {\PRL} {\bf 42} {1117}
\item[] Drees M and Nojiri M M {1993} {\PR} D{\bf 47} {376}
\item[] Dreiner H and Ross G G {1993} {\NP} B{\bf 410} {188}
\item[] Dubbers D {1991} {\PPNP} {\bf 26} {173} 
\item[] Duncan D K, Lambert D L and Lemke M {1992} {\ApJ} {\bf 401} {584}
\item[] Dvali G {1995} {\preprint} {IFUP-TH 09/95 (hep-ph/9503259)} ?


\item[] Edmunds M G {1994} {\MN} {\bf 270} {L37}
\item[] Efstathiou G P {1990} {\it Physics of the Early Universe} 
         {ed J~Peacock \etal (Bristol: Adam Hilger) p~361}
\item[] Elliott T, King S F and White P L {1994} {\PR} D{\bf 49} {2435}
\item[] Ellis J {1987} {\PS} T{\bf 15} {61}
\item[]  \dash {1993} {\preprint} {CERN-TH-7083-93} [to appear in {\it Les 
                     Houches Session 60: Cosmology and Large Scale Structure}] 
\item[]  \dash {1994} {\NC} {\bf 107A} {1091} 
\item[]  \dash {1995} {\it International Symposium on Lepton Photon 
          Interactions, Beijing} {ed Z-P Zheng and H-S Chen (Singapore: 
          World Scientific)}
\item[] Ellis J and Olive K A {1983} {\NP} B{\bf 223} {252}
\item[] Ellis J and Steigman G {1979} {\PL} {\bf 89B} {186}
\item[] Ellis J, Nanopoulos D V and Rudaz S {1982} {\NP} B{\bf 202} {43}
\item[] Ellis J, Linde A D and Nanopoulos D V {1983} {\PL} {\bf 118B} {59}
\item[] Ellis J, Hagelin J S, Nanopoulos D V, Olive K A and Srednicki M
         {1984a} {\NP} B{\bf 238} {453}
\item[] Ellis J, Kim J E and Nanopoulos D V {1984b} {\PL} {\bf 145B} {181}
\item[] Ellis J, Gelmini G B, Jarlskog C, Ross G G and Valle J W F {
         1985a} {\PL} {\bf 150B} {142}
\item[] Ellis J, Nanopoulos D V and Sarkar S {1985b} {\NP} B{\bf 259} {175}
\item[] Ellis J, Enqvist K, Nanopoulos D V, Olive K A, Quir\'os M and Zwirner F
         {1986a} {\PL} B{\bf 176} {403}
\item[] Ellis J, Enqvist K, Nanopoulos D V and Sarkar S 
         {1986b} {\PL} {\bf 167B} {457}
\item[] Ellis J, Nanopoulos D V and Quir\'os M {1986c} {\PL} B{\bf 174} {176}
\item[] Ellis J, Enqvist K, Nanopoulos D V and Olive K A 
         {1987} {\PL} B{\bf 191} {343}
\item[] Ellis J, Gunion J, Haber H, Roszkowski L and Zwirner F 
         {1989} {\PR} D{\bf 39} {844}
\item[] Ellis J, Kelly S and Nanopoulos D V {1991} {\PL} B{\bf 249} {441}
\item[] Ellis J, Gelmini G B, Nanopoulos D V, Lopez J and Sarkar S 
         {1992} {\NP} B{\bf 373} {399}
\item[] Ellis J, Nanopoulos D V, Olive K A and Rey S-J {1996} {\AP}
         {\bf 4} {371}
\item[] Ellwanger U {1983} {\PL} {\bf 133B} {187}
\item[] Elmfors P, Enqvist K and Vilja I {1994} {\PL} B{\bf 326} {37}
\item[] Enqvist K {1986} {\it Proc. 2nd ESO-CERN Symp. on Cosmology, Astronomy
         and Fundamental Physics} {ed G~Setti and L~van~Hove (Garching: ESO) 
         p~137}
\item[] Enqvist K and Eskola K J {1990} {\MPL} A{\bf 5} {1919}
\item[] Enqvist K and Nanopoulos D V {1986} {\PPNP} {\bf 16} {1}
\item[] Enqvist K and Sirkka J {1993} {\PL} B{\bf 314} {298}
\item[] Enqvist K and Uibo H {1993} {\PL} B{\bf 301} {376}
\item[] Enqvist K, Kainulainen K and Maalampi J 
         {1990a} {\PL} B{\bf 244} {186}
\item[]  \dash {1990b} {\PL} B{\bf 249} {531}
\item[]  \dash {1991} {\NP} B{\bf 349} {754}
\item[] Enqvist K, Kainulainen K and Semikoz V {1992a} {\NP} B{\bf 374} {392}
\item[] Enqvist K, Kainulainen K and Thomson M {1992b} {\NP} B{\bf 373} {498}
\item[]  \dash {1992c} {\PRL} {\bf 68} {744}
\item[]  \dash {1992d} {\PL} B{\bf 280} {245}
\item[] Epstein R, Lattimer J and Schramm D N {1976} {\Nature} {\bf 263} {198}
\item[] Escribano R, Mass\'o E and Toldr\`a R {1995} {\PL} B{\bf 356} {313}
\item[] Esmailzadeh R, Starkman G D and Dimopoulos S 
         {1991} {\ApJ} {\bf 378} {504}


\item[] Faber S and Gallagher J S {1979} {\ARAA} {\bf 17} {135}
\item[] Faraggi A and Nanopoulos D V {1991} {\MPL} A{\bf 6} {61}
\item[] Fargion D and Roos M {1984} {\PL} {\bf 147B} {34}
\item[] Fargion D and Shepkin M G {1981} {\PL} {\bf 146B} {46}
\item[] Farhi E and Susskind L {1981} {\PRep} {\bf 74} {277}
\item[] Farrar G and Fayet P {1978} {\PL} {\bf 76B} {575}
\item[] Faul D D, Berman B L, Meyer P and Olson D L {1981} {\PR} {\bf 24} {84}
\item[] Fayet P {1979} {\PL} {\bf 86B} {272}
\item[]  \dash {1982} {\it Electroweak Interactions and Grand Unified
          Theories} {ed J~Tran~Thanh~Van (Gif Sur Yvette: Editions 
          Fronti\'eres) p~161}
\item[]  \dash {1984} {\PRep} {\bf 105} {21}
\item[]  \dash {1987} {\PS} T{\bf 15} {46}
\item[] Fayet P and Ferrara S {1975} {\PRep} {\bf 32} {249}
\item[] Felten J E and Steigman G {1995} {\SSR} {\bf 74} {245}
\item[] Ferrara S, Kounnas C, Porrati M and Zwirner F 
         {1989} {\NP} B{\bf 318} {75}
\item[] Fields B D and Olive K A {1996} {\PL} B{\bf 368} {103}
\item[] Fields B D, Dodelson S and Turner M S {1993} {\PR} D{\bf 47} {4309}
\item[] Fields B D, Kainulainen K and Olive K A {1995} {\preprint}
         {CERN-TH/95-335 (hep-ph/9512321)} [{\AP}, in press]
\item[] Fields B D, Kainulainen K, Olive K A and Thomas D {1996} {\preprint}
         {CERN-TH/96-59 (astro-ph/9603009)}
\item[] Fischler W {1994} {\PL} B{\bf 332} {277} 
\item[] Fischler W, Giudice G F, Leigh R G and Paban S {1991} {\PL} 
         B{\bf 258} {45}
\item[] Foot R and Volkas R R {1995a} {\PRL} {\bf 75} {4350}
\item[]  \dash {1995b} {\PR} D{\bf 52} {6595} 
\item[] Foot R, Thomson M and Volkas R R {1996} {\PR} D{\bf 53} {5349}
\item[] Fowler W A and Hoyle F {1964} {\Nature} {\bf 203} {1108} 
\item[] Fowler W A, Caughlan G R and Zimmerman B A {1975} {\ARAA} {\bf 13} {69}
\item[] Freedman S {1990} {\CNPP} {\bf 19} {209}
\item[] Freese K, Kolb E W and Turner M S {1983} {\PR} D{\bf 27} {1689}
\item[] Frieman J A and Giudice G {1989} {\PL} B{\bf 224} {125}
\item[] Frieman J A, Kolb E W and Turner M S {1990} {\PR} D{\bf 41} {3080}
\item[] Fukugita M, N\"otzold D, Raffelt G and Silk J {1988} {\PRL}
         {\bf 60} {879}
\item[] Fukugita M, Hogan C J and Peebles P J E {1993} {\Nature} {\bf 366} 
         {309}
\item[] Fukugita M and Yanagida T {1986} {\PL} {\bf 174B} {45}
\item[]  \dash {1994} {\it Physics and Astrophysics of 
          Neutrinos} {ed A~Suzuki (Tokyo: Springer Verlag) p~248}
\item[] Fukugita M and Yazaki {1987} {\PR} D{\bf 36} {3817}
\item[] Fuller G M, Boyd R N and Kalen J D {1991} {\ApJ} {\bf 371} {L11}
\item[] Fuller G M and Malaney R A {1991} {\PR} D{\bf 43} {3136}


\item[] Galli D, Palla F, Ferrini F and Penco U {1995} {\ApJ} {\bf 443} {536}
\item[] Gari M and Hebach H {1981} {\PRep} {\bf 72} {1}
\item[] Geiss J {1993} {\it Origin and Evolution of the Elements} {ed 
         N~Prantzos \etal (Cambridge: Cambridge University Press) p~89}
\item[] Geiss J and Reeves H {1972} {\AA} {\bf 18} {126}
\item[] Gelmini G, Nussinov S and Peccei R {1992} {\IJMP} A{\bf 7} {3141} 
\item[] Gelmini G and Roncadelli M {1981} {\PL} {\bf 99B} {411}
\item[] Gelmini G and Roulet E {1995} {\RPP} {\bf 58} {1207}
\item[] German G and Ross G G {1986} {\PL} B{\bf 172} {305}
\item[] Gershte\v{\ii}n S and Zel'dovich Ya B {1966} {\JETPL} {\bf 4} {120}
\item[] Gherghetta T {1996} {\preprint} {UM-TH-96-10 (hep-ph/9607448)}
\item[] Gibbons G, Hawking S W and Siklos S T C (ed) {1983} {\it The Very 
         Early Universe} {(Cambridge: Cambridge University Press)}
\item[] Gilmore G, Gustafsson B, Edvardsson B and Nissen P E 
         {1992} {\Nature} {\bf 357} {379}
\item[] Giudice G {1990} {\PL} B{\bf 251} {460}
\item[] Giudice G and Masiero A {1988} {\PL} B{\bf 206} {480}
\item[] Giveon A, Porrati M and Rabinovici E {1994} {\PRep} {\bf 244} {77}
\item[] Gloeckler G and Geiss J {1996} {\Nature} {\bf 381} {210}
\item[] Gnedin N Y and Ostriker J P {1992} {\ApJ} {\bf 400} {1}
\item[] Goldberg H and Hall L J {1986} {\PL} B{\bf 174} {151}
\item[] Goncharov A S, Linde A D and Vysotski\v{\ii} M I {1984} {\PL}
         {\bf 147B} {279}
\item[] Gondolo P and Gelmini G {1991} {\NP} B{\bf 360} {145}
\item[] Gondolo P, Gelmini G and Sarkar S {1993} {\NP} B{\bf 392} {111}
\item[] Gonzalez-Garcia M C and Valle J W F {1990} {\PL} B{\bf 240} {163}
\item[] Gott J R, Gunn J E, Schramm D N and Tinsley B M {1974} {\ApJ} 
         {\bf 194} {543}
\item[] Govaerts J, Lucio J L, Martinez A and Pestiau J {1981} {\NP} 
         A{\bf 368} {409}
\item[] Grasso D and Kolb E W {1996} {\PR} D{\bf 54} {1374}
\item[] Gratsias J, Scherrer R J and Spergel D N {1991} {\PL} B{\bf 262} {298}
\item[] Green M, Schwarz J H and Witten E {1987} {\it Superstring Theory} 
         {(Cambridge: Cambridge University Press)}
\item[] Grifols J A and Mass\'o E {1987} {\MPL} A{\bf 2} {205}
\item[] Gronau M and Yahalom R {1984} {\PR} D{\bf 30} {2422}
\item[] Gross D J, Pisarski R D and Yaffe, L G {1981} {\RMP} {\bf 53} {43}
\item[] Gross D J, Harvey J A, Martinec E and Rohm R {1985} {\PRL}
         {\bf 54} {502}
\item[] Guth A {1981} {\PR} D{\bf 23} {347}
\item[] Gyuk G and Turner M S {1993} {\preprint} {FERMILAB-Pub-93/181-A 
         (astro-ph/9307025)}
\item[]  \dash {1994} {\PR} D{\bf 50} {6130}


\item[] Haber H and Kane G {1985} {\PRep} {\bf 117} {75}
\item[] Haber H and Weldon A {1981} {\PR} D{\bf 25} {502}
\item[] Hagelin J S and Parker R J D {1990} {\NP} B{\bf 329} {464}
\item[] Hall L J {1988} {\it Proc. 16th SLAC Summer Institute on
         Particle Physics} {ed E~C~Brennan (Stanford: SLAC) p~85}
\item[] Hall L and Hsu S {1990} {\PRL} {\bf 64} {2848}
\item[] Hall L and Suzuki M {1984} {\NP} B{\bf 231} {419}
\item[] Hall L, Lykken J and Weinberg S {1983} {\PR} D{\bf 27} {2359}
\item[] Halm I {1987} {\PL} B{\bf 188} {403}
\item[] Hannestad S and Madsen J {1995} {\PR} D{\bf 52} {1764}
\item[]  \dash {1996} {\PRL} {\bf 76} {2848}; erratum hep-ph/9606452
\item[] Harris M J, Fowler W A, Caughlan G R and Zimmerman B A 
         {1983} {\ARAA} {\bf 21} {165}
\item[] Harrison E R {1973} {\ARAA} {\bf 11} {155}
\item[]  \dash {1993} {\ApJ} {\bf 403} {28}
\item[] Harvey J A, Kolb E W, Reiss D B and Wolfram S {1981} {\NP} B{\bf 177} 
         {456}
\item[] Hata N, Scherrer R J, Steigman G, Thomas D, Walker T P, Bludman S 
         and Langacker P {1995} {\PRL} {\bf 75} {3977}
\item[] Hata N, Scherrer R J, Steigman G, Thomas D and Walker T P 
         {1996a} {\ApJ} {\bf 458} {637}
\item[] Hata N, Steigman G, Bludman S and Langacker P {1996b} {\preprint}
         {OSU-TA-6/96 (astro-ph/9603087)}
\item[] Hawking S and Ellis G F R {1973} {\it The Large-scale Structure of
         Space-time} {(Cambridge: Cambridge University Press)}
\item[] Hayashi C {1950} {\PTP} {\bf 5} {224} 
\item[] Hecht H F {1971} {\ApJ} {\bf 170} {401}
\item[] Hemmick T K \etal {1990} {\PR} D{\bf 41} {2074}
\item[] Herrera M A and Hacyan S {1989} {\ApJ} {\bf 336} {539}
\item[] Hewett, J L and Rizzo T G {1989} {\PRep} {\bf 183} {193}
\item[] Hime A {1992} {\MPL} A{\bf 7} {1301}
\item[] Hime A and Jelley N {1991} {\PL} B{\bf 257} {441}
\item[] Hime A, Phillips R J N, Ross G G and Sarkar S {1991} {\PL} B{\bf 260} 
         {381}
\item[] Hinchliffe I (ed) {1987} {\it Cosmology and Particle Physics:
         Proc. Theoretical Workshop, Berkeley} {(Singapore: World Scientific)}
\item[] Hindmarsh M B and Kibble T W B {1995} {\RPP} {\bf 58} {477} 
\item[] Hobbs L and Pilachowski C {1988} {\ApJ} {\bf 334} {734}
\item[] Hobbs L and Thorburn J A {1991} {\ApJ} {\bf 375} {116}
\item[]  \dash {1994} {\ApJ} {\bf 428} {L25}
\item[] Hodges H and Turner M S {1988} {\preprint} {FERMILAB-PUB-88/115-A-Rev}
\item[] Hogan C J {1994} {\it Advances in Astrofundamental Physics} 
         {ed N~Sanchez and A~Zichichi (Singapore: World Scientific) p~38} 
\item[]  \dash {1995a} {\it Cosmic Abundances} {ed S~Holt and G. Sonneborn 
          (San Francisco: PASP) p~67} 
\item[]  \dash {1995b} {\ApJ} {\bf 441} {L17}
\item[]  \dash {1996} {\PR} D{\bf 54} {112}
\item[] Hogan C J and Rees M {1984} {\Nature} {\bf 311} {109}
\item[] Holman R, Ramond P and Ross G G {1984} {\PL} {\bf 137B} {343}
\item[] Hosotani Y {1981} {\NP} B{\bf 191} {411}
\item[] Hoyle F and Tayler R J {1964} {\Nature} {\bf 203} {1108}
\item[] Hu W and Silk J {1993} {\PRL} {\bf 70} {2661}
\item[] Huchra J P {1992} {\Science} {\bf 256} {321}


\item[] Ib\'a\~nez L {1994} {\it Proc. Intern. EPS Conf. on High
         Energy Physics, Marseilles} {ed J~Carr and M~Perrottet (Paris: 
         Editions Fronti\'eres) p~911} 
\item[] Ib\'a\~nez L and Ross G G {1993} {\it Perspectives on Higgs
         Physics} {ed G~Kane (Singapore: World Scientific) p~229}
\item[] Iben I and Truran J W {1978} {\ApJ} {\bf 220} {980}
\item[] Illarianov A F and Sunyaev R A {1975} {\SA} {\bf 18} {413}
\item[] Izotov Y I, Thuan T X and Lipovetsky V A {1994} {\ApJ} {\bf 435} {647} 
\item[]  \dash {1996} {\preprint} {(submitted to \ApJ)}


\item[] Jacoby G H, Branch D, Ciardullo R, Davies R, Harris W E, Pierce M J, 
         Pritchet C J, Tonry J L and Welch D L {1992} {\PASP} {\bf 104} {599}
\item[] Jakobsen P, Boksenberg A, Deharveng J M, Greenfield P, Jedrzejewski 
         and Paresce F {1994} {\Nature} {\bf 370} {35}
\item[] Jedamzik K, Fuller G M and Mathews G J {1994a} {\ApJ} {\bf 423} {50} 
\item[] Jedamzik K, Fuller G M, Mathews G J and Kajino T
         {1994b} {\ApJ} {\bf 422} {423}
\item[] Juszkiewicz R, Silk J and Stebbins A {1985} {\PL} {\bf 158B} {463}


\item[] Kainulainen K {1990} {\PL} B{\bf 244} {191}
\item[] Kajino T, Toki H and Austin S M {1987} {\ApJ} {\bf 319} {531}
\item[] Kajantie K, Laine M Rummukainen K and Shaposhnikov M {1996} {\NP}
         B{\bf 466} {189}; {\preprint} {CERN-TH-96-126 (hep-ph/9605288)}
\item[] Kane G {1987} {\it Modern Elementary Particle Physics} {(Redwood City, 
         CA: Addison-Wesley)}
\item[] Kang H-S and Steigman G {1992} {\NP} B{\bf 372} {494}
\item[] Kapusta J {1988} {\it Finite Temperature Field Theory} {(Cambridge: 
         Cambridge University Press)}
\item[] Kaul R K {1983} {\RMP} {\bf 55} {449}
\item[] Kawano L {1988} {\preprint} {FERMILAB-Pub-88/34-A}
\item[]  \dash {1992} {\preprint} {FERMILAB-Pub-92/04-A}
\item[] Kawano L, Schramm D N and Steigman G {1988} {\ApJ} {\bf 327} {750}
\item[] Kawano L, Fuller G M, Malaney R A and Savage M J {1992} {\PL}
         B{\bf 275} {487}
\item[] Kawasaki M and Moroi T {1995a} {\PTP} {\bf 93} {879}
\item[]  \dash {1995b} {\ApJ} {\bf 452} {506}
\item[]  \dash {1995c} {\PL} B{\bf 346} {27}
\item[] Kawasaki M and Sato K {1987} {\PL} B{\bf 189} {23}
\item[] Kawasaki M, Terasawa N and Sato K {1986} {\PL} B{\bf 178} {71}
\item[] Kawasaki M, Kernan P, Kang H-S, Scherrer, R J, Steigman G and 
         Walker T P {1994} {\NP} B{\bf 419} {105}
\item[] Kayser B, Gibrat-Debu F and Perrier F {1989} {\it The Physics
         of Massive Neutrinos} {(Singapore: World Scientific)}
\item[] Kennicutt R C, Freedman W L and Mould J R {1995} {\AJ} {\bf 110} {1476}
\item[] Kernan P J {1993} {\it PhD thesis} {(Ohio State University)}
\item[] Kernan P J and Krauss L M {1994} {\PRL} {\bf 72} {3309}
\item[] Kernan P J and Sarkar S {1996} {\PR} D{\bf 54} {3681} 
\item[] Khlopov M Yu and Linde A D {1984} {\PL} {\bf 138B} {265}
\item[] Khlopov M Yu and Petcov S {1981} {\PL} {\bf 99B} {117}
         (erratum {\bf 100B} {520}) 
\item[] Kibble T W B {1976} {\JPA} {\bf 9} {1387}
\item[] Kim J E {1987} {\PRep} {\bf 150} {1}
\item[] Kim J E and Nilles H P {1984} {\PL} {\bf 138B} {150}
\item[] King S F {1995} {\RPP} {\bf 58} {263}
\item[] Kingdon J and Ferland G J {1995} {\ApJ} {\bf 442} {714}
\item[] Kofman L, Linde A D and Starobinsky A A {1996} {\PRL} {\bf 76} {1011}
\item[] Kolb E W {1980} {\it Proc. Telemark Workshop on Neutrino Mass} 
         {ed V~Barger and D~Cline (Madison: University of Wisconsin) p~61}
\item[] Kolb E W and Goldman T {1979} {\PRL} {\bf 43} {897}
\item[] Kolb E W and Peccei R D (ed) {1995} {\it Particle and Nuclear
         Astrophysics and Cosmology in the Next Millenium: Snowmass '94}
         {(Singapore: World Scientific)}
\item[] Kolb E W and Scherrer R J {1982} {\PR} D{\bf 25} {1481}
\item[] Kolb E W and Turner M S {1983} {\ARNPS} {\bf 33} {645}
\item[]  \dash {1990} {\it The Early Universe} {(Redwood City, CA: 
          Addison-Wesley)}
\item[] Kolb E W, Seckel D and Turner M S {1985} {\Nature} {\bf 314} {415}
\item[] Kolb E W, Perry M J and Walker T P {1986a} {\PR} D{\bf 33} {869}
\item[] Kolb E W, Turner M S, Lindley D, Olive K A and Seckel D (ed)
         {1986b} {\it Inner Space--Outer Space} {(Chicago: University of 
         Chicago Press)}
\item[] Kolb E W, Turner M S and Walker T P {1986c} {\PR} D{\bf 34} {2197}
\item[] Kolb E W, Turner M S and Schramm D N {1989} {\it Neutrino Physics} 
         {ed K~Winter (Cambridge: Cambridge University Press) p~239}
\item[] Kolb E W, Turner M S, Chakravorty A and Schramm D N 
         {1991} {\PRL} {\bf 67} {533}
\item[] Kounnas C, Zwirner F and Pavel I {1994} {\PL} B{\bf 335} {403}
\item[] Krakauer D A \etal {1990} {\PL} B{\bf 252} {177}
\item[]  \dash {1991} {\PR} D{\bf 44} {R6}
\item[] Krauss L M {1983a} {\NP} B{\bf 227} {1303}
\item[]  \dash {1983b} {\PL} {\bf 128B} {37}
\item[]  \dash {1984} {\PRL} {\bf 53} {1976}
\item[]  \dash {1985} {\preprint} {HUTP-85-A040}
\item[] Krauss L M and Kernan P {1994} {\ApJ} {\bf 432} {L79} 
\item[]  \dash {1995} {\PL} B{\bf 347} {347} 
\item[] Krauss L M and Rey S-J (ed) {1992} {\it Baryon Number
         Violation at the Electroweak Scale} {(Singapore: World Scientific)}  
\item[] Krauss L M and Romanelli P {1990} {\ApJ} {\bf 358} {47}
\item[] Krauss L M, Terning J and Appelquist T {1993} {\PRL} {\bf 71} {823}  
\item[] Krauss L M, Guth A H, Spergel D N, Field G B and Press W H
         {1986} {\Nature} {\bf 319} {748}
\item[] Kundic T \etal {1995} {\ApJ} {\bf 455} {L5}
\item[] Kurki-Suonio H, Matzner R A, Olive K A and Schramm D N 
         {1990} {\ApJ} {\bf 353} {406}
\item[] Kuzmin V A, Rubakov V A and Shaposhnikov M E {1985} {\PL} {\bf 155B} 
         {36}
\item[] \dash {1987} {\PL} B{\bf 191} {171}


\item[] Lahanas A B and Nanopoulos D V {1987} {\PRep} {\bf 145} {1}
\item[] Lam W P and Ng K-W {1991} {\PR} D{\bf 44} {3345}
\item[] Landau L D and Lifshitz E M {1982} {\it Statistical Physics} 
         {(Oxford: Pergamon Press)}
\item[] Lane K {1993} {\it The Building Blocks of Creation (TASI'93)} 
         {ed S~Raby and T~Walker (Singapore: World Scientific) p~381}
\item[] Lang K R {1992} {\it Astrophysical Data: Planets and Stars} {(Berlin: 
         Springer-Verlag)}
\item[] Langacker P {1981} {\PRep} {\bf 72} {185}
\item[]  \dash {1988} {\it Neutrino Physics} {ed H~V~Klapdor (Berlin:
         Springer-Verlag) p~71}
\item[]  \dash (ed) {1994} {\it Precision Tests of the Standard Electroweak 
         Model} {(Singapore: World Scientific)} 
\item[] Langacker P and Luo M {1991} {\PR} D{\bf 44} {817}
\item[] Langacker P, Segr\`e G and Soni S {1982} {\PR} D{\bf 26} {3425}
\item[] Langacker P, Sathiapalan B and Steigman G {1986} {\NP} 
         B{\bf 266} {669}
\item[] Langacker P, Petcov S T, Steigman G and Toshev S {1987} {\NP} 
         B{\bf 282} {589}
\item[] Laurent C {1983} {\it Proc. ESO Workshop on Primordial 
         Helium} {ed P~Shaver \etal (Garching: ESO) p~335}
\item[] Lazarides G, Schaefer R, Seckel D and Shafi Q {1990} {\NP}
         B{\bf 346} {193}
\item[] Lee B W and Schrock R E {1977} {\PR} D{\bf 16} {1444}
\item[] Lee B W and Weinberg S {1977} {\PRL} {\bf 39} {165}
\item[] Leigh R G and Rattazzi R {1995} {\PL} B{\bf 352} {20}
\item[] LEP Electroweak Working Group {1995} {\preprint} {CERN PPE/95-172}
\item[] Liddle A and Lyth {1993} {\PRep} {\bf 231} {1}
\item[] Linde A D {1979} {\RPP} {\bf 42} {389}
\item[]  \dash {1984} {\RPP} {\bf 47} {925}
\item[]  \dash {1990} {\it Particle Physics and Inflationary 
          Cosmology} {(New York: Harwood Academic)}
\item[]  \dash {1996} {\PR} D{\bf 53} {R4129}
\item[] Lindley D {1979} {\MN} {\bf 188} {15p} 
\item[]  \dash {1985} {\ApJ} {\bf 294} {1}
\item[] Linsky J L \etal {1993} {\ApJ} {\bf 402} {694}
\item[]  \dash {1995} {\ApJ} {\bf 451} {335}
\item[] Longair M S {1981} {\it High Energy Astrophysics} {(Cambridge:
         Cambridge University Press)}
\item[] Lopez J L and Nanopoulos D V {1990} {\PL} B{\bf 241} {392}
\item[] Lynden-Bell D and Gilmore G (eds) {1990} {\it Baryonic Dark Matter} 
         {(Dordrecht: Kluwer)}
\item[] Lyth D H and Stewart E D {1995} {\PRL} {\bf 75} {201}
\item[]  \dash {1996} {\PR} D{\bf 53} {1784} 
\item[] Lyttleton R A and Bondi H {1959} {\PRS} A{\bf 252} {313}


\item[] Maalampi J and Roos M {1990} {\PRep} {\bf 186} {53}
\item[] Mahajan S {1986} {\PR} D{\bf 33} {338}
\item[] Malaney R A and Fowler W A {1988} {\ApJ} {\bf 333} {14}
\item[] Malaney R A and Mathews G J {1993} {\PRep} {\bf 229} {145}
\item[] Mampe W, Ageron P, Bates C, Pendlebury J M and Steyerl A 
         {1989} {\PRL} {\bf 63} {593}
\item[] Manohar A {1987} {\PL} B{\bf 186} {370}
\item[] Marciano W J {1991} {\ARNPS} {\bf 41} {469}
\item[] Marciano W J and Sanda A {1977} {\PL} {\bf 67B} {303}
\item[] Mass\'o E and Pomarol A {1989} {\PR} D{\bf 40} {2519}
\item[] Mass\'o E and Toldr\`a R {1994a} {\PL} B{\bf 333} {132}
\item[]  \dash {1994b} {\PR} D{\bf 52} {1755}
\item[] Mather J C \etal (COBE collab.) {1994} {\ApJ} {\bf 420} {439}
\item[] Mathews G J, Alcock C R and Fuller G M {1990a} {\ApJ} {\bf 349} {449}
\item[] Mathews G J, Meyer B M, Alcock C and Fuller G M {1990b} {\ApJ} 
         {\bf 358} {36}
\item[] Mathews G J, Boyd R N and Fuller G M {1993} {\ApJ} {\bf 403} {65}
\item[] McCullough P R {1992} {\ApJ} {\bf 390} {213}
\item[] McKellar B H J and Pakvasa S {1983} {\PL} {\bf 122B} {33}
\item[] McLerran L {1986} {\RMP} {\bf 58} {1021}
\item[] Melnick J, Heydani-Malyeri M and Leisy P {1992} {\AA} {\bf 253} {16}
\item[] Michaud G and Charbonneau P {1991} {\SSR} {\bf 57} {1}
\item[] Mihalas D and Binney J {1981} {\it Galactic Astronomy} {(San 
         Francisco: Freeman)}
\item[] Milne E A {1935} {\it Relativity, Gravitation and World
         Structure} {(Oxford: Clarendon Press)}
\item[] Miyama S and Sato K {1978} {\PTP} {\bf 60} {1703} 
\item[] Mohapatra R N and Marshak R {1980} {\PRL} {\bf 44} {1316} 
\item[]  \dash {1992} {\it Unification and Supersymmetry} {(New York: 
          Spinger-Verlag)}
\item[] Mohapatra R N and Pal P B {1991} {\it Massive Neutrinos in Physics
         and Astrophysics} {(Singapore: World Scientific)}
\item[] Molaro P, Primas F and Bonifacio P {1995} {\AA} {\bf 295} {L47}
\item[] Morgan J A {1981a} {\PL} {\bf 102B}{247}
\item[]  \dash {1981b} {\MN} {\bf 195} {173}
\item[] Moroi T, Murayama H and Yamaguchi M {1993} {\PL} B{\bf 303} {289}
\item[] Mukhanov V F, Feldman H A and Brandenberger R H 
         {1982} {\PRep} {\bf 215} {203}


\item[] Nanopoulos D V (ed) {1991} {\it Proc. Intern. School of 
         Astroparticle Physics} {(Singapore: World Scientific)}
\item[] Nanopoulos D V, Olive K A and Srednicki M {1983} {\PL} {\bf 127B} {30}
\item[] Nath P, Arnowitt R and Chamseddine A {1984} {\it Applied N=1 
         Supergravity} {(Singapore: World Scientific)}
\item[] Nelson A E and Seiberg N {1994} {\NP} B{\bf 416} {46}
\item[] Nieves J F {1983} {\PR} D{\bf 28} {1664}
\item[] Nilles H P {1984} {\PRep} {\bf 110} {1}
\item[]   \dash {1990} {\IJMP} A{\bf 5} {4199} 
\item[] Nilles H P, Srednicki M and Wyler D {1983} {\PL} {\bf 120B} {346}
\item[] Nilsson J S, Gustafson B and Skagerstam B -S (ed) {1991} {\it The Birth
        and Early Evolution of Our Universe} {(Singapore: World Scientific)}
\item[] Norris J E, Ryan S G and Stringfellow G S {1994} {\ApJ} {\bf 423} {386}
\item[] N\"otzold D and Raffelt G {1988} {\NP} B{\bf 307} {924}
\item[] Nussinov S, {1985} {\PL} {\bf 165B} {55} 


\item[] Oberauer L and von Feilitzsch F {1992} {\RPP} {\bf 55} {1093}
\item[] Oberauer L and von Feilitzsch F and M\"ossbauer R L {1987}
         {\PL} B{\bf 198} {113}
\item[] Olive K A {1990a} {\PRep} {\bf 190} {309}
\item[]  \dash {1990b} {\Science} {\bf 251} {1194}
\item[] Olive K A and Schramm D N {1992} {\Nature} {\bf 360} {439}
\item[] Olive K A and Scully S T {1996} {\IJMP} A{\bf 11} {409}
\item[] Olive K A and Steigman G {1995a} {\ApJS} {\bf 97} {49}
\item[]  \dash {1995b} {\PL} B{\bf 354} {357}
\item[] Olive K A and Thomas D {1991} {\NP} B{\bf 335} {192}
\item[] Olive K A and Turner M S {1982} {\PR} D{\bf 25} {213}
\item[] Olive K A, Schramm D N and Steigman G {1981a} {\NP} B{\bf 180} {497}
\item[] Olive K A, Schramm D N, Steigman G, Turner M S and Yang J 
 {1981b} {\ApJ} {\bf 246} {557}
\item[] Olive K A, Schramm D N, Steigman G and Walker T P {1990} {\PL} 
         B{\bf 236} {454}
\item[] Olive K A, Schramm D N, Thomas D and Walker T P {1991} {\PL} 
         {\bf 65B} {239}
\item[] Olive K A, Rood R T, Schramm D N, Truran J and Vangioni-Flam E 
         {1995} {\ApJ} {\bf 444} {680} 
\item[] Ostriker J, Thompson C and Witten E {1986} {\PL} B{\bf 180} {231}


\item[] Paczy\'{n}ski B {1986} {\ApJ} {\bf 304} {1}
\item[] Padmanabhan T {1993} {\it Structure Formation in the
         Universe} {(Cambridge: Cambridge University Press)}
\item[] Padmanabhan T and Vasanti R {1982} {\PL} {\bf 89A} {327}
\item[] Pagel B E J {1982} {\PRS} A{\bf 307} {19}
\item[]  \dash {1987} {\it An Unified View of the Macro- and Micro-Cosmos} 
          {ed A~De~R\'{u}jula \etal (Singapore: World Scientific) p~399}
\item[]  \dash {1992} {\it Observational and Physical Cosmology} {ed
          F~Sanchez \etal (Cambridge: Cambridge University Press) p~116}
\item[]  \dash {1993} {\PNAS} {\bf 90} {4789}
\item[] Pagel B E J and Kazlaukas A {1992} {\MN} {\bf 256} {49p}
\item[] Pagel B E J, Simonson E A, Terlevich R J and Edmunds M G 
         {1992} {\MN} {\bf 255} {325}
\item[] Pagels H and Primack J R {1982} {\PRL} {\bf 48} {223}
\item[] Pal P B {1983} {\NP} B{\bf 277} {237}
\item[]  \dash {1992} {\IJMP} A{\bf 7} {5387}
\item[] Pal P B and Wolfenstein L {1982} {\PR} D{\bf 25} {766}
\item[] Palla F, Galli D and Silk J {1995} {\ApJ} {\bf 451} {44} 
\item[] Particle Data Group {1990} {\PL} B{\bf 239} {1}
\item[]  \dash {1992} {\PR} D{\bf 45} {S1}
\item[]  \dash {1994} {\PR} D{\bf 50} {1173} 
\item[]  \dash {1996} {\PR} D{\bf 54} {1}
\item[] Pasachoff J M and Vidal-Madjar A {1989} {\CA} {\bf 14} {61}
\item[] Peacock J A, Heavens A F and Davies A T (ed) {1990} {\it Physics of
         the Early Universe} {(Bristol: Adam Hilger)}
\item[] Peccei R D {1989} {\it CP Violation} ed C~Jarlskog {(Singapore:
         World Scientific) p~503}
\item[] Peebles P J E {1966a} {\PRL} {\bf 16} {411}
\item[]  \dash {1966b} {\ApJ} {\bf 146} {542}
\item[]  \dash {1971} {\it Physical Cosmology} {(Princeton: Princeton 
          University Press)}
\item[]  \dash {1980} {\it The Large-Scale Structure of the Universe} 
          {(Princeton: Princeton University Press)}
\item[]  \dash {1987} {\ApJ} {\bf 315} {L73}
\item[]  \dash {1993} {\it Principles of Physical Cosmology} {(Princeton: 
          Princeton University Press)}
\item[] Peimbert M and Torres-Peimbert S {1974} {\ApJ} {\bf 193} {327}
\item[] Penrose R {1979} {\it General Relativity: An Einstein Centenary 
         Survey} ed S~W~Hawking and W~Israel {(Cambridge: Cambridge 
         University Press) p~581}
\item[]  \dash {1989} {\it Proc. 14th Texas Symp. on Relativistic Astrophys.} 
          {ed E~Fenyves (New York: New York Academy of Sciences) p~249}
\item[] Perkins D {1984} {\ARNPS} {\bf 34} {1}
\item[]  \dash {1993} {\NP} B{\bf 399} {3}
\item[] Persic M and Salucci P {1992} {\MN} {\bf 258} {14p}
\item[] Petcov S {1977} {\SJNP} {\bf 25} {344} (erratum {\bf 25} {698})
\item[] Philipsen O, Teper M and Wittig H {1996} {\NP} B{\bf B469} {445}
\item[] Pinsonneault M H, Deliyannis C P and Demarque P 
         {1992} {\ApJS} {\bf 78} {179}
\item[] Piran T and Weinberg S (ed) {1986} {\it Intersection between Particle 
         Physics and Cosmology: Proc. 1st Jerusalem Winter School for
         Theoretical Physics} {(Singapore: World Scientific)}
\item[] Pl\"umacher M {1996} {\preprint} {hep-ph/9604229}
\item[] Polchinski J {1996} {\preprint} {NSF-ITP-96-60 (hep-th/9607050)}
\item[] Prantzos N, Cass\'{e} and Vangioni-Flam E {1993} {\ApJ} {\bf 403} {630}
\item[] Preskill J {1984} {\ARNPS} {\bf 34} {461}
\item[] Primack J, Seckel D and Sadoulet B {1988} {\ARNPS} {\bf 38} {751}
\item[] Protheroe R J, Stanev T and Berezinsky V S {1995} {\PR} D{\bf 51} 
         {4134}


\item[] Quir\'os M {1991} {\PL} B{\bf 267} {27}


\item[] Raffelt G {1990} {\PRep} {\bf 198} {1}
\item[] Rai B and Senjanovi\'c G {1994} {\PR} D{\bf 49} {2729}
\item[] Rana N C {1991} {\ARAA} {\bf 29} {129}
\item[] Rana N C and Mitra B {1991} {\PR} D{\bf 44} {393}
\item[] Randall L and Thomas S {1995} {\NP} B{\bf 449} {229}
\item[] Rangarajan R {1995a} {\NP} B{\bf 454} {357}
\item[]  \dash {1995b} {\NP} B{\bf 454} {369}
\item[] Rauscher T, Applegate J H, Cowan J J, Thielemann F-K and Wiescher M 
         {1994} {\ApJ} {\bf 429} {499}
\item[] Rebolo R, Molaro P and Beckman J E {1988} {\AA} {\bf 192} {192}
\item[] Reeves H {1991} {\PRep} {\bf 201} {335}
\item[] \dash {1994} {\RMP} {\bf 66} {193}
\item[] Reeves H, Audouze J, Fowler W A and Schramm D N {1973} {\ApJ} 
         {\bf 179} {909}
\item[] Reno M H and Seckel D {1988} {\PR} D{\bf 37} {3441}
\item[] Rephaeli Y {1990} {\it The Cosmic Microwave Background:
         25 Years Later} {ed N~Mandolesi and N~Vittorio (Dordrecht: Kluwer
         Academic) p~67}
\item[] Rich J, Lloyd-Owen D, and Spiro M {1987} {\PRep} {\bf 151} {239}
\item[] Riley S P and Irvine J M {1991} {\JPG} {\bf 17} {35}
\item[] Rindler W {1977} {\it Essential Relativity} {(Berlin: Springer-Verlag)
\item[] Rohm R {1984} {\NP} B{\bf 237} {553}
\item[] Rom\~ao J C and Freire F (ed) {1994} {\it Electroweak Physics and the 
         Early Universe} {(London: Plenum Press)}
\item[] Roncadelli M and Senjanovi\'c G {1981} {\PL} {\bf 107B} {59}
\item[] Rood R T, Steigman G and Tinsley B {1976} {\ApJ} {\bf 207} {L57}
\item[] Rood R T, Bania T M and Wilson T L {1992} {\Nature} {\bf 355} {618}
\item[] Ross G G {1984} {\it Grand Unified Theories} {(New York: 
         Benjamin/Cummings)}
\item[] Ross G G and Roberts R G {1992} {\NP} B{\bf 377} {571}
\item[] Ross G G and Sarkar S {1996} {\NP} B{\bf 461} {597}
\item[] Rowan-Robinson M {1985} {\it The Cosmological Distance Ladder} {(New
         York: Freeman)}
\item[]  \dash {1988} {\SSR} {\bf 48} {1}
\item[] Rubakov V A and Shaposhnikov M E {1996} {\preprint}
         {CERN-TH/96-13 (hep-ph/9603208)} [{\SPU}, in press]
\item[] Rugers M and Hogan C J {1996a} {\ApJ} {\bf 459} {L1}
\item[]  \dash {1996b} {\AJ} {\bf 111} {2135}
\item[] Ryan S G, Beers T C, Deliyannis C P and Thorburn J {1996} {\ApJ} 
         {\bf 458} {543}


\item[] Sadoulet B {1992} {\it Research Directions for the Decade:
         Proc. 1990 Summer Study on High-Energy Physics, Snowmass}
         {ed E~L~Berger (Singapore: World Scientific) p~3}
\item[] Sakharov A D {1967} {\JETPL} {\bf 5} {24}
\item[] Salati P, Delbourgo-Salvador P and Audouze J {1987} {\AA} {\bf 173} {1}
\item[] Sanchez N and Zichichi A (ed) {1992} {\it Current Topics in 
         Astrofundamental Physics} {(Singapore: World Scientific)}
\item[] Sarkar S {1985} {\it Superstrings, Supergravity and Unified 
         Theories} {ed G~Furlan \etal (Singapore: World-Scientific) p~465}
\item[]  \dash {1986} {\it Proc. XXVI Intern. Symp. on Multiparticle Dynamics}
         {ed M~Markatyan \etal (Singapore: World Scientific) p~863} 
\item[]  \dash {1988} {\ApLC} {\bf 27} {293}
\item[]  \dash {1991} {\it Observational Tests of Cosmological 
          Inflation} {ed T~Shanks \etal (Dordrecht: Kluwer Academic) p~91}
\item[]  \dash {1996} {\it Proc. Intern. EPS Conf. on High Energy Physics, 
          Brussels} {ed J~Lemonne \etal (Singapore: World Scientific) p~95} 
\item[]  \dash {1997} {\it The Big Bang: A Laboratory for Particle 
          Physics} {(Cambridge: Cambridge University Press) to appear}
\item[] Sarkar S and Cooper A M {1984} {\PL} {\bf 148B} {347}
\item[] Sasselov D and Goldwirth D S {1995} {\ApJ} {\bf 444} {L5}
\item[] Sato K and Audouze J (ed) {1991} {\it Primordial Nucleosynthesis and 
         Evolution of Early Universe: Proc. Intern. Conf., Tokyo} 
         {((Dordrecht: Kluwer Academic)}
\item[] Sato K and Kobayashi M {1977} {\PTP} {\bf 58} {1775} 
\item[] Satz H {1985} {\ARNPS} {\bf 35} {245}
\item[] Savage M J, Malaney R A and Fuller G M {1991} {\ApJ} {\bf 368} {1}
\item[] Scalo J M {1986} {\FCP} {\bf 11} {1}
\item[] Schechter J and Valle J W F {1982} {\PR} D{\bf 25} {774} 
\item[] Scherrer R J {1983} {\MN} {\bf 205} {683}
\item[]  \dash {1984} {\MN} {\bf 210} {359}
\item[] Scherrer R J and Turner M S {1985} {\PR} D{\bf 31} {681} 
\item[]  \dash {1986} {\PR} {\bf D33} {1585} (erratum D{\bf 34} {3263})
\item[]  \dash {1988a} {\ApJ} {\bf 331} {19}
\item[]  \dash {1988b} {\ApJ} {\bf 331} {33} 
\item[] Scherrer R J, Cline J, Raby S and Seckel D {1991} {\PR} D{\bf 44} 
         {3760}
\item[] Schramm D N and Wagoner R V {1977} {\ARNS} {\bf 27} {37} 
\item[] Schreckenbach K and Mampe W {1992} {\JPG} {\bf 18} {1}
\item[] Schrock R E {1981} {\PR} D{\bf 24} {1232,1275} 
\item[] Sciama D W {1982} {\PL} {\bf 118B} {327}
\item[]  \dash {1993} {\it Modern Cosmology and the Dark Matter Problem} 
          {(Cambridge: Cambridge University Press)}
\item[] Scully S T, Cass\'e M, Olive K A, Schramm D N, Truran J and
         Vangioni-Flam E {1996} {\ApJ} {\bf 462} {960}
\item[] Seckel D {1993} {\preprint} {BA-93-16 (hep-ph/9305311)} 
\item[] Sengupta S and Pal P B {1996} {\PL} B{\bf 365} {175}
\item[] Senjanovi\'c G {1986} {\it Inner Space--Outer Space} {eds 
         E~W~Kolb \etal (Chicago: University of Chicago Press) p~554}
\item[] Setti G and Van Hove L (ed) {1984} {\it First ESO-CERN Symp.: 
         Large-scale Structure of the Universe, Cosmology and Fundamental 
         Physics} {(Geneva: CERN)}
\item[] Shapiro P and Wasserman I {1981} {\Nature} {\bf 289} {657}
\item[] Shapiro P, Teukolsky S A and Wasserman I {1980} {\PRL} {\bf 45} {669}
\item[] Shaposhnikov M E {1991} {\PS} T{\bf 36} {183}
\item[]  \dash {1992} {\it Electroweak Interactions and Unified Theories} 
          {ed J~Tran~Thanh~Van (Gif Sur Yvette: Editions Fronti\'eres) p~201}
\item[] Shaver P A, Kunth D and Kj\"{a}r K (ed) {1983} {\it Primordial
         Helium} {(Garching: European Southern Observatory)}
\item[] Sher M {1989} {\PRep} {\bf 179} {273}
\item[] Shi X, Schramm D N and Fields B {1993} {\PR} D{\bf 48} {2563}
\item[] Shields G {1987} {\it 13th Texas Symp. on Relativistic 
         Astrophysics} ed P~Ulmer {(Singapore: World Scientific Press) p~192}
\item[] Shu F H {1981} {\it The Physical Universe: An Introduction to 
         Astronomy} {(Mill Valley, CA: University Science Books)}
\item[] Shuryak E {1980} {\PRep} {\bf 61} {2}
\item[] Shvartsman V F {1969} {\JETPL} {\bf 9} {184}
\item[] Sigl G, Jedamzik K, Schramm D N and Berezinsky V S 
         {1995} {\PR} D{\bf 52} {6682}
\item[] Simpson J J {1985} {\PRL} {\bf 54} {1891}
\item[] Skillman E and Kennicutt R C {1993} {\ApJ} {\bf 411} {655}
\item[] Skillman E, Terlevich R J, Kennicutt R C, Garnett D R and Terlevich E 
         {1993} {\ANYAS} {\bf 688} {739}
\item[] Smilga A V {1995} {\preprint} {SUNY-NTG-95-304 (hep-ph/9508305)}
\item[] Smith M S, Kawano L H and Malaney R A {1993} {\ApJS} {\bf 85} {219}
\item[] Smith P F {1988} {\it Contemp. Phys.} {\bf 29} {159}
\item[] Smith P F and Lewin J D {1990} {\PRep} {\bf 187} {203}
\item[] Smith P F \etal {1982} {\NP} B{\bf 206} {333} 
\item[] Smith V V, Lambert D L and Nissen P E {1992} {\ApJ} {\bf 408} {262}
\item[] Smits D P {1991} {\MN} {\bf 248} {20}
\item[]  \dash {1996} {\MN} {\bf 278} {683}
\item[] Smoot G F \etal (COBE collab.) {1992} {\ApJ} {\bf 396} {L1}
\item[] Songaila A, Cowie L L, Hogan C J and Rugers M {1994} {\Nature}
         {\bf 368} {599}
\item[] Spite F and Spite M {1982} {\AA} {\bf 115} {357}
\item[] Spite M, Spite F, Peterson R C and Chafee F C {1987} {\AA} {\bf 172} 
         {L9}
\item[] Srednicki M (ed) {1990} {\it Particle Physics and Cosmology:
         Dark Matter} {(Amsterdam: North-Holland)}
\item[] Srednicki M, Watkins W and Olive K A {1988} {\NP} B{\bf 310} {693}
\item[] Starkman G D {1992} {\PR} D{\bf 45} {476}
\item[] Starobinsky, A A {1993} {\JETPL} {\bf 57} {622}
\item[] Steigman G {1976} {\ARAA} {\bf 14} {33}
\item[]  \dash {1979} {\ARNPS} {\bf 29} {313}
\item[] Steigman G and Tosi M {1992} {\ApJ} {\bf 401} {150}
\item[]  \dash {1995} {\ApJ} {\bf 453} {173}
\item[] Steigman G, Schramm D N and Gunn J {1977} {\PL} {\bf 66B} {202}
\item[] Steigman G, Olive K A and  Schramm D N {1979} {\PRL} {\bf 43} {239}
\item[] Steigman G, Olive K A, Schramm D N and Turner M S 
         {1986} {\PL} B{\bf 176} {33}  
\item[] Steigman G, Fields B D, Olive K A, Schramm D N and Walker T P 
         {1993} {\ApJ} {\bf 415} {L35}
\item[] Steinhardt P J {1995} {\IJMP} A{\bf 10} {1091}
\item[] Stevens D, Scott D and Silk J {1993} {\PRL} {\bf 71} {20}
\item[] Sugiyama N and Silk J {1994} {\PRL} {\bf 73} {509}
\item[] Szalay A S and Marx G {1976} {\AA} {\bf 49} {437}


\item[] Terasawa N and Sato K {1987} {\PL} B{\bf 185} {412}
\item[]  \dash {1988} {\PTP} {\bf 80} {468}
\item[]  \dash {1991} {\PS} T{\bf 36} {60}
\item[] Terasawa N, Kawasaki M and Sato K {1988} {\NP} B{\bf 302} {697}
\item[] Thomas D, Schramm D N, Olive K A and Fields B D 
         {1993} {\ApJ} {\bf 406} {569}
\item[] Thomas S {1995} {\PL} B{\bf 351} {424}
\item[] Thorburn J A {1994} {\ApJ} {\bf 421} {318}
\item[] Thuan T X, Izotov Y I and Lipovetsky V A {1996} {\it Interplay
         between Massive Star Formation, the ISM and Galaxy Evolution} 
         {ed D.~Kunth \etal (Gif-Sur-Yvette: Edition Fronti\'eres)}
\item[] Tinsley B M {1980} {\FCP} {\bf 5} {287}
\item[] Toussaint D {1992} {\NPBPS} {\bf 26} {3}
\item[] Trimble V {1989} {\ARAA} {\bf 25} {425}
\item[] Turner M S {1990} {\PRep} {\bf 197} {67}
\item[]  \dash {1991} {\PS} T{\bf 36} {167}
\item[]  \dash {1993} {\PNAS} {\bf 90} {4827}
\item[] Tytler D and Burles S {1996} {\preprint} {astro-ph/9606110}
\item[] Tytler D, Fan X and Burles S {1996} {\Nature} {\bf 381} {207}


\item[] Unruh W G and Semenoff G W (ed) {1988} {\it The Early 
         Universe} {(Dordrecht: Reidel)}


\item[] Valle J W F {1991} {\PPNP} {\bf 26} {91}
\item[] Van den Bergh S {1992} {\PASP} {\bf 104} {861}
\item[]  \dash {1994} {\PASP} {\bf 106} {1113}
\item[] Van Nieuwenhuizen P {1981} {\PRep} {\bf 68} {189}
\item[] Vangioni-Flam E and Casse\'e M {1995} {\ApJ} {\bf 441} {471}
\item[] Vangioni-Flam E, Olive K A and Prantzos N {1994} {\ApJ} {\bf 427} {618}
\item[] Vauclair S {1988} {\ApJ} {\bf 335} {971}
\item[] Vauclair S and Charbonnel C {1995} {\AA} {\bf 295} {715}
\item[] Verkerk P \etal {1992} {\PRL} {\bf 68} {1117}
\item[] Vidal-Madjar A {1986} {\it Space Astronomy and Solar System 
  Exploration} {[ESA SP-268] (Noordwijk: ESA) p~73}
\item[] Vilenkin A {1985} {\PRep} {\bf 121} {263}
\item[] Vilenkin A and Shellard E P S {1994} {\it Cosmic Strings and Other
         Topological Defects} {(Cambridge: Cambridge University Press)}
\item[] Vysotski\v{\ii} M I, Dolgov A D and Zel'dovich Ya B {1977} {\JETPL} 
         {\bf 26} {188}


\item[] Wagoner R V {1969} {\ApJS} {\bf 18} {247}
\item[]  \dash {1973} {\ApJ} {\bf 179} {343}  
\item[]  \dash {1980} {\it Les Houches Session XXXII: Physical 
          Cosmology} {ed R~Balian \etal (Amsterdam: North-Holland) p~395} 
\item[]  \dash {1990} {\it Modern Cosmology in Retrospect} {ed R~Bertotti 
          \etal (Cambridge: Cambridge University Press) p~159}
\item[] Wagoner R V, Fowler W A and Hoyle F {1967} {\ApJ} {\bf 148} {3}
\item[] Walker T P, Steigman G, Schramm D N, Olive K A and Kang H-S {1991} 
         {\ApJ} {\bf 376} {51}
\item[] Wampler E J \etal {1995} {\preprint} {astro-ph/9512084} 
         [{\AA}, in press]
\item[] Wasserburg G J, Boothroyd A I and Sackmann I {1995} {\ApJ} 
  {\bf 447} {L37}
\item[] Webb J K, Carswell R F, Irwin M J and Penston M V {1991} {\MN} 
         {\bf 250} {657}
\item[] Weinberg S {1972} {\it Gravitation and Cosmology} {(New York: Wiley)}
\item[]  \dash {1977} {\it The First Three Minutes} {(New York: Basic Books)}
\item[]  \dash {1980} {\PS} {\bf 21} {773}
\item[]  \dash {1982} {\PRL} {\bf 48} {1303}
\item[]  \dash {1989} {\RMP} {\bf 61} {1}
\item[] Wess J and Bagger J {1993} {\it Supersymmetry and Supergravity} 
         {(Princeton: Princeton University Press)}
\item[] White M, Viana P T P, Liddle A R and Scott D {1996} {\preprint} 
         {SUSSEX-AST 96/5-2 (astro-ph/9605057)} [submitted to {\MN}]
\item[] White S D M {1994} {\preprint} {astro-ph/9410043}
         [to appear in {\it Les Houches Session 60: Cosmology and Large 
          Scale Structure} 
\item[] White D A and Fabian A {1995} {\MN} {\bf 273} {72}
\item[] White S D M, Navarro J F, Evrad A E and Frenk C S {1993} {\Nature} 
         {\bf 366} {429}
\item[] Wilczek F {1991} {\PS} T{\bf 36} {281}
\item[] Wilkinson D H {1982} {\NP} A{\bf 377} {474}
\item[] Wilson T L and Rood R T {1994} {\ARAA} {\bf 32} {191}
\item[] Witten E {1981a} {\CMP} {\bf 80} {381}
\item[]  \dash {1981b} {\NP} B{\bf 188} {513}
\item[]  \dash {1984} {\PR} D{\bf 30} {272}
\item[]  \dash {1985a} {\NP} B{\bf 249} {557}
\item[]  \dash {1985b} {\NP} B{\bf 258} {75}
\item[] Wolfram S {1979} {\PL} {\bf 82B} {65}
\item[] Wright E L \etal {1994} {\ApJ} {\bf 420} {450}



\item[] Yaffe L G {1995} {\preprint} {hep-ph/9512265}
\item[] Yahil A and Beaudet G {1976} {\ApJ} {\bf 206} {26}
\item[] Yamazaki T \etal {1984} {\it Proc. Neutrino '84} {eds K~Kleinknecht and
         E~A~Paschos (Singapore: World Scientific Press) p~183}
\item[] Yang J, Schramm D N, Steigman G and Rood R T {1979} {\ApJ} {\bf 227} 
         {697} 
\item[] Yang J, Turner M S, Steigman G, Schramm D N and Olive K A {1984} 
         {\ApJ} {\bf 281} {493}
\item[] Yoshimura M, Totsuka Y, Nakamura K and Lim C S (ed) {1988} {\it Proc. 
         2nd Workshop on Elementary Particle Picture of the Universe} 
         {(Tsukuba: KEK)}


\item[] Zdziarski A {1988} {\ApJ} {\bf 335} {786}
\item[] Zdziarski A and Svensson R {1989} {\ApJ} {\bf 344} {551}
\item[] Zee A (ed) {1982} {\it Unity of Forces in the Universe} {(Singapore:
         World Scientific)}
\item[] Zel'dovich Ya B {1965} {\AAA} {\bf 3} {241}
\item[] Zel'dovich Ya B, Kobzarev I Yu and Okun L B 
         {1975} {\JETP} {\bf 40} {1}
\item[]  Zwirner F {1996} {\it Proc. Intern. EPS Conf. on High Energy Physics, 
          Brussels} {ed J~Lemonne \etal (Singapore: World Scientific) p~743} 
}
\endrefs

\Tables

\fulltable{Thermodynamic history of the RD era\label{tabthermhist}}
\br
$T$ &Threshold (GeV) &Particle Content &$g_{\R}(T)$
 &$\case{N_\gamma(T_0)}{N_\gamma(T)}$\\ 
\mr
$< m_{e}$ &$0.511 \times 10^{-3}$ &$\gamma$ (+ 3 decoupled $\nu$'s) &2 &1 \\
$m_{e} - T_{\D}(\nu)$ &\# &add $e^{\pm}$ &11/2 &2.75 \\
$T_{\D}(\nu) - m_{\mu}$ &0.106 &$\nu$'s become interacting&43/4 &2.75 \\
$m_{\mu} - m_{\pi}$ &0.135 &add $\mu^{\pm} $ &57/4 &3.65 \\
$m_{\pi} - T_{\c}^{\qh}$ &\$ &add $\pi^{\pm}$, $\pi^{0}$ & 69/4 &4.41 \\
$T_{\c}^{\qh} - m_{s}$ &0.194 &$\gamma$, $3\nu$'s, $e^{\pm}$, $\mu^{\pm}$ 
 &205/4 &13.1 \\
& &\ $u$, $\bar{u}$, $d$, $\bar{d}$, 8 $g$'s & & \\ 
$m_{s} - m_{c}$ &$1.27 \pm 0.05$ &add $s$, $\bar{s}$ &247/4 &15.8 \\
$m_{c} - m_{\tau}$ &1.78 &add $c$, $\bar{c}$ &289/4 &18.5 \\
$m_{\tau} - m_{b}$ &$4.25 \pm 0.10$ &add $\tau^{\pm}$ &303/4 &19.4 \\
$m_{b} - m_{W}$ &$80.3 \pm 0.3$ &add $b$, $\bar{b}$ &345/4 &22.1 \\
$m_{W} - m_{t}$ &$180\pm12$ &add $W^{\pm}$, $Z^{0}$ &381/4 &24.4 \\
$m_{t} - m_{H^0}$ &\dag &add $t$, $\bar{t}$ &423/4 &27.1 \\
$m_{H^0} - T_{\c}^{\EW}$ &\ddag &add $H^{0}$ &427/4 &27.3 \\ 
\br
\tabnotes

\noindent \#\ Neutrinos decouple from the thermal plasma at 
 $T_{\D}(\nu)\approx2.3-3.5\MeV$.\\ 
\noindent \$\ $T_{\c}^{\qh}\approx150-400\MeV$ characterizes the quark-hadron 
 phase transition (assumed to be adiabatic).\\
\noindent \dag\ We have assumed that the Higgs boson is heavier than
 both the $W^{\pm},Z^{0}$ bosons and the $t$ quark.\\
\noindent \ddag\ Note that $g_{\R}$ does {\em not} change when the
 $SU(2)_{\L}\,\otimes\,U(1)_{Y}$ symmetry is restored at
 $T_{\c}^{\EW}\sim300$ GeV since the total number of degrees of
 freedom in the gauge plus Higgs fields is invariant.

\endtabnotes

\Figures

\epsfxsize\hsize\epsffile{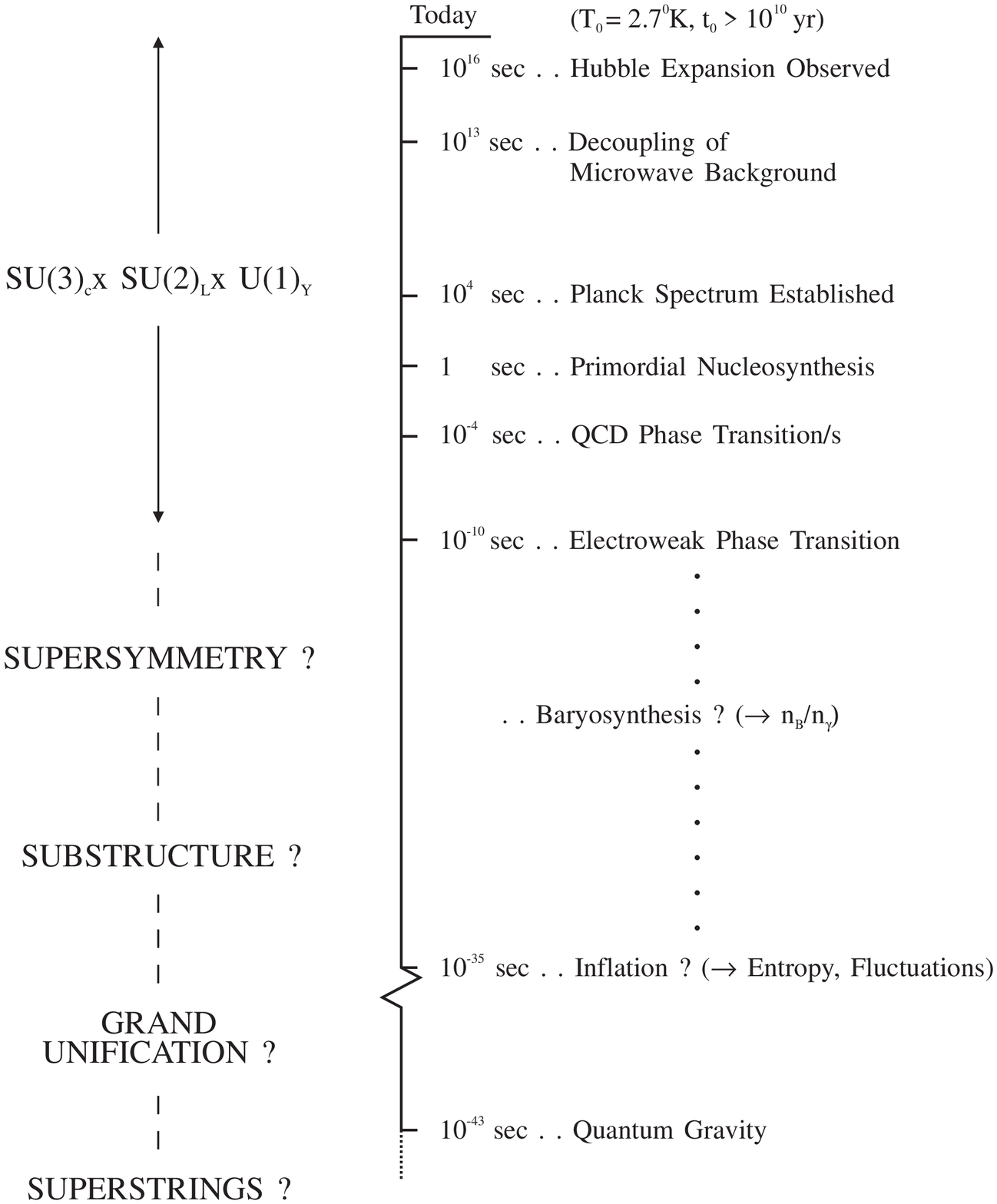}
\Figure{The cosmological history of the universe\label{cosmohist}}

\epsfxsize\hsize\epsffile{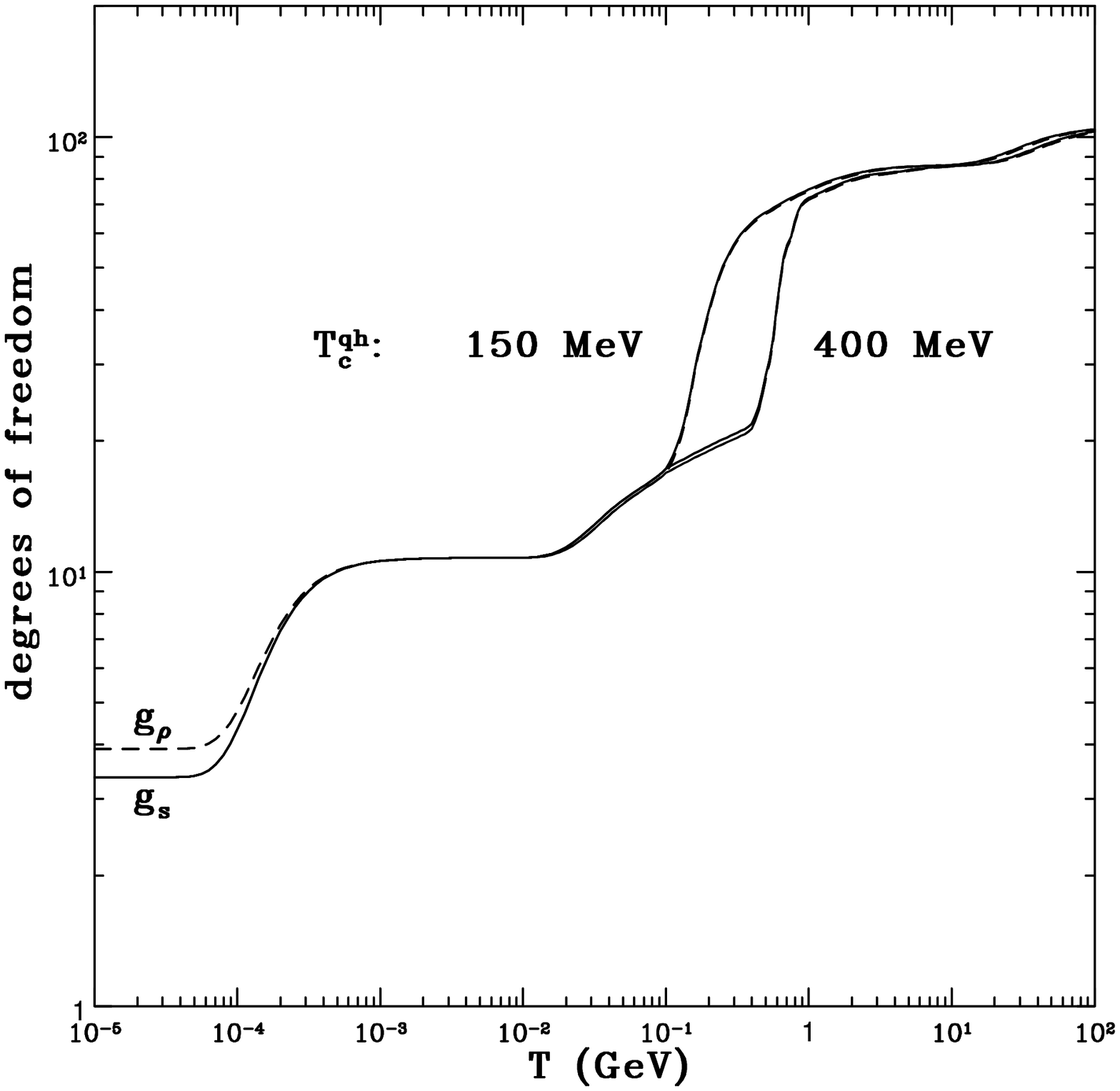}
\Figure{The number of relativistic degrees of freedom characterizing
 the entropy density $g_{s}$ (dashed line) and the energy density
 $g_{\rho}$ (solid line), as a function of temperature in the Standard
 $SU(3)_{\c}{\otimes}SU(2)_{\L}{\otimes}U(1)_{Y}$
 Model.\label{degfreedom}}

\epsfxsize\hsize\epsffile{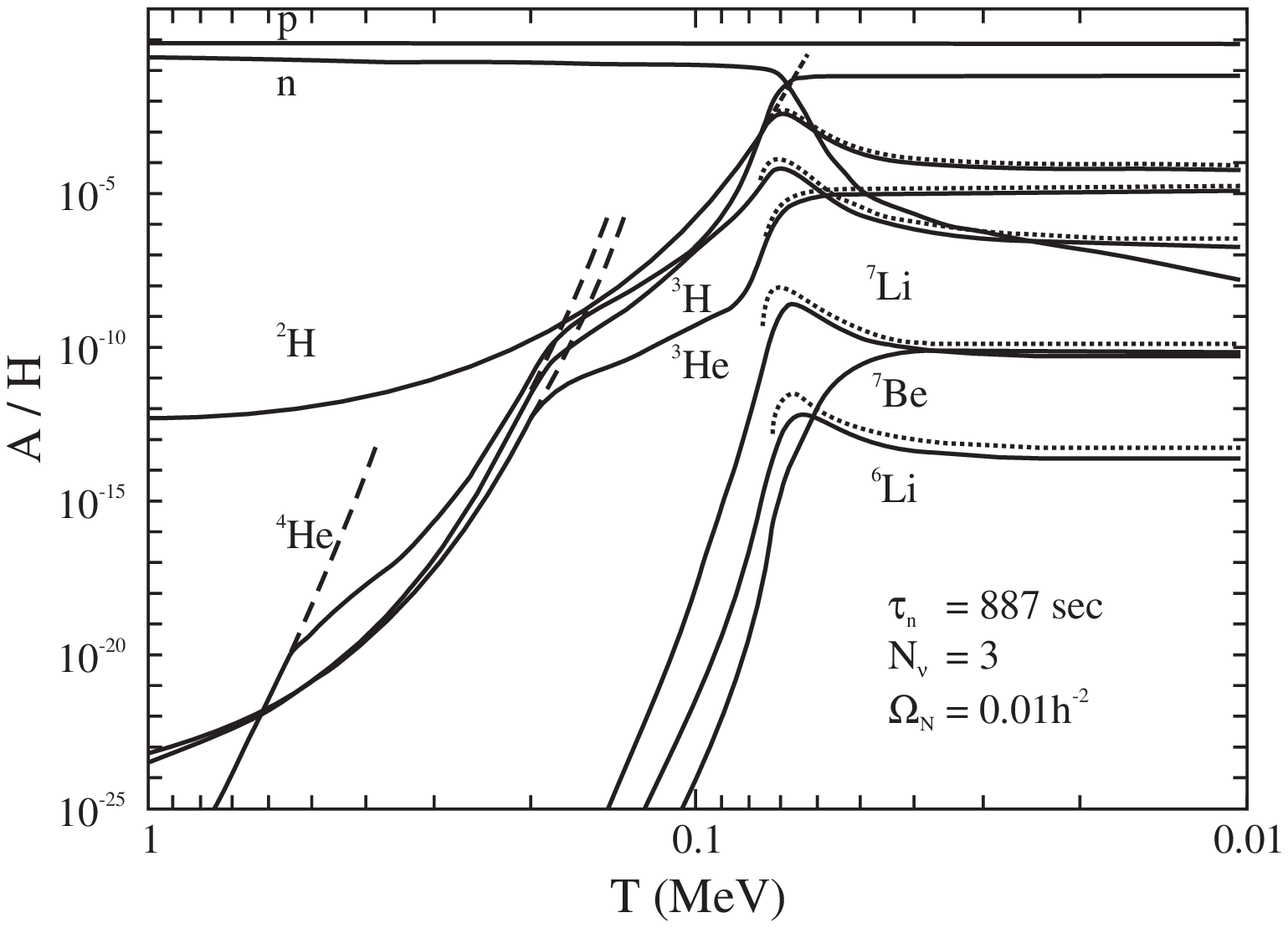}
\Figure{Evolution of the abundances of primordially synthesized light
 elements with temperature according to the Wagoner (1973) numerical
 code as upgraded by Kawano (1992). The dashed lines show the values
 in nuclear statistical equilibrium while the dotted lines are the
 `freeze-out' values as calculated analytically by Esmailzadeh \etal
 (1991).\label{abunevol}}

\epsfxsize\hsize\epsffile{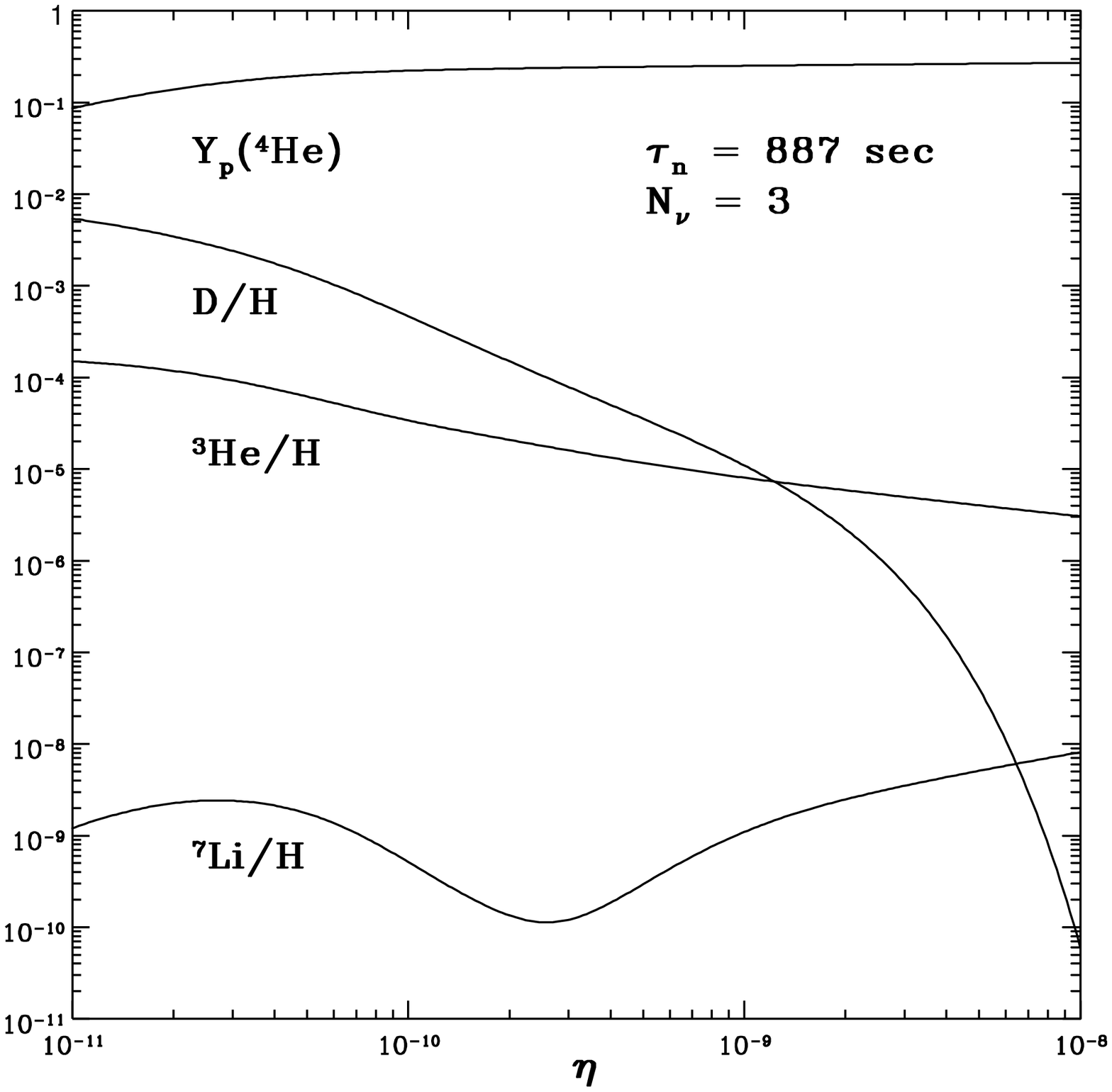}
\Figure{Dependence of primordially synthesized light element
 abundances on the nucleon-to-photon ratio $\eta$, calculated using
 the upgraded Wagoner code.\label{abuneta}}

\epsfxsize\hsize\epsffile{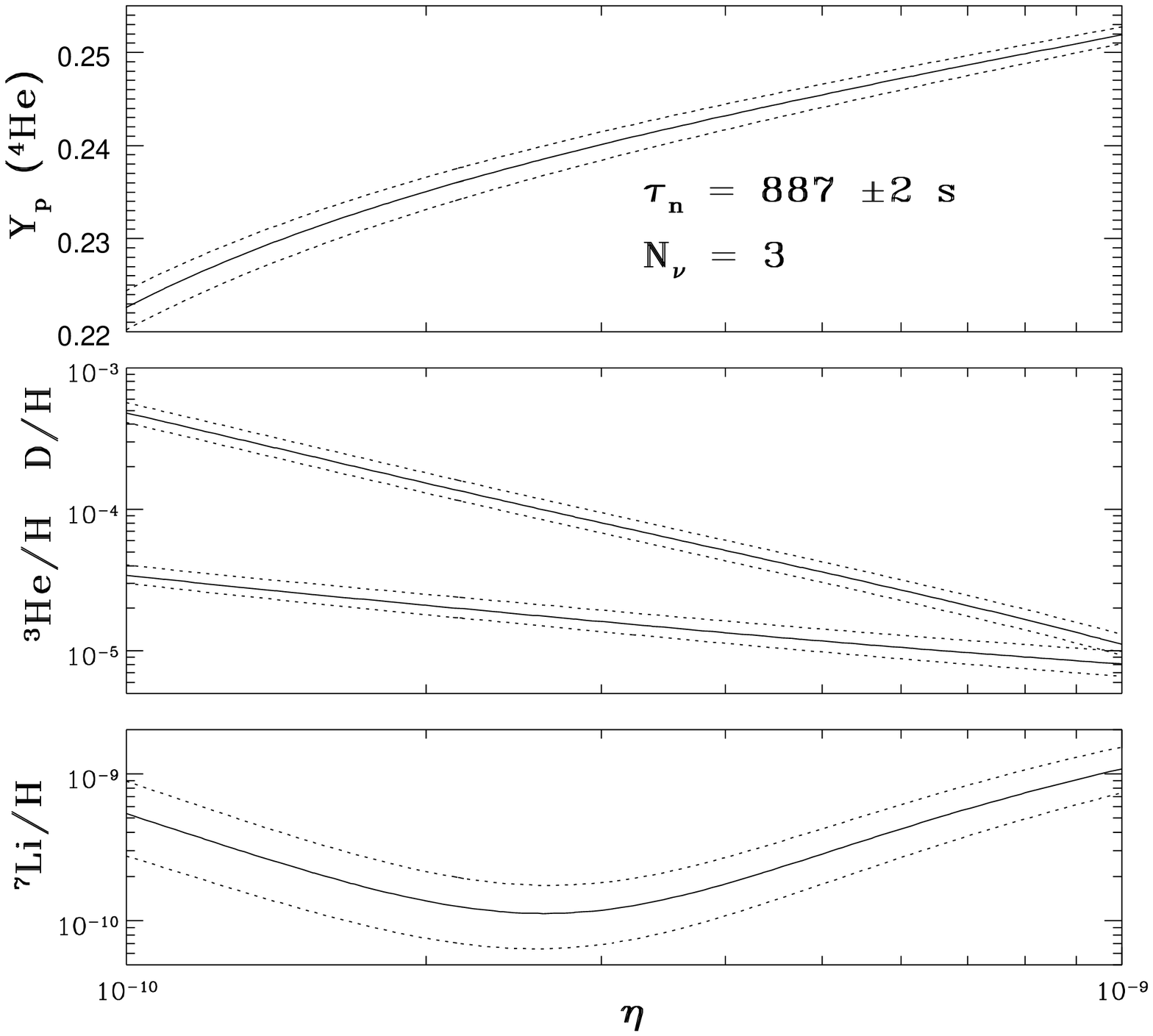}
\Figure{Monte Carlo results (Krauss and Kernan 1995) for the $\95cl$
 limits (dashed lines) on primordially synthesized elemental
 abundances, along with their central values (full lines). Note that
 the $\Hefour$ mass fraction $Y_{\pr}$ is shown on a linear
 scale.\label{abunetamc}}

\epsfxsize7cm\epsffile{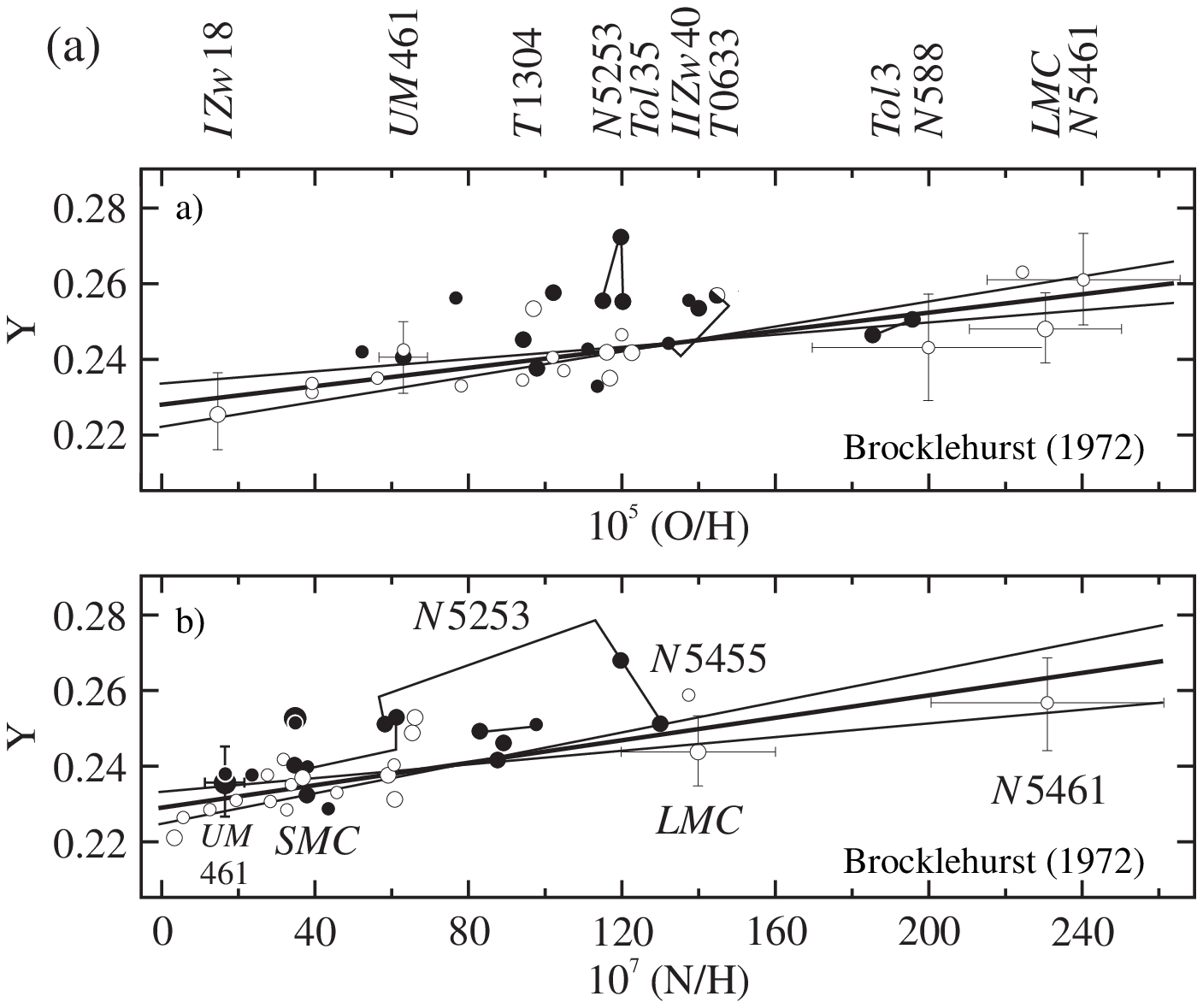}
\epsfxsize7cm\epsffile{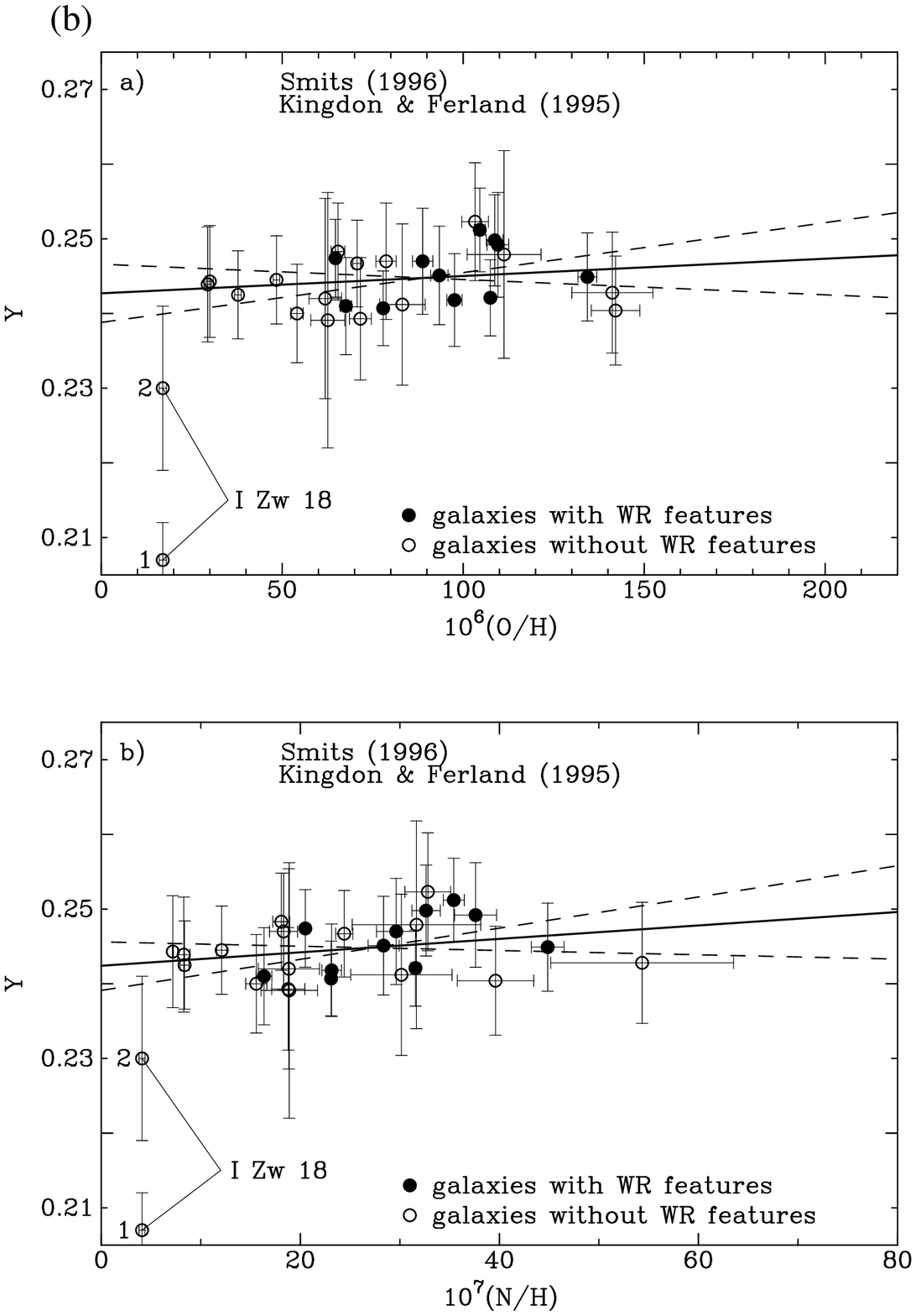} 
\Figure{Regressions of the helium mass fraction against the oxygen and
 nitrogen abundances in extragalactic low-metallicity HII regions,
 with (filled circles) and without (open circles) broad Wolf-Rayet
 features. Panel (a) shows abundances for 33 objects obtained using
 the Brocklehurst (1972) emissivities by Pagel \etal (1992), with the
 maximum-likelihood linear fits (with $\pm1\sigma$ limits) for the
 latter category. Panel (b) shows abundances for 27 objects obtained
 using the Smits (1996) emissivities by Izotov \etal (1996) along with
 the maximum-likelihood linear fits (with $\pm1\sigma$
 limits).\label{Ypreg}}

\epsfxsize\hsize\epsffile{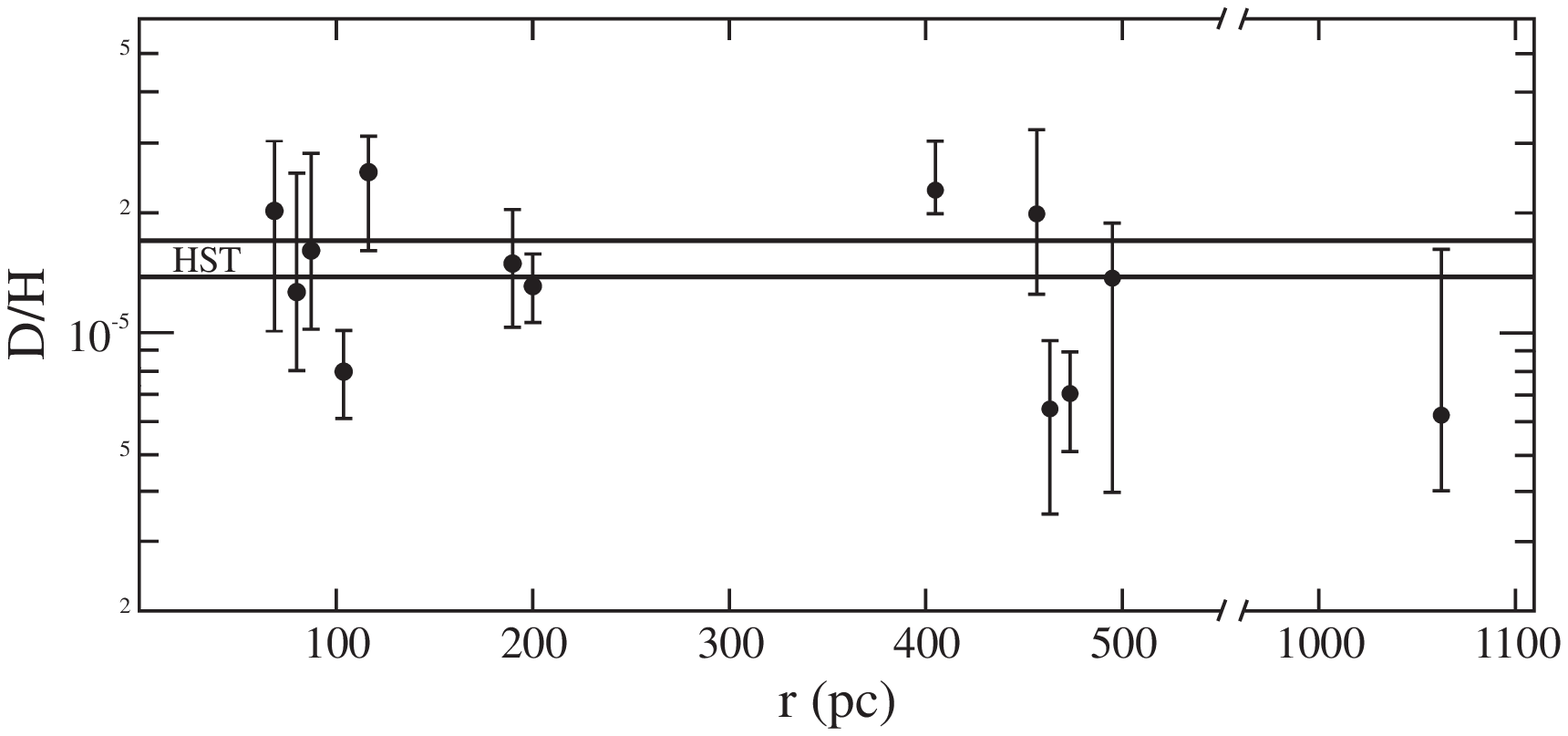}
\Figure{The interstellar deuterium abundance as observed by {\sl
 Copernicus} and {\sl IUE} towards distant hot stars. The band shows
 the value measured towards the nearby star {\sl Capella} (at 12.5
 kpc) by the {\sl Hubble Space Telescope} (Linsky \etal
 1995).\label{Dmeas}}

\epsfxsize18cm\epsffile{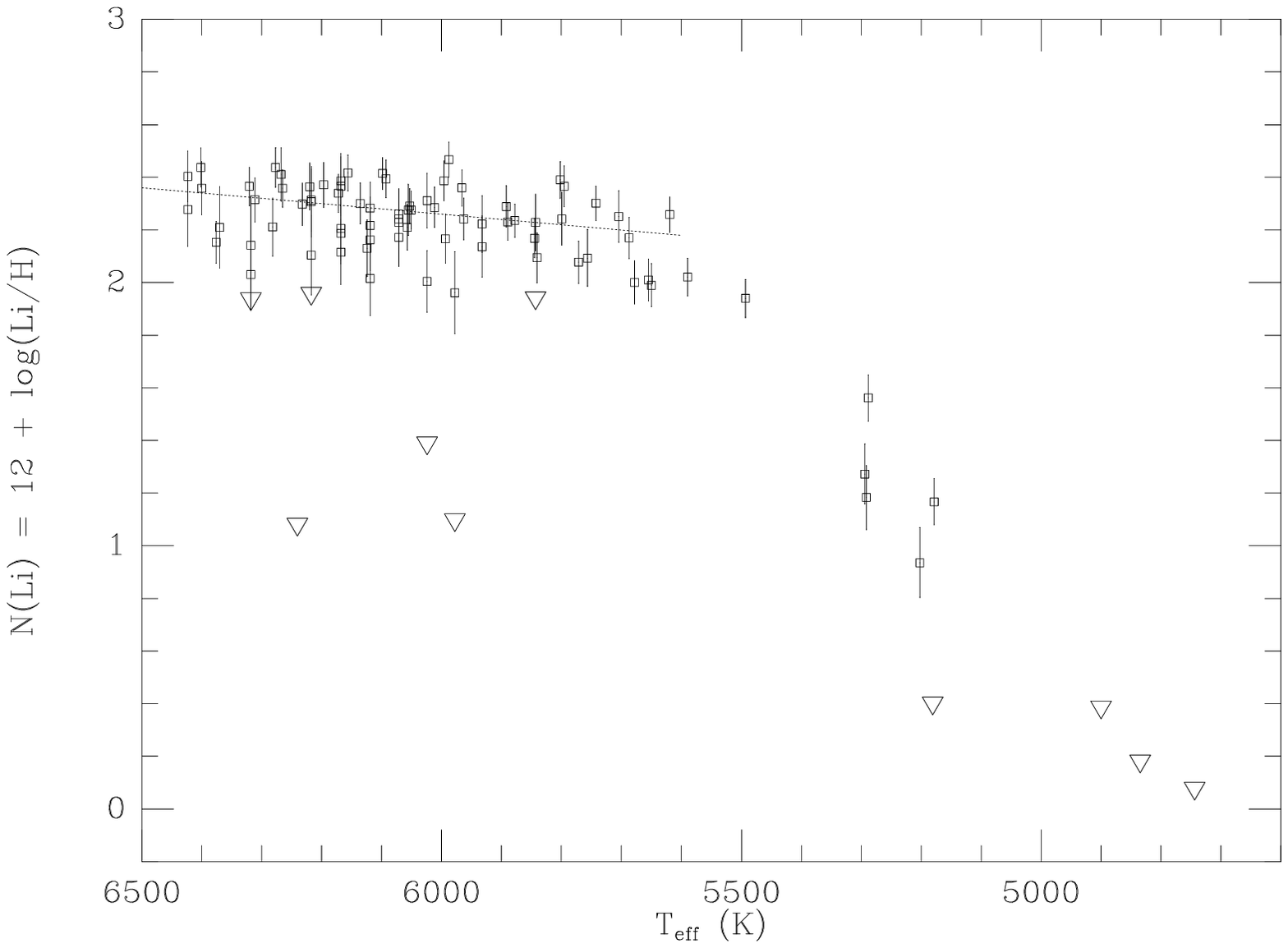}
\Figure{The Lithium abundance in 90 halo dwarf and main-sequence
 turnoff stars versus their effective surface temperature. Error bars
 indicate the $1\sigma$ interval for detections while triangles denote
 $3\sigma$ upper limits for non-detections; the dotted line is a fit
 which minimizes the absolute deviation of the detections (from
 Thorburn 1994).\label{Li7meas}}

\epsfxsize\hsize\epsffile{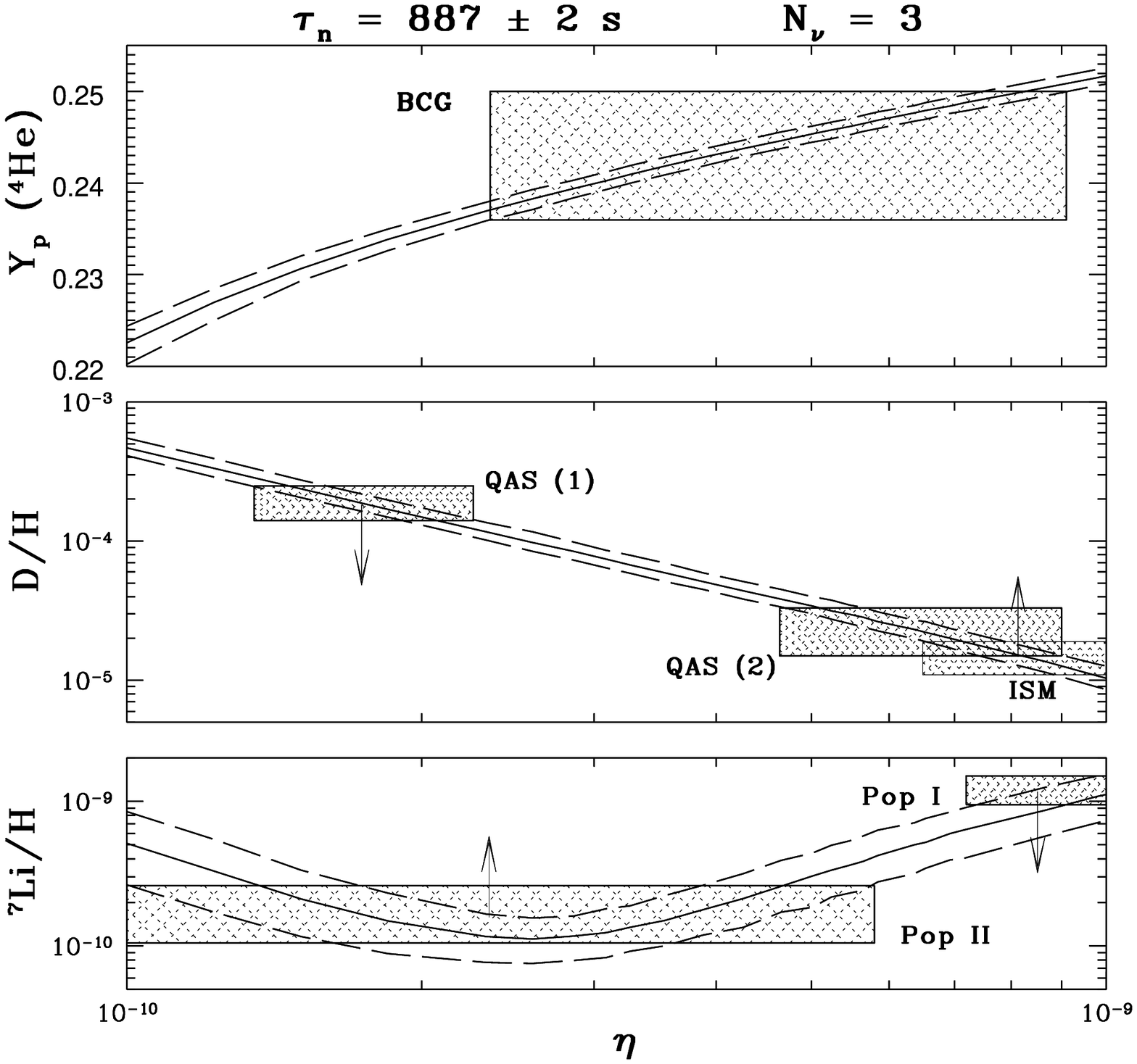}
\Figure{Concordance of the predicted abundances with present
 observational bounds (from Kernan and Sarkar 1996). Note that only
 the $\Hefour$ abundance inferred from BCG is established to be
 primordial. The two conflicting measurements of the $\Htwo$ abundance
 in QAS are shown along with its ISM value, and both the Pop\,I and
 Pop\,II $\Liseven$ abundances are shown.\label{concord}}

\epsfxsize13cm\epsffile{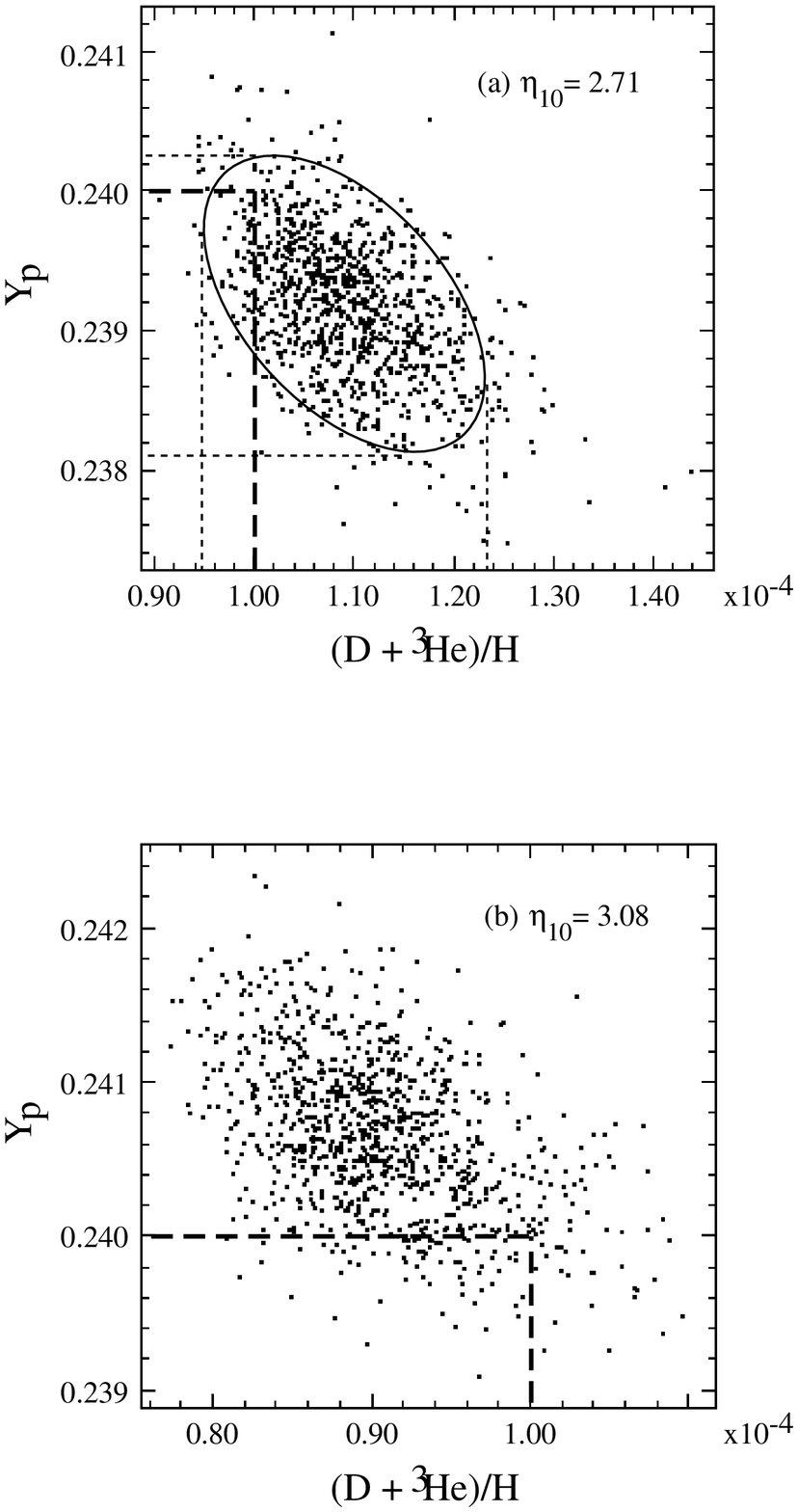}
\Figure{Monte Carlo predictions for the $\Hefour$ versus the
 ${\Htwo}+{\Hethree}$ abundances (taking $N_{\nu}=3$ and
 $\tau_{\n}=889\pm2.1$ sec) for (a) $\eta=2.71\times10^{-10}$ and (b)
 $\eta=3.08\times10^{-10}$. The dashed lines indicate the adopted
 ``reasonable'' observational upper bounds. In panel (a), a gaussian
 contour with $\pm2\sigma$ limits (dotted lines) on each variable is
 also shown (from Kernan and Krauss 1994).\label{He4vsHe3Dmc}}

\epsfxsize12cm\epsffile{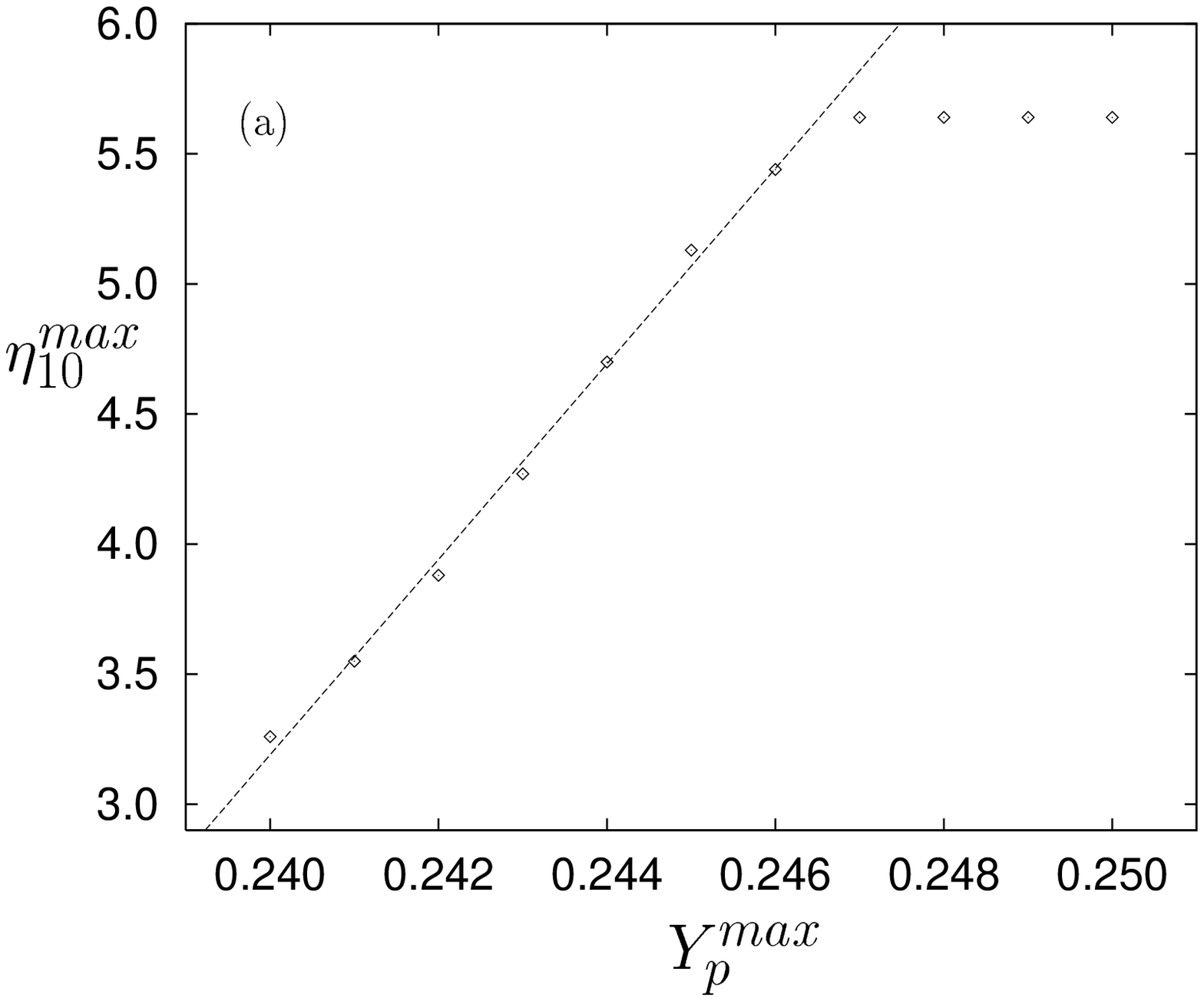}
\epsfxsize12cm\epsffile{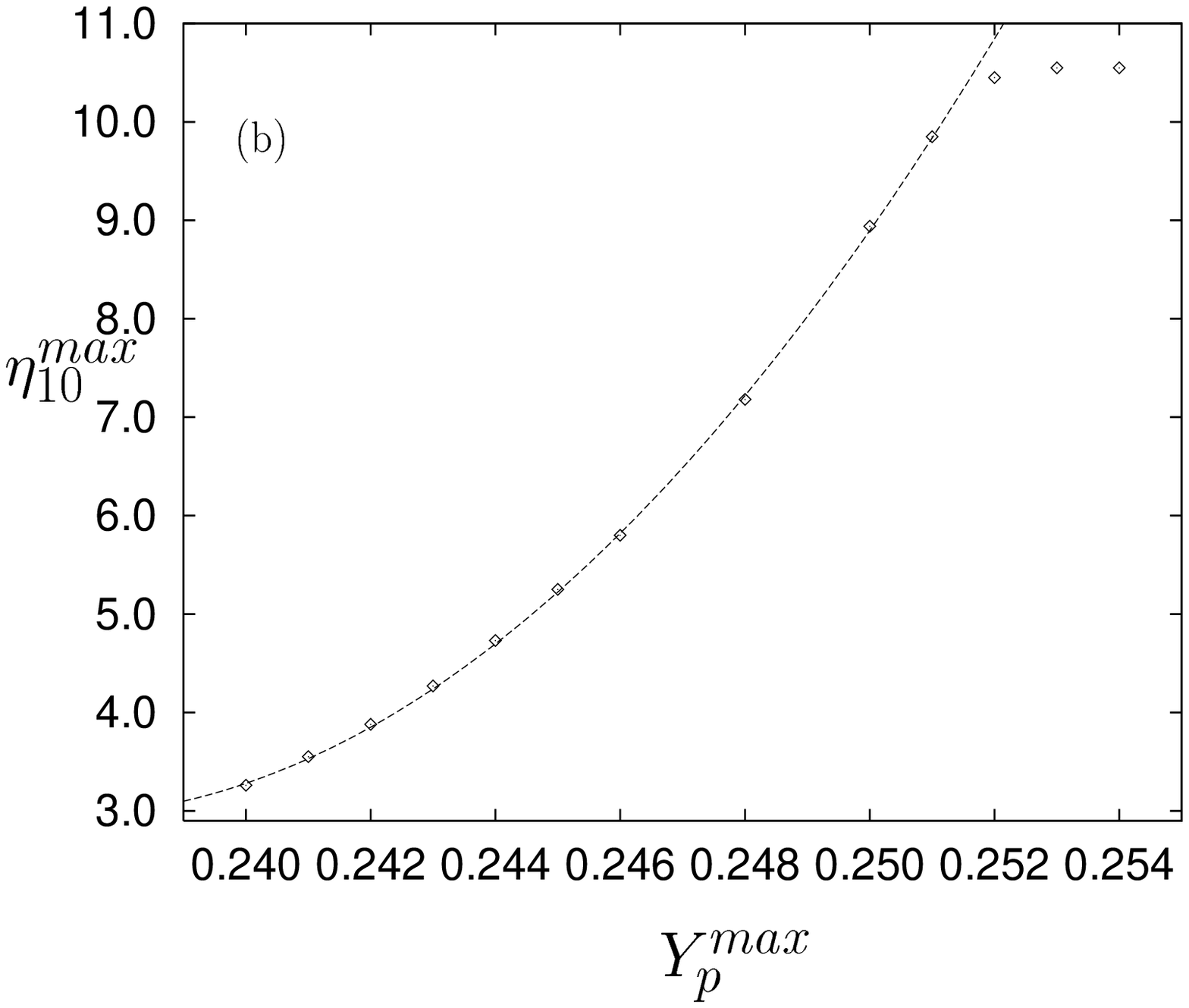}
\Figure{The upper limit to the nucleon-to-photon ratio (in units of
 $10^{-10}$) implied by the ISM bound ${\Htwo}/{\H}>1.1\times10^{-5}$
 combined with (a) the Pop\,II bound
 ${\Liseven}/{\H}<2.6\times10^{-10}$, and (b) the Pop\,I bound
 ${\Liseven}/{\H}<1.5\times10^{-9}$, as a function of the maximum
 ${\Hefour}$ mass fraction.\label{etalim}}

\epsfxsize10cm\epsffile{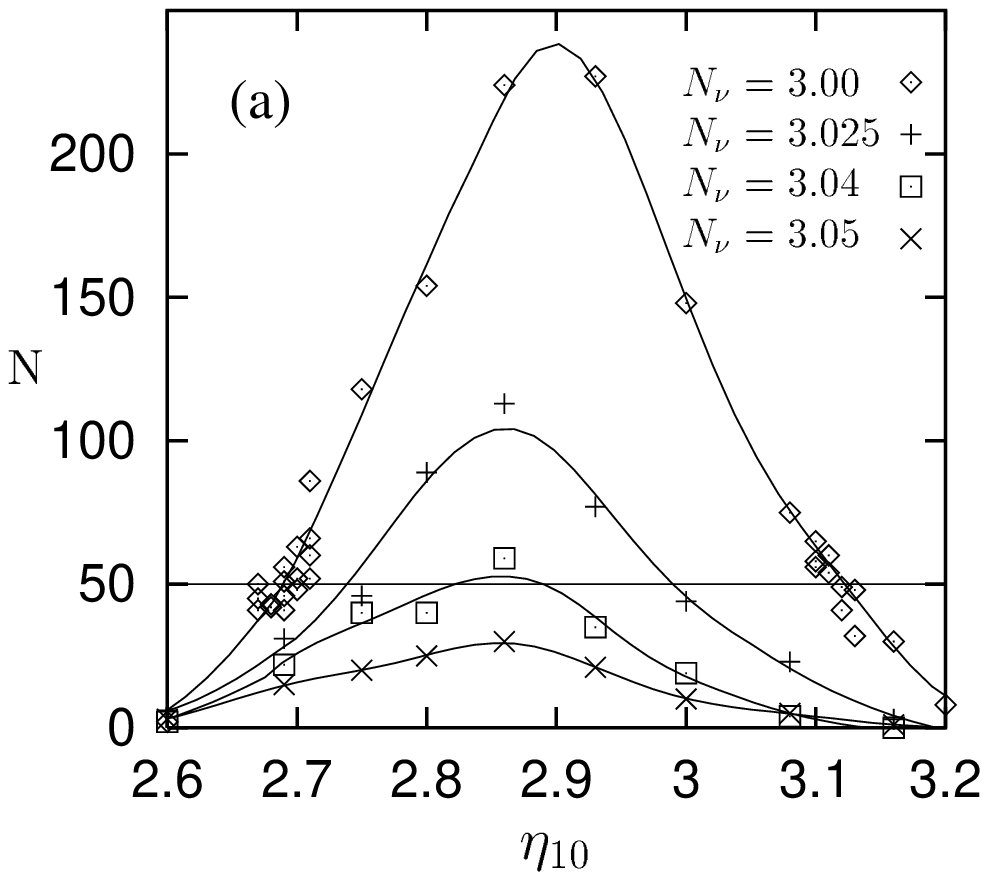}
\epsfxsize10cm\epsffile{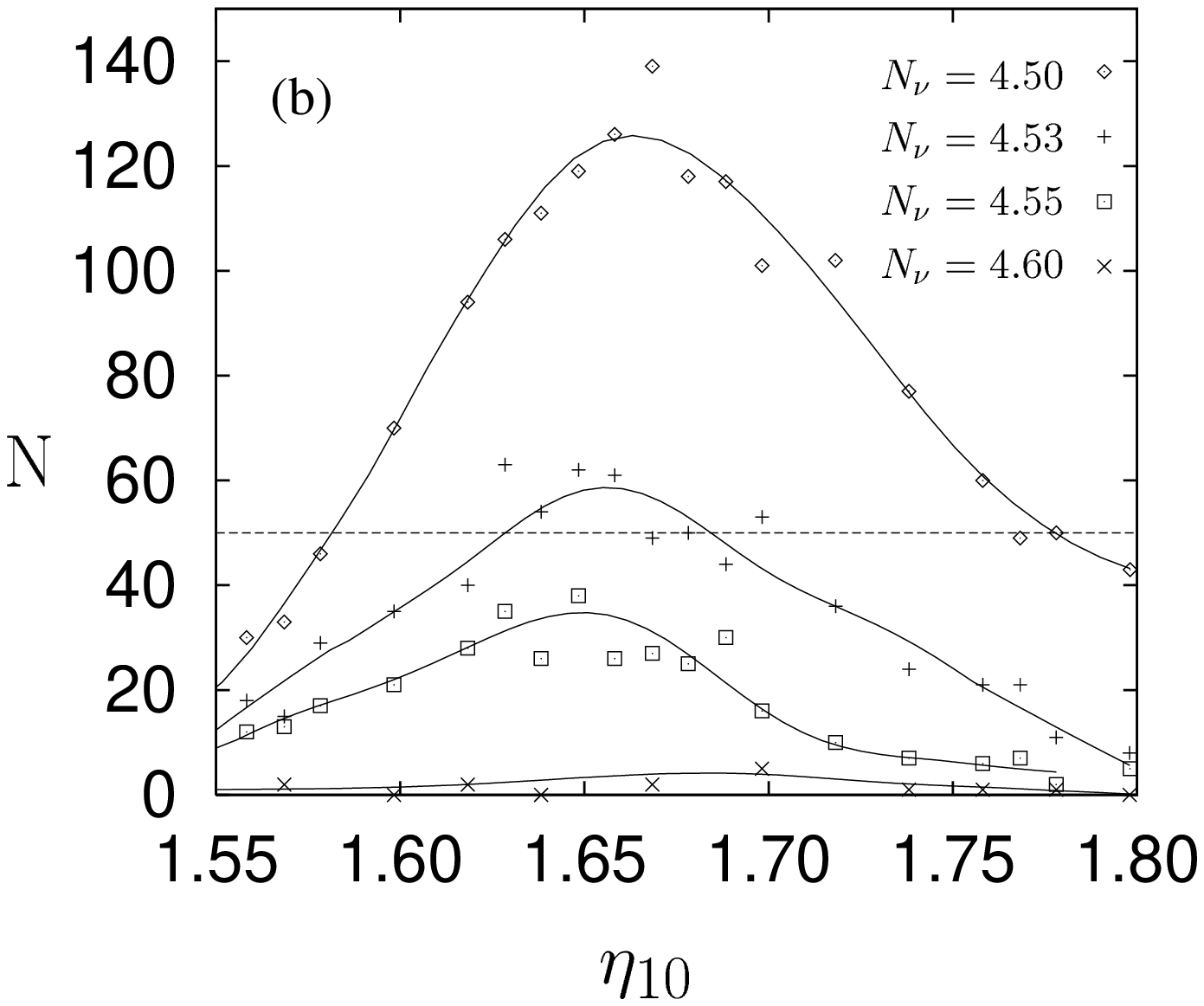}
\Figure{Number of Monte Carlo runs (out of 1000) which simultaneously
 satisfy the assumed abundance bounds, as a function of $\eta$ (in
 units of $10^{-10}$), for various values of $N_{\nu}$. Panel (a) is
 obtained adopting $Y_{\pr}({\Hefour})\leq0.24$ and
 $[({\Htwo}+{\Hethree})/{\H}]_{\pr}\leq10^{-4}$ (from Kernan and
 Krauss 1994) while panel (b) is obtained taking
 $Y_{\pr}({\Hefour})\leq0.25$,
 $[{\Htwo}/{\H}]_{\pr}\leq2.5\times10^{-4}$ and
 $[{\Liseven}/{\H}]_{\pr}\leq2.6\times10^{-10}$ (from Kernan and
 Sarkar 1996).\label{He4vsHe3DNnu}}

\epsfxsize9cm\epsffile{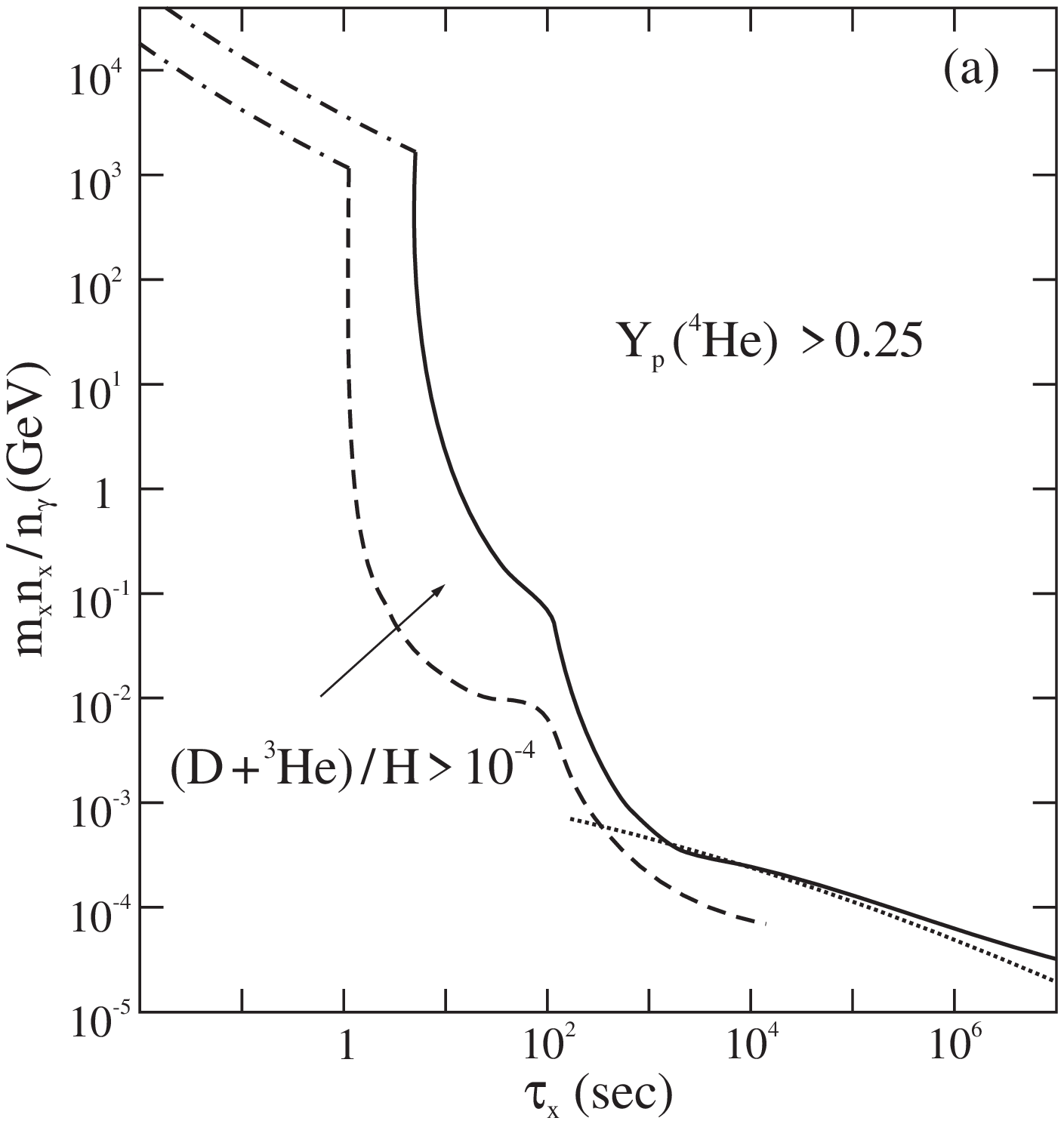}
\epsfxsize9cm\epsffile{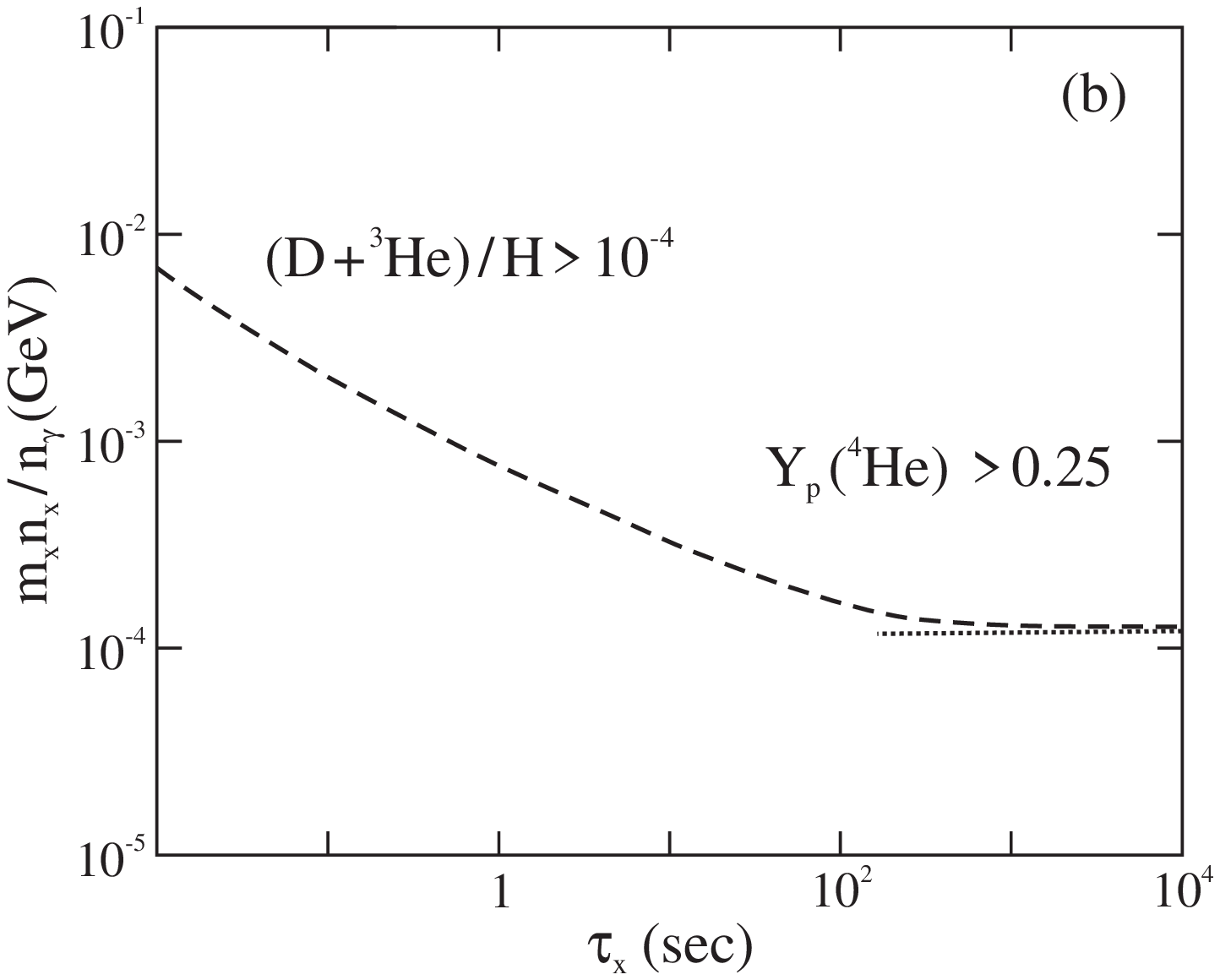}
\Figure{Upper bounds on the decaying-particle abundance as a function
 of its lifetime obtained from considerations of (a) entropy
 generation and increase in the expansion rate and (b) increase in the
 expansion rate alone (for `invisible' decays). Above the full lines
 $\Hefour$ is overproduced whereas above the dashed lines
 ${\Htwo}+{\Hethree}$ is overproduced; the dot-dashed lines in
 panel~(a) assume in addition that the initial nucleon-to-photon ratio
 is less than $10^{-4}$ (Scherrer and Turner 1988a,b). The dotted
 lines indicate the approximate bounds given by Ellis \etal
 (1985b).\label{nxlimentexp}}

\epsfxsize\hsize\epsffile{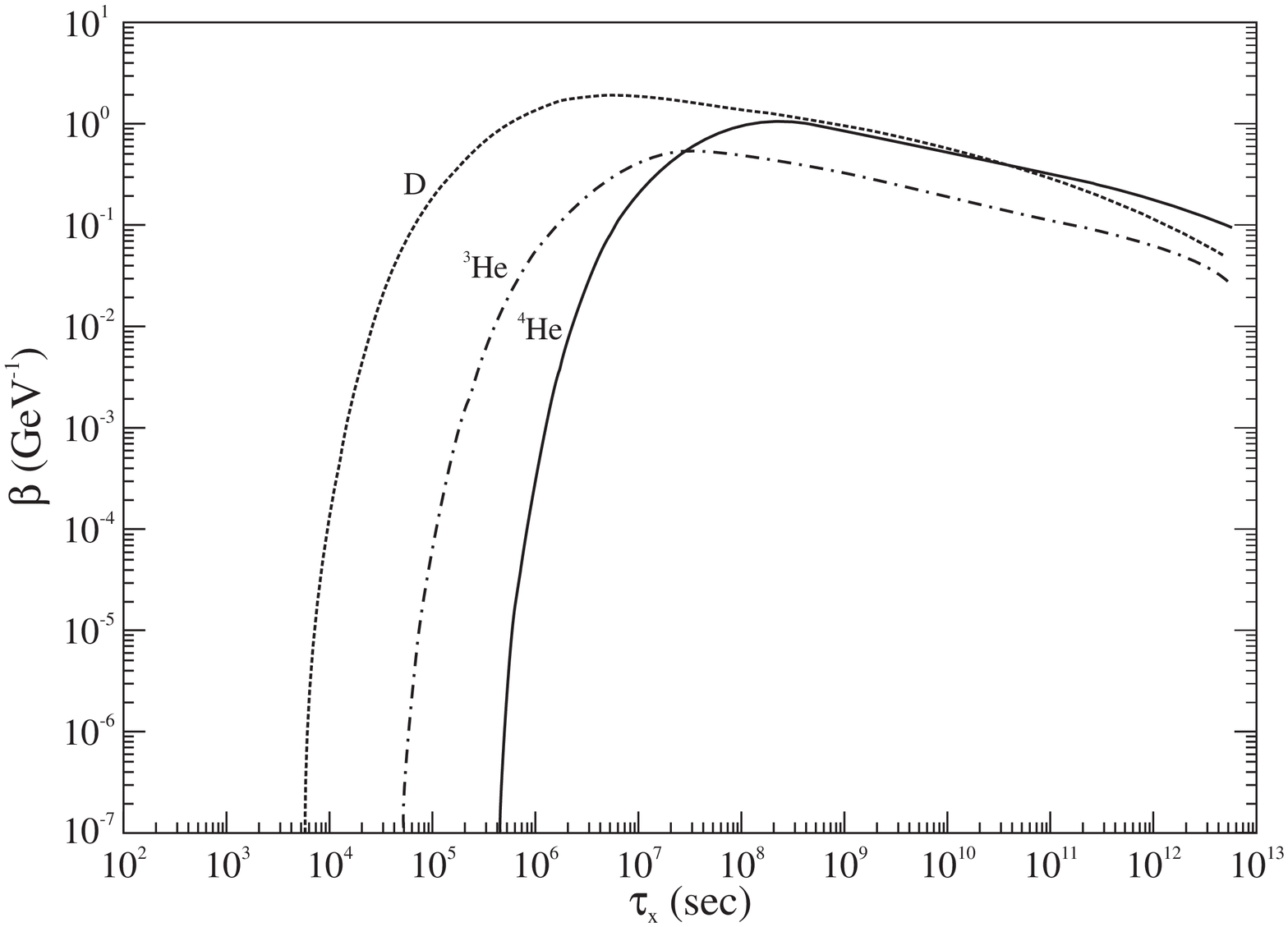}
\Figure{Normalized rates for photodissociation of light elements (as
  labelled) by electromagnetic cascades generated by massive unstable
  particles, as a function of the particle lifetime (Ellis \etal
  1992).\label{photorates}}

\epsfxsize\hsize\epsffile{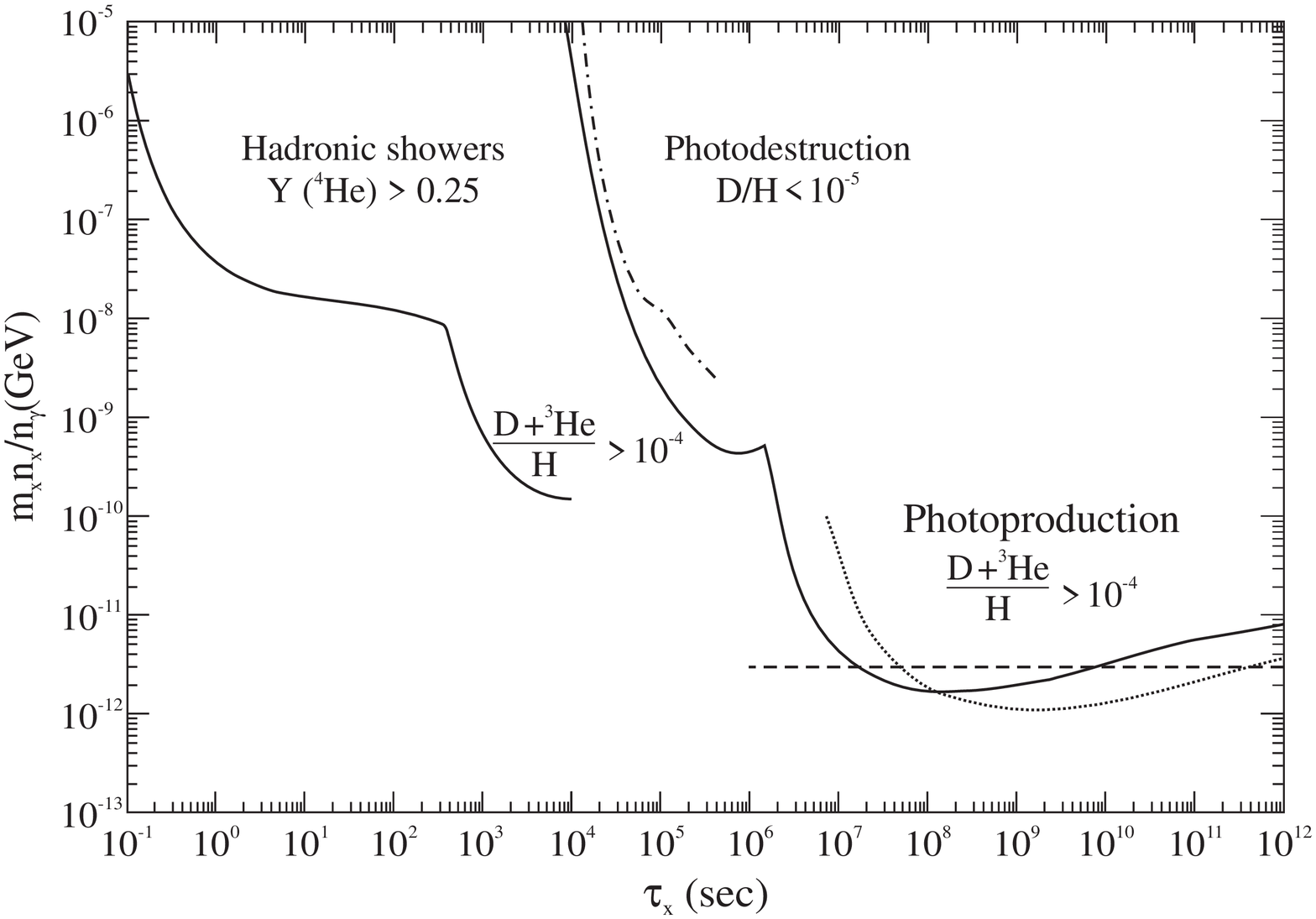}
\Figure{Upper bounds on the abundance of an unstable particle as a
  function of its lifetime from the effects of electromagnetic (Ellis
  \etal 1992) and hadronic cascades (Reno and Seckel 1988) on the
  primordially synthesized abundances. Other results shown are from
  Dimopoulos \etal (1989) (dot-dashed line), Ellis \etal (1985b)
  (dashed line) and Protheroe \etal (1995) (dotted
  line).\label{nxlimcascade}}

\epsfxsize\hsize\epsffile{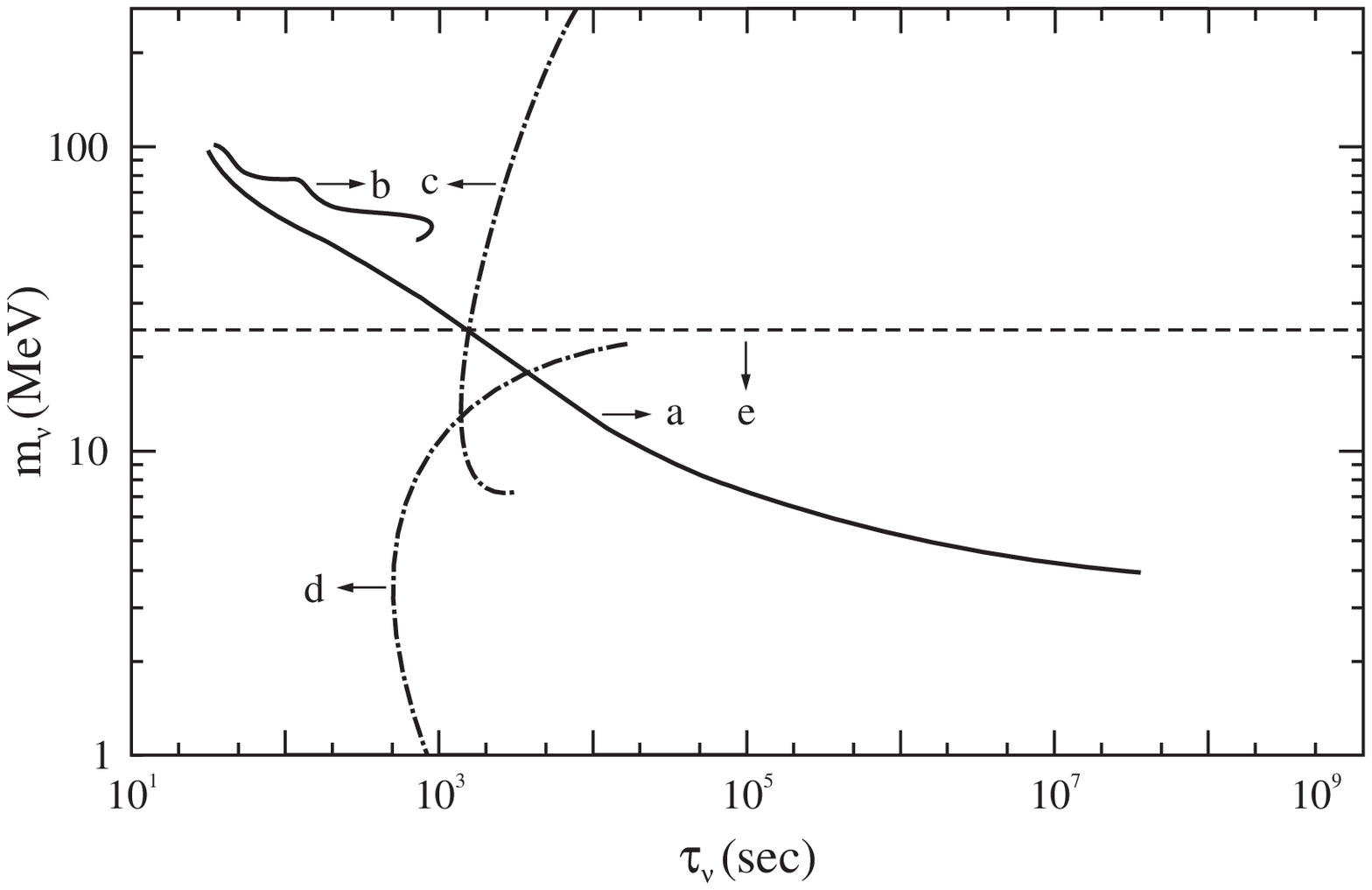}
\Figure{Upper bounds (dot-dashed curves) on the lifetime for
 $\nu_{\tau}\to{\el}^-{\el}^+\nu_{\el}$ (or
 $\nu_{\tau}\to\nu_{\el,\mu}\gamma$) from nucleosynthesis compared
 with lower bounds (full lines) from laboratory experiments (updated
 from Sarkar and Cooper 1984). Curves (a) and (b) are calculated from
 limits on the mixing angle $|U_{{\el}3}|^2$ obtained from,
 respectively, searches for additional peaks in $\pi\to{\el}\nu$ decay
 (Britton \etal 1992, De Leener-Rosier 1991) and measurement of the
 branching ratio $\pi\to{\el}\nu/\pi\to\mu\nu$ (Britton \etal
 1994). Curve (d) is the bound from entropy production and speed-up of
 the expansion rate, while curve (e) is obtained from consideration of
 deuterium photofission. Curve (c) is the present experimental bound
 on the $\nu_{\tau}$ mass (Busculic \etal 1995). Note that there is
 {\em no} allowed region for an unstable tau
 neutrino with mass exceeding 1 MeV.\label{nutaudecay}}

\epsfxsize\hsize\epsffile{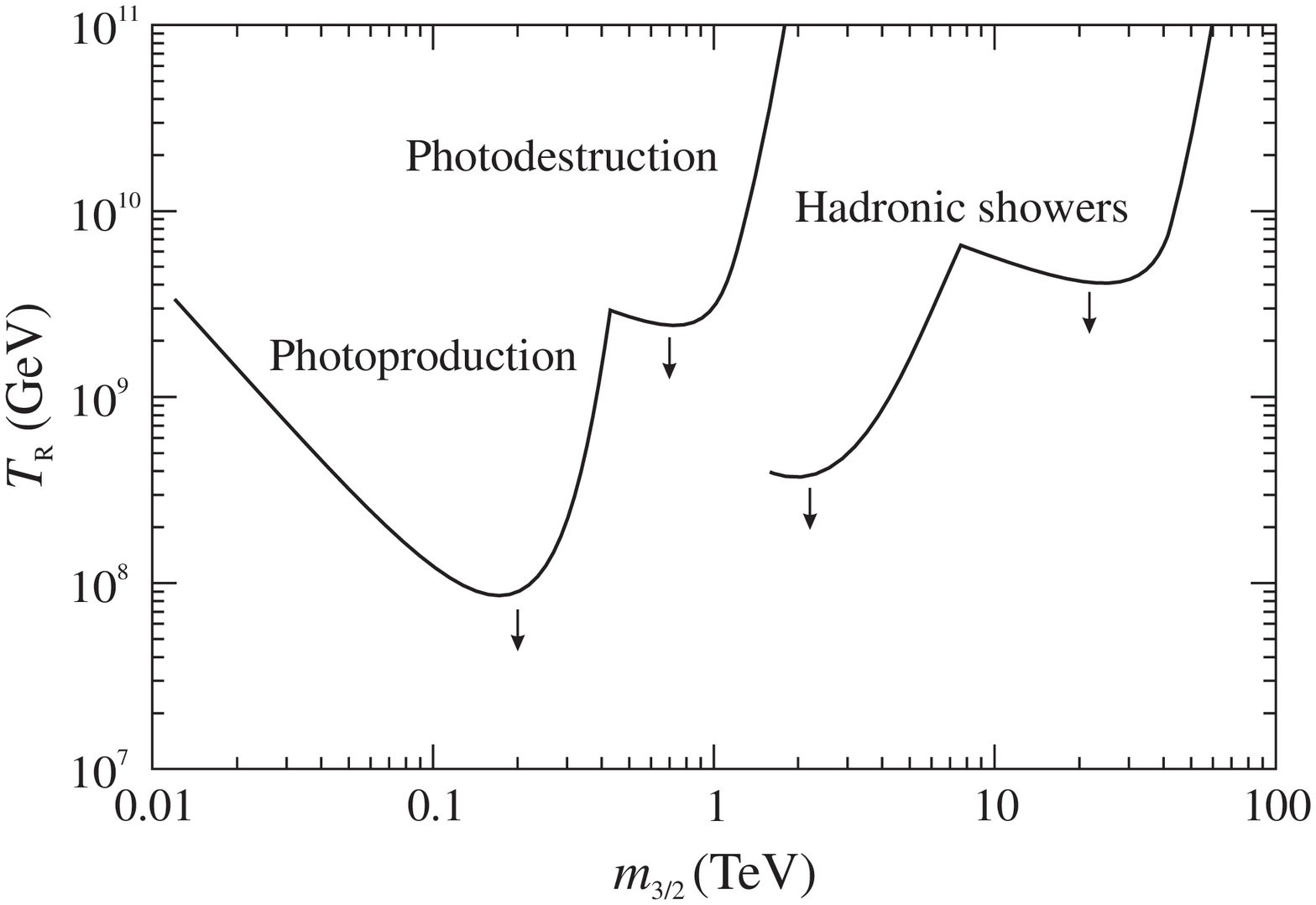}
\Figure{Nucleosynthesis bounds on the reheating temperature after
  inflation, from consideration of the generation of massive unstable
  gravitinos (Ellis \etal 1984b) and the effects of their hadronic
  (Reno and Seckel 1988) and radiative (Ellis \etal 1992) decays on
  elemental abundances.\label{TR}}

\epsfxsize\hsize\epsffile{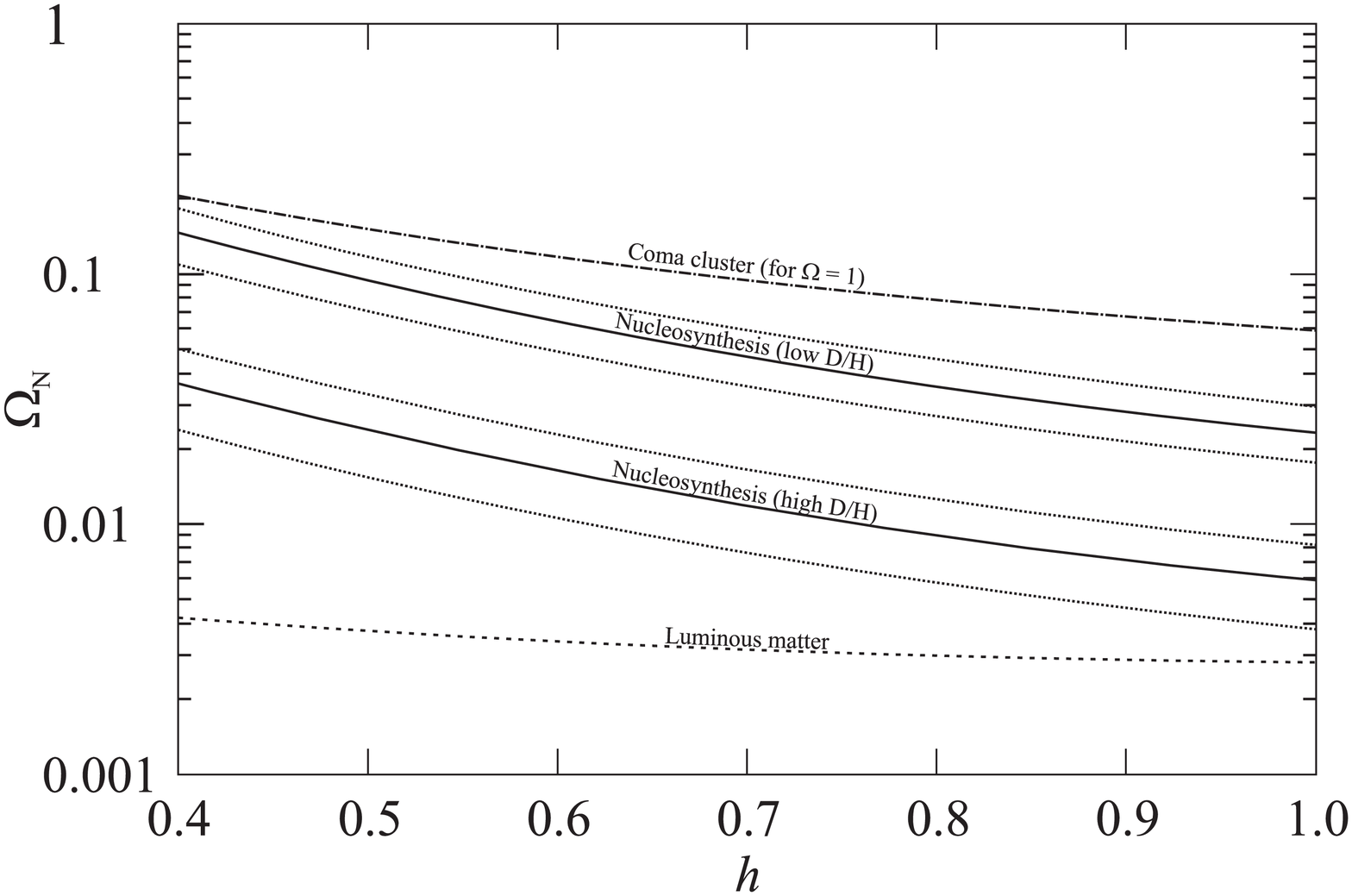}
\Figure{The contribution of nucleons to the cosmological density
  parameter as a function of the assumed Hubble parameter (after Hogan
  1994). The full lines (with dotted `$2\sigma$' error bands) show the
  standard BBN values according as whether the primordial $\Htwo$
  abundance is taken to be the high value (Songaila \etal 1994, Rugers
  and Hogan 1996a,b) or the low value (Tytler \etal 1996, Burles and
  Tytler 1996) measured in QAS. The dashed line is the lower limit
  from an audit of luminous matter in the universe (Persic and Salucci
  1992). The dot-dashed line indicates the value deduced from the
  observed luminous matter in the {\sl Coma} cluster (White \etal
  1993) for $\Omega=1$; it should be lowered by a factor of
  $\Omega^{-1}$ for $\Omega<1$.\label{omegan}}

\end{document}